\documentclass[12pt,letterpage]{amsart}
\usepackage{amsmath}
\newcommand{\be}{\begin{equation}}
\newcommand{\ee}{\end{equation}}
\newcommand{\A}{\mathcal{A}}
\newcommand{\D}{\mathcal{D}}
\usepackage{color}
\definecolor{orange}{rgb}{1,0.5,0}
\usepackage{graphics}
\usepackage{wrapfig}
\usepackage{amsmath}
\usepackage{amssymb}
\usepackage{subfigure}
\usepackage{float}
\usepackage{floatflt}
\usepackage{hyperref}
\usepackage{amssymb,amsmath}
\usepackage{graphicx}
\usepackage{epstopdf}
\usepackage{fancyhdr}
\begin{document}

\title{Deriving Finite Sphere Packings} 
\author{Natalie Arkus, Vinothan N. Manoharan and Michael P. Brenner}
\address{School of Engineering and Applied Sciences,
Harvard University, Cambridge, MA 02138 }

\begin{abstract}
Sphere packing problems have a rich history in both mathematics and physics;  yet, relatively few analytical analyses of sphere packings exist, and answers to seemingly simple questions are unknown.  Here, we present an analytical method for deriving all packings of $n$ spheres in $\mathbb{R}^3$ satisfying minimal rigidity constraints ($\geq 3$ contacts per sphere and $\geq 3n-6$ total contacts).  We derive such packings for $n \leq 10$, and provide a preliminary set of maximum contact packings for $10 < n \leq 20$.  The resultant set of packings has some striking features, among them: 
(i) all minimally-rigid packings for $n \leq 9$ have exactly $3n-6$ contacts,  
(ii) non-rigid packings satisfying minimal rigidity constraints arise for $n \geq 9$, (iii) the number of ground states (\begin{frenchspacing}i.e.\end{frenchspacing} packings with the maximum number of contacts) oscillates with respect to $n$, (iv) for $10 \leq n \leq 20$ there are only a small number of packings with the maximum number of contacts, and for $10 \leq n < 13$ these are all commensurate with the HCP lattice. The general method presented here may have applications to other related problems in mathematics, such as the Erd\"os repeated distance problem and Euclidean distance matrix completion problems.
\end{abstract}

\maketitle

\section{Introduction}

%now I think can be streamlined more and shorter - the sentence I added back in the last paragraph beginning with "Mathematically" might be a little redundant given the last sentence of the 2nd paragraph.  Also, the paragraph going into problems in materials science etc. beginning with "Furthermore" is now stuck in there - should make it more cohesive.  And, the sentence describing the motivating problem might not be clear - make sure the motivating problem is clear and this being the 1st step is clear...
We consider all configurations of $n$ identical impenetrable spheres in $\mathbb{R}^3$ that maximize the number of contacts between the spheres.

Our interest in this problem arose through the following question: can one direct the self-assembly of colloidal particles into any desired structure simply by imposing a spherically symmetric binding specificity on those colloidal particles?\footnote{By `spherically symmetric binding specificity' we are referring to a situation in which colloidal particles bind isotropically, but not all colloidal particles can bind to one another.  For example, imagine assigning labels to the colloidal particles; $A$, $B$, $C$, and so on.  Particles sharing common labels will be able to bind to one another, but colloidal particles that do not share a common label can not bind.}  Colloidal particles are small,  (nanometer to micron sized) spherical particles in aqueous solution. In recent years a myriad of methods have been developed to control the binding of particles to each other, thus causing them to self-assemble into clusters \cite{whitesides02b,manoharan03,wang04,crock05,dins98,grier98}.

The interactions between the particles typically have a range much smaller than the particle size; the potential energy of such a cluster is then simply proportional to the number of contacts. Structures with maximum numbers of contacts are thus the minima of the potential energy, and correspond to what will form in thermodynamic equilibrium.

Controlling which structures will form thus becomes a problem of controlling which structures correspond to energetic minima, or equivalently to contact maxima.  Although minimal-energy clusters have been catalogued for many different potentials, \begin{frenchspacing}e.g.\end{frenchspacing} the Lennard-Jones potential \cite{freeman_computational_1996}, they had not previously been calculated for hard (impenetrable) spheres.  Thus there was no detailed understanding of what structures could self-assemble in this colloidal system.

The motivating question of how to control the self-assembly of a colloidal system could thus be broken up into 2 parts: (i) Determine what structures can self-assemble, (ii) Determine how to direct the self-assembly of the system such that only the desired structure(s) form.  In this paper, we address (i).  In \cite{thesis, specificity_paper}, we address (ii).

It should be noted that, in addition to self-assembly, the structures of small clusters of atoms or particles bear directly on problems central to materials science and condensed matter physics, including nucleation, glass formation, and the statistical mechanics of clusters~\cite{doye_controlling_2007, sheng_atomic_2006, reichert_observation_2000}.  
%maybe delete this last sentence
For instance, the first step towards understanding the thermodynamics of a particular cluster system is to calculate the ground states as a function of particle number $n$~\cite{doye_entropic_2002}.

%The beginning of this paragraph could still flow/be worded a little better
The structures that can self-assemble in this colloidal system can be determined by solving the mathematical problem of which structures globally or locally maximize the number of contacts between $n$ spheres in $\mathbb{R}^3$ -- \begin{frenchspacing}i.e.\end{frenchspacing} all  structures in which either (i) no additional contacts between spheres can exist, or (ii) a contact must first be broken in order  to form an additional contact.  We will refer to such structures as \textit{packings}.  We formulate our problem by focusing on enumerating only \textit{minimally-rigid} sphere packings, which we define as packings with at least 3 contacts per particle and at least $3n-6$ total contacts.  Minimal rigidity is necessary, but not sufficient, for a structure to be rigid.  We previously detailed the ground states of these packings as well as some of their interesting features in \cite{short_paper}.  Here, we present the results and method more completely.   %put footnote here also about code being modified and the results differing a little...?
The method we introduce combines graph theory and geometry to analytically derive all minimally-rigid packings of $n$ spheres.  We perform this enumeration for $n \leq 10$ spheres.  
%Here put a note about computational sources of error that might cause some packings to be missed.
Due to the large number of packings that must be evaluated, this analytical method is implemented computationally, and near $n = 10$ we reach the method's computational limitations.  Finding scalable methods for enumerating packings at higher $n$ is a significant challenge for the future.

Already by $n = 10$, a number of interesting features set in.  For $n\le 9$, all minimally-rigid packings
have exactly $3n-6$ contacts.  The first instance of  a non-rigid sphere packing that satisfies minimal rigidity constraints occurs at $n = 9$, and more such non-rigid packings arise at $n = 10$.  The first instance of packings with greater than $3n-6$ contacts occurs at $n = 10$.  We discuss the geometrical manner in which these maximum contact packings arise, and conjecture that maximum contact packings for all $n$ in this system must contain octahedra.  We provide preliminary evidence for this maximum contact conjecture for $n \leq 20$, and we show that the putative maximum contact packings of $10 \leq n \leq 13$ are commensurate with the hexagonal close packing (HCP) lattice, but that maximum contact packings of $14 \leq n \leq 20$ are not. 
Furthermore, we show that the number of packings containing the maximum number of contacts is oscillatory with $n$, and we discuss the origins of these oscillations.

The set of packings we enumerate includes, as a subset, structures previously observed and described in the literature: for example, it includes all minimal-second-moment clusters reported by \cite{sloane95}, packings observed experimentally through capillary driven assembly of colloidal spheres \cite{lauga04,manoharan03}, as well as the Janus particle structures observed by Hong \textit{et al.}~\cite{hong_clusters_2008}.

This problem is  closely related to several unsolved problems in mathematics, such as the Erd\"os unit distance problem (\begin{frenchspacing}a.k.a.\end{frenchspacing} the Erd\"os repeated distance problem), Euclidean distance matrix and positive semidefinite matrix completion problems, and 3-dimensional graph rigidity.  Thus the method and results presented here may have direct bearing on these problems.

The organization of this paper is as follows: In the next section we outline our mathematical
formulation of the problem and describe our methodology for finding all minimally-rigid packings of a fixed $n$.  We combine graph theoretic enumeration of adjacency matrices with solving for their corresponding distance matrices.  The elements of these distance matrices correspond to the relative distances between spheres in 3-dimensional Euclidean space, and thus yield the packings that correspond to those adjacency matrices.  Analytical methods for solving such adjacency matrices that scale efficiently with $n$ do not exist. Sections \ref{Geom Solution Section} and \ref{gen rule section} thus derive a method with improved scaling:  Section \ref{Geom Solution Section} derives geometrical rules that map patterns in adjacency matrices either (i) to a configuration of spheres, in which case the adjacency matrix is solved for its corresponding distance matrix or matrices; or (ii) to an unrealizable configuration, in which case no real-valued embedding in 3-dimensional Euclidean space of that adjacency matrix exists in which spheres do not overlap.  We show how these geometrical rules,  combined with adjacency matrix enumeration, can lead to a complete set of minimally-rigid packings.  Each time a new pattern is encountered, a new geometrical rule must be derived; thus this part of the method requires new derivations at each $n$, and does not scale efficiently\footnote{This method allowed us to enumerate packings up to $n=8$, while standard methods in algebraic geometry (using the package \textit{SINGULAR}) allowed us to enumerate up to $n=7$.}.  In section \ref{gen rule section}, we derive a single geometrical rule (the \textit{triangular bipyramid rule}) that can solve all iterative adjacency matrices.  An iterative adjacency matrix is an $n \times n$ matrix in which all minimally-rigid $m < n$ subgraphs also correspond to packings.  This part of the method applies to any $n$, and thus scales efficiently for all $n$.  Most adjacency matrices at small $n$ are iterative; therefore, this greatly reduces the number of geometrical rules necessary to derive a complete set of packings.  Section \ref{the packings} describes the set of sphere packings we find from our study.  We provide analytical formulas for packings up to $n=7$, and the set of packings for $n=8,9,10$ are included in the supplementary information \cite{supp_info}.  Section \ref{packing properties section} summarizes notable properties of the packings, including how the number of contacts changes with $n$, the emergence of minimally-rigid structures that are not rigid, and the emergence of maximum contact packings that are commensurate with lattice packings.  Section \ref{future directions section} summarizes the main roadblocks towards obtaining results at higher $n$ and contains several ideas and conjectures therein, and also discusses extensions to dimensions other than 3, as well as relevance to other problems of mathematical interest.  Section \ref{concluding remarks} provides some concluding remarks.

\section{Mathematical Formulation}\label{mathematical formulation section}

We begin by presenting a two-step method for enumerating a complete set of sphere packings that satisfies minimal rigidity constraints.  The set of all packings of $n$ spheres is a subset of all possible configurations of $n$ spheres.  Thus, to enumerate a complete set of sphere packings we (i) use graph theory to enumerate all $n$ sphere configurations, and (ii) determine which of those configurations correspond to sphere packings.  The sphere packings we consider here correspond to maxima (local or global) in the number of contacts\footnote{\begin{frenchspacing}i.e.\end{frenchspacing} either the configuration of spheres can not form an extra contact (global maxima), or if an extra contact can be formed, it first requires the breaking of an existing contact (local maxima).}.  Provided step (ii) is exact, this method will produce a complete set of packings.  However, current analytical methods do not scale efficiently with $n$, and are therefore ill-suited for step (ii).  Thus, in sections  \ref{Geom Solution Section} and \ref{gen rule section}, we use basic geometry to derive an analytical method that can exactly determine which configurations correspond to sphere packings.  Because the number of possible configurations grows exponentially with $n$,  this analytical process must be implemented computationally.  We focus our search to only those sphere packings satisfying minimal rigidity constraints ($\geq 3$ contacts per sphere, $\geq 3n-6$ total contacts), because doing this guarantees that if a graph has an embedding in 3-dimensional Euclidean space, it corresponds to a sphere packing; whereas graphs that are not minimally-rigid can have 3-dimensional embeddings without corresponding to packings due to a continuous degree of freedom that allows for the formation of another contact.  %Thus, enforcing minimal rigidity constraints greatly reduces the search through graphs that do not correspond to packings; yet even after the imposition of this constraint, only a small subset of graphs correspond to packings. %%%I think this is unnecessary - perhaps the last part of this sentence should be included somewhere else, if it isn't already.

We also note that spheres can be thought of as points (Fig.~\ref{AdjToDistMatFig}a,b), where points correspond to the centers of spheres, and we measure the distance between spheres as the distance between their centers.  Throughout this paper we will use the words sphere, point, and particle interchangeably.

\subsection{Graph Theory Produces the Set of Possible Packings}

A configuration of $n$ spheres can be described by an $n \times n$ \textit{adjacency matrix}, $\A$, detailing which spheres are in contact: $\A_{ij}=1$ if the $i^{th}$ and $j^{th}$ particles touch, and $\A_{ij} = 0$ if they do not.  A system of $n$ spheres has $n(n-1)/2$ interparticle distances; the 2 possibilities (touching or not touching) per distance thus leaves $2^{n(n-1)/2}$ different ways of arranging contacts amongst the distances. There are thus $2^{n(n-1)/2}$ possible adjacency matrices, each of which potentially corresponds to a packing.

\begin{figure}[h]
 \begin{center}
 \includegraphics[width=6in]{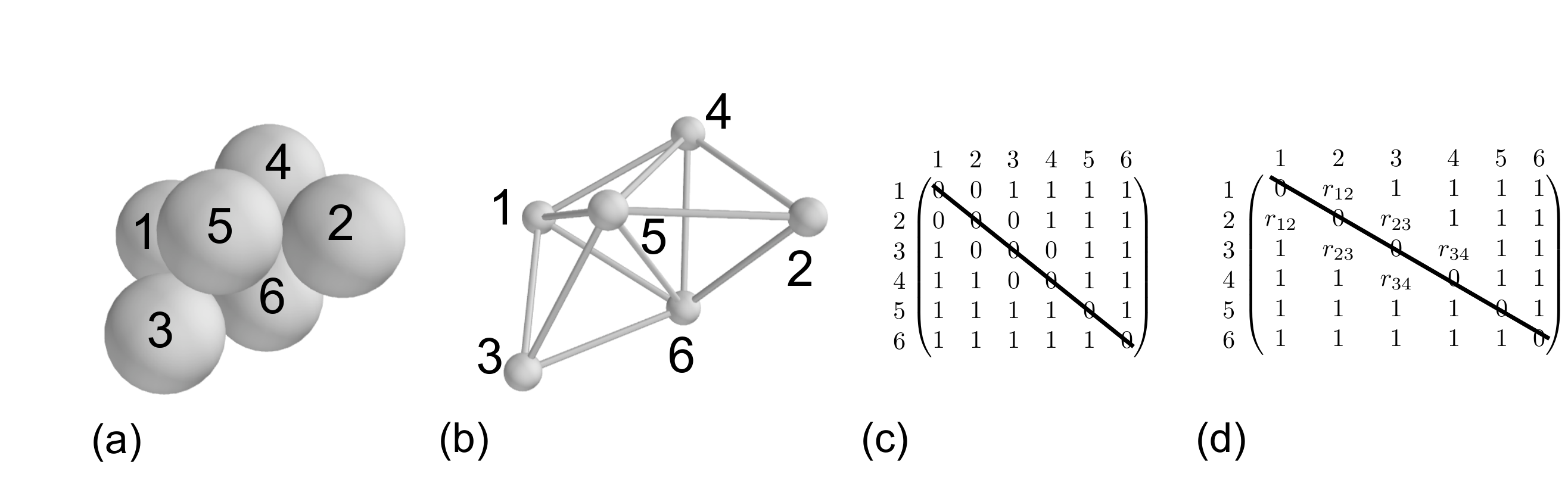}
% (a)%\includegraphics[width=1.3in]{AdjMatDistMatExPicSphere.pdf}
% (b)%\includegraphics[width=1.3in]{AdjMatDistMatExPicBallStick.pdf}
% (c)%\includegraphics[width=1.3in]{AdjMatDistMatExPicAdjMat.pdf}
% (d)%\includegraphics[width=1.3in]{AdjMatDistMatExPicDistMat.pdf}
 \end{center}
 \caption{\textbf{Adjacency and Distance Matrix Representation of Packings.}\newline(a) 6 particle polytetrahedral sphere packing. (b) The same 6-particle packing shown in point/line representation. (c) Corresponding 6 particle adjacency matrix, $\A$.  (d) Corresponding relative distance matrix, $\D$.}\label{AdjToDistMatFig}
 \end{figure}

Figure \ref{AdjToDistMatFig} shows a packing of 6 particles, both as a sphere packing (Fig.~\ref{AdjToDistMatFig}a) and as points connected by line segments (Fig.~\ref{AdjToDistMatFig}b). The adjacency matrix corresponding to this packing is shown in figure ~\ref{AdjToDistMatFig}c.  
The set of possible packings can be enumerated by considering {\sl all adjacency matrices}. For $n=6$, there are $2^{15}=$ 32,768 different adjacency matrices. Table \ref{AdjMatTable} shows that the number of adjacency matrices grows rapidly with $n$, reaching $3.5184\times 10^{13}$ by $n=10$; however, many of these correspond to the same structure due to particle labeling degeneracy.  For example, switching labels 5 and 2 of figure \ref{AdjToDistMatFig} yields another adjacency matrix corresponding to the same structure.\footnote{Note that this is purely an exercise in switching particle \textit{labels}, it is not a statement about the symmetry of the structure - we are not saying that particles 2 and 5 have the same contact distribution.  The point is that many matrices can lead to the same structure, because how you label the particles is an arbitrary factor.}  Adjacency matrices corresponding to the same structure are \textit{isomorphic} to one another -- meaning there will exist a permutation of rows and columns that can translate one matrix into the other\footnote{See \cite{hoare} for a nice example of such a permutation.}.  

To generate the complete set of possible packings, we need only enumerate \textit{nonisomorphic} $\A$'s.  Such algorithms exist; examples include \textit{nauty} and the \textit{SAGE} package called \textit{nice} \cite{nauty, sage}. The number of nonisomorphic matrices is much smaller but still grows exponentially with $n$. Table \ref{AdjMatTable} shows this growth also; for example at $n=6$ the number of potential structures is $156$, and at $n=10$ it is 12,005,168.

The set of $\A$'s (potential packings) can be further reduced by imposing rigidity constraints.  Most structures with less than $3n-6$ total contacts or less than 3 contacts per particle will not correspond to a packing because there will exist a continuous degree of freedom through which the structure can form one or more bonds.  Rigidity requires (i) there  be at least 3 contacts per particle, and  (ii)  there be at least as many contacts as internal degrees of freedom -- thus there must be at least $3n-6$ contacts\footnote{Note that in restricting to structures with exactly $3n-6$ contacts, we will also find
structures with more than $3n-6$ contacts. This is because when we solve for the packings as outlined below, the solutions can end up having more contacts than are assumed in the algebraic formulation.} 

Table \ref{AdjMatTable} shows how imposing minimal rigidity constraints restricts the number of adjacency matrices. For $n\leq$ 5 particles, this eliminates all but 1 adjacency matrix, thus identifying a unique packing for each of these $n$; the doublet, triangle, tetrahedron, and triangular bipyramid, respectively (section \ref{the packings}).  For $n \leq 4$, all relative distances within these packings are touching and are thus known.  However, for $n = 5$, the packing contains one unknown relative distance, which must be determined.  For $n \geq 6$, more than one minimally-rigid $\A$ exists, and thus rigidity constraints alone are insufficient to identify the number of sphere packings (or to solve for them, as more than 1 distance in each minimally-rigid $\A$ is unknown).

\subsection{Solving Potential Packings: Algebraic Formulation}\label{Solving For Packings Section}

To make further progress, we reformulate our problem algebraically.
Each  adjacency matrix element,  $\A_{ij}$, is associated with an interparticle distance,
\begin{eqnarray}
\nonumber r_{ij}=\sqrt{(x_i - x_j)^2 + (y_i - y_j)^2 + (z_i - z_j)^2},
\end{eqnarray} 
which is the distance between particles $i$ and $j$ whose centers are located at $(x_i,y_i,z_i), (x_j,y_j,z_j)$, respectively.  The distances are constrained by the adjacency matrix as follows:
\begin{eqnarray}
\A_{ij} &=& 1 \; \; \Longrightarrow r_{ij} = 2r\\
\A_{ij} &=& 0 \; \; \Longrightarrow r_{ij} \geq 2r
\end{eqnarray}
where $r$ is the sphere radius\footnote{Strictly speaking, $A_{ij} = 0$ implies that particles are not touching, and thus that $r_{ij} > 2r$.  However, it is convenient to consider all solutions of $\A$, which thus include the possibility $r_{ij} = 2r$ for $\A_{ij} = 0$; in other words, a packing can be represented by multiple $\A$'s with different numbers of 1's if the solution to $A_{ij} = 0$ forces it to a 1.}.

For adjacency matrices with $3n-6$ contacts,  this leads to precisely as many equations as unknowns\footnote{Without loss of generality, we can set one particle to reside at the origin, another to reside along a single axis (for example the $y$-axis), and a third to reside in one plane (such as the $xy$-plane).  6 coordinates are then fixed, leaving $3n-6$ instead of $3n$ coordinates.}. 
The task is thus to solve for the $r_{ij} \geq 2r$ given a particular set of $r_{ij} = 2r$.  The particle configuration encoded by each $\A$ is thus specified by a \textit{distance matrix}, $\D$, whose elements $\D_{ij}=r_{ij}$.  

The fundamental question is to find an efficient, exact method for mapping $\A\to \D$.  If any  $\D_{ij} < 2r$, this implies that particles $i,j$ overlap; thus any $\D$ with $\D_{ij} < 2r$ is unphysical.
Figure \ref{AdjToDistMatFig}d shows the distance matrix corresponding to an $n=6$ packing. The interparticle distance between each of the particles that are touching is normalized to $1$; for this packing, this leaves three distances that need to be determined; $r_{12}, r_{23}$, and $r_{34}$.

In solving $\A$ for $\D$, the following scenarios are possible:

\begin{enumerate}
\item Continuous set(s) of real-valued $\D$ correspond to a given $\A$, in which case the structure(s) are not rigid.  

\item No real-valued $\D$ exists that solves $\A$, in which case the structure is unphysical.

\item Finite, real-valued $\D$ exist that solve $\A$.  In this case, the structure(s) correspond to rigid sphere packing(s) provided that all $\D_{ij} \geq 2r$.

\end{enumerate}

For every nonisomorphic adjacency matrix, whether there exists a corresponding packing
requires solving for $\D$ and asking whether the resulting $r_{ij}$'s satisfy these constraints.

\begin{table}
 \begin{tabular}{|c|c|c|c|c|c|}
\hline
& & & & &\\
 $n$ & $\A$'s & Non-Isomorphic $\A$'s & Minimally & Iterative $\A$'s & Non-Iterative $\A$'s\\
& & & rigid $\A$'s & &\\
\hline
%& & & & &\\
1 & 1 & 1 & 1 & 1 &0 \\
\hline
2 & 2 & 2 & 1 & 1 & 0\\
\hline
3 & 8 & 4 & 1 & 1 & 0\\
\hline
4 & 64 & 11 & 1 & 1 & 0\\
\hline
5 & 1,024 & 34 & 1 & 1 & 0\\
\hline
6 & 32,768 & 156 & 4 & 3 & 1\\
\hline
7 & 2,097,152 & 1,044 & 29 & 26 & 3\\
\hline
8 & 268,435,456 & 12,346 & 438 & 437 & 1\\
\hline
9 & 6.8719 $\cdot 10^{10}$ & 274,668 & 13,828 & 13,823 & 5\\
\hline
10 & 3.5184 $\cdot 10^{13}$ & 12,005,168 & 750,352 & 750,258 & 94\\
\hline
\end{tabular}
\caption{\textbf{The Growth of Adjacency Matrices with $n$}.\newline The number of adjacency matrices (constructed by \cite{nauty}) decreases rapidly as isomorphism and rigidity constraints are imposed.  Iterative and non-iterative are defined in the text.  The classification of whether an $\A$ is iterative or not is here shown after all rules for $n-1$ particles are applied; thus the non-iterative column shows $n$-particle non-iterative structures only, and does not include non-iterative structures of less than $n$ particles.  Also note that the classification of whether an $\A$ is iterative or not is thus sensitive to which geometrical rules are used.  The number of iterative and non-iterative $\A$'s at $n=10$ is different from \cite{short_paper} because some modifications were made to the code since the publication of that paper.  Please see the supplemental information \cite{supp_info} for exactly which modifications have been made.}\label{AdjMatTable}
\end{table}

\subsection{Limitations of Existing Solution Methods}

The issue now becomes one of solving a system of $n(n-1)/2$ constraints ($3n-6$ equations and $n(n-1)/2 - (3n-6)$ inequality constraints). Numerical approaches for solving such a system cannot be guaranteed to converge; for example, Newton's method requires an accurate initial guess for guaranteed convergence. When a solution does not converge, we do not know if it is because a solution does not exist, or because the initial guess is not sufficiently accurate\footnote{Furthermore, convergence to an unphysical solution lying within numerical error is also a problem.}.   Algebraic geometric methods (\begin{frenchspacing}e.g.\end{frenchspacing} Gr\"obner bases) \cite{buch01} are effective, but these algorithms 
 do not scale efficiently with $n$.  Our own implementation\footnote{At $n = 8$, using the software package SINGULAR \cite{singular}, one matrix takes several hours to solve, and there are a total of $438$ minimally-rigid non-isomorphic $\A$'s.} of this method using the package {\sl SINGULAR} \cite{singular} was only able to solve for structures up to $n = 7$.

In the following section, we use another approach and derive a different geometrical method to efficiently solve for all sphere packings given a set of nonisomorphic, minimally-rigid $\A$'s.  We implement this method up to $n = 10$, at which point we begin to hit some practical roadblocks; these are discussed at the end of the paper, where we outline potential ways to overcome them.

\subsubsection{Chiral Structures}

Before proceeding further, it is worth remarking that structures with different handedness will correspond to the same distance matrix.  We can analyze each $\D$, and determine whether it corresponds to a structure that has a chiral counterpart (see section \ref{Chiral Section}).  We refer to chiral structures as the same packing but different states -- thus, a distance matrix having a left- and right-handed counterpart corresponds to 1 packing with 2 different states.  Note that according to our definition, a different packing necessarily corresponds to a different state.  Thus, the total number of states is equal to the total number of packings plus the total number of chiral counterparts.

\section{Geometrical rules solve for sphere packings} \label{Geom Solution Section}

We now show how geometrical rules can be used to effectively and analytically solve the class of polynomial equations that are generated by adjacency matrices.  We use basic geometry to construct rules associating patterns of 1's and 0's in $\A$'s with either a given relative distance, $\D_{ij}$, or an unphysical conformation (in which case no $\D \geq 2r$ exists).  There thus exist two types of rules: {\it Elimination rules} eliminate an $\A$ as unphysical, and {\it distance rules} solve an $\A_{ij}$ for its corresponding $\D_{ij}$.

\subsection{Neighbor Spheres and Intersection Circles}

\begin{figure}[h]
\begin{center}
%(a) %\includegraphics[width = 2.75in]{NeighbSphereW4TouchingSpheresBW4.pdf}
%(b) %\includegraphics[width = 2.75in]{IntCircGreyWLengths.pdf}
%(a) %\includegraphics[width = 2.75in]{NeighbSphereW4TouchingSpheresBW4Cropped.png}
(a) \includegraphics[width = 2.75in]{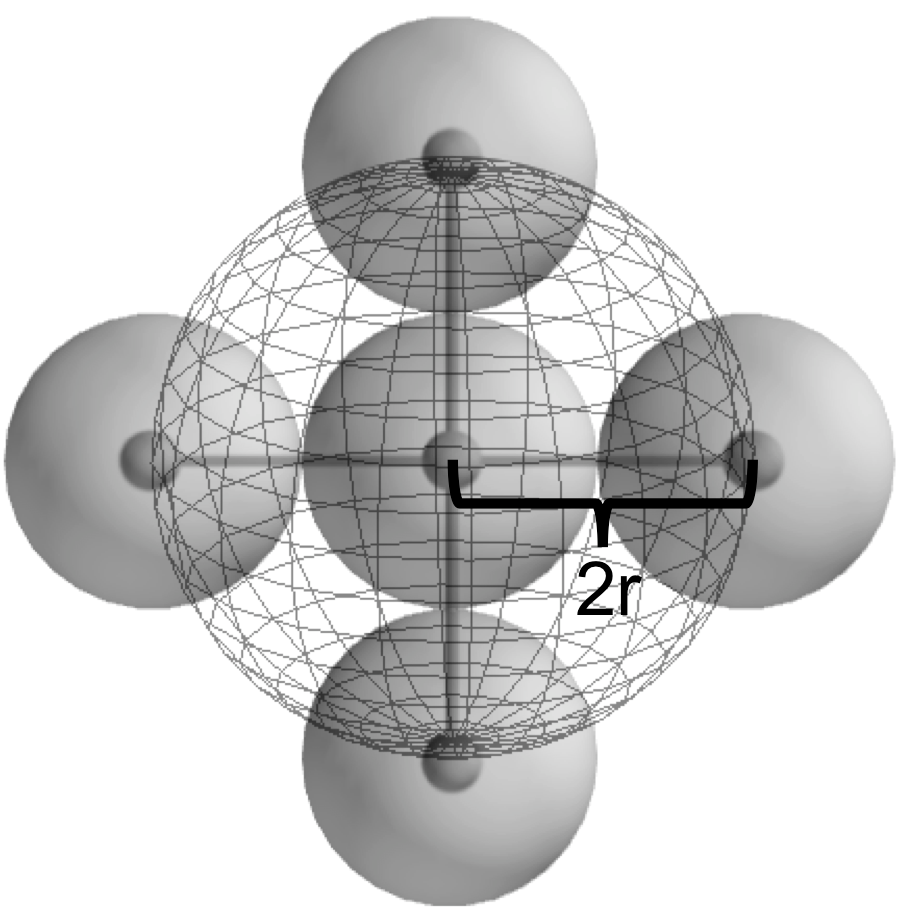}
(b) \includegraphics[width = 2.75in]{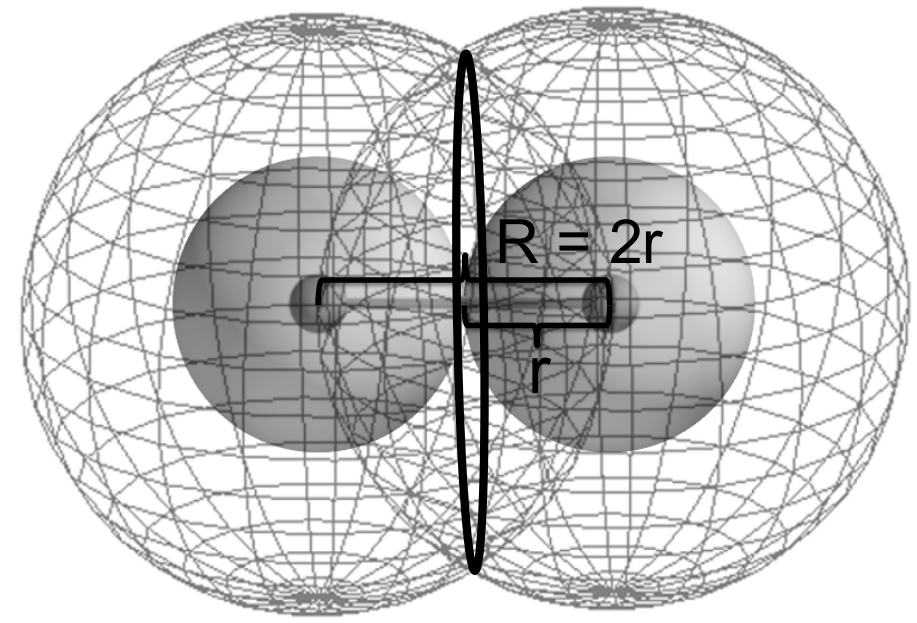}
\caption{\textbf{The Neighbor Sphere.}\newline(a) A particle and its neighbor sphere.  4 particles are shown touching the center particle, and it is seen that their centers lie on the surface of the particle's neighbor sphere.  (b) 2 touching particles.  The associated neighbor spheres of the particles intersect in an \textit{intersection circle} of radius $(\sqrt{3}/2)R$.}\label{neighbor sphere}
\end{center}
\end{figure}

With each sphere of radius $r$, we can associate a \textit{neighbor sphere} of radius $R = 2r$, whose surface defines where another sphere must lie if it is to touch the original sphere in question (Fig.~\ref{neighbor sphere}a).  When 2 spheres touch, their neighbor spheres intersect in an \textit{intersection circle} (Fig. \ref{neighbor sphere}b). The radius of the intersection circle follows from straightforward geometry, and is $\frac{\sqrt{3}}{2} R$ (see supplemental information for derivation \cite{supp_info}).

Now we can interpret each $\A_{ij} = 1$, in geometrical terms, as an intersection circle between spheres $i$ and $j$.  Minimal rigidity constraints imply  that each particle is associated with at least 3 intersection circles.    Intersection circles can be used to derive geometrical rules because, in general, a packing of $n$ particles involves intersections of intersection circles, and intersections of intersection circles define points in space.  A particle touching $m$ other particles will lie at the intersection of those $m$ neighbor spheres.  For example, a particle touching the dimer depicted in figure \ref{neighbor sphere}b will lie on the circumference of the associated intersection circle.  The intersection of $m \geq 3$ neighbor spheres are points -- and by defining points in space, basic trigonometry can be used to calculate the distances between those points, thus solving $\A$'s for $\D$'s.

\subsection{Individual Geometrical Rules}
Using intersection circles, we now derive several  representative geometrical rules for eliminating and solving adjacency matrices.  The supplementary information \cite{supp_info} contains  the complete set of rules used to derive the results of this paper.

\subsubsection{Rule 1}\label{Elim Rule 1 Section}
The simplest rule arises from the fact that intersection circles can intersect in either 2, 1 or 0 points, but never in more than 2 points.  This geometrical fact implies that any $\A$ with the following property is unphysical: any 2 of the set $\left\lbrace \A_{jk},\A_{jp}, \A_{kp} \right\rbrace$ equal 1, and there exist more than 2 $i$'s for which $\A_{ij} = \A_{ik} = \A_{ip} = 1$.\\

\begin{figure}[htbp!]
 \begin{center}
 \includegraphics[width=3.2in]{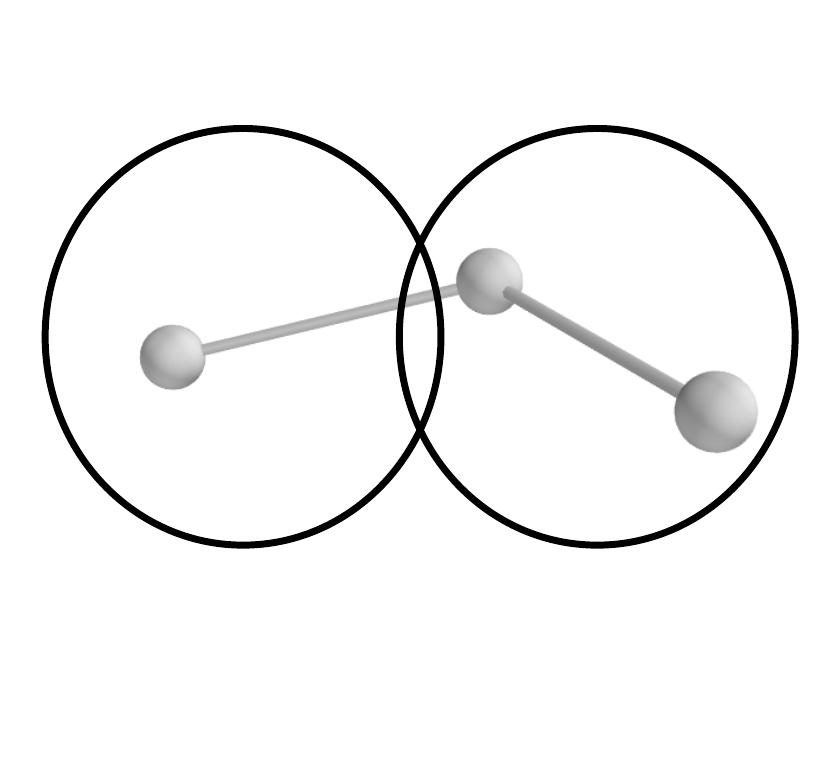}
 \end{center}
 \caption{\textbf{No More than 2 Particles can Touch 3 Connected Particles.} Schematic of 3 linearly connected particles and their associated intersection circles.  The center of each intersection circle lies at the midpoint of the line segment connecting the associated points.  There can never be more than 2 intersection points of these intersection circles, indicating that no more than 2 particles can touch the same 3 linearly connected particles.}\label{NoMoreThan2PointsTouching3Fig}
 \end{figure}

\begin{figure}[htbp!]
 \begin{center}
(a) \includegraphics[width=1.7in]{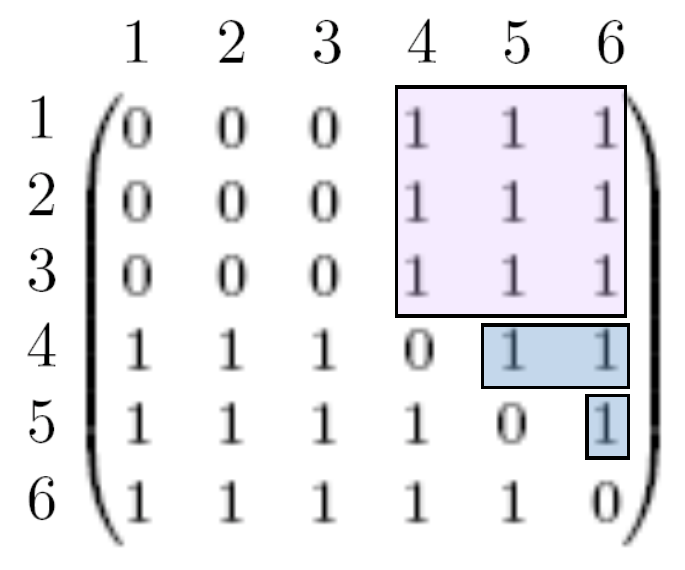}
(b) \includegraphics[width=3.5in]{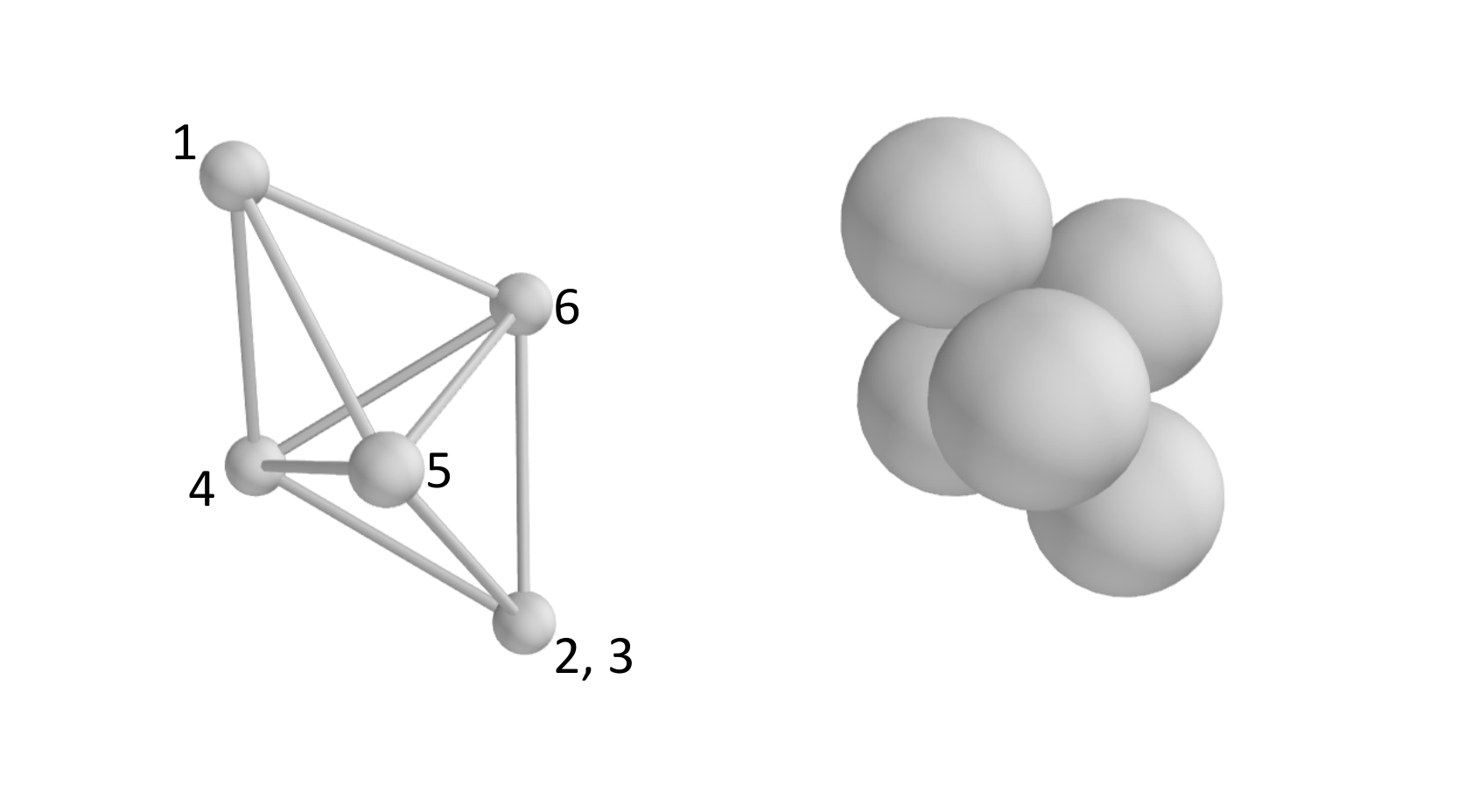}
 \end{center}
 \caption{\textbf{Example of an Unphysical Adjacency Matrix.} (a) An adjacency matrix that is unphysical because it implies more than 2 intersections of intersection circles.  The blue highlights show that particles 4,5,6 make up a trimer.  The purple highlighted part shows that particles 1, 2, and 3 all touch the same trimer, 4,5,6.  (b) A sphere packing corresponding to this unphysical adjacency matrix (shown in both sphere and point/line representations).  For it to be realized, 2 particles must occupy the same point in space.}\label{EliminationRuleUnphysicalExFig}
 \end{figure}

Physically, this implies that no more than 2 spheres can simultaneously touch 3 connected spheres.  Three spheres are connected if at least 2 contacts exist between them (Fig.~\ref{NoMoreThan2PointsTouching3Fig}).  This in turn tells us how many identical spheres can mutually touch a trimer: 2.  Figure \ref{EliminationRuleUnphysicalExFig} shows an example of an adjacency matrix that is unphysical for this reason: the blue highlighted section shows that spheres 4,5,6 make up a trimer; but the purple highlighted section shows that spheres 1,2,3 all touch the same trimer. This is impossible given the argument outlined above and hence this adjacency matrix does not correspond to a packing.

\subsubsection{Rule 2}\label{Dist Rule 1 Section}

\begin{figure}[htbp!]
 \begin{center}
 (a)\includegraphics[width = 2.75in]{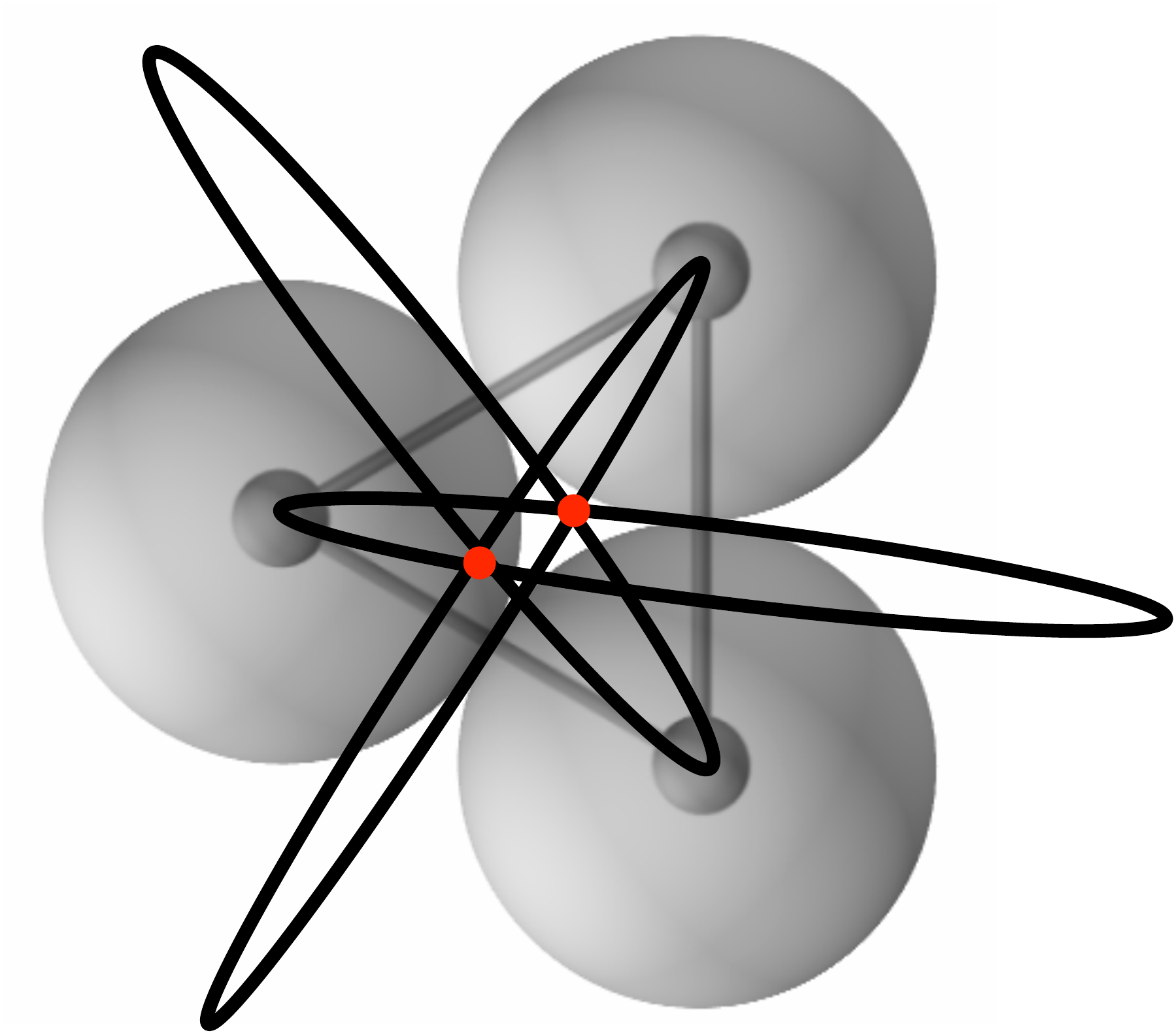}
 (b)\includegraphics[width = 2.5in]{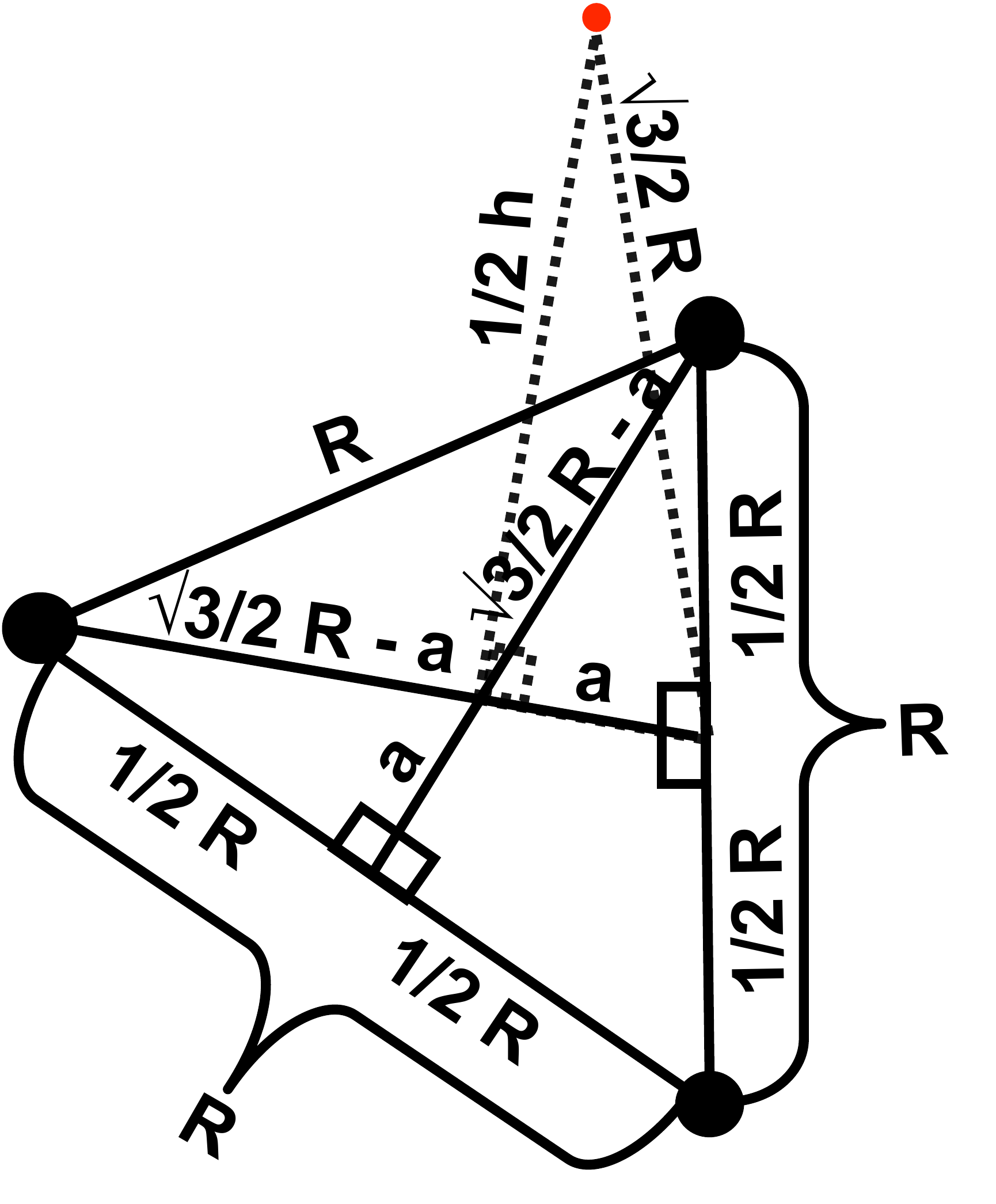}
\caption{\textbf{The Intersections of 3 Intersection Circles.}\newline(a) A trimer and its corresponding intersection circles.  The 3 intersection circles mutually intersect at 2 points, shown in red.  (b) The triangles that relate the trimer to one of the points of intersection (red dot).  The distance between this point and the center of the triangle is equivalent to half the distance between the 2 points of intersection (here denoted as $h$).}\label{3 int circs and triangle}
 \end{center}
\end{figure}

A \textit{trimer}, a configuration of 3 spheres forming an equilateral triangle, is associated with 3 mutually intersecting intersection circles (Fig.~\ref{3 int circs and triangle}a).  These 3 intersection circles intersect at 2 points (shown in red).  Here we calculate the distance between these 2 intersection points.

\begin{figure}[htbp!]
 \begin{center}
  %%\includegraphics[width = 5in]{5PartPackingForTrimer.pdf}
   % %\includegraphics[width = 5in]{5PartPackingForTrimer2.pdf}
   %%\includegraphics[width = 5in]{5PartPackingForTrimer3.pdf}
  % %\includegraphics[width = 5in]{5PartPackingForTrimer4.pdf}
     %(a)%\includegraphics[width = 2.75in]{5PartPackingForTrimer2Sphere.pdf}
       %(b)%\includegraphics[width = 2.4in]{5PartPackingForTrimer2BallStick.pdf}
       % %\includegraphics[width = 5in]{5PartPackSphereBallStickDoubTransWLength.pdf}
         \includegraphics[width = 5in]{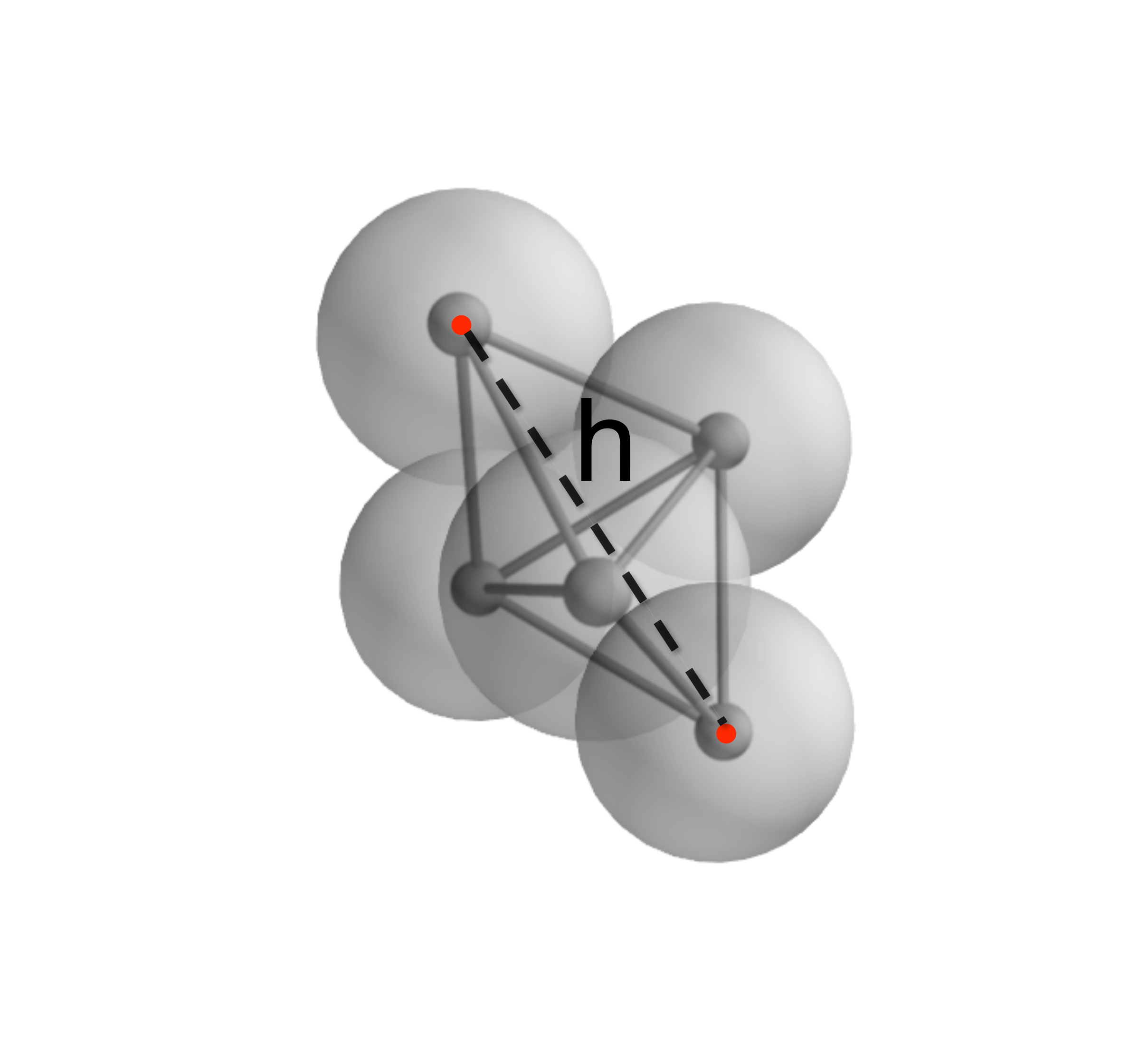}
\caption{\textbf{The 2 Intersection Points of a Trimer Correspond to the 5-Particle Packing.} A 5-particle packing is shown with its point representation overlain.  The center triangle of the point representation corresponds to a trimer, and the 2 points that contact the trimer correspond to the 2 intersection points of the trimer's 3 intersection circles.  The 2 intersection points are shown in red, and $h$ corresponds to the distance between them.}\label{5 Part Pack For Trimer}
 \end{center}
\end{figure}

Note that a particle lying at one of the intersection points forms the 4-particle packing (the tetrahedron).  And that 2 particles, lying at each intersection point, form the 5-particle packing (the 5-point polytetrahedron). The distance between these 2 intersection points, $h$, is the only distance that is greater than R in the 5-particle packing (Fig.~\ref{5 Part Pack For Trimer}).

To calculate this distance, we note that the trimer and its associated intersection circles form the set of triangles shown in figure \ref{3 int circs and triangle}b (where the dashed line indicates an out-of-plane triangle).  We calculate $a$ by considering the right triangle with sides $\sqrt{3}/2R - a, a, 1/2 R$.  Trigonometry then implies that $a=R/(2\sqrt{3})$, and $h=2 \sqrt{2/3} R$.

\raggedbottom

This implies that the solution to an adjacency matrix corresponding to the 5-particle packing is
\begin{eqnarray}
\nonumber \left( \begin{matrix} 0	&1	&1	&1	&1\\
1	&0	&1	&1	&1	\\
1	&1	&0	&1	&1	\\
1	&1	&1	&0	&0	\\
1	&1	&1	&0	&0	\end{matrix} \right)
\longrightarrow \left( \begin{matrix} 0	&1	&1	&1	&1\\
1	&0	&1	&1	&1	\\
1	&1	&0	&1	&1	\\
1	&1	&1	&0	&2 \sqrt{\frac{2}{3}}	\\
1	&1	&1	&2 \sqrt{\frac{2}{3}}	&0	\end{matrix} \right)
\end{eqnarray}
where the right matrix is the corresponding distance matrix, $\D$, and without loss of generality we have let $R=1$.  For $n = 5$, there is only 1 non-isomorphic minimally-rigid $\A$.\\

We can formalize this construction as a distance rule, which can be used whenever
a submatrix of some $\A$ has the same structure as the 5-particle packing.
Such submatrices can be identified with the following pattern: $\A_{ij} = \A_{ik} = \A_{kj} = 1$, and there exist 2 points $p$ for which $\A_{pi} = \A_{pj} = \A_{pk} = 1$.  Whenever this pattern exists, the distance submatrix between the associated points corresponds to $\D$ for the 5-particle packing.
In particular, this rule solves for the distance between the 2 points $p$, for example if $p = l, m$, then $\D_{lm} = 2\sqrt{\frac{2}{3}}R$.

\subsubsection{Rule 3}\label{Elim Rule 2 Section}

Another elimination rule follows directly from any distance rule, including 
Rule 2 derived above. Suppose we determine that for a given pattern of $\A_{ij}$, the contact distribution implies that $\D_{kp} > R$. If it then happens that $\A_{kp} = 1$, then this implies that all of the geometrical constraints cannot be satisfied simultaneously, so that $\A$ is unphysical.

For example, if an $\A$ contained the intersection circle construction discussed in the previous section, that would imply that $\D_{lm} = 2\sqrt{2/3}R$, but if the adjacency matrix also stated that $\A_{lm} = 1$, then that $\A$ would be unphysical.

\subsubsection{Rule 4}\label{Dist Rule 2 Section}

\begin{figure}[htbp!]
 \begin{center}
 \includegraphics[width = 6in]{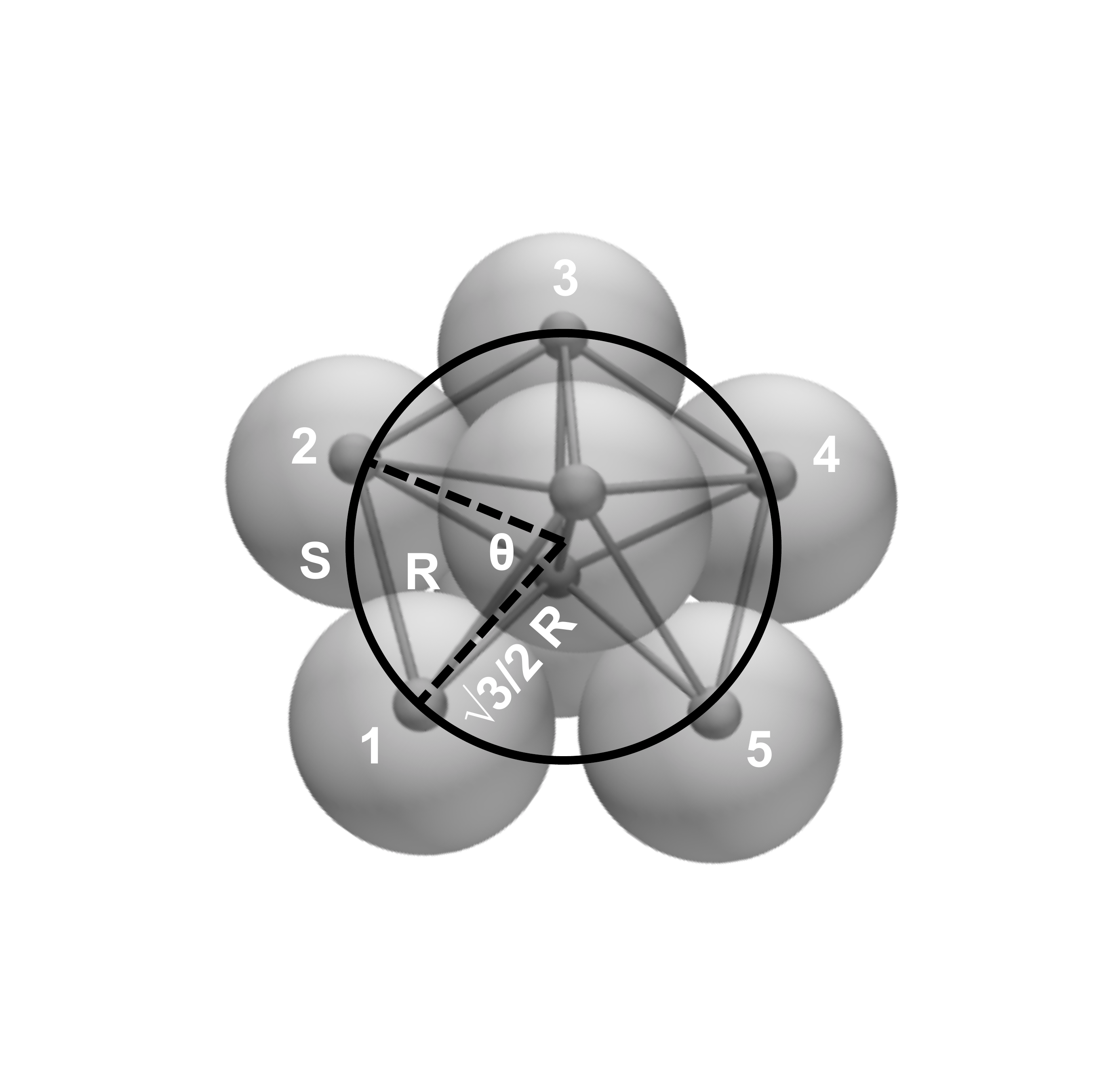}
 \caption{\textbf{5 Points on an Intersection Circle.}\newline The intersection circle shown in black corresponds to the dimer in the center.  5 points are shown lying on the intersection circle; this corresponds to 5 particles touching the center dimer.  The radius of the intersection circle is $(\sqrt{3}/2)R$ (shown as dashed black lines), and connects the center of the dimer (which is the origin of the intersection circle) to points 1--5 on the intersection circle.  The arc length swept out by one pair of particles on the intersection circle is $S$.  It can be seen that the 1st and 5th particles nearly touch.  The space between them is not big enough to fit another particle, and thus it can be seen that no more than 5 particles can touch a dimer.}\label{5PointsOnIntCircPaper}
\end{center}
\end{figure}

We can derive another set of geometrical rules by finding the maximum number of points that
can lie on an intersection circle -- this corresponds to the maximum number of spheres that can touch a dimer.  Figure \ref{5PointsOnIntCircPaper} shows the dimer (top and bottom spheres), as well as points lying on their intersection circle.

The maximum number of spheres that can lie on an intersection circle is 5, and this can be calculated as follows: we divide the circumference of the entire intersection circle by the arc length swept out by 2 spheres lying a unit distance apart\footnote{Without loss of generality, we refer to the distance between two touching spheres, $R$, as the unit distance.} (see figure \ref{5PointsOnIntCircPaper}).  This arc length is given by
$S = r\theta$,
where $r$ is the radius of the intersection circle, and $\theta$ is the angle between 2 radial line segments.  The law of cosines then implies that
\begin{eqnarray}\label{theta eqn}
\theta = \cos^{-1}\left(\frac{1}{3}\right),
\end{eqnarray}
so that the number of points a distance $R$ apart that can fit on an intersection circle is given by
 \begin{eqnarray}
\frac{2\pi \frac{\sqrt{3}}{2} R}{\frac{\sqrt{3}}{2} R \cos^{-1}\left( \frac{1}{3} \right)} \approx 5.1043.
\end{eqnarray}
This indicates that (i) any $\A$ implying that more than 5 points lie on an intersection circle, and (ii) any $\A$ implying a unit distance between all $m \leq 5$ points lying on an intersection circle is unphysical.  We can identify 5 points lying on an intersection circle by the following adjacency matrix pattern: $\A_{ij} = 1$, such that there are 5 points $k$ for which $\A_{ik} = \A_{jk} = 1$.\\

To solve for the structure of $m \leq 5$ points lying on an intersection circle, we must compute the distances between the non-touching particles on the intersection circle (Fig.~\ref{5PointsOnIntCircPaper}).  Of these distances, we have already calculated that between points 1 and 3 and shown that it is $= 2 \sqrt{2/3} R$ (section \ref{Dist Rule 1 Section}).  All of these distances can be obtained by the isosceles triangle with equivalent lengths $\sqrt{3}/2R$ (corresponding to the dashed black lines in figure \ref{5PointsOnIntCircPaper} -- note that only 2 such lines are shown, but that they exist between the midpoint of the dimer and every point along the intersection circle).  The unique length of the isosceles triangle will be the unknown distance, $r_{ij}$, and the angle between the two $\sqrt{3}/2R$ sides connecting particles $i$ and $j$ will be called $\phi_{ij}$.  Thus, the unknown distances will all be given by 
\begin{eqnarray}
\label{base eqn int circ} \sin \left( \frac{1}{2} \phi_{ij} \right) = \frac{\frac{1}{2}r_{ij}}{\frac{\sqrt{3}}{2}R}
\end{eqnarray}
where
\begin{eqnarray}
\nonumber \phi_{13} &=& 2\theta\\
\nonumber \phi_{14} &=& 2\pi - 3\theta\\
\nonumber \phi_{15} &=& 2\pi - 4\theta
\end{eqnarray}
where $\theta$ is given by equation \ref{theta eqn}, thus yielding
\begin{eqnarray}
\label {2sqrt(2/3) dist rule 2} r_{13} &=& 2 \sqrt{\frac{2}{3}} R\\
\label{5/3 dist rule} r_{14} &=& \frac{5}{3} R\\
\label{4sqrt(6)/9 dist rule} r_{15} &=& \frac{4 \sqrt{6}}{9} R
\end{eqnarray}

%fix this sentence - 'fixed' at least once
These calculations apply to any adjacency matrix in which
$\A_{ij} = 1$, there exist $n$ points, $k$, for which $\A_{ik} = \A_{jk} = 1$, and there also exist $n-1$ instances where $\A_{pq} = 1$ amongst the $n$ points, $k$; where $n = 3, 4, 5$ for $r_{13},r_{14},r_{15}$, respectively.  Then the distance between the two endpoints of the $n$ particles is given by $\D_{pk} = r_{13},r_{14},r_{15}$, respectively.\\
That is for\\
$n = 2$: if $k = p, q, l$, then $\A_{pq} = \A_{ql} = 1$, and the distance $\D_{pl} = 2\sqrt{\frac{2}{3}}R$.\\
$n = 3$: if $k = p, q, l, m$, then $\A_{pq} = \A_{ql} = \A_{lm} = 1$, and the distance $\D_{pm} = \frac{5}{3}R$.\\
$n = 4$: if $k = p, q, l, m, z$, then $\A_{pq} = \A_{ql} = \A_{lm} = \A_{mz} = 1$, and $\D_{pz} = \frac{4\sqrt{6}}{9}R$.\\

Note for $n = 3$, we have already identified this $\A$ pattern in section \ref{Dist Rule 1 Section}, whereas the $n=2,4$ structures are new.  Also note that, by symmetry, $r_{13} = r_{24} = r_{35}$, and $r_{14} = r_{25}$, and that these equivalences are identified by the above patterns in $\A$.

\raggedbottom
\subsection{Using Geometrical Rules to Derive a Complete Set of Sphere Packings}

\begin{figure}[htdp!]
 \begin{center}
 \includegraphics[width = 6.25in]{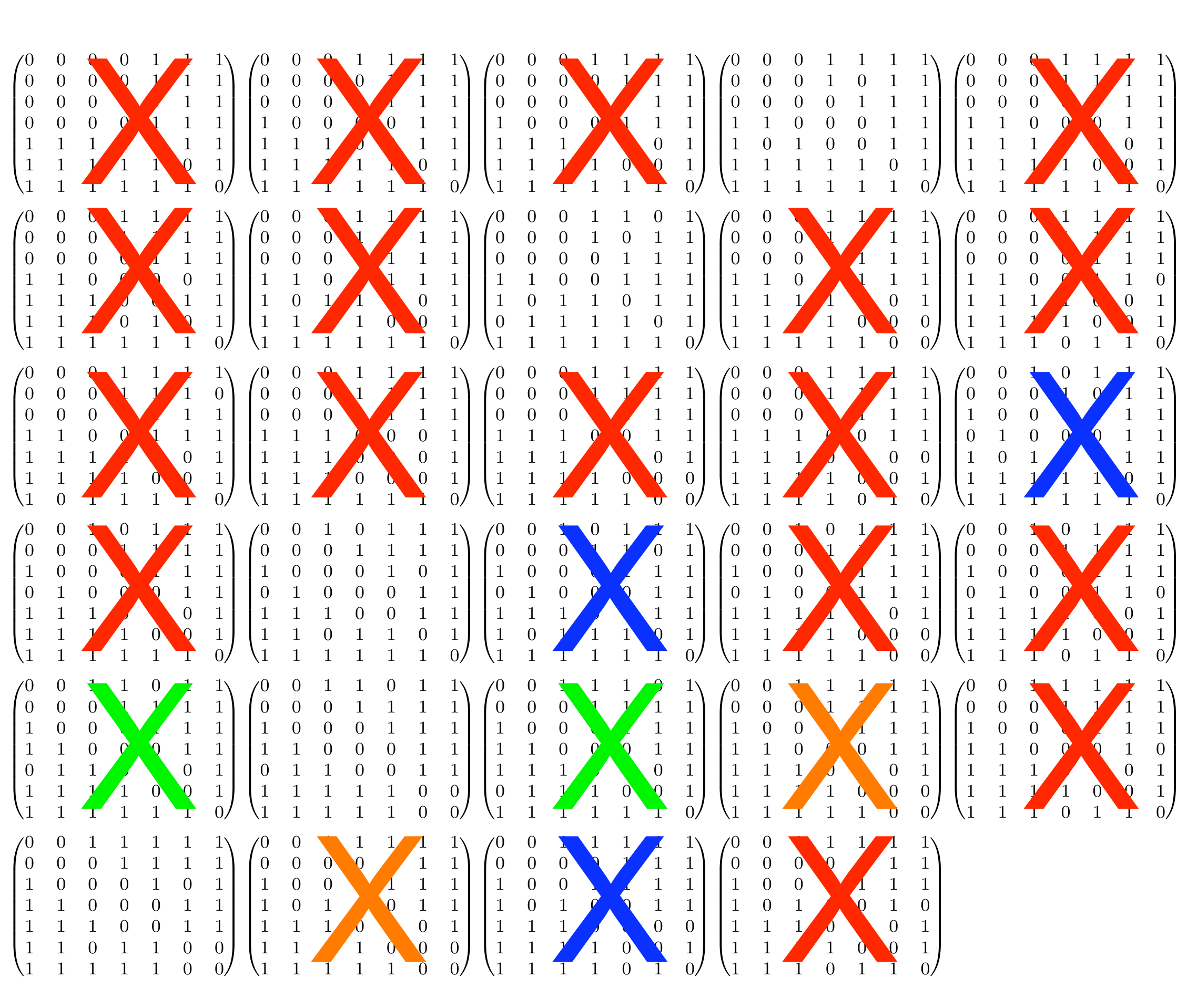}
 \caption{\textbf{Eliminating Adjacency Matrices.} There are 29 non-isomorphic adjacency matrices satisfying minimal rigidity constraints for 7 particles.  24 out of the 29 $\A$'s are eliminated by geometrical rules, which are shown here as color-coded \textbf{X}'s -- see table \ref{7PartElimTable} and appendix A \cite{supp_info} for the corresponding rules.  See figure \ref{7 packings fig} for solutions to the 5 physically realizable $\A$'s (which correspond to the matrices that appear here without an \textbf{X}).}\label{7PartAdjMatElim}
\end{center}
\end{figure}

The aforementioned geometrical rules can be used, along with the set of nonisomorphic adjacency matrices, to derive a complete set of packings for a given $n$. To explain more clearly how this is done, we show as an example the derivation for $n = 7$ particle packings. For this $n$,  there exist 29 minimally-rigid $\A$'s, all of which potentially correspond to packings (table \ref{AdjMatTable}). 

\begin{figure}[htbp]
\begin{center}
\includegraphics[width = 1.0 \textwidth]{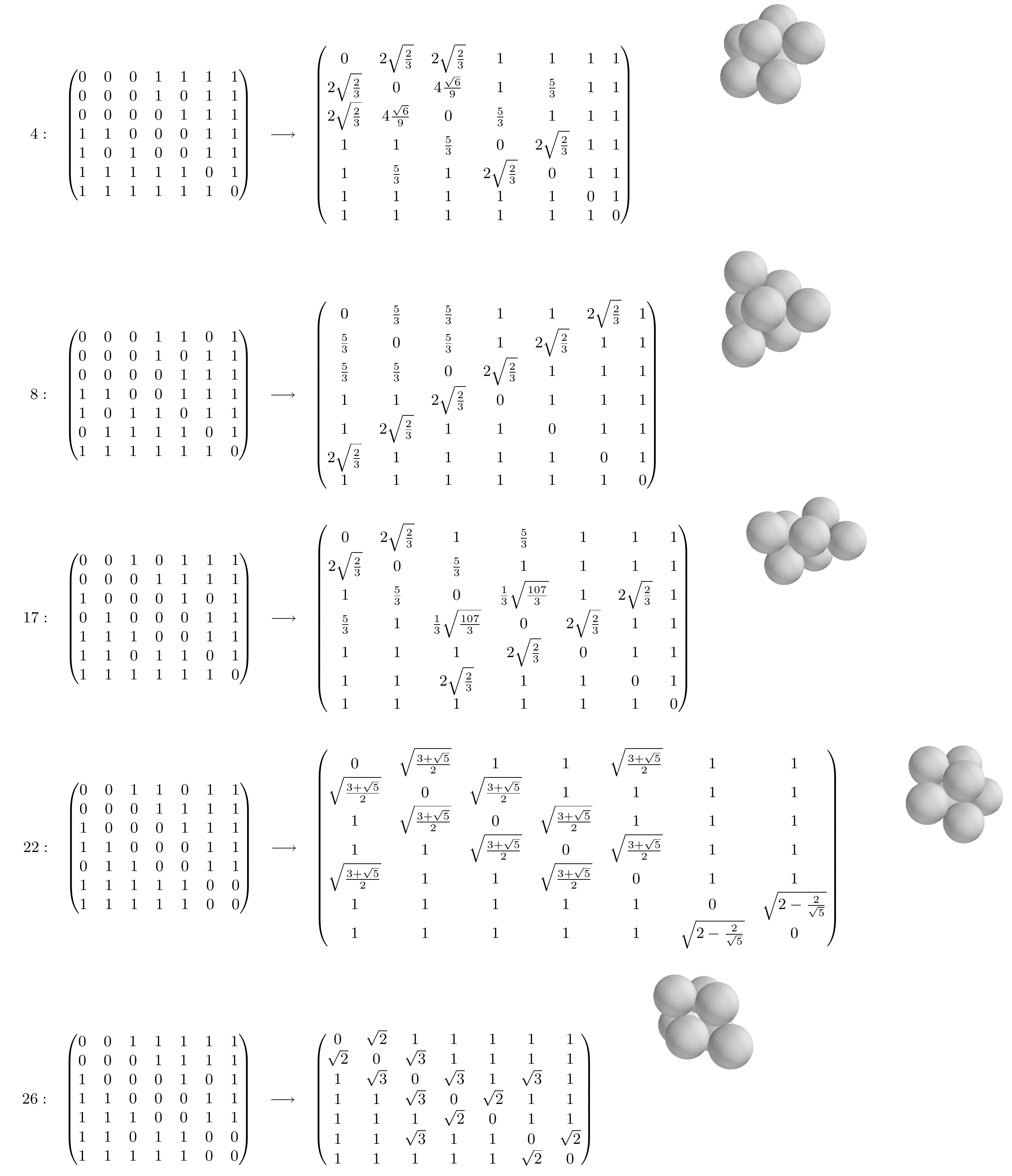}
\caption{7-Particle Packings.\newline These are the 5 possible packings of 7 particles.  Graph numbers appear to the left -- corresponding to a sequential numbering of the graphs appearing in figure \ref{7PartAdjMatElim} from left to right and top to bottom, respectively.  Following the graph numbers are $\A$, $\D$, and a picture of the corresponding packing, respectively.  Note that graph 17 is chiral -- it has both a left-handed and a right-handed form (this can be seen by moving the top particle, currently shown on the left side, over to the right side).}
\label{7 packings fig}
\end{center}
\end{figure}

%I don't know why the 7PartSolveTable is being referenced as 3.3 instead of 3!
To these matrices, we apply the elimination rules just outlined, as well as those that appear in the supplementary information \cite{supp_info}. This immediately eliminates  24/29 $\A$'s  as unphysical. Figure \ref{7PartAdjMatElim} shows which of the matrices are eliminated.  Table \ref{7PartElimTable} shows which rules are used to eliminate the $\A$'s.  17/24 of the matrices are eliminated because they imply more than 2 intersections of intersection circles. Three of the matrices are eliminated because of the relative distance rule for three points on an intersection circle. Two matrices each are eliminated by rules acting on 5 rings and 4 rings, respectively. For the remaining
five adjacency matrices, we apply distance rules to the $\A$'s to solve for the corresponding $\D$'s.  Table \ref{7PartSolveTable} details which distance rules are used to solve for the packings corresponding to each $\A$.  The analytical solutions for the distance matrices as well as the associated packings are shown in figure \ref{7 packings fig}; to each $\A$  (numbered by the order in which it appears in figure \ref{7PartAdjMatElim}, in ascending order from left to right, and top to bottom, respectively), we apply the rules outlined in Table \ref{7PartSolveTable} to analytically solve for the packing.

\begin{table}[htdp!]
\caption{\textbf{Elimination Rules Used for 7-Particle Packings.}\newline Each rule appears in its own section either in the text or in appendix A \cite{supp_info}, where the complete set of rules is included.  The rule column thus lists in what section(s) the relevant rules can be found.}
\begin{center}
\begin{tabular}{|c|l|l|}
\hline
Color: & Unphysical Because: & Rule: \\
\hline
\color{red}{X} & 2 or more intersection circles intersect at more than 2 points & A.1 \\
&&section \ref{Elim Rule 1 Section}\\
\hline
\color{blue}{X} & All relative distances between 3 points lying on an & A.3 \\
& intersection circle $=R$  & sections \ref{Dist Rule 1 Section}, \ref{Elim Rule 2 Section} and \ref{Dist Rule 2 Section} \\
\hline
\color{green}{X} & A closed 5 ring surrounds a circle of intersection & A.14 \\
\hline
\color{orange}{X} & 2 points on opposite sides of a closed 4 ring touch & A.15 \\
\hline
\end{tabular}
\end{center}
\label{7PartElimTable}
\end{table}

\begin{table}[htdp]
\caption{\textbf{Rules Needed to Solve 7-Particle Packings.}\newline The rules listed here correspond to distance rules.  Rule `A.\#' corresponds to rule \# in appendix A \cite{supp_info}, otherwise the relevant equation and section numbers are listed for rules found within the paper.  (Note that rule 4, found in section \ref{Dist Rule 2 Section} (eqn \ref{2sqrt(2/3) dist rule 2}), is the same as rules 2 (section \ref{Dist Rule 1 Section}) and A.1, rule 4 (eqn \ref{5/3 dist rule}) is the same as rule A.2, and rule 4 (eqn \ref{4sqrt(6)/9 dist rule}) is the same as rule A.4).  Graphs are numbered in ascending order from left to right and top to bottom as they appear in figure \ref{7PartAdjMatElim}.  These graphs correspond to the ones without \textbf{X}'s.}
\begin{center}
\begin{tabular}{|c|l|}
\hline
Graph Number: & Rules Used: \\
\hline
4 & section \ref{Dist Rule 2 Section} eqns \ref{2sqrt(2/3) dist rule 2}, \ref{5/3 dist rule}, and \ref{4sqrt(6)/9 dist rule} \\
\hline
8 & section \ref{Dist Rule 2 Section} eqns \ref{2sqrt(2/3) dist rule 2} and \ref{5/3 dist rule} \\
\hline
17 & section \ref{Dist Rule 2 Section} eqns \ref{2sqrt(2/3) dist rule 2}, \ref{5/3 dist rule}; and A.11 \\
\hline
22 & A.7 and A.6 \\
\hline
26 & A.3 and A.9 \\
\hline
\end{tabular}
\end{center}
\label{7PartSolveTable}
\end{table}%

%
%
%{\bf Graph number isnt defined--do we need this? I guess we do but then should say that the numbers follow from nauty output-though then we need to be very clear about how the nauty calculation was set up.}
%
% MICHAEL - I DON'T THINK WE NEED TO HAVE THE GRAPH NUMBERS HERE, AS IT IS JUST A SEQUENTIAL NUMBERING OF THE ORDER IN WHICH THEY APPEAR ON THE PAGE (WHICH WE ALREADY MENTION), AND THERE ARE FEW ENOUGH THAT I BELIEVE THEY'RE EASY TO SPOT OUT (WE ONLY REFER TO NUMBERS OF GRAPHS WITHOUT X'S).  ALSO, I THINK THE FIGURE MIGHT BE MORE CLEAR WITHOUT THE GRAPH NUMBERS, AS PUTTING IN THE NUMBERING WILL MAKE THE GRAPHS THEMSELVES SMALLER IN ORDER TO FIT IN THE SAME SPACE... - NA
%
%
%\begin{footnotesize}
%\[
%\begin{array}{cccc}
%\text{Packing 7.1 (graph 4):} & & &\\ 
%\left( \begin{matrix} 0&0&0&1&1&1&1\\
%0&0&0&1&0&1&1\\
%0&0&0&0&1&1&1\\
%1&1&0&0&0&1&1\\
%1&0&1&0&0&1&1\\
%1&1&1&1&1&0&1\\
%1&1&1&1&1&1&0 \end{matrix} \right) & \longrightarrow & \left( \begin{matrix} 0&2\sqrt{\frac{2}{3}}&2\sqrt{\frac{2}{3}}&1&1&1&1\\
%2\sqrt{\frac{2}{3}}&0&4\frac{\sqrt{6}}{9}&1&\frac{5}{3}&1&1\\
%2\sqrt{\frac{2}{3}}&4\frac{\sqrt{6}}{9}&0&\frac{5}{3}&1&1&1\\
%1&1&\frac{5}{3}&0&2\sqrt{\frac{2}{3}}&1&1\\
%1&\frac{5}{3}&1&2\sqrt{\frac{2}{3}}&0&1&1\\
%1&1&1&1&1&0&1\\
%1&1&1&1&1&1&0 \end{matrix} \right) & %\includegraphics[width = 2in]{7Graph4Sphere.png}
%\end{array}
%\]
%\end{footnotesize}

This is the set of 7 particle sphere packings.  Note that the packing corresponding  to graph 17 (row 4 from the top) is the only one where distinct left and right handed structures are possible; thus it corresponds to  2 distinct states.

\subsection{Non-Uniqueness of Geometrical Rules}

Note that the geometrical rules described here are not unique in that (i) the rules themselves can be derived in different ways, and (ii) a different set of rules altogether could be derived/applied to solve for the same packings.  This is simply one set of rules that works.  One example of this is that either Rule 2 or Rule 4, equation \ref{2sqrt(2/3) dist rule 2} can be used to determine the unknown distance in a 5-particle packing.  Another example of this is that all iterative packings can be solved using the {\sl triangular bipyramid rule} that will be introduced in section \ref{gen rule section} instead of using the aforementioned rules.  There are undoubtedly many such examples, and the list of rules just presented were not derived with the goal of conciseness.  They are complete in the sense that they allow one to solve \textit{all} adjacency matrices for the $n$ presented here.  Beyond this however, they can only solve adjacency matrices containing the structure the rule identifies.  If an $\A$ contains an identifiable structure as well as a non-identifiable (not previously encountered) structure, then the rules will solve the identifiable part and only partially solve that $\A$ for its corresponding $\D$.  If an $\A$ contains no identifiable structure, then new rules must be derived to solve any part of it.  It is because of these latter two cases that we continued to derive new geometrical rules as we increased in $n$.  The \textit{triangular bipyramid rule} in section \ref{gen rule section} is a rule that we derived in order to have one general rule that could recognize a certain class of adjacency matrix structures -- the iterative class -- for all $n$.  Because it is a general rule, applicable to all $n$, its introduction makes the set of rules much more concise.
 
\subsection{Chirality}\label{Chiral Section}

Once all packings have been derived by solving all $\A$'s for their corresponding $\D$'s, we must determine how many states each packing has.  If a packing is chiral, it will have more than one state.  This will show up by a packing having a non-superimposable mirror image; for example a packing having different `handedness,' such as distinct left and right-handed structures.

One can calculate whether a packing is chiral as follows:  The automorphism group of a packing, $\{\alpha\}$, gives the set of self-isomorphisms -- \begin{frenchspacing}i.e. all\end{frenchspacing} possible permutations of the structure into itself.  Each element of the automorphism group will thus correspond either to a rotation or to a reflection.  Rotations are transformations with determinant $=1$, and reflections are transformations with determinant $=-1$.  Thus, one can construct the set of all isomorphic graphs, $\{\D\}$, and construct the automorphism group for any one $\D_i$ within the isomorphic set.  If $\exists$ any matrix $\D_j \in \{\D\}$ that is isomorphic to $\D_i$ \textit{only} through reflections and not through rotations -- \begin{frenchspacing}i.e. if the isomorphism group of $\D_j$ to $\D_i$ corresponds to transformations that all have determinant -1\end{frenchspacing}, then the packing $\D$ is chiral.

A property related to chirality is the symmetry number, $\sigma$.  This corresponds to the number of ways that a structure can be rotated into itself.  The symmetry number is necessary for calculating the equilibrium probability distribution of packings \cite{thesis, thermo_paper}.  The symmetry number of a packing, $\D$, will be equal to the number of transformations within the automorphism group of $\D$ that have determinant 1.  Thus, if a packing has no reflections, then the symmetry number will simply equal the size of the automorphism group.  If a packing has reflections, then the symmetry number will equal the size of the automorphism group divided by 2 (dividing by 2 will remove all automorphism mappings corresponding to reflections, and not rotations).

Related to both symmetry numbers and chirality are point groups.  A point group is a group of symmetry operations which all leave at least one point unmoved.  Point groups have been calculated for many structures -- and there exist programs that allow one to enter in a set of coordinates and retrieve the point group corresponding to those coordinates \cite{RotConstCalc}.  Symmetry numbers and chirality can alternatively be calculated directly from the point group of a structure.  For example, compounds in the $C_m$ point group, where $C_m$ is the cyclic group consisting of rotations by $360^\circ/m$ and all integer multiples (where $m$ is an integer), are always chiral \cite{PointGroupWeb}.

Point groups, symmetry numbers, and chirality of packings are included in the lists of packings appearing in section \ref{the packings} and in the supplementary information \cite{supp_info}.  The growth of chiral structures with $n$ is interesting -- surprisingly, over half of all 9-particle packings are chiral -- see table \ref{PackingResultsTable}.

%\begin{table}[htdp]
%\caption{Number of Enantiomers}
%\begin{center}
%\begin{tabular}{|c|c|c|}
%\hline
%$n$ & Packings & Enantiomers\\
%\hline
%6&2&0\\
%7&5&1\\
%8&13&3\\
%9&50&27?\\
%\hline
%\end{tabular}
%\end{center}
%\label{ChiralTable}
%\end{table}%

\section{One Geometrical Rule That Solves for all Iterative Packings: The Triangular Bipyramid Rule}\label{gen rule section}

In principle, these types of geometrical rules can be used to derive a  complete set of sphere packings for any number of particles, $n$.  However, in practice, the number of rules used here grows too quickly with $n$ for this to be a practical method: at $n=5$ spheres, only 1 rule is required, 3 rules are needed at $n=6$, 12 rules at $n=7$, and at $n=8$, 14 rules solve 435/438 minimally-rigid non-isomorphic $\A$'s.  This leaves 3 unsolved $\A$'s for which more geometrical rules must be derived; looking ahead at the 13,828 and 750,352 $\A$'s that must be solved at $n = 9,10$, respectively, it becomes clear that deriving a rule or set of rules that does not grow significantly with $n$ is a necessary step.  Here, we derive one geometrical rule that can solve one class of packings for any $n$, thereby greatly reducing the number of rules needed to derive a complete set of $n$ sphere packings.  In section \ref{new seed gen rule section}, we discuss how one geometrical rule can also be used to solve the other class of packings for all $n$.

\begin{figure}
 \begin{center}
  (a) \includegraphics[width = 2.75in]{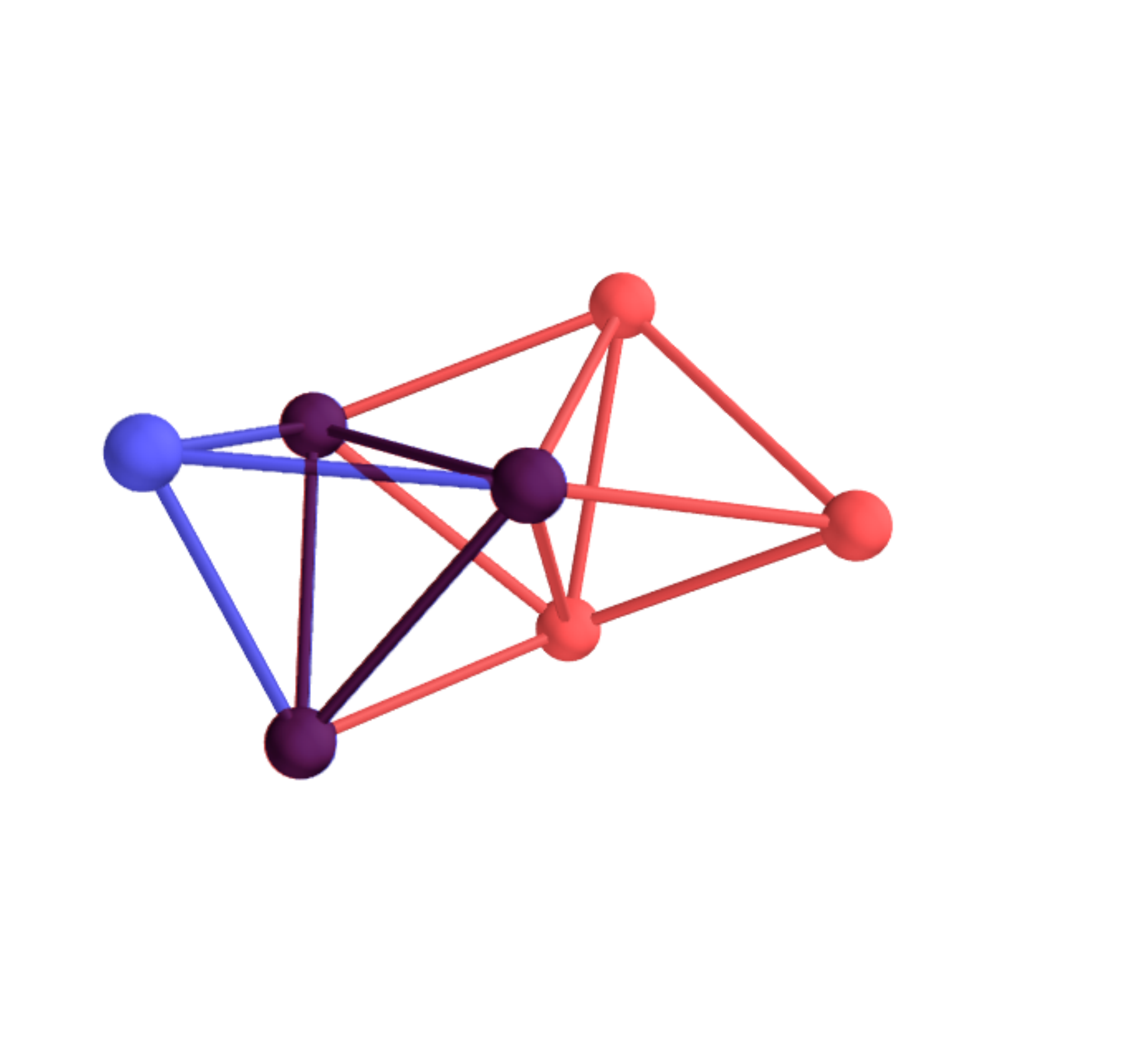}
 (b) \includegraphics[width = 2.75in]{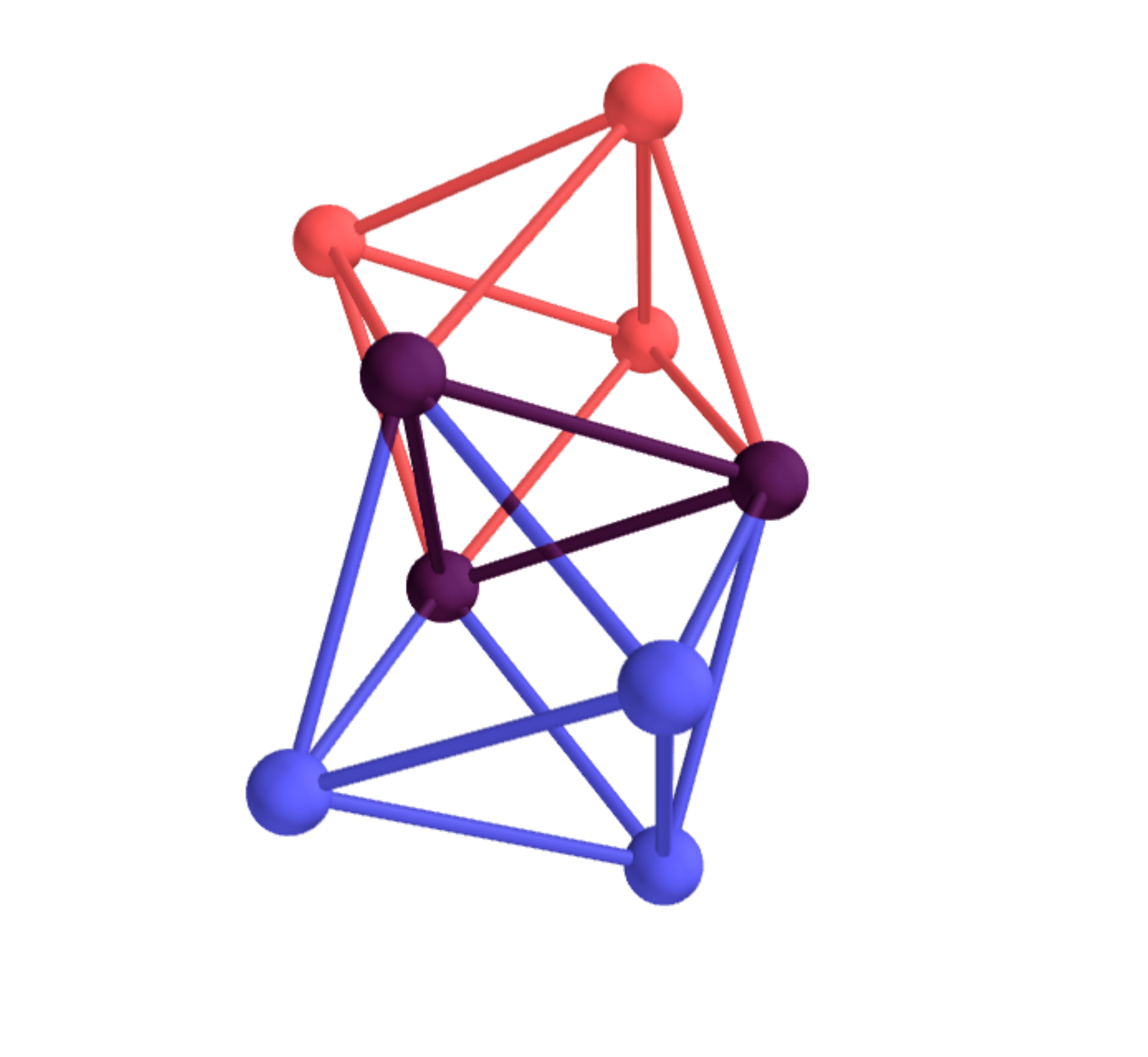}
 %(a) %\includegraphics[width = 2.6in]{7Graph17RedBluMergeCropped}
 %(b) %\includegraphics[width = 1.75in]{2TouchingOcts9PartsRedAndBlueMergeCropped}
 %(a) %\includegraphics[width = 3.25in]{7Graph17RedBluMerge}
 %(b) %\includegraphics[width = 1.7in]{2TouchingOcts9PartsRedAndBlueMergeCropped}
  \caption[c]{\textbf{Iterative Packings.} Two examples of iterative packings.  (a) A 6 particle polytetrahedron (red) with one particle added to it (blue).  This decomposes into a tetrahedron (blue) plus a 6 particle polytetrahedron (red), with a shared triangular base (purple).  (b) 2 joined octahedra (one red and one blue, with a shared purple triangular base) forming a 9-particle packing. }\label{it_packs_ex_fig}
  \end{center}
 \end{figure}
 
\begin{figure}
 \begin{center}
  (a) \includegraphics[width = 2.95in]{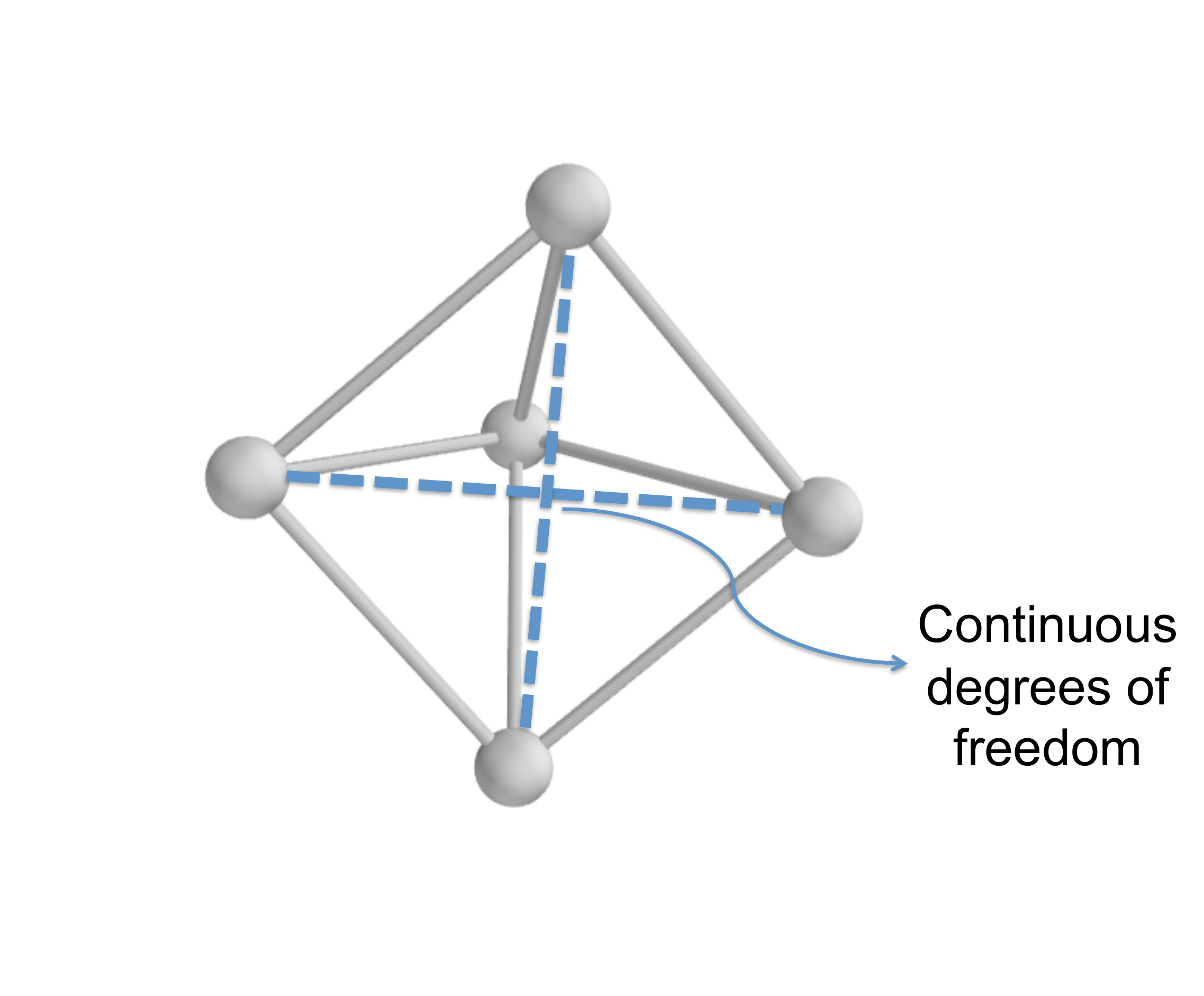}
 (b) \includegraphics[width = 2.55in]{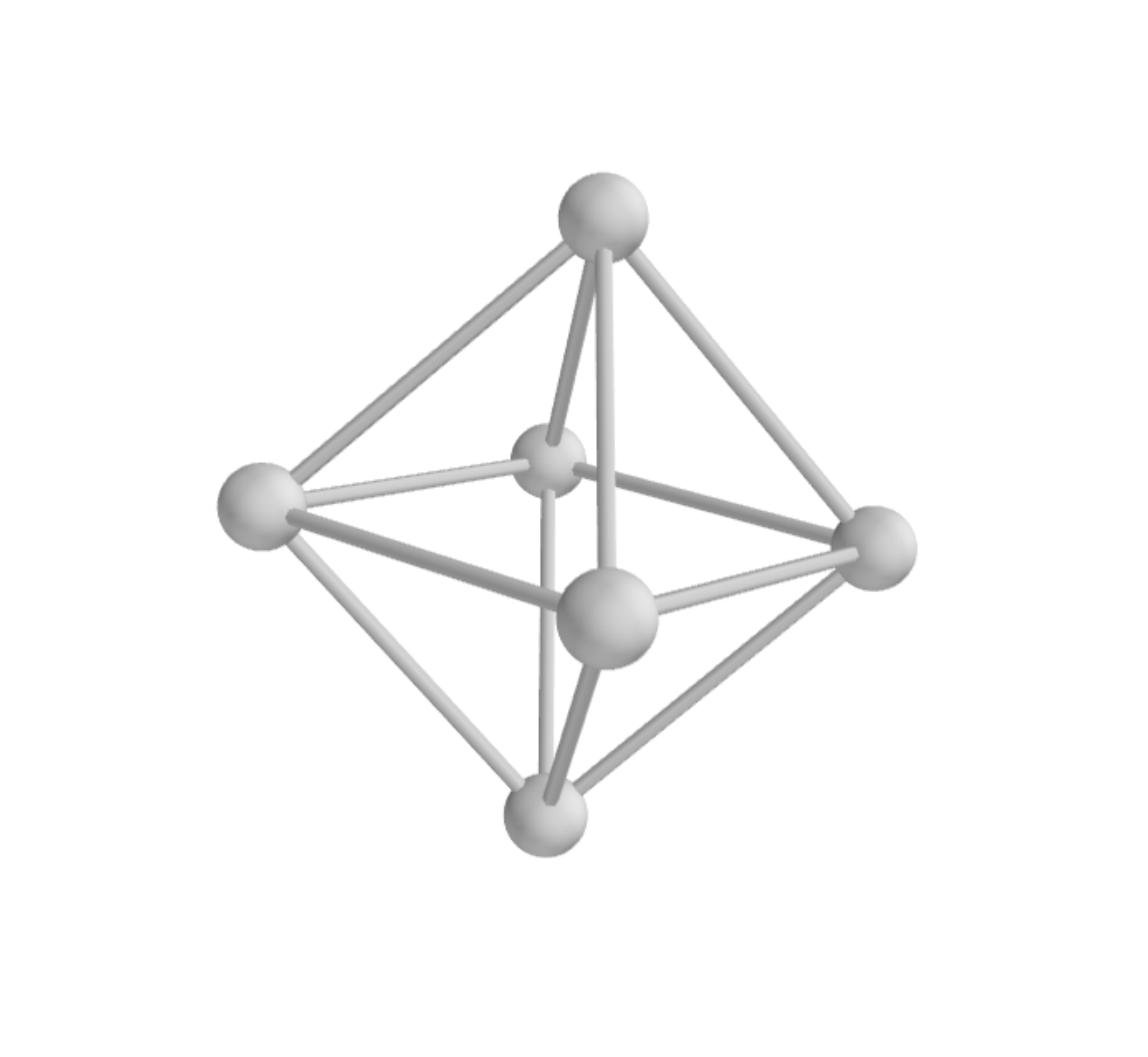}
  \caption[c]{\textbf{New Seeds.} The octahedron is an example of a new seed.  (a) The `base' of an octahedron has a continuous degree of freedom through which a 5 particle polytetrahedron can form, and thus is not a packing.  The continuous degrees of freedom are shown as dashed lines; bringing either of the pairs of particles connected by these dashed lines into contact forms the 5 particle polytetrahedron.  (Note that the 5 particle structure shown in (a) is not minimally-rigid as it has fewer than $3n-6 = 9$ contacts.) (b) Once a 6th particle is added, the octahedral structure can be stabilized, thereby forming a \textit{new seed}.  This `new seed' is an inherently new structure.}\label{new_seeds_ex_fig}
  \end{center}
 \end{figure}

Packings can be broken up into two types or classes: \textit{iterative} and non-iterative, or \textit{new seeds}.  Iterative packings are $n$-particle packings that are solely combinations of packings of less than $n$ particles (see Fig.~\ref{it_packs_ex_fig}).   New seeds are $n$-particle packings that cannot be constructed solely out of packings of less than $n$ particles -- \begin{frenchspacing}i.e.\end{frenchspacing} they contain within them (in part or in whole) an inherently new structure (Fig.~\ref{new_seeds_ex_fig}).  Put another way, iterative packings correspond to $\A$'s for which all minimally-rigid $m$ x $m$ subgraphs, $m < n$, correspond to packings that have been identified at lower $n$.

\clearpage
%\newpage
\subsection{Solving iterative structures}

An iterative packing is a polyhedron containing \textit{solely} packings of less than $n$ particles\footnote{An iterative $\A$ is an $n \times n$ graph composed \textit{solely} of $m \times m$ ($m < n$) subgraphs, each of  which correspond to minimally-rigid $\A$'s of less than $n$ spheres.}. Thus any iterative packing can be decomposed into 2 joined polyhedra (see Fig.~\ref{it_packs_ex_fig} -- the red and blue packings are the joined polyhedra).  The 2 polyhedra are joined via a common base of particles (shown in purple)\footnote{Given the minimal rigidity constraints we have imposed, this common base will always consist of at least 3 particles.}.  Because the joined polyhedra are less than $n$-particle packings, all of their \textit{intra}polyhedral distances are known from lower-order packings. 

Thus, deriving one geometrical rule that can solve for \textit{all} iterative packings requires solving the following geometrical problem:  Given 2 joined polyhedra, where all \textit{intra}polyhedral distances are known, derive a general formula for the \textit{inter}polyhedral distances.  Note that the solution to this problem immediately extends to unphysical iterative structures as well, as they are composed of structures of less than $n$ particles, where either (i) one or more of the joined structures is unphysical, or (ii) the particular combination of the structures is unphysical.

%BTW - This property below is not related to spherical trigonometry.  The only part of this rule that is related to spherical trigonometry is the rule that relates dihedral angles to regular angles (this ends up being the same as the spherical rule of cosines) -- and this is stated later on in the text.  This relation of dihedral to regular angles is one component of the entire rule; thus, as a whole, the rule is not related to spherical trigonometry, it makes use of something that is related to sphere trig. -- NA
The geometrical problem is solved with the following observation: An explicit analytical formula for the distance between any 2 points can be derived if those 2 points can be related to a common triangular base. Let there exist two particles $i,j$ whose interparticle distance, $r_{ij}$, is unknown.  If there also exist 3 particles, $k,p,q$, with known interparticle distances ($r_{kp}$, $r_{kq}$, $r_{pq}$), and if the distances between $i$, $j$ and the 3 particle base ($r_{ip}, r_{ik}, r_{iq}, r_{jp},r_{jk},r_{jq}$) are also known; then there exists an analytical relationship for the resulting $r_{ij}$.  We call this the {\sl triangular bipyramid rule} because the 5 points $i,j,k,p,q$ together form a (potentially irregular) ditetrahedron or triangular bipyramid (see figure \ref{triang dipyramid}).  We show the rule here, while a complete derivation can be found in the supplemental information (Appendix A \cite{supp_info}).

The dihedral angles $A_2$ and $A_3$ of figure \ref{triang dipyramid} are given by:
\begin{eqnarray}
\nonumber A_3 &=& \cos^{-1}\left( \frac{\cos a_3 - \cos b_3 \cos c_3}{\sin c_3 \sin b_3} \right)\\
\nonumber A_2 &=& \cos^{-1}\left( \frac{\cos a_2 - \cos b_2 \cos c_2}{\sin c_2 \sin b_2} \right)
\end{eqnarray}
These formulas are essentially obtained using the spherical rule of cosines - see Appendix A for details \cite{supp_info}.

The dihedral angle $A_1$ will be either the sum or the difference of the dihedral angles $A_2$ and $A_3$ (see Fig.~\ref{triang dipyramid}), depending on whether the points $i,j$ lie on the same or on opposite sides of the base $p,k,q$.  If $i,j$ lie on the same side, then $A_1 = |A_2 - A_3|$, and if $i,j$ lie on opposite sides then $A_1 = A_2 + A_3$.

The angle $a_1$ is then given by
\begin{eqnarray}\label{a1 eqn}
a_1 = \cos^{-1}(\sin c_1 \sin b_1 \cos A_1 + \cos b_1 \cos c_1)
\end{eqnarray}
and from the law of cosines, we can then calculate $r_{ij}$:
\begin{eqnarray}\label{gen rule eqn rij}
r_{ij} = \sqrt{r_{ip}^2 + r_{pj}^2 - 2r_{ip}r_{pj}\cos a_1}
\end{eqnarray}

Associated with each $r_{ij}$ we have 2 possible $A_1$, and thus 2 possible solutions (similar, in principle, to one having 2 possible solutions to a quadratic equation). 

\begin{figure}
 \begin{center}
  \includegraphics[width = 6.5in]{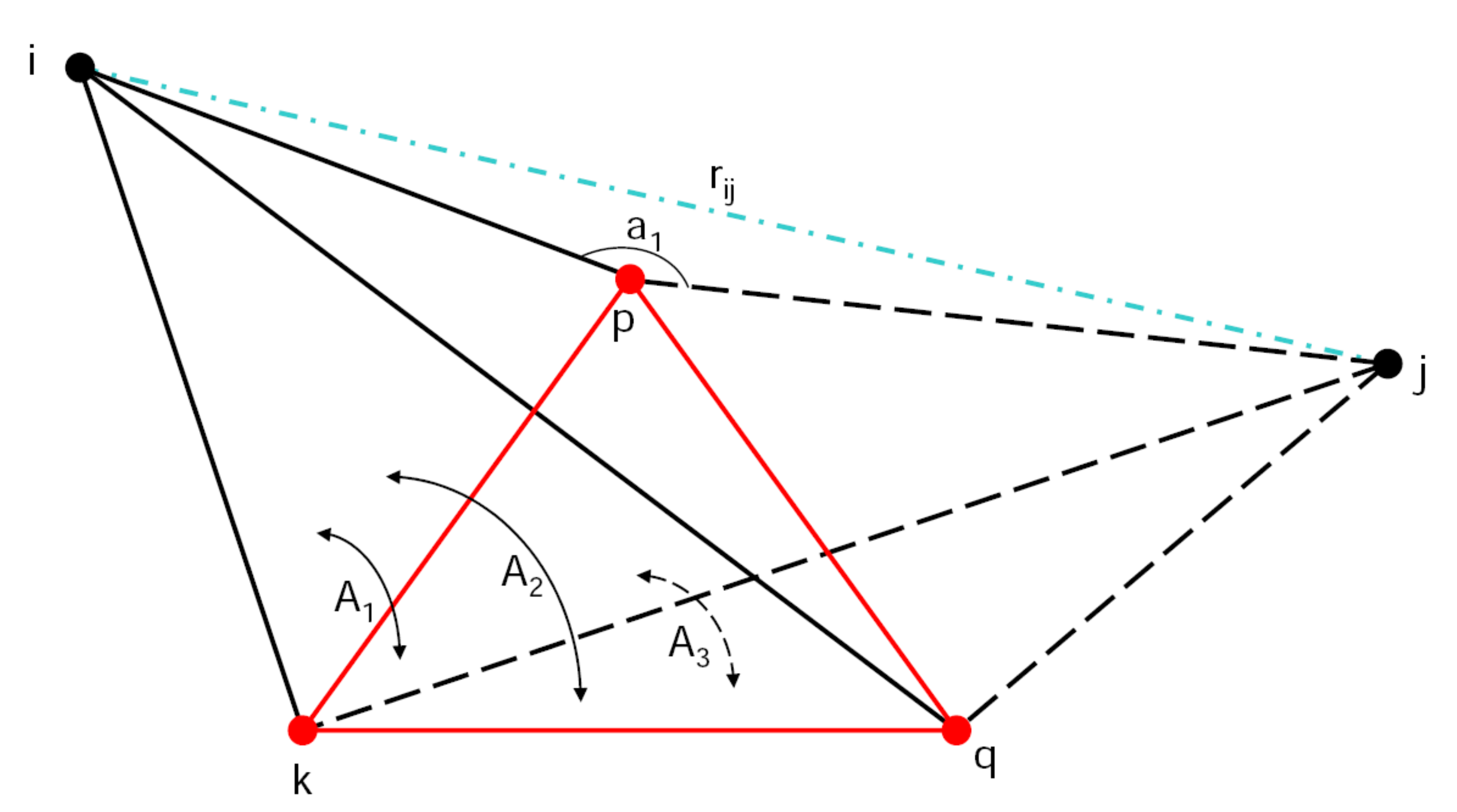}
  \caption[c]{\textbf{The Triangular Bipyramid.}\newline The triangular bipyramid (or ditetrahedron) constructed in the triangular bipyramid rule.  The center triangle ($kpq$), shown in red, corresponds to the common 3 particle base.  Particles $i$ and $j$ are related to one another through the common base.  The distance between $i$ and $j$, $r_{ij}$, shown as the dash-dot blue line, is unknown.  $a_1$ corresponds to $\angle jpi$.  $A_1$ is the dihedral angle between $\triangle ipk$ and $\triangle jpk$, $A_2$ is the dihedral angle between $\triangle ipk$ and $\triangle kpq$, and $A_3$ is the dihedral angle between $\triangle kpq$ and $\triangle jpk$.  Points $i$ and $j$ can either both lie on the same side of the base $kpq$ or each lie on opposite sides of the base (indicated by the dashed lines, that can either go into or come out of the plane).  If $i,j$ lie on the same side, then $A_1$ is equal to the difference of $A_2$ and $A_3$, and if $i,j$ lie on opposite sides of the base then $A_1$ is equal to the sum of $A_2$ and $A_3$.  When all distances other than $r_{ij}$ are known, then an explicit analytical formula can be derived to solve $r_{ij}$.}\label{triang dipyramid}
  \end{center}
 \end{figure}

\subsection{Applying the triangular bipyramid rule}\label{app triang bipyramid rule section}

The triangular bipyramid rule can be used to solve all iterative packings as follows.  We first search for subgraphs of $\A$ corresponding to lower-$n$ seeds.  The elements of $\D$ corresponding to these lower-order structures are known and inserted appropriately.  If $\A$ is iterative, all minimally-rigid subgraphs of $m < n$ particles (\begin{frenchspacing}i.e.\end{frenchspacing} $m$ subgraphs with at least $3m-6$ contacts and at least 3 contacts per particle) will correspond to $m$-particle packings.  Once all lower-order seeds are inserted as appropriate, all unknown $r_{ij}$ correspond to the distances between the spheres of different known lower-order subpackings.  The triangular bipyramid rule is then applied to each unknown element\footnote{If $\A$ is not iterative, there will exist unknown $r_{ij}$ that do not correspond to distances between spheres of known subpackings.  In this case, equation \ref{gen rule eqn rij} will contain at least 1 unknown element on the right hand side and can not be applied directly.} of $\D$.  For each unknown distance, $r_{ij}$, both solutions are potentially stored, as are all possible sets of unknown distances $\{r_{ij}\}$.  Along the way, each $r_{ij}$ solution is tested for consistency, and it is always possible that both, 1, or neither solution will be consistent.  Once all locally consistent $r_{ij}$ are stored, the resultant $\{r_{ij}\}$ are tested for global consistency.

A solution will be inconsistent, and thus unphysical, for one of the following reasons:

\renewcommand{\labelenumi}{\arabic{enumi}.}
\begin{enumerate}
\item \label{local and global inconsistency} It violates the triangle inequality (meaning that no real solution exists -- this shows up as the absolute value of the argument of the inverse cosine being greater than 1).

\item \label{overlap inconsistency} One or more distance(s) are less than R.

\item \label{global inconsistency} Different triangular bases lead to different $r_{ij}$; this indicates that a structure  is in conflict with itself. One part of it implies it should have one structure, whereas another part implies a different structure. Such a structure is physically and mathematically inconsistent.
\end{enumerate}

%Michael, the first violation can be a measure of both local and global inconsistency (depending on which points are tested), and it can register inconsistency by considering only 3 points, or different combinations of 5 points for the entire triangular bipyramid... -- NA
Violation \ref{overlap inconsistency} arises within the 5 particles of one triangular bipyramid, and thus registers a physical local inconsistency.  Violation \ref{local and global inconsistency} occurs within individual triangular bipyramids, as each $r_{ij}$ is determined, in which case it registers a local inconsistency; as well as over the entire set of triangular bipyramids, once all $\{r_{ij}\}$ have been determined, in which case it registers a global inconsistency\footnote{In this case, some triangular bipyramids are locally consistent, whereas others are not.  All possible triangular bipyramids of a structure need not be tested to solve for all $r_{ij}$, thus it is important to check all bipyramids to ensure global consistency once $\{r_{ij}\}$ have been determined.  This violation is related to violation \ref{global inconsistency}, except that here the violation is registered between the angles associated with the line segments, and in violation \ref{global inconsistency} the violation occurs within the dihedral angles.}.  Violation \ref{global inconsistency} occurs when solutions are consistent within individual triangular bipyramids, and thus locally consistent, but inconsistent within combinations of triangular bipyramids -- these solutions are thus \textit{globally inconsistent}.  This violation can be checked as follows:  Figure \ref{triang dipyramid} shows that the dihedral angle, $A_1$, is given by either the sum or the difference of $A_2$ and $A_3$, if particles $i$ and $j$ lie on opposite sides or on the same side of the triangular base, respectively (to within a $2\pi$ modulation of course).  Test all possible 5 particle combinations of triangular bipyramids within the $n$ particle structure, and if there exists a triangular bipyramid that does not satisfy
\begin{eqnarray}\label{global consistency check}
A_1 = \begin{matrix} &A_2 + A_3\\ &|A_2 - A_3|\\ &2\pi - (A_2 + A_3)\\ &2\pi - |A_2 - A_3|\end{matrix}
\end{eqnarray}
%\[
%A_1 = \left\{
%\begin{array}{l}
%A_2 + A_3\\
%|A_2 - A_3|\\
%2\pi - (A_2 + A_3)\\
%2\pi - |A_2 - A_3|
%\end{array}
%\right.
%\]
the solution is globally inconsistent.  (In calculating $r_{ij}$ in equation \ref{gen rule eqn rij}, we need not consider the latter two $A_1$ solutions, as $\cos(2\pi - x) = \cos(x)$). %Figure \ref{} shows an example of this.

There is one scenario in which 5 points need not satisfy this global consistency check, and that is when 3 or more of the 5 points lie in a line.  In this case, the 3 or more points define a line and not a plane, and thus the dihedral angle is not defined, and equation \ref{global consistency check} can not be applied.  This situation is encountered in certain structures that contain octahedra (for example, see graphs 5416 and 10664 at $n=9$ in the supplementary information \cite{supp_info}).  In this scenario, the following global check can be performed: for $m$ points lying in a line, there must exist the following number of line segments with a given minimum distance:
\begin{eqnarray}
\begin{matrix} \text{Number of Line Segments} & \text{Minimum Length of Line Segment}\\
m-1 & R\\
 m-2 & 2R\\
. & .\\
. & .\\
. & .\\
m - (m -1) & (m-1)R
\end{matrix} 
\end{eqnarray}
Note that the first constraint, having $m-1$ line segments with a minimum length of $R$, is automatically satisfied by the fact that we eliminate all solutions containing a distance $< R$.  Up to $n=10$, it turns out that performing this check for $m=3$ alone is sufficient to find all globally inconsistent solutions.  We evaluate that 3 points lie in a line by identifying that the angle associated with those 3 points is 0 or $\pi$.  We then check that there is at least one distance amongst the 3 points that is $\geq 2$ (where once again, $R = 1$ without loss of generality).  If this is not satisfied, then the packing is globally inconsistent and is eliminated.  Note that the fact that some dihedral angles may not be defined was overlooked in \cite{short_paper}, and is the primary source of the difference in numbers reported here in tables \ref{AdjMatTable} and \ref{PackingResultsTable} versus the numbers previously reported in \cite{short_paper}.

\subsection{The Growth of New Seeds}

For new seeds, we have a structure that contains an inherently new polyhedron, and thus some or all \textit{intra}polyhedral distances are also unknown, \begin{frenchspacing}i.e. one\end{frenchspacing} or more of the 9 distances within $\{r_{ip}, r_{ik}, r_{iq}, r_{jp}, r_{jk}, r_{jq}, r_{pk}, r_{pq}, r_{kq}\}$ are unknown.  Thus, deriving the equation for $r_{ij}$, as is done for iterative packings, will yield one equation with more than one unknown.  The triangular bipyramid rule, therefore, can not be applied directly to new seeds, and new geometrical rules must be derived to analytically solve non-iterative $\A$'s.

Using a general rule to solve for iterative $\A$'s, and deriving individual geometrical rules for non-iterative $\A$'s is feasible so long as the non-iterative $\A$'s do not grow too quickly with $n$.  Table \ref{AdjMatTable} shows that this is the case for $n \leq 9$, where the number of non-iterative $\A$'s is 5 or less.  However, at $n = 10$, there are 94 non-iterative $\A$'s\footnote{Note that the non-iterative and iterative $\A$'s listed in this table are constructed \textit{after} applying the geometrical rules for $n \leq 8$ that appear in the text and in the supplementary information \cite{supp_info}.  Thus, this reflects the number of iterative and non-iterative $\A$'s with respect to these geometrical rules, and not the absolute number of iterative and non-iterative $\A$'s}.  To sift through the 94 potential seeds at $n=10$ requires inventing new geometrical rules, and the growth of such rules demonstrates that the method we have described does not scale efficiently with $n$.  In section \ref{new seed gen rule section}, we will discuss a potential extension of the triangular bipyramid rule, which might be able to break this bottleneck, at least computationally.  Here, we present the packing results derived from a combination of the triangular bipyramid rule and individual geometrical rules.  For $n \leq 9$, we have analytically solved for all packings.  At $n = 10$, we analytically solve for all iterative packings, and produce a preliminary list of new seeds, found by solving the non-iterative $\A$'s numerically using Newton Iterations\footnote{Our runs of Newton Iterations used random initial guesses (between -5 and 5 for the coordinates of the particles).  We performed 20 iterations of each initial guess, and approximately 150,000 total runs.  Every matrix for which a solution was found, was found multiple times.  We believe this to be a reasonably thorough search and that this preliminary list of new seeds at $n=10$ might be complete.  It is worth noting that the preliminary list of new seeds reported here found by Newton Iterations is the same as the list reported in \cite{short_paper} found by constructing the non-iterative $\A$'s manually with the construction toy {\it Geomags}.}.

%\newpage
\section{The Set of Sphere Packings}\label{the packings}

Here we present the list of sphere packings derived by this method\footnote{The number of packings presented here for $n = 9,10$ differs from the number we reported in \cite{short_paper}.  This is primarily because our previous code did not run a check to ensure that the dihedral angle was well defined when checking for global consistency.  As mentioned in section \ref{app triang bipyramid rule section}, this can occur when the 3 points used to define a plane are collinear, occurring in some of the packings that contain octahedra.   In our original code (used for \cite{short_paper}), the dihedral angle check was still being performed in such an instance and erroneously deemed some packings globally inconsistent.  It was a personal communication with Rob Hoy, who extended this work in \cite{Hoy_O?Hern_2010}, that brought it to our attention that 2 packings were missed in our $n = 9$ list.  In examining this discrepancy, we discovered that this issue with the dihedral angles was what caused these packings to be missed, and we have since made the relevant correction in the code.  This caused 52 packings to be realized at $n=9$ and many more packings to be realized at $n=10$.  We also made 2 more modifications to the code, such as further correcting for numerical round-off error, which further caused several more packings to be registered at $n=10$.  Please see supplemental information, Appendix C \cite{supp_info} for a complete list of the changes that were made to the code.}.  In principle, the analytical method outlined here will yield a complete set of minimally-rigid packings.  However, we have not implemented the triangular bipyramid rule symbolically, leading to the following practical issues, which could lead to numerical errors:

\textit{Numerical round-off error:} All calculations involving the triangular bipyramid rule are subject to numerical precision.  Our algorithm eliminates many packings by finding situations where the argument of the $\cos^{-1}$ term is larger than unity; this can also occur erroneously due to round-off error, causing  packings to erroneously be recognized as unphysical.  Similarly, round-off errors are possible when checking for the consistency of a packing; in checking the equivalence between dihedral angles, or checking that there exists at least one distance greater than or equal to 2 when 3 points lie in a line.   Round-off issues could be improved by using general precision libraries such as \textit{gmp} and \textit{mpfr} \cite{gmp,mpfr}, or altogether avoided by doing all calculations symbolically.  Thus, while the analytical method presented here should in principle yield a complete set of sphere packings, practical issues such as these are a source of error.  

We present a complete set of sphere packings of $n \leq 9$, save round-off error.  At $n = 10$, we present a complete set of iterative packings and a preliminary list of new seeds.  Packings of $n \leq 7$ particles are included here, and packings of $n \leq 8 \leq 10$ particles are included in supplementary information \cite{supp_info}.  All results are summarized in table \ref{PackingResultsTable}.

In the list presented here, $\phi$ corresponds to the point group, and $\sigma$ to the symmetry number.  We have included the 2nd moment of each packing, and a `*' appears next to the 2nd moment that corresponds to the minimum of the 2nd moment of $n$ particles.  The `special properties' column denotes whether a structure is convex, a new seed, chiral, or non-rigid.  If the special properties column is blank, then that packing contains none of these properties.

\newpage

\raggedbottom
\[
\begin{array}{l}
\includegraphics[width = 0.75\textwidth]{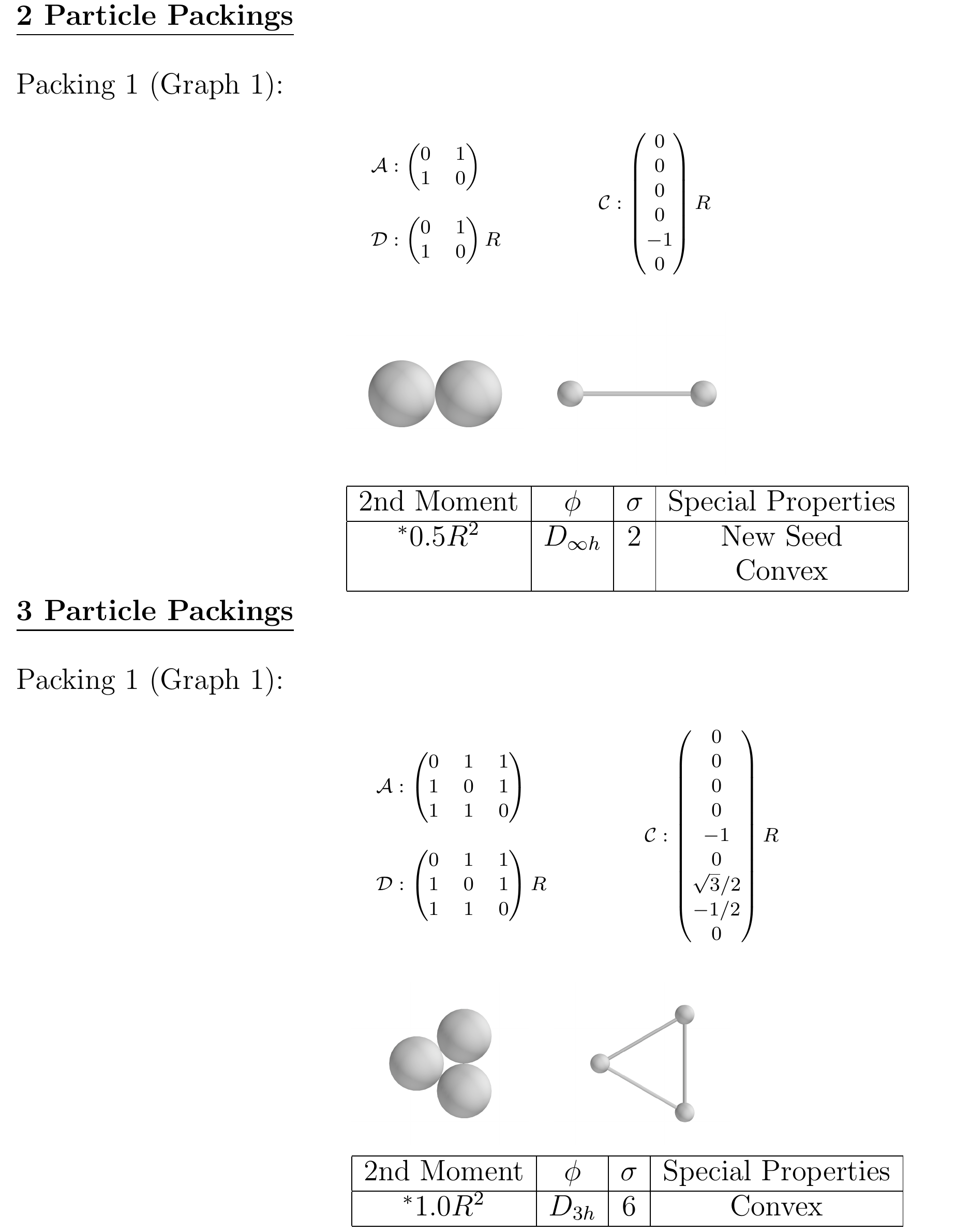}
\end{array}
\]
\newpage
\[
\begin{array}{l}
\includegraphics[width = 0.7\textwidth]{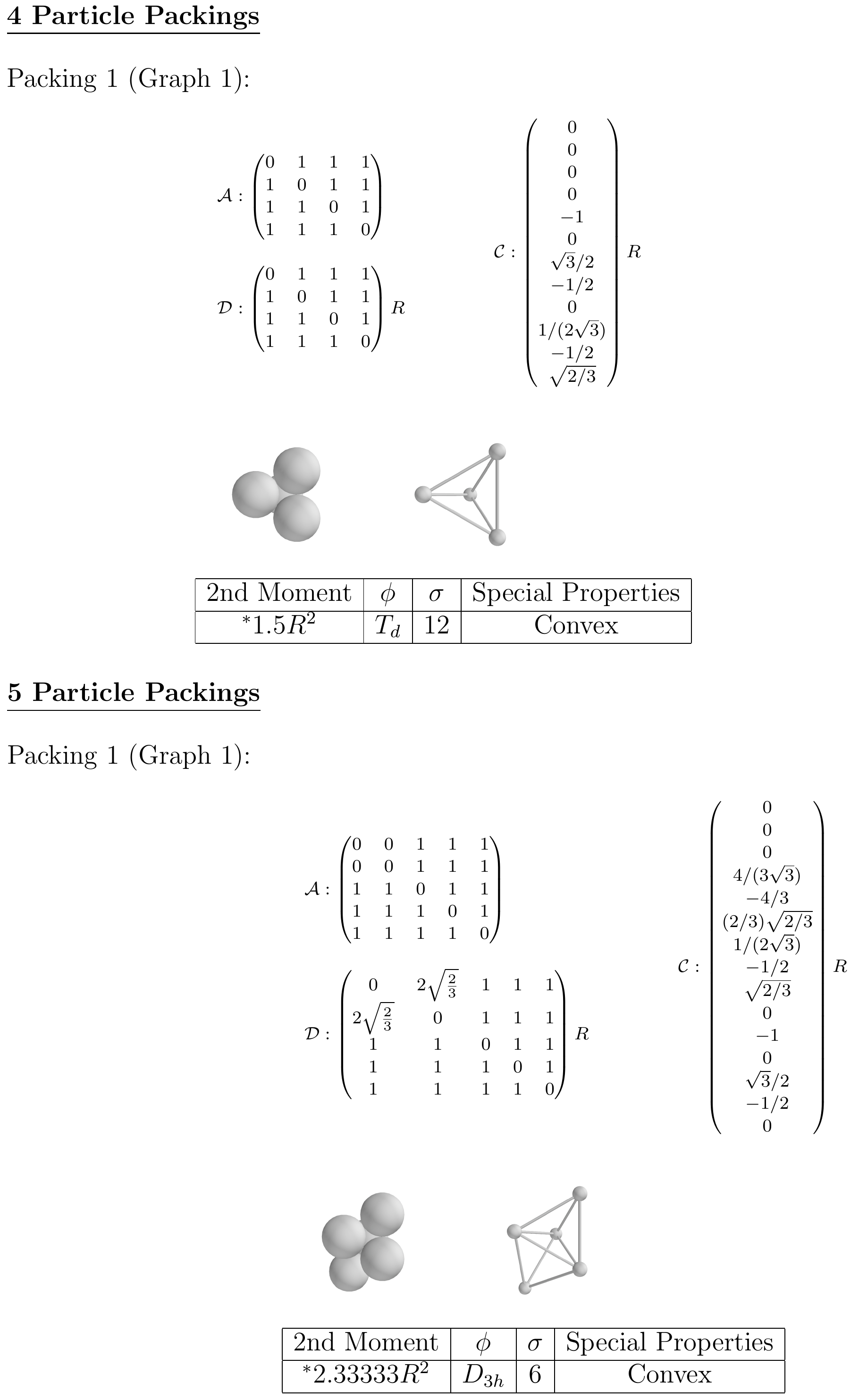}
\end{array}
\]
\newpage
 \[
 \begin{array}{l}
 \includegraphics[width = 0.8\textwidth]{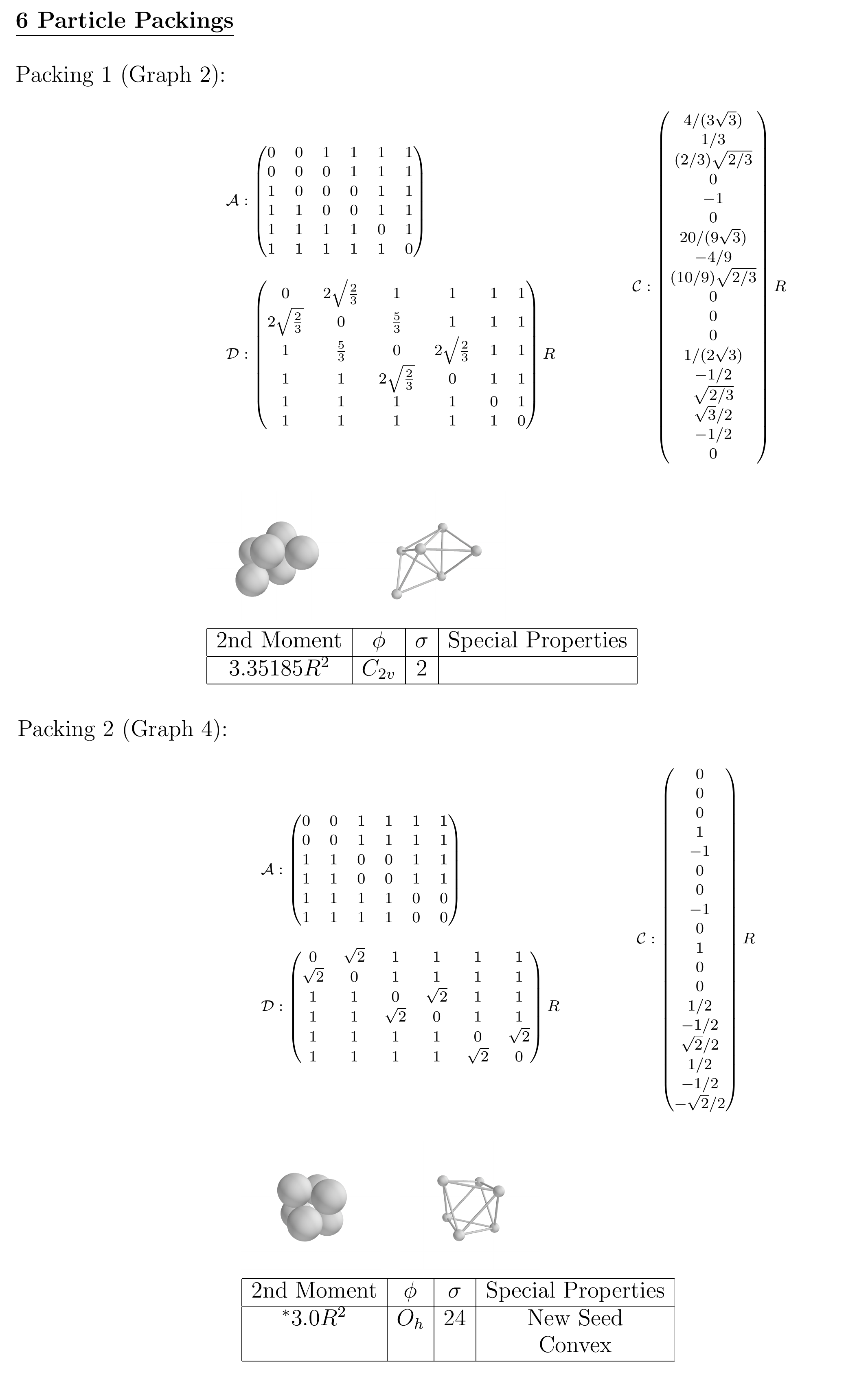}
 \end{array}
 \]
\newpage
\[
\begin{array}{l}
\includegraphics[width = 0.75\textwidth]{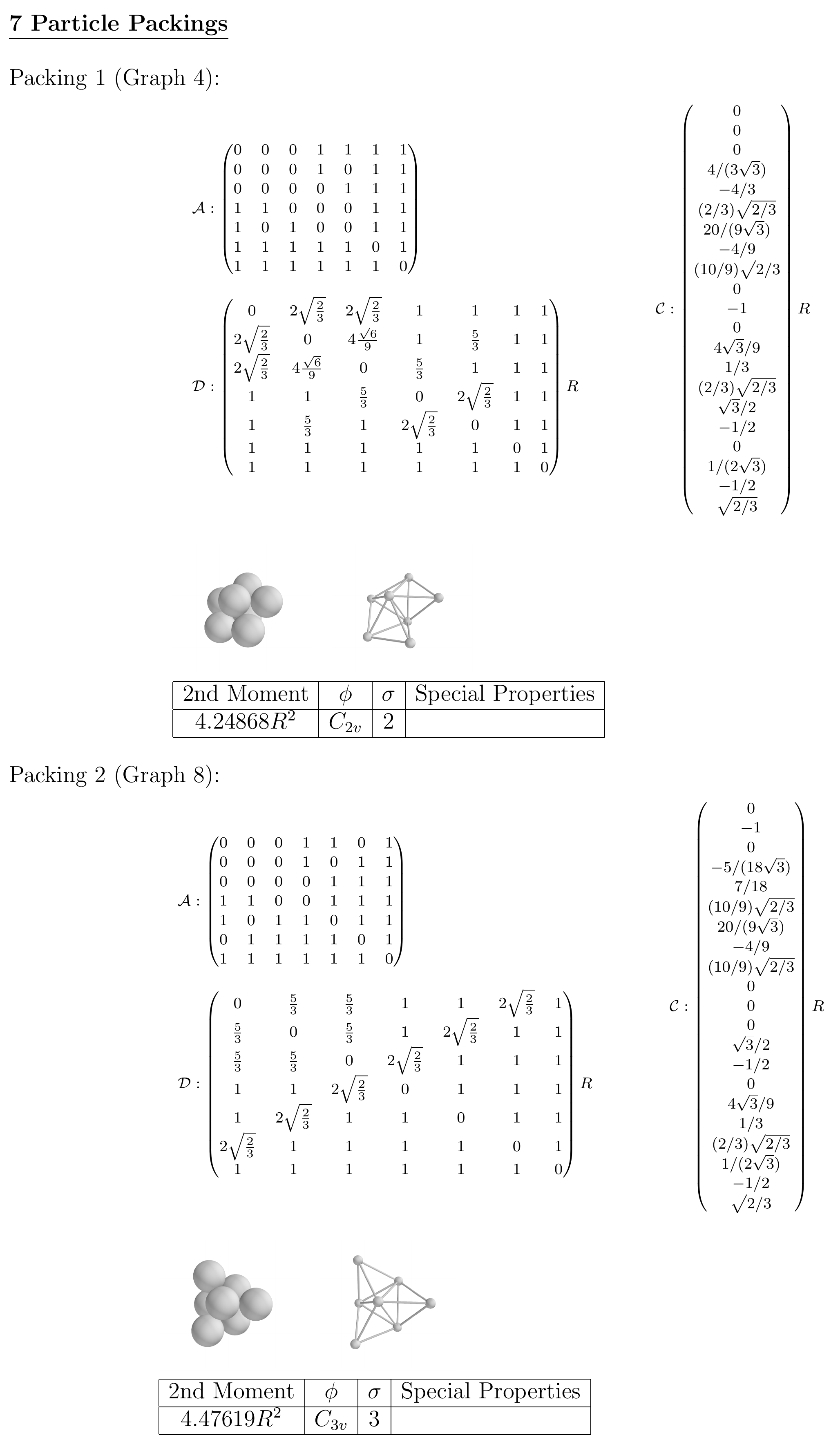}
\end{array}
\]
\newpage
\[
\begin{array}{l}
\includegraphics[width = 0.9\textwidth]{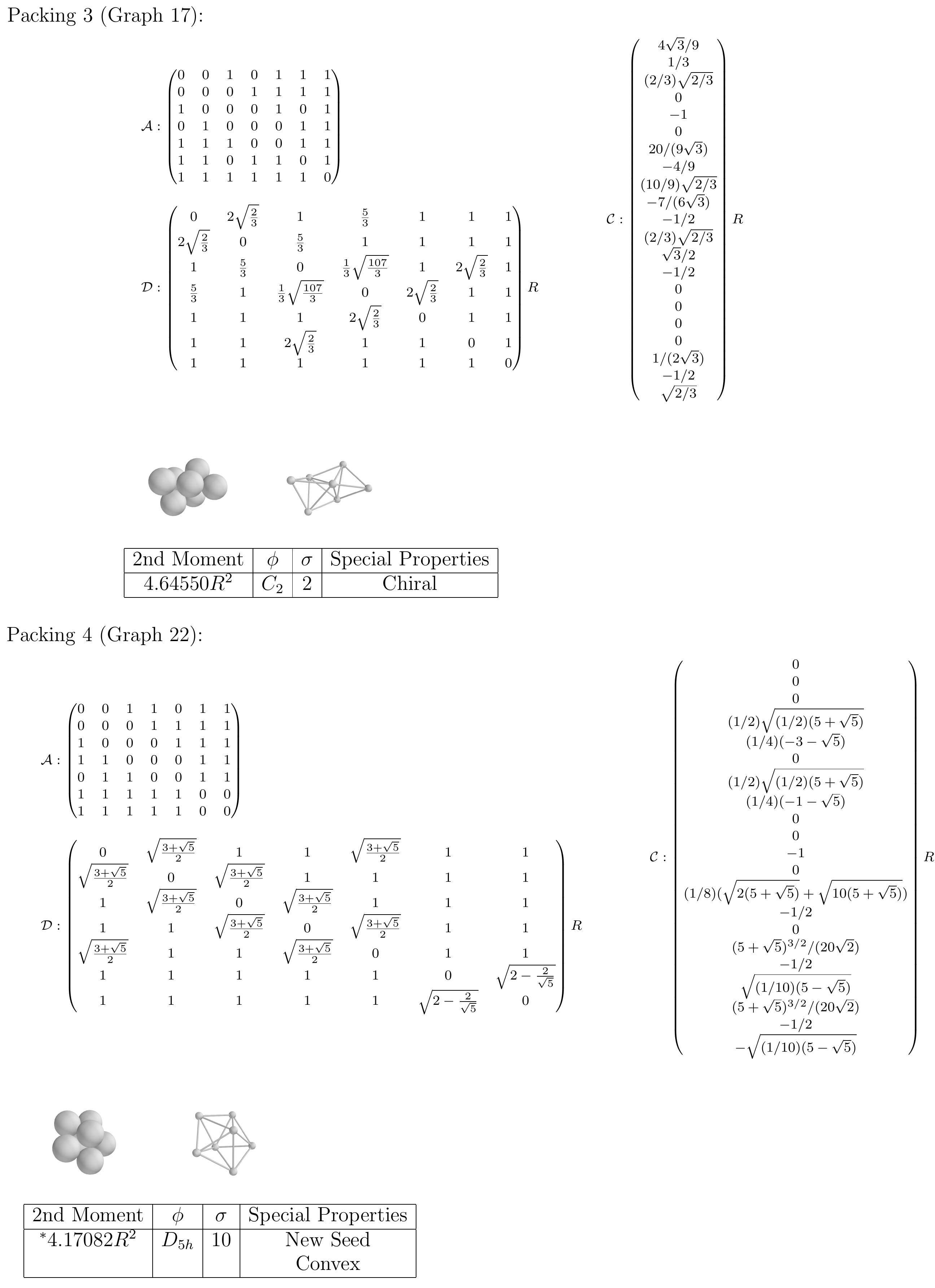}
\end{array}
\]
\newpage
\[
\begin{array}{l}
\includegraphics[width = 0.75\textwidth]{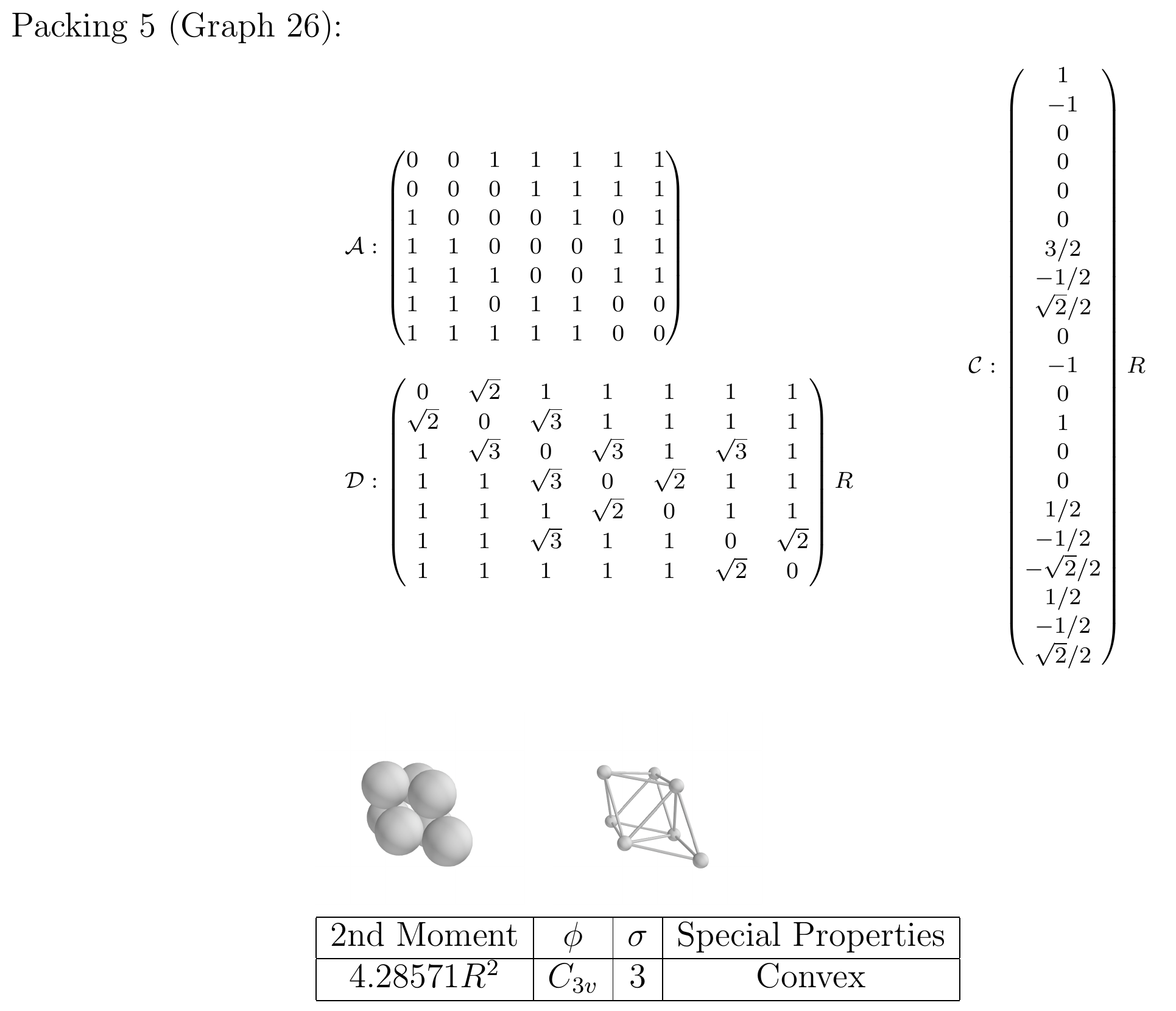}
\end{array}
\]

\begin{table}
 \begin{tabular}{|c|c|c|c|c|c|c|}
  \hline
& & & & & &\\
$n$ & Packings & Total Packings & New Seeds & Non-Rigid & Chiral & Total States\\
&  from \cite{hoare} & (Current Study) & & Packings & &\\
& & & & & &\\
\hline
%& & & &\\
2 & 1 & 1 & 0 & 0 & 0 & 1\\
\hline
3 & 1 & 1 & 0 & 0 & 0 & 1\\
\hline
4 & 1 & 1 & 0 & 0 & 0 & 1\\
\hline
5 & 1 & 1 & 0 & 0 & 0 & 1\\
\hline
6 & 2 & 2 & 1 & 0 & 0 & 2\\
\hline
7 & 4 & 5 & 1 & 0 & 1 & 6\\
\hline
8 & 10 & 13 & 1 & 0 & 3 & 16\\
\hline
9 & 32 & 52 & 4 & 1 & 28 & 80\\
\hline
10 & 113 & 262 & 8 & 4 & 201 & 463\\ 
\hline
 \end{tabular}
\caption{\textbf{Packings.}\newline Total number of packings found by the current study, compared to those found by Hoare and McInnes \cite{hoare}, who used only the tetrahedral ($n=4$) and octahedral ($n = 8$) seeds to iteratively calculate hard sphere packings by adding one particle at a time to $(n-1)$-particle packings.  They include chiral structures within their list of packings, so that a left and right-handed structure are considered to be 2 packings.  We distinguish between chiral structures and packings, such that a left and right-handed structure is considered to be 1 packing with 2 distinct states (for $n \leq 10$, left/right-handedness is the only type of chiral packing encountered). The number of packings having chiral counterparts is included in the column marked `chiral.'  The total number of states per $n$ is equal to the number of packings plus the number of chiral structures.  This is included in the table, along with the number of packings corresponding to new seeds and to non-rigid structures.  The number of packings and total states for $n=9,10$ differ from the numbers we previously reported in \cite{short_paper} primarily because in \cite{short_paper} we did not run a check to confirm that the dihedral angle was defined.  Here we report 2 more packings (and 3 more states) at $n=9$; and 39 more packings (and 70 more states) at $n=10$.  In a paper extending our work \cite{Hoy_O?Hern_2010}, Hoy and O'Hern reported 52 packings at $n=9$ and $279$ packings at $n=10$.  However, the 279 packings that they reported multiply counted the three 25 contact packings.  When all instances of the same packing being counted more than once are removed, it yields 261 packings that were reported by Hoy and O'Hern in \cite{Hoy_O?Hern_2010}.  Please see supplementary information, Appendix B \cite{supp_info} for a list of exactly which packings are reported here but were not reported in \cite{short_paper} or in \cite{Hoy_O?Hern_2010}.}\label{PackingResultsTable}
\end{table}

\section{Properties of Packings}\label{packing properties section}

Here we highlight some interesting properties of packings.

\subsection{New Seeds}

\begin{figure}[htbp]
\begin{center}
\includegraphics[width = 0.8\textwidth]{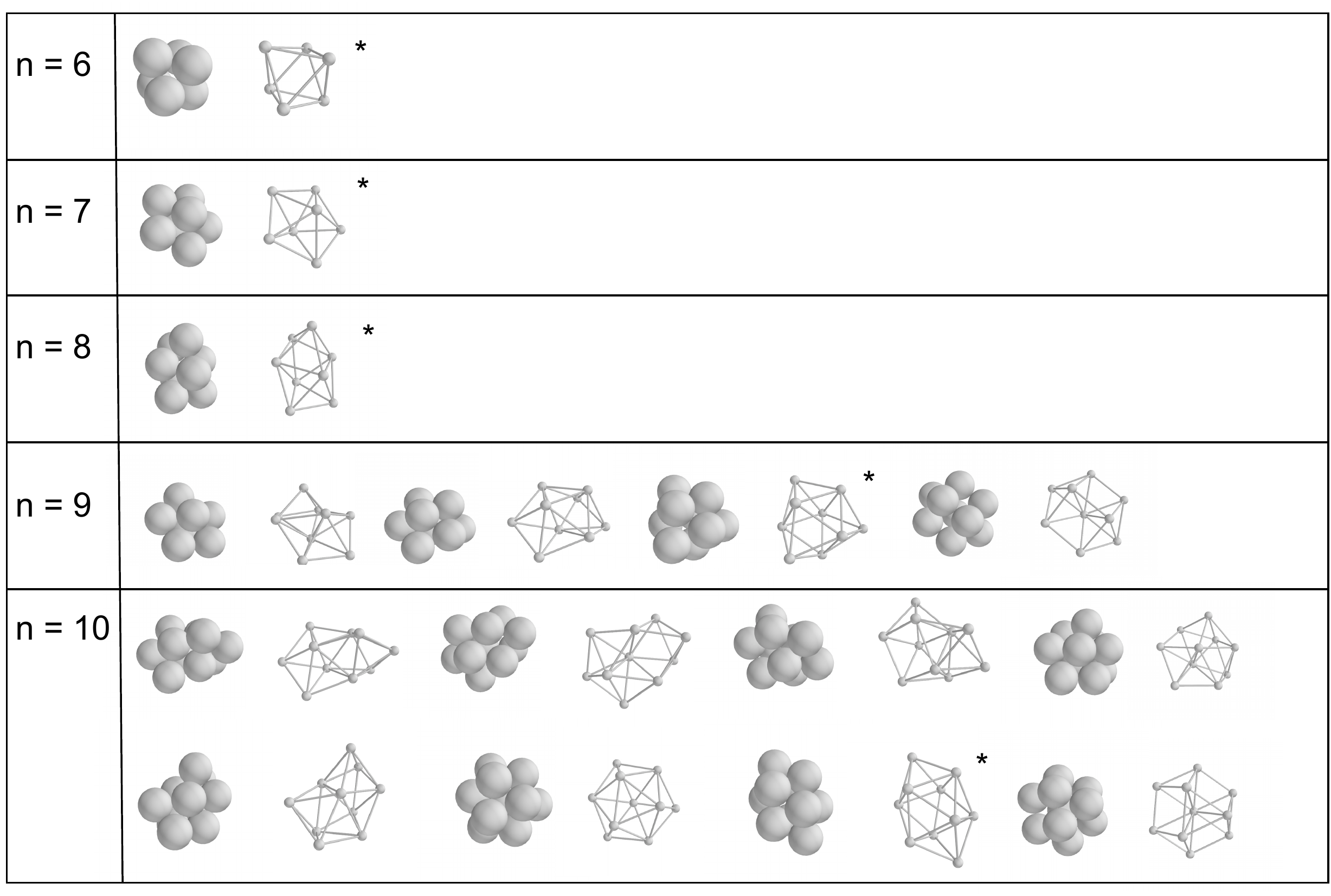}
\caption{\textbf{New Seeds.}\newline All new seeds of $n \leq 10$ particles shown in both sphere and point/line representation.  There exists only 1 packing for each $n$ of $n \leq 5$ particles, that can each be constructed iteratively from a dimer; thus there exist no new seeds for $n \leq 5$.  $n = 6$ is the first instance of a new seed.  The set of new seeds reported for $n = 10$ is putative and thus represents a lower bound.  New seeds with a $^*$ appearing to the right correspond to minima of the second moment out of all packings for that $n$.  It can be seen here that, for the packings we have analyzed, when more than 1 packing exists, the minimum of the second moment happens to correspond to a new seed.}
\label{NewSeeds6to8Fig}
\end{center}
\end{figure}

New seeds are interesting because they are inherently new structures of $n$ particles.  They are also `generating sets,' \begin{frenchspacing}i.e. once\end{frenchspacing} they exist at a given starting $n = m$, they are propagated iteratively for all $n > m$.  Geometrically, new seeds are inherently global structures, stabilized \textit{exactly} by the $n$ particles for which that new seed arises.  Iterative packings are geometrically locally stable, in that subsets of less than $n$ particles within the packing also correspond to packings.  Thus, new seeds are unique events for a given number of particles, $n$. Figure \ref{NewSeeds6to8Fig} shows all new seeds of $n \leq 10$ particles (where the set of new seeds at $n = 10$ is putative).  The proportion of new seeds to total packings is relatively small for small $n$, which can be seen in table \ref{PackingResultsTable}.

\subsection{Rigidity}

\begin{figure}[h]
 \begin{center}
  \includegraphics[width=1.0\textwidth]{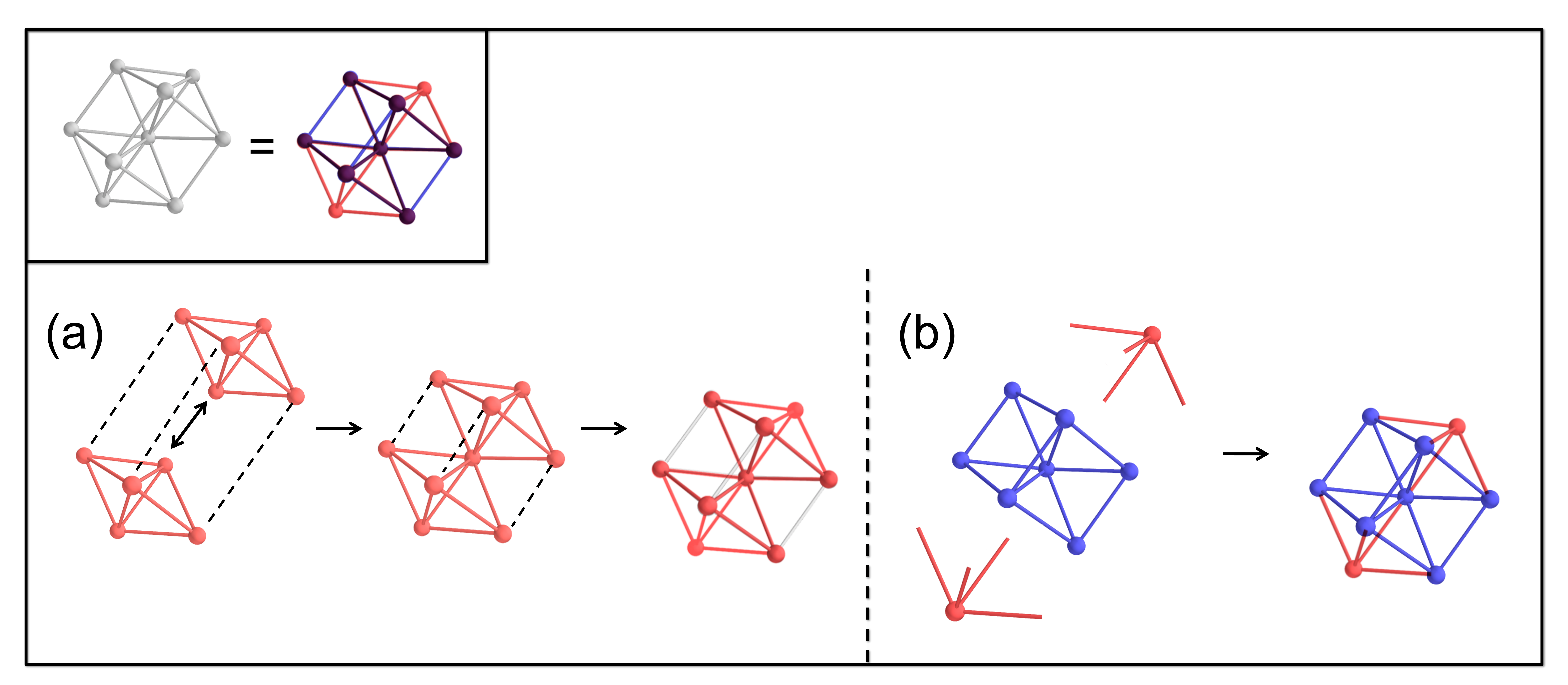}
 \end{center}
 \caption{The 9 particle non-rigid new seed (grey) is shown in the top left-hand box.  It is 
 composed
of two joined 5-point polytetrahedra (red) attached to two joined half octahedra (blue). The substructures are shown overlain -- purple bonds and particles are shared by both substructures; whereas
only red or blue bonds and particles belong to the polytetrahedral or octahedral structures alone, respectively. Two
representative ways of forming this non-rigid structure are shown in (a) and (b). (a) Two 5-point polytetrahedra are
joined by sharing one common point (on bottom). The two polytetrahedra are then fully attached to one another
via the remaining 3 bonds first shown in dashed black lines, as potential bonds, and then in solid white lines, as
actualized bonds. These bonds form the two connected half octahedra. (b) 2 particles (red) are attached to the
concave side faces of the 2 joined half octahedra (blue). The 2 red particles form the two 5-point polytetrahedral
substructures once they are attached to the joined octahedra.}\label{9NonRigidFig}
 \end{figure}

\begin{figure}[htbp]
\begin{center}
\includegraphics[width=1.0\textwidth]{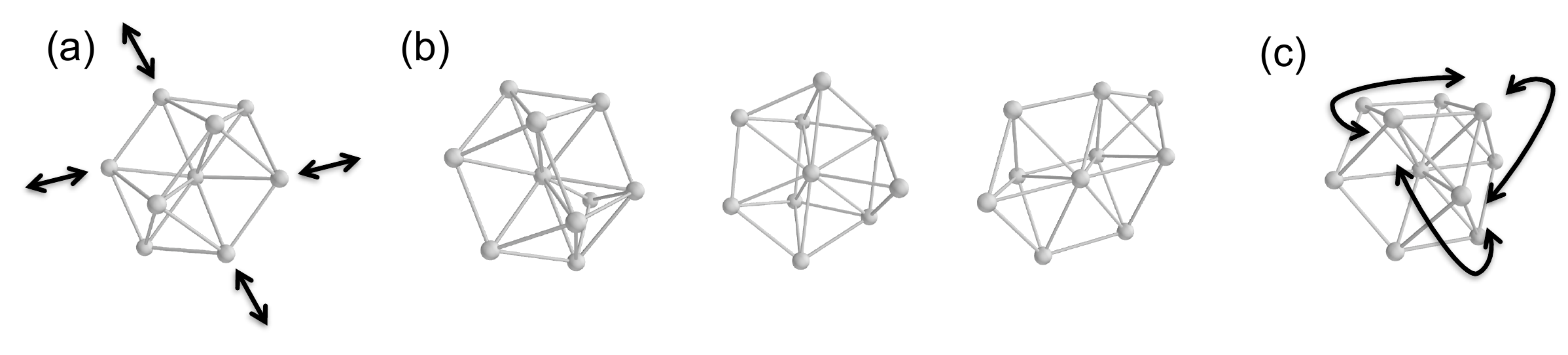}
\caption{Non-rigid packings of 9 and 10 spheres.  (a) The non-rigid $n = 9$ packing, with non-rigid motion, corresponding to a twisting of the square faces, shown by black arrows.  (b) Non-rigid $n = 10$ packings formed iteratively by adding one particle to the non-rigid $n = 9$ seed.  The non-rigid motion of these structures is the same as in (a).  (c) Non-rigid $n = 10$ seed.  Non-rigid motion, corresponding to a twisting of the radially connected square faces, connected by the bottom particle, is shown by the black arrows.  This twisting motion can be accomplished by, for example, twisting the top triangle.}
\label{9and10NonRigidFig}
\end{center}
\end{figure}

We have enumerated packings satisfying minimal rigidity constraints; these constraints are necessary but not sufficient for rigidity, and thus we can find non-rigid packings that satisfy these constraints.  For these packings, there exists a degree of freedom in which particles can move without breaking or forming additional contacts.  The first instance of a non-rigid packing occurs at $n=9$, at which there is one.  At $n=10$, there are 4 non-rigid packings: 1 non-rigid new seed, and 3 iterative non-rigid packings that derive from the $n=9$ non-rigid new seed (Fig.~\ref{9NonRigidFig} and \ref{9and10NonRigidFig}).

These 10 particle non-rigid packings will iteratively produce at least $m \geq 1$ non-rigid packings at $n = 11$, and so on.  All non-rigid packings enumerated thus far contain at least 2 deformable open square faces.  We do not know whether or not  at least 2 open square faces are a requisite of non-rigid packings that satisfy minimal rigidity constraints.  The open square faces must be `connected' for the extra degree of freedom to exist -- in the packings encountered thus far, this manifests itself by the existence of half-octahedra sharing at least 1 vertex.

\subsection{The Tree Nature of Packings}

\begin{figure}[htbp]
\begin{center}
\includegraphics[width = 0.5\textwidth]{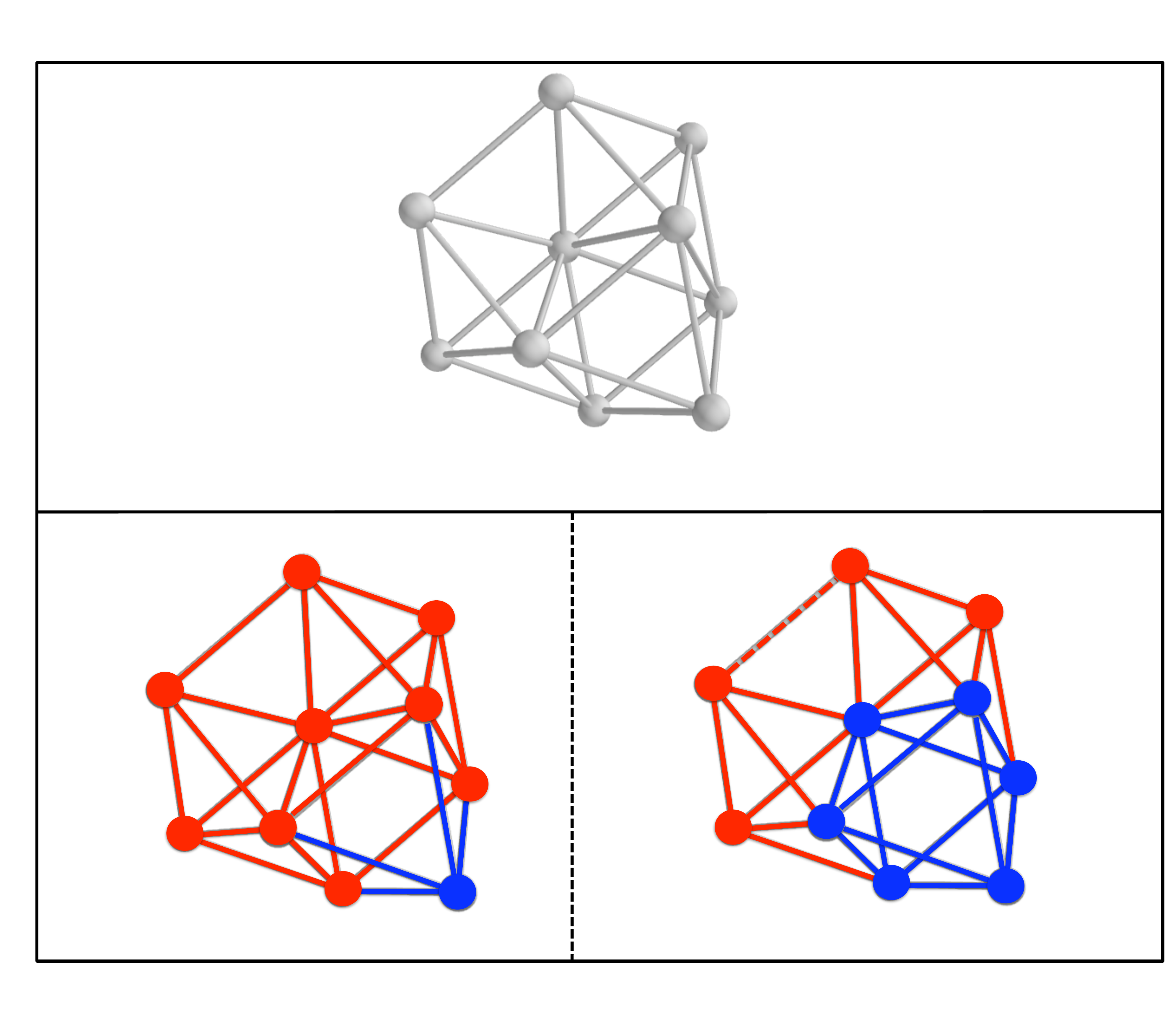}
\caption{\textbf{Tree Convergence in a 10-Particle Packing.}\newline An example of tree convergence in one of the 25 contact packings of 10 particles.  This packing, shown in grey (top panel), can be decomposed into (bottom left panel) the 9 particle non-rigid new seed (red) plus one particle (blue) that rigidifies the structure, or into (bottom right panel) an octahedron (blue) with 2 attached polytetrahedra (red).  The red dashed line indicates an `implicit contact,' a contact that automatically forms once the other contacts are in place (this corresponds to the 25th contact).}
\label{TreeMergeFig25Conts}
\end{center}
\end{figure}

%for now I've put this in - but I think it's probably too much to have this in the main text, and so probably should move to supp info -- NA.
\begin{figure}[htbp]
\begin{center}
\includegraphics[width = 1.0\textwidth]{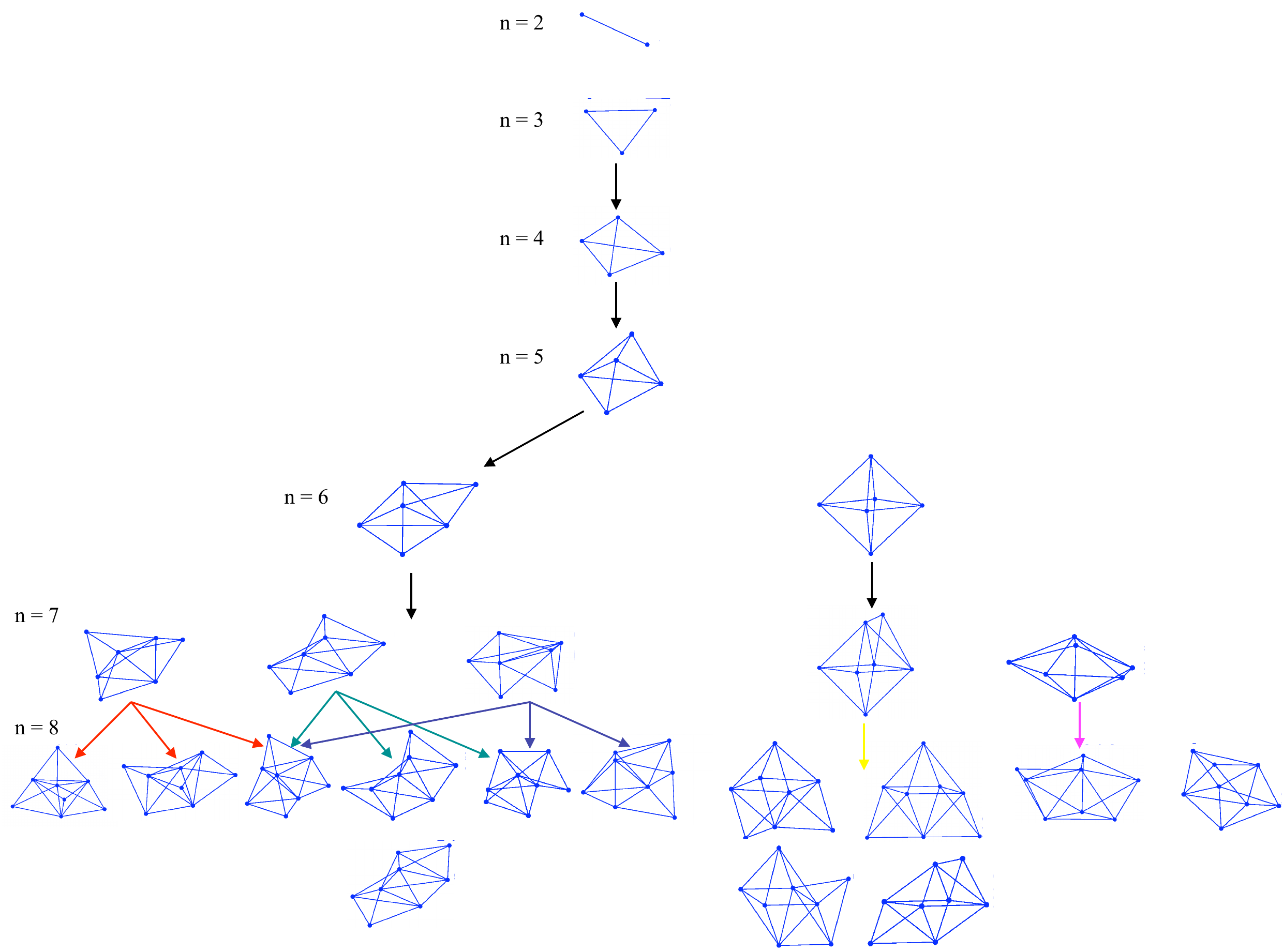}
\caption{\textbf{Tree Convergence for $n \leq 8$.}\newline An example of tree convergence for $2 \leq n \leq 8$.  Packings above which there is no arrow correspond to new seeds, and thus to the beginning of a new branch.  The arrows then point to the $n$-particle packing or group of packings that form iteratively by adding one particle.  It can be seen that even for the iterative case of adding one particle, there is tree convergence from $n = 7$ to $n = 8$ (shown by multiple arrows feeding into the same packing).}
\label{TreeMerge1to8Fig}
\end{center}
\end{figure}

There is a distinct tree nature to packings.  New seeds are the origin of a branch in the tree.  Iterative packings continue the branch.  All $n$ particle iterative packings can be decomposed into combinations of less than $n$-particle packings, and this decomposition is often not unique.  When the decomposition is not unique, the branches of the tree converge.  Figures \ref{TreeMergeFig25Conts} and \ref{TreeMerge1to8Fig} show examples of tree convergence.  Figure \ref{TreeMergeFig25Conts} shows the 2 possible decompositions of one of the 10 particle 25 contact packings.  This packing can be formed either by adding 1 particle to the 9 particle non-rigid new seed, or by combining two polytetrahedra with a 6 particle octahedron.

Figure \ref{TreeMerge1to8Fig} shows the tree structure of $2 \leq n \leq 8$ packings.  It can be seen that tree convergence occurs from $n = 7$ to $n = 8$, where multiple 7-particle packings produce the same 8-particle packing under the addition of another particle.

\subsection{Minima of the Second Moment}

The second moment measures the deviation of particles from their collective center (centroid), and is given by
\begin{eqnarray}
M = \sum_{i = 1}^n|\mathbf{r}_i - \mathbf{c}|^2 = \sum_{i = 1}^n (x_i - c_x)^2 + (y_i - c_y)^2 + (z_i - c_z)^2
\end{eqnarray}
where $\mathbf{r}_i$ is the $x,y,z$ position of particle $i$; and $\mathbf{c}$ is the centroid, the average $x,y,z$ position over all particles, given by
\begin{eqnarray}
c_x = \frac{1}{n}\sum_{i=1}^nx_i
\end{eqnarray}
and analogously given for $c_y$ and $c_z$.

%\begin{figure}[htbp]
%\begin{center}
%\includegraphics[width = 0.8\textwidth]{2ndMomentFig2to10}
%\caption{\textbf{Minima of the 2nd Moment.}\newline Here shown for $6 \leq n \leq 10$.  We did not include $n \leq 5$, as there only exists one packing per $n$ in this regime, and thus the minimum of the 2nd moment is trivial.}
%\label{2ndMomFig}
%\end{center}
%\end{figure}

The minimum of the second moment corresponds to the packing with the smallest $M$.  The 2nd moment is listed within the list of packings in section \ref{the packings} and in appendix B \cite{supp_info}, and a `$^*$' signifies the minimum of the 2nd moment for each $n$ in figure \ref{NewSeeds6to8Fig}.  We confirm that the minima of the 2nd moment reported by Sloane \textit{et al.} \cite{sloane95} are correct (they proved the 2nd moment minima for $n\leq4$, but for $n > 4$ these were putative structures).  For $n \leq 10$, each minimum of the second moment corresponds to a new seed, if a new seed exists.

\subsection{Ground States and the Maximum Number of Contacts}

A fundamental question related to sphere packings is what is the maximum number of contacts that a packing of $n$ spheres can have?  Not only is this question of mathematical interest in its own right, but it is also of significant physical interest as such packings correspond to ground states.  The number of packings that contain the maximum number of contacts in turn corresponds to the ground state degeneracy.  Table \ref{NumContactTable} shows how the ground state degeneracy changes with $n$.  Interestingly, this relation appears to be oscillatory.

% A reviewer recommended we delete this figure as it's redundant with Table 6.5
% \begin{figure}[htbp]
%  \centering
%     \includegraphics[width = 0.75\textwidth]{GroundStateWNumContsExtended2_2.eps}
%  \caption{Number of packings versus $n$, for $3n-m$ contacts, where $-3 \leq m \leq 6$.  The ground state degeneracy as a function of $n$ is shown by the solid curve.  Degeneracy for $n\geq11$ is conjectured.  Dashed curves are shown for $m = 6,5,4,3$ before and after $3n-m$ becomes the ground state number of contacts, illustrating the exponential growth of these packings that is primarily due to iterative growth and not to the formation of $3n-m$ seeds.}
% \label{GroundStateDegeneracyFig}
% \end{figure}

%this was commented out in my thesis -- see if i have a replacement fig somewhere else, or if should be using this one...
%\begin{figure}[h]
% \begin{center}
% %%\includegraphics[width=15cm]{SpecialPackingsNonPoster}
% %%\includegraphics[width=15cm]{SpecialPackings}
% %%\includegraphics[width=15cm]{SpecialPackings2}
%  \includegraphics[width=15cm]{SpecialPackings3}
% \end{center}
% \caption{\textit{Special Packings}. The top part of the figure denotes new seeds for $n=9$ and $n=10$. The 4th 9 particle new seed and the 8th 10 particle new seed are nonrigid (despite having $3n-6$ total contacts and $\geq 3$ contacts per sphere). At $n=10$ there are three non-rigid packings that derive from the nonrigid $n=9$ seed, and there are also two packings with $25$ contacts.}\label{SpecialPacksFig}
% \end{figure}
 
 \begin{figure}
 \centering
\includegraphics[width = 1.0\textwidth]{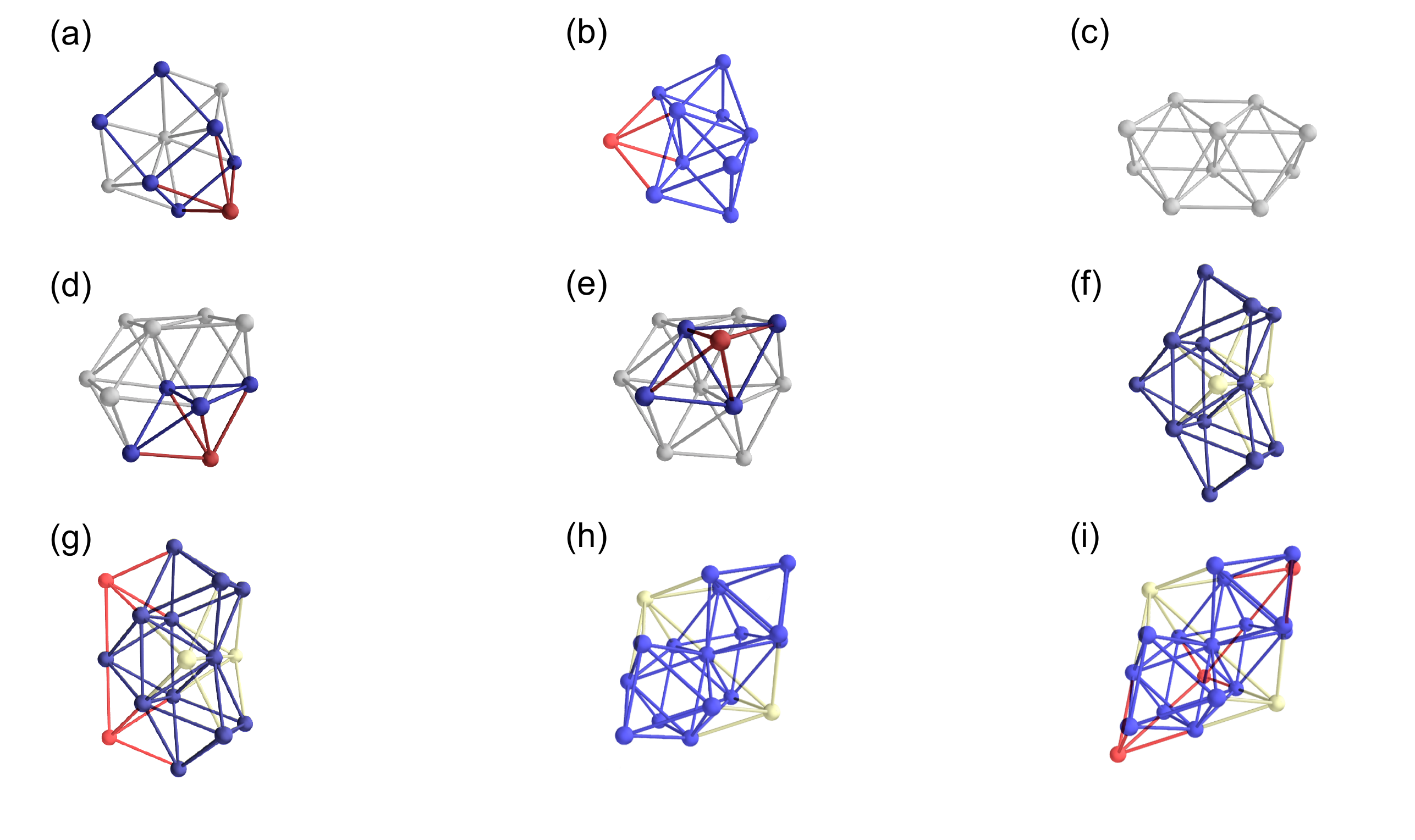}
 \caption{Maximum contact packings for $10 \leq n \leq 20$.  We are reasonably confident that the $n = 10$ packings shown here correspond to the maximum contact packings, but maximum contact packings of $n > 10$ are conjectured.  (a)-(c) 10-particle packings with $3n-5 = 25$ contacts.  (a) can be formed either by adding one particle (red) to one of the open square faces (blue) or as otherwise detailed in figure \ref{TreeMergeFig25Conts}.  (b) can be formed by adding one particle (red) to the concave 4 particle face created by 2 joined octahedra (blue).  (c) is formed by connecting 2 octahedra by one edge.  (d) 11 particle maximum contact packing ($3n-4$ = 29 contacts).  This can be formed either by adding one particle (red) to the concave 4 particle face created by the 2 joined octahedra of (b), or by adding one particle to the one remaining open square face of (a).  (e) 12 particle maximum contact packing ($3n-3 = 33$ contacts).  This is formed by adding one particle (red) to the concave 4 particle face created by 2 joined octahedra (blue).  13 particle maximum contact packings are constructed iteratively from this packing.  (f) 14 particle maximum contact packing ($3n-2 = 40$ contacts).  This is formed by 3 radially connected octahedra (blue) and 2 particles (yellow) added to each of the concave 5 particle faces created by the 3 joined octahedra.  (g) 15 and 16 particle maximum contact packings containing $3n-1$ and $3n$ contacts respectively.  The 15-particle packing corresponds to the addition of only one of the red particles to the concave 4 particle face of (f), and the 16-particle packing includes both red particles.  (h) 17 particle $3n$ contact packing formed by 4 radially connected octahedra (blue) and 2 particles (yellow) connected to the concave 6 particle faces created by the 4 joined octahedra.  At 17 particles, $3n$ contact packings correspond to this packing as well as packings constructed iteratively from (g).  (i) 18,19,20 particle maximum contact packings corresponding to $3n+1$, $3n+2$, and $3n+3$ contacts respectively.  Each of these packings is constructed by adding a particle (red) to one of the concave 4 particle faces created by the joined octahedra -- the 18-particle packing is constructed by adding one such particle to (h), 19 by adding 2, and 20 by adding 3.}
 \label{max_contact_packings}
\end{figure}
 
For $n \leq 9$, every packing has exactly $3n-6$ contacts.  Thus, at each fixed $n$, $n \leq 9$, all packings have the same potential energy.  Table \ref{NumContactTable} shows that, for $n \le 9$, the ground state degeneracy increases exponentially.  But at $n = 10$ this trend changes due to a small number of packings that can have $25 = 3n-5$ contacts (all other 10-particle packings have $3n-6$ contacts).  There exist 3 such packings, each containing octahedra (Fig.~\ref{max_contact_packings}a-c).  These three structures are the ground states at $n=10$.

For $n \geq 11$, we conjecture as to what the maximum number of contacts are, and provide examples of such structures.  The maximum contact packings at $n = 10$ arise because it becomes possible to add 4 contacts to minimally-rigid 9-particle packings; whereas all other iterative packings of $n \leq 10$ spheres are formed by the addition of $3m$ contacts to a minimally-rigid $n-m$ sphere packing.  All maximum contact packings found thus far correspond to iterative packings.  We have not determined whether this is true for all $n$, but we conjecture that it is, because new seeds tend to contain more empty space, and thus less contacts.  We have found three types of structures that allow for the addition of more than $3m$ contacts: (i) $m$ octahedra, where each pair of octahedra share one edge (as in Fig.~\ref{max_contact_packings}c), (ii) an open square face created by half an octahedra (as shown in blue in figure \ref{max_contact_packings}a), and (iii) the concave $m$ point face created by octahedra sharing 3 edges (as shown in blue in figure \ref{max_contact_packings}d,e).  4 point concave faces are shown, for example, in figure \ref{max_contact_packings}e; 5 point concave faces in figure \ref{max_contact_packings}f,g; and 6 point concave faces in figure \ref{max_contact_packings}h,i.  To each $m$ point concave face, it is possible to add one particle with $m$ contacts; this is evidenced, for example, by the red particle with 4 contacts in figure \ref{max_contact_packings}e and the yellow particles with either 5 or 6 contacts in figure \ref{max_contact_packings}f-i.  All structures leading to maximum contact packings that we have found thus far are related to octahedra, and we conjecture that this will be the case for all $n$.  (Interestingly all non-rigid packings found thus far are also related to octahedra, in that they contain the open square faces created by half-octahedral structures; they all contain half-octahedra sharing at least 1 point).

There are many fewer ways of adding greater than $3m$ contacts to an $n-m$ sphere packing than there are of adding $3m$ contacts, thus leading to a relatively small number of ground state packings when a new maximum number of contacts, as a function of $n$, is reached.  Thus, each time a packing with a greater maximum number of contacts, as a function of $n$, is possible, we expect the ground state degeneracy to either decrease or to remain small.  When the functional form for the maximum number of contacts remains constant, we expect the ground state degeneracy to grow rapidly, due in large part to the iterative growth of adding a particle with 3 contacts to packings.  Table \ref{NumContactTable} for example, shows that for $n \le 9$ the ground state degeneracy increases exponentially because all packings have $3n-6$ contacts.  But at $n=10$ this trend changes because packings with more than $3n-6$ contacts become possible.  At $n = 13$ and at $n = 17$, rapid growth in the ground state degeneracy resumes as the functional form for the maximum number of contacts remains constant, but for all other $n$, the ground state degeneracy either decreases or remains constant, because a new maximum number of contacts, as a function of $n$, is possible.

At low temperatures, we expect the experimentally observable packings to be dominated by the maximum contact packings.  Thus, for $n = 10$, and for any $n$ in which either (i) the maximum number of contacts increases as a function of $n$ or (ii) the maximum number of contacts does not remain constant for too long ($\leq 3$ particles), we expect that there are only a small number of packings that will be observable at low temperature.  This trend can be seen in the conjectured ground state degeneracy for $9 < n \leq 20$ in Table \ref{NumContactTable}, and we conjecture that a similar low amplitude oscillatory trend might also continue for $n > 20$.

\begin{table}[htdp]
\begin{center}
\caption[c]{\textbf{Number of Contacts \begin{frenchspacing}vs.\end{frenchspacing} Number of Particles.}\newline  The first column corresponds to the number of particles, $n$, the 2nd column to the number of contacts that the ground state packing(s) have, and the 3rd column to the number of packings in the ground state.  Note that for $n \geq 11$ these are putative results and thus the number of packings with $3n-m$ contacts represents a lower bound, and the ground state number of contacts is conjectured (we have not encountered packings with more contacts, but we have not proved that they don't exist).\newline}
\begin{tabular}{|c|c|c|}
\hline
$n$ & Contacts & Ground State Packings \\
\hline
2 & $3n-6$ & 1\\
\hline
3 & $3n-6$ & 1  \\
\hline
4 & $3n-6$ &1 \\
\hline
5 & $3n-6$ & 1 \\
\hline
6 & $3n-6$ & 2 \\
\hline
7 & $3n-6$ & 5 \\
\hline
8 & $3n-6$ & 13\\
\hline
9 & $3n-6$ & 52\\
\hline
10 & $3n-5$ & 3\\
\hline
11 & $3n-4$ & 1\\
\hline
12 & $3n-3$ & 1\\
\hline
13 &  $3n-3$ & 7 \\ %the 1 new 3n-3 packing plus about 6 new ones from adding 3 contacts to the 1 3n-3 packing of 11 particles
\hline
14 & $3n-2$ & 1 \\
\hline
15 & $3n-1$ & 1 \\
\hline
16 & $3n$ & 1 \\
\hline
17 & $3n$ & 8 \\
\hline
18 & $3n+1$ & 1 \\
\hline
19 & $3n+2$ & 1 \\
\hline
20 & $3n + 3$ & 1\\
\hline
\end{tabular}
\end{center}
\label{NumContactTable}
\end{table}

\subsection{Lattice Structure}

The non-rigid new seeds at $n = 9$ and $n = 10$, as well as the maximum contact packings of $n < 13$ are all subunits of the hexagonally close-packed (HCP) lattice, being combinations of face-sharing tetrahedra and octahedra.  Additionally, the structure shown in figure \ref{max_contact_packings}c is a subunit of either the HCP or face-centered cubic (FCC) lattice.  The non-rigid packings are entropically favored, and we thus expect these to form with higher probability at higher temperatures; while the maximum contact packings are energetically favored, corresponding to the structures that will form with higher probability as the temperature, $T \rightarrow 0$.  For $14 \leq n \leq 20$, the maximum contact packings are not HCP subunits (Fig.~\ref{max_contact_packings} f-i). %I believe these are commensurate with a different type of lattice structure, but I don't know what it's called...

Frank predicted \cite{frank_supercooling_1952} that icosahedral short-range order would be a hallmark of liquid structure, and experimental studies have shown local cluster-like order in bulk atomic liquids and glasses \cite{reichert_observation_2000,sheng_atomic_2006}.  Results from a recent study suggest that structural arrest in condensed phases may be related to geometrical constraints at the scale of a few particles \cite{royall_direct_2008}.  The propensity for icosahedra \cite{hoare_physical_1975,doye_structural_1997} in longer-range systems is absent in ours.  We have proven that the icosahedron is \emph{not} the ground state at $n=12$, nor is an icosahedron with a central sphere the ground state at $n=13$.  A $12$-sphere icosahedron has only $3n-6=30$ contacts, and in a $13$-sphere icosahedron the outer spheres would not be close enough to interact with each other.  %Thus, we do not expect icosahedra to be a structural motif in hard sphere gels or fluid cluster phases governed by short-ranged attractions and local clusters of the order of 13 particles.

It is possible, and perhaps even likely, that the lattice structures corresponding to ground state packings will be periodic with $n$.  For example, although the ground states for $14 \leq n \leq 20$ are not commensurate with HCP, the ground states for a finite range of higher $n$ may be, and may then subsequently return to the lattice structure commensurate with $14 \leq n \leq 20$.  Detailing the structures of ground state packings for all $n$, and geometrical patterns contained therein, is a subject of future work.  Furthermore, the appearance of crystalline order, such as HCP, at very low $n$ may influence nucleation.

\section{Extensions and Conjectures}\label{future directions section}

\subsection{The major roadblock for reaching higher $n$}
The main roadblock to the \textit{analytical} enumeration of sphere packings at higher $n$ in the current work is deriving one analytical geometrical rule that can solve for all new seeds.  In the next section, we outline a numerical method, based on the triangular bipyramid rule, that is capable of finding \textit{all} solutions of $\A \longrightarrow \D$, and which can thus solve for all new seeds. However, the implementation of either this numerical scheme or the derivation of an analytical rule would only allow us to enumerate packings of up to about $n = 14$ spheres.  This is because the real limitation of the current method arises from the enumeration of minimally-rigid, non-isomorphic adjacency matrices.  For $n < 10$, the enumeration of such adjacency matrices using \textit{nauty} \cite{nauty} takes on the order of seconds.  For $n = 10, 11$, this enumeration takes minutes.  For $n = 12$, enumerating all minimally-rigid, non-isomorphic $\A$'s takes approximately 2 hours.  Extrapolating, we expect the enumeration at $n = 13, 14$ to take on the order of 2 days and 2 weeks, respectively.  Thus, around $n = 14$, we begin to reach the computational limitations of this method, which is due to the enumeration of $\A$'s.

Only a very small fraction of adjacency matrices correspond to sphere packings; for example, at $n = 10$, out of the 750,226 $\A$'s, only 262 correspond to sphere packings.  Thus, the enumeration of all $\A$'s really is a brute force and wasteful step.  Further advances in enumerating sphere packings will require overcoming this roadblock.  In section \ref{bond breakage section}, we propose one method that might be able to overcome this limitation.

\subsection{Applying the Triangular Bipyramid Rule to New Seeds}\label{new seed gen rule section}
The Triangular Bipyramid rule solves for iterative structures but does not work for noniterative ones, which we also showed increase rapidly starting at $n=10$.  Here, we discuss how the triangular bipyramid rule might also be applied to new seeds.  In this case, the equations for the unknown interparticle distances, $r_{ij}$, are implicit and thus must be solved numerically.

\begin{figure}
 \begin{center}
  \includegraphics[width = 5in]{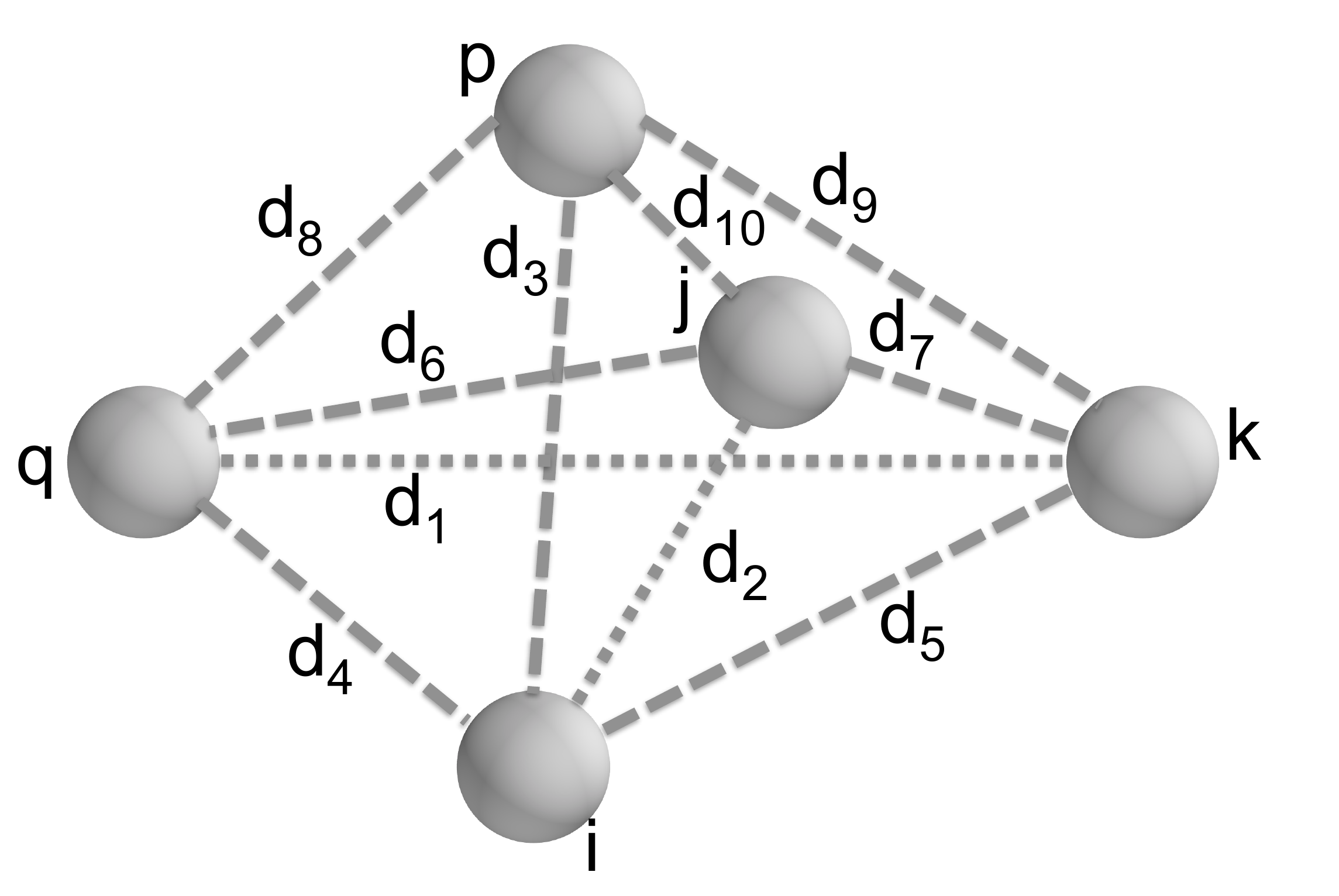}
\caption[c]{\textbf{General Triangular Bipyramid.} A triangular bipyramid where all distances, labeled $d_1 - d_{10}$, are potentially unknown.  The particles, $i,j,k,p,q$ correspond to the same labeling as in figure \ref{triang dipyramid}.  This triangular bipyramid is used to derive the general equation for an unknown relative distance, $r_{ij}$ (appendix A \cite{supp_info}).}\label{Gen ditet fig}
%eqn \ref{GenDistEqnNewSeeds}
\end{center}
\end{figure}

%\begin{figure}
% \begin{center}
%  %\includegraphics[width = 5in]{OctahedronGenDitetEx}
%\caption[c]{\textbf{A New Seed and the General Triangular Bipyramid.}\newline Here we show an example of how to apply the general triangular bipyramid to a new seed in order to solve its unknown distances.  This new seed corresponds to the 6 particle octahedron.  It has 3 unknown distances, $r_{12}$, $r_{34}$, and $r_{56}$.  The first triangular bipyramid constructed consists of particles $1,2,4,5,6$, and has unknown distances $d_1$ and $d_2$, corresponding to $r_{12}$ and $r_{56}$ ($d_3$ to $d_{10} = R$).  The second triangular bipyramid consists of particles $1,2,3,5,6$ and has unknown distances $d_1', d_2'$ corresponding to $r_{56}, r_{12}$, respectively.}\label{New seed ex w gen ditet}
%\end{center}
%\end{figure}

\begin{figure}
 \begin{center}
  \includegraphics[width = 5in]{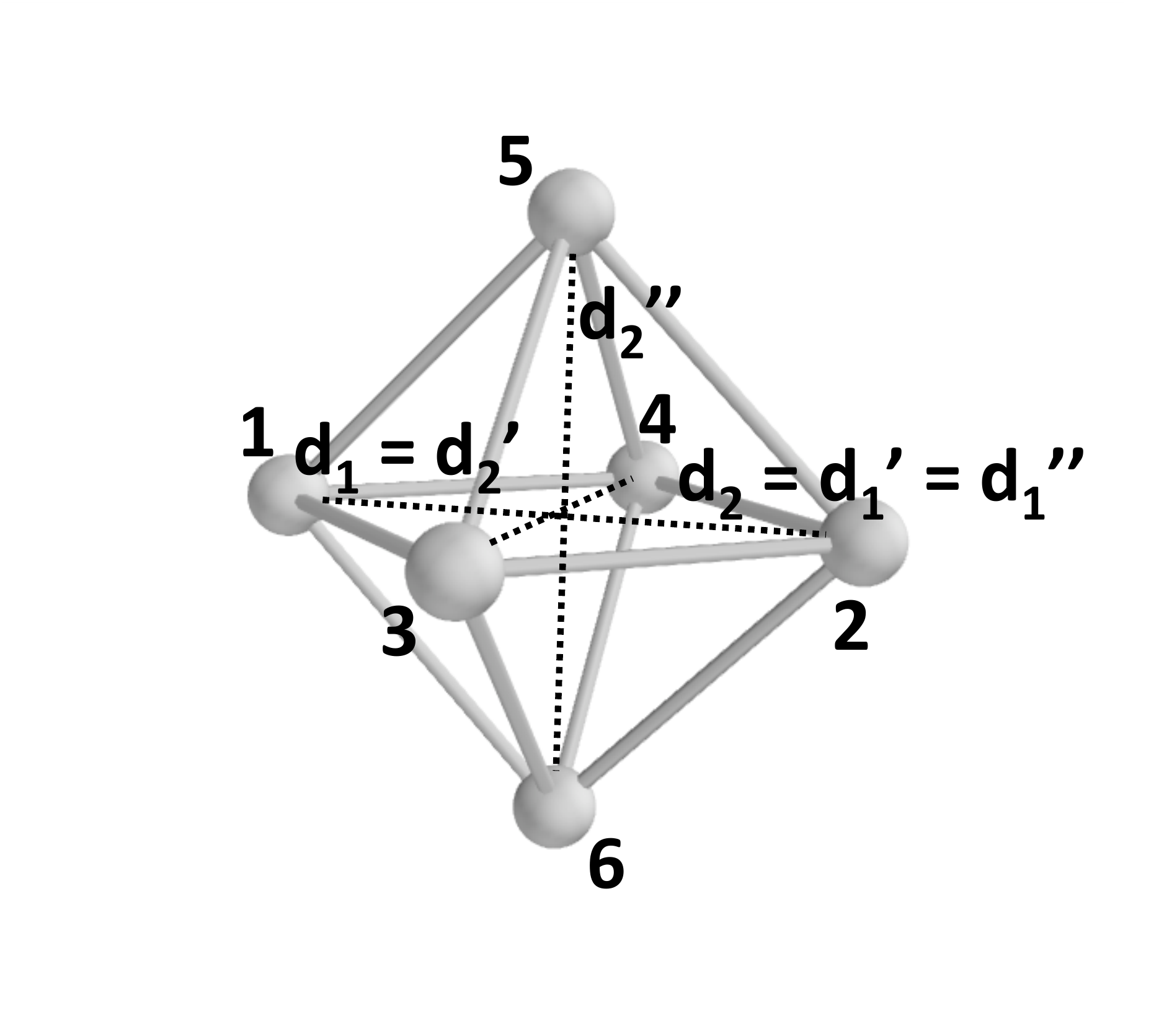}
\caption[c]{\textbf{A New Seed and the General Triangular Bipyramid.}\newline Here we show an example of how to apply the general triangular bipyramid to a new seed in order to determine its unknown distances.  This new seed corresponds to the 6 particle octahedron.  It has 3 unknown distances, $r_{12}$, $r_{34}$, and $r_{56}$ (dashed black lines).  The first triangular bipyramid constructed consists of particles $1,2,3,4,5$, and has unknown distances $d_1$ and $d_2$, corresponding to $r_{34}$ and $r_{12}$, respectively.  Note that ($d_3$ to $d_{10} = R$).  The second triangular bipyramid consists of particles $1,2,3,4,6$ and has unknown distances $d_1', d_2'$ corresponding to $r_{12}, r_{34}$, respectively.  The third triangular bipyramid consists of particles $1,2,4,5,6$ and has unknown distances $d_1'', d_2''$ corresponding to $r_{12}, r_{56}$, respectively.  Note that the first two triangular bipyramids comprise 2 equations and 2 unknowns, and thus are alone sufficient to solve $r_{12}$ and $r_{34}$.  Once these 2 distances are known, applying the third triangular bipyramid involves only 1 unknown distance, $r_{56}$, and thus follows as applying the triangular bipyramid to an iterative packing.}\label{New seed ex w gen ditet}
\end{center}
\end{figure}

In a new seed, any triangular bipyramid that contains an unknown distance will contain more than one unknown distance.  Thus, $r_{ij}$ can not be determined one at a time, as with iterative packings; instead we must construct a system of equations to solve for the unknown distances\footnote{As can be seen in section \ref{gen rule section} or in Appendix A of the supplemental information \cite{supp_info}, the equation for an unknown distance is quadratic.  Thus, this will be a quadratic system of equations.}.  Let us consider a general triangular bipyramid, like that shown in figure \ref{triang dipyramid}, but here all 10 distances are potentially unknown (Fig.~\ref{Gen ditet fig}).  For $m$ unknown $r_{ij}$, we construct $m$ triangular bipyramids to solve for the $r_{ij}$ (see Fig.~\ref{New seed ex w gen ditet}, for example).  Each triangular bipyramid yields one equation for an unknown distance, thus yielding $m$ equations with $m$ unknowns in total, making the system well defined.

We assign a label, $d_i$, to each of the 10 distances within the triangular bipyramid (Fig.~\ref{Gen ditet fig}).  Explicit formulas can always be obtained for $d_2$ and $d_3$.  Thus, over all $m$ triangular bipyramids, we place each unknown distance in location $d_2$ or $d_3$ at least once.  New seeds are inherently global structures, thus the $m$ triangular bipyramids should contain all $n$ points amongst them in order to ensure that the solution space is sufficiently constrained.  Also, to avoid redundancy, each triangular bipyramid must contain a unique combination of 5 particles.

The equations for the $r_{ij}$ are derived from the triangular bipyramid property that the dihedral angle $A_1$ must equal either the sum or the difference of the dihedral angles $A_2$ and $A_3$ (as was detailed in section \ref{gen rule section} and in Fig.~\ref{triang dipyramid}).  This equation is cumbersome, and it as well as its derivation can be found in Appendix A of the supplemental information \cite{supp_info}.  As always with the triangular bipyramid rule, due to the 2 possibilities of $A_1$ equaling either the sum or the difference of $A_2$ and $A_3$, each unknown distance has 2 possible solutions.

For each set of $r_{ij}$ that is to be solved, construct initial guesses between the bounds of the triangle inequality and no-overlap constraint, and iterate the initial guess with a step size less than or equal to the minimum difference between different solutions (for rigid structures).  There will always exist unknown $r_{ij} \leq 2R$ because each particle has at least 3 contacts\footnote{This also holds true when each particle has at least 2 contacts.}.  These are the $r_{ij}$ for which there exists a $k$ satisfying $\A_{ik} = \A_{jk} = 1$, $\A_{ij} = 0$.  Thus, first solve the set of $R \leq r_{ij} \leq 2R$.  If unknown $r_{ij}$ remain, then solve the set of $r_{ij}$ that now have known triangle inequality bounds, due to the previously solved set of $r_{ij}$, repeat until all $r_{ij}$ have been solved\footnote{We have tested this method on the new seed $\A$'s for $n \leq 8$, and have shown that it works; however, we have not implemented it for up to $n = 14$.}.

\subsection{The Bond Breakage Conjecture}\label{bond breakage section}
A packing of $n$ spheres can be formed by (i) taking an $n-m$ sphere packing, breaking a contact (or bond), adding $m$ new spheres, and forming the appropriate contacts to complete the packing, %(fig \ref{}), 
or by (ii) breaking one bond of an $n$ sphere packing and reforming another.  %Both of these principles can shed insight into sphere packings, here we will focus on mechanism (ii).
%From this property, we arrive at the following conjecture.
%From this property, we propose the following `bond breakage theory':
From this property, we propose the following theory.\\

\underline{\textit{Bond Breakage Conjecture:}}  All packings of $n$ spheres can be obtained by breaking one bond and reforming another in every possible way.  %(See figure \ref{}).  
For any packing, there exists an $m$ step path, of breaking one contact and reforming another, that will form that packing out of another packing with $3n-6$ contacts.  Each of the $m$ steps will correspond to an $n$-particle packing.  The end points of the path (\begin{frenchspacing}i.e. which\end{frenchspacing} $3n-6$ packing one begins with and which packing one ends up with) determine what value $m$ takes.  For every packing, there exists at least 1 other packing for which $m = 1$.\\

This suggests an alternative method for enumerating all packings of $n$ spheres: construct just one $3n-6$ contact packing of $n$ particles (this can easily be done, simply construct an $n$ particle polytetrahedron for example), and then break and reform bonds in all possible ways.  For each packing, it is important to explore every combination of breaking and reforming a bond -- \begin{frenchspacing}i.e. to\end{frenchspacing} go down all paths, and not just one path. 

We have confirmed this conjecture up to as high as we have enumerated packings ($n=10$) using the following algorithm:\\
For every $\A$ that corresponds to a packing
\begin{enumerate}
%\item One at a time, remove each 1 from the $\A$ and place it in all possible 0 locations
\item For each 1 that appears in the $\A$:
\begin{enumerate}
%\item Remove the 1 and place it in a location that has a 0, constructing a new $\A$ with a 0 in the location that originally had the 1 and a 1 in the location that originally had the 0. -- only examine replacements of the 1 that lead to graphs that are non-isomorphic to the $\A$ that is being examined.
\item Swap the 1 with an existing 0.  Do this in every possible non-isomorphic way.  (This is the mathematical analogue of physically breaking an existing bond and reforming a different bond that was not present in that packing).\\
\\
For each new $\A$ that is generated by swapping a 1 and a 0 (these are 1 bond away $\A$'s, \begin{frenchspacing}i.e.\end{frenchspacing} where $m = 1$), 
\begin{enumerate}
\item Test for an isomorphism with the $\A$'s of all other packings (\begin{frenchspacing}i.e.\end{frenchspacing} all $\A$'s other than the one being examined).

\item If an isomorphism is found, stop the examination of this $\A$, as it has been shown that there exists a 1 bond bath between the packing being examined and another packing\footnote{If one is interested in examining all 1 bond paths that exist between all packings, then this same algorithm can be executed without this termination step to yield all possible 1 bond paths.}.
\end{enumerate}
\end{enumerate}
\end{enumerate}

Thus implementing this algorithm computationally, we have proven that, for each $n \leq 10$, every packing is a 1 bond distance away from at least 1 other packing of the same $n$.  We have not proven this for $n > 10$, as we have not enumerated all packings of $n > 10$, but we suspect that this conjecture holds for all $n$.

Mapping out all possible 1 bond distances might be able to shed insight into the kinetic pathways of packings.

To implement the bond breakage conjecture into an improved method for enumerating sphere packings, the bonds must be broken and reformed intelligently, such that unphysical conformations are not explored.  If all physical \textit{and unphysical} conformations were to be explored, then we would simply return to the same computational problem we had with enumerating all non-isomorphic $\A$'s.  One should be able to break a bond and reform it only as is physically possible by calculating the 1 degree of freedom motion that is left over from the one broken bond.  Furthermore, one can take advantage of symmetry to, a priori, not explore redundant (\begin{frenchspacing}i.e.\end{frenchspacing} isomorphic) pathways of breaking and reforming bonds.

\subsection{Extensions to Other Dimensions}

The method presented here is, in large part, not dimension specific.  The only step which is dimension specific is the set of geometrical rules used to solve $\A$ for $\D$.  However, at least for the triangular bipyramid rule (which solves for most of the packings), the geometrical rules can easily be modified to account for a different number of dimensions, $d$.  Once this is done, the same method can be used to solve for sphere packings of $d \neq 3$ dimensions.

\subsection{`Lower Dimensional Packings'}

While a packing of $n$ spheres depends on the dimensionality of its embedding space, there exists a cutoff number, $m$, for which the packing of $n$ spheres remains constant for all $d + i$ dimensions, $i \geq 1$.  For this $m$, the $n$ spheres have accessed all dimensions possible to them, and so the embedding space becomes irrelevant.  For example, for all $d$, the unique packing of 1 and 2 spheres is the singlet and doublet, respectively.  For 3 spheres, the unique packing in 1 dimension is a linear connected chain of 3 spheres; whereas in 2 dimensions it is the equilateral triangle.  For $d \geq 2$ however, the unique packing of 3 spheres remains the same, it is always the triangle.  For 4 spheres, the unique packing in 2 dimensions is the di-triangle; whereas, in 3-dimensions it is the tetrahedron.  It is generally true that packings of $d+1$ particles remain the same for d or more dimensions\footnote{We originally posed this as a question, but one of the reviewers of this paper pointed out that this was true due to the fact that any $d+1$ points are always contained in an affine subspace of at most $d$ dimensions (without loss of generality, one point can be taken to be the origin).  Thus, the $d+1$ points can only access $d$ dimensions, and considering a higher than $d$ dimensional space will not change the structures that they can form.}.

\subsection{Patterns in Adjacencies and Distances}

Does there exist a signature pattern in the $\A$'s or $\D$'s that signifies a packing?  In other words, is there a pattern in the distribution of adjacencies (\begin{frenchspacing}i.e.\end{frenchspacing} number of contacts per particle and connections therein) and/or the distribution of distances that corresponds to packings?  If such a pattern does exist, it could illuminate a more general method for solving for packings.  It also might shed light on the spectrum of allowed solutions for the system of quadratic equations corresponding to the adjacency matrices -- detailing which do and do not have real valued solutions in $\mathbb{R}^3$ satisfying $\D \geq R$.

\subsection{Related Mathematical Problems}

\subsubsection{Erd\"os Unit Distance Problem}  

The Erd\"os unit distance problem (\begin{frenchspacing}a.k.a.\end{frenchspacing} Erd\"os Repeated Distance Problem)\footnote{Without loss of generality, a \textit{repeated distance} can be called a \textit{unit distance}, because one can always uniformly rescale all distances such that the repeated distance is the unit distance.  Put another way, `a unit' can be given any value -- here, the unit is simply given the value of the repeated distance.} was posed in 1946 by the Hungarian mathematician, Paul Erd\"os.  It asks what the maximum number of unit (or repeated) distances that can connect $n$ points in $d$ dimensions is \cite{Erdos1970, Erdos1988}.  This problem is still unsolved.  Even in 2 and 3 dimensions, only upper and lower bounds are known \cite{croftUnsolvedProbs,Guth_Katz}.  The solution to this problem in 3 dimensions, where the unit distance is also the minimum distance, would answer what the maximum number of contacts in any sphere packing is; thus giving the number of contacts corresponding to the ground state packing(s).

\subsubsection{3-Dimensional Rigidity}

In solving adjacency matrices for both rigid and non-rigid packings that satisfy minimal rigidity constraints in 3-dimensions, this problem is directly related to determining whether a graph is rigid in 3-dimensions.  Much work has been done in this field \cite{Reports03thedress, Connelly02infinitesimallylocked, Laurent97cuts, CONNELLY91ongeneric, Jackson03connectedrigidity, gortler-2007}, as well as in other dimensions \cite{Lovasz82ongeneric}.  The existing work on rigidity may help to further inform sphere packings, and the work presented here may in addition be applicable to rigidity theory.  In particular, it may allow for the development of a simple method for reading off whether a 3-dimensional graph is rigid or not.  By the method presented here, a graph is determined to be non-rigid if there exists a continuum of solutions to $\A$.  However, if we can determine a signature pattern that corresponds to all non-rigid (but minimally-rigid) $\A$'s, this would allow for a very simple determination of whether a 3D graph is rigid.

\subsubsection{Solutions to Systems of Polynomial Equations}

The method presented here is inherently solving a system of quadratic equations.  Thus presenting an alternative analytical solution to this class of problems.  Current standards in the field for analytically solving systems of polynomial equations include Gr\"obner bases \cite{buch01}; however, these are time-consuming and thus do not scale efficiently with the number of equations.  The method presented here solves a certain class of polynomial equations efficiently for a relatively large number of equations.  Is it possible to extend this method in order to more efficiently solve large systems of polynomial equations?

\subsubsection{Euclidean Distance Matrix Completion Problems}

Given a symmetric matrix, $M$, where only certain elements are specified, the Euclidean distance matrix completion problem is to find the unspecified elements of $M$ that make $M$ a Euclidean distance matrix.  Euclidean distance matrix and positive semidefinite matrix completion problems are closely linked \cite{Laurent98connectionsEuclidean, Laurent98polynomialinstances, Laurent01matrixcompletion, Laurent95connectionsbetween, huang_607918, Alfakih99solvingeuclidean}.  In solving adjacency matrices for distance matrices, the method presented here is directly related and potentially directly applicable to the euclidean distance matrix completion problem and, by extension, to the positive semidefinite matrix completion problem.

\section{Concluding Remarks}\label{concluding remarks}

In this work, we present an analytical method for deriving all packings of $n$ spheres.  We carry out this derivation for $n \leq 10$; where the set of $n = 10$ new seeds is preliminary, and all iterative packings of $n = 10$ spheres and all packings of $n \leq 9$ spheres are potentially complete, save the numerical round-off error present from implementing this analytical method computationally.

We consider the derivation of these sphere packings to be the first step in directing the self-assembly of spherical colloidal particles; where we have divided this problem up into 2 parts: (i) understanding what the system of colloids can self-assemble, (ii) deriving a mechanism to control that self-assembly.  The derivation of all packings of $n$ spheres gives us everything that a system of $n$ colloidal particles can self-assemble, thus taking care of this first step.  Future work will detail the second step, which is the derivation of a mechanism that directs the self-assembly of the packings such that only one packing forms.

Beyond the problem of self-assembly, the results reported here are interesting in their own right.  We find many interesting properties from the sphere packings enumerated up to $n = 10$, as well as from the conjectured maximum contact packings of $11 \leq n \leq 20$.  Furthermore, the results are directly related to the physics of colloidal clusters, and may have applications to glassy systems and the nucleation of crystals.  They are also directly related to unsolved problems in mathematics, such as the Erd\"os unit distance problem.
\\
\\

We thank John Lee for consultations in coding, and for writing code that (i) converted the packing output to latex format for appropriate display in appendix B, and (ii) interfaced the packing output with the rotational constant calculator website and with the mayavi2 graphing package so that the 10 particle point groups, symmetry numbers, and figures did not have to be generated manually.  We also thank David Roach, Noam Elkies, and Marcus Roper for helpful discussions.  We acknowledge support from the MRSEC program of the National Science Foundation under award number DMR-0820484, the NSF Division of Mathematical Sciences under grant number DMS-0907985, and DARPA under contract BAA 07-21.
\newpage
\bibliographystyle{siam}
\bibliography{long_paper_bib}

%\appendix
%\title{Appendix A}
%\maketitle
\vspace{1in}
%{\Large \begin{center}Appendix A: Individual Geometrical Rules\end{center}}
\textbf{{\large \begin{center}APPENDIX A: INDIVIDUAL GEOMETRICAL RULES\end{center}}}
%\baselineskip
\vspace{.5in}

In this appendix, we derive the individual geometrical rules that were used to solve $\A$ for $\D$.

\setcounter{section}{0}
\setcounter{figure}{0}
\section{Rules Sufficient to Solve All 6 Particle Packings}

\noindent \textbf{\underline{Rule 1:}}

The first rule is the most simple; it states that any 2 intersection circles can not intersect in $> 2$ points (fig 2).
\begin{figure}[h]
\begin{center}
\subfigure[2 intersections of intersection circles]{
	\includegraphics[scale = .33]{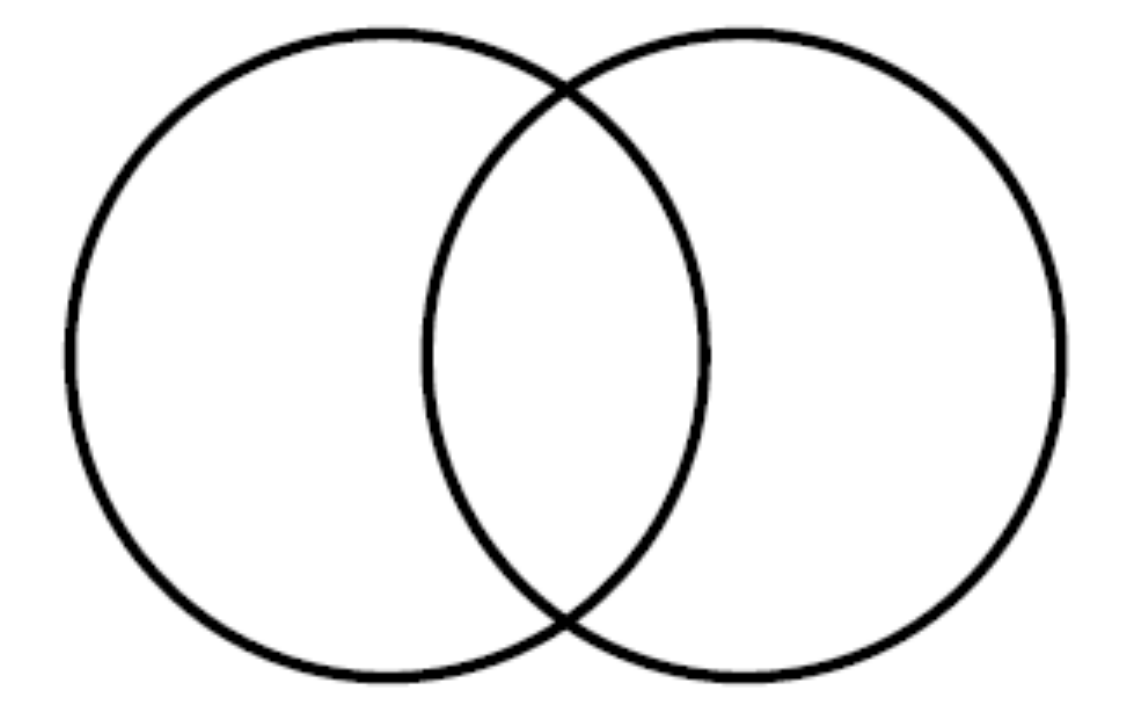}
}
\subfigure[1 intersection of 2 intersection circles]{
	\includegraphics[scale = .33]{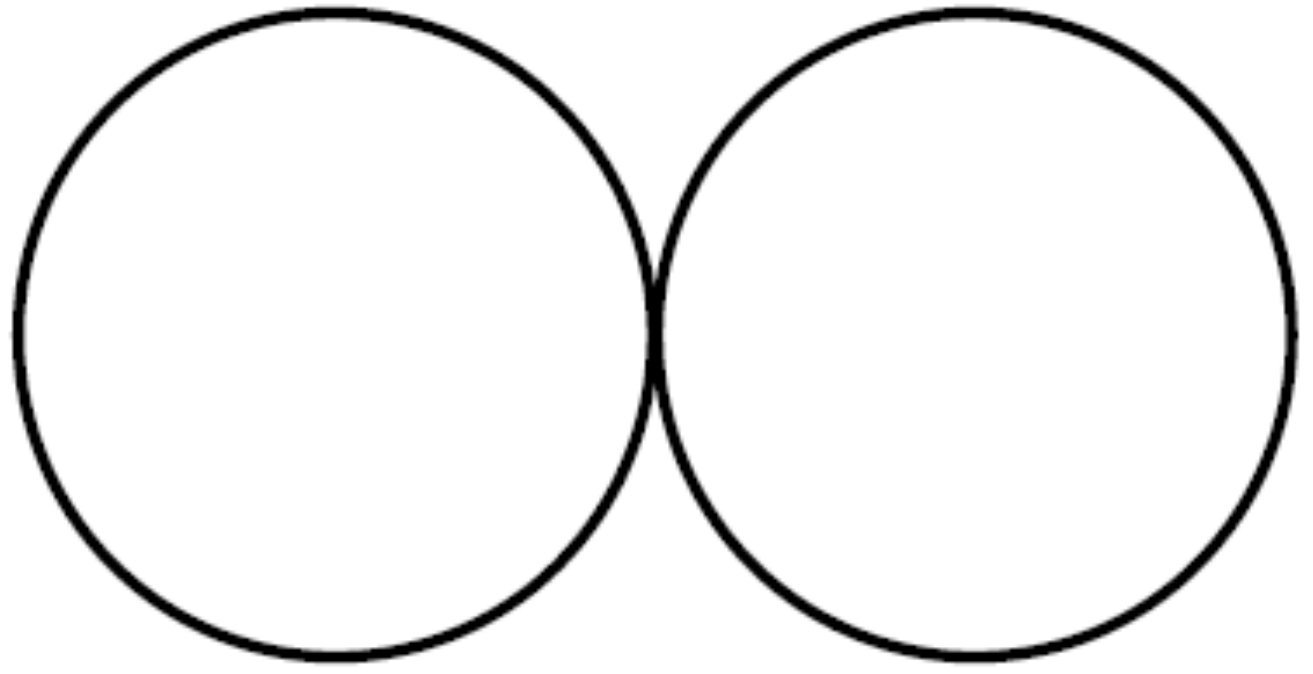}
}
\subfigure[0 intersections of intersection circles]{
	\includegraphics[scale = .33]{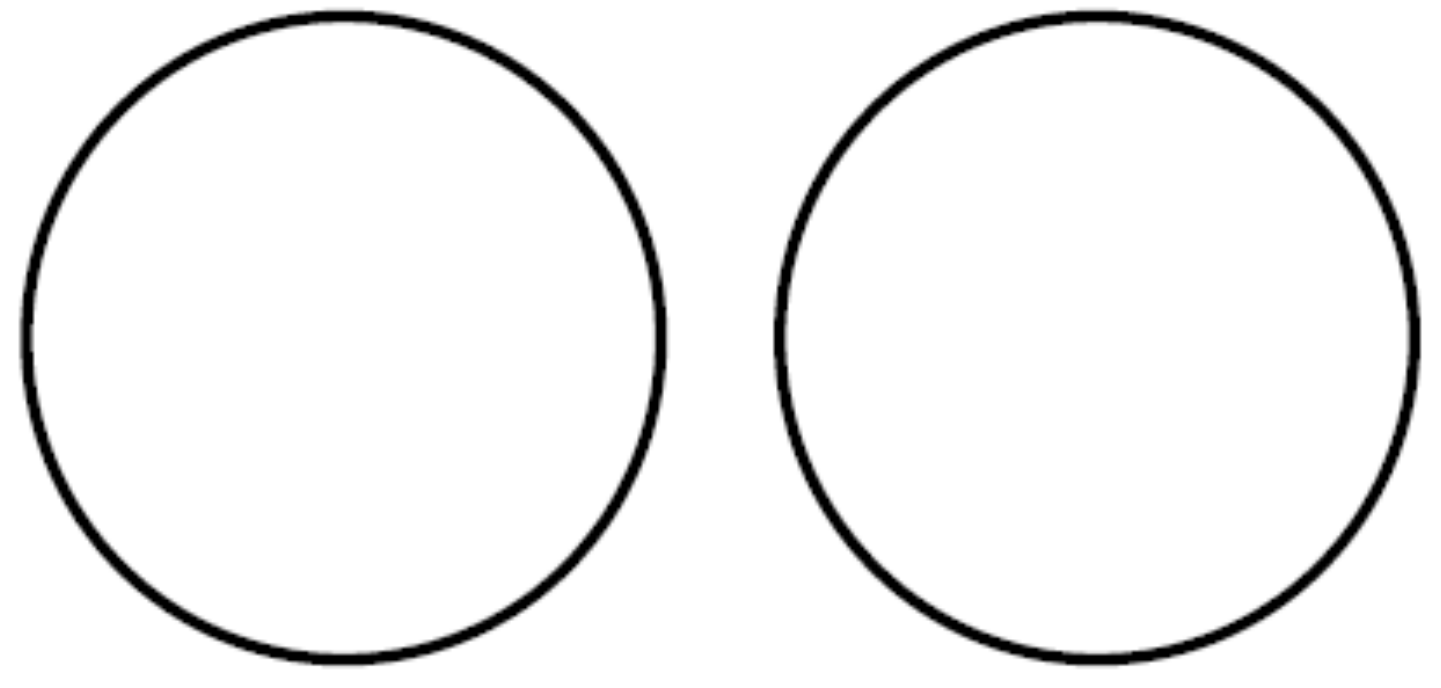}
}
\caption{Intersecting Intersection Circles}\label{Ints of int circs}
\end{center}
\end{figure}
\\
\underline{This rule is identified by the following adjacency matrix pattern:}\\ Any 2 of the set $\left\lbrace \A_{jk},\A_{jp}, \A_{kp} \right\rbrace$ equal 1, and $\exists > 2 \; i$'s for which $\A_{ij} = \A_{ik} = \A_{ip} = 1$.
\\
\\
\textbf{\underline{Rule 2:}}

\begin{figure}[h]
 \begin{center}
  \includegraphics[width = 5in]{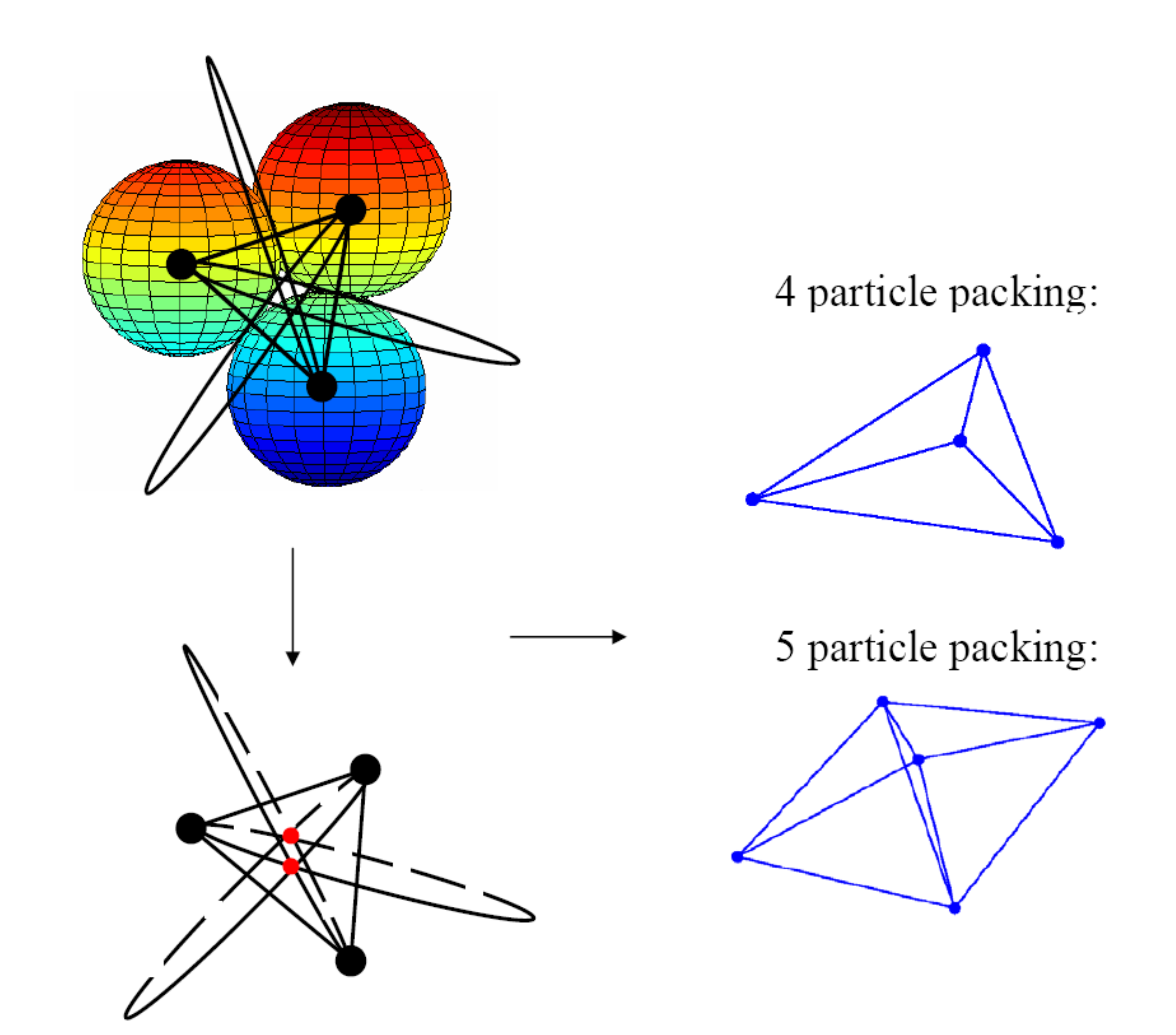}
\caption{The Intersections of 3 Intersection Circles}\label{3 int circs full}
 \end{center}
\end{figure}

The 2nd rule calculates the distance between the 2 points of intersection of 3 mutually intersecting intersection circles associated with a triplet (fig \ref{3 int circs full}).  If a particle lies at one of the intersection points, this forms a 4 particle packing (the tetrahedron), for which all distances $= R$. If 2 particles lie at each of the intersection points, this forms a 5 particle packing -- the distance between these 2 points is the only distance $> R$.

The non-unit distance (where the `unit distance' is $= R$) of the 5 particle packing is calculated as follows (fig 4): A trimer forms an equilateral triangle with sides of length $R$.  The intersection circles have a radius of $(\sqrt{3}/2) R$, and the points, corresponding to the particles' centers, lie on the circumference of these intersection circles.  The centers of the intersection circles lie at the midpoint of each of the triangle sides (of length $R$).  Thus, we can draw a line segment of length $(\sqrt{3}/2) R$ emanating from the points of the equilateral triangle to the midpoints of the opposite line segment.  (See figure \ref{3 int circs to triangles}).  The intersection of these line segments corresponds to the in-plane location of where the intersection circles intersect -- \begin{frenchspacing}i.e.\end{frenchspacing} projecting the intersection points to the face of the equilateral triangle.  For example, in the pentamer in figure \ref{3 int circs to triangles}b (\begin{frenchspacing}i.e.\end{frenchspacing} the 5 particle packing), the center triangle corresponds to the in plane triangle we are considering, and the top and bottom points correspond to the 2 intersection points of the 3 intersection circles.  As all of these line segments = $R$ (as they represent particles touching each other), all of the triangles formed in this pentamer are equivalent equilateral triangles oriented differently in 3-dimensional space.  Let us consider the 3rd triangle in figure \ref{3 int circs to triangles}a; using right triangle math, we can calculate what $a$ is:

\begin{figure}[h]
 \begin{center}
(a)\includegraphics[width = 5.9in]{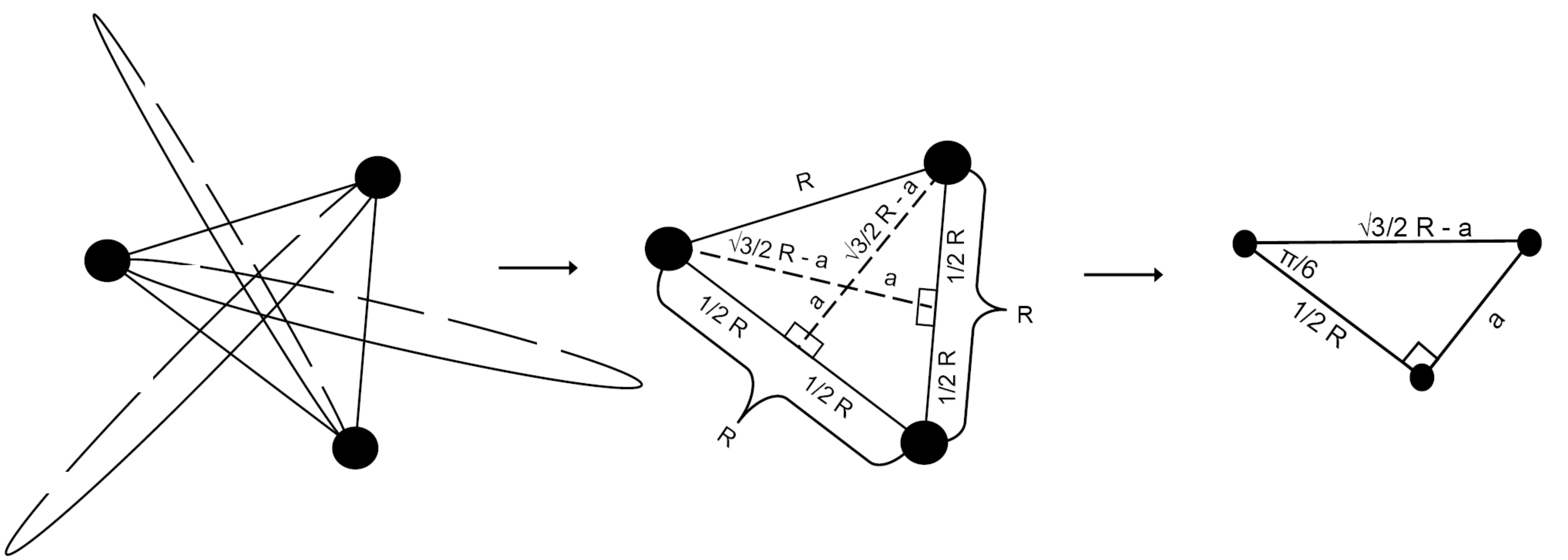}
(b)\includegraphics[width = 5in]{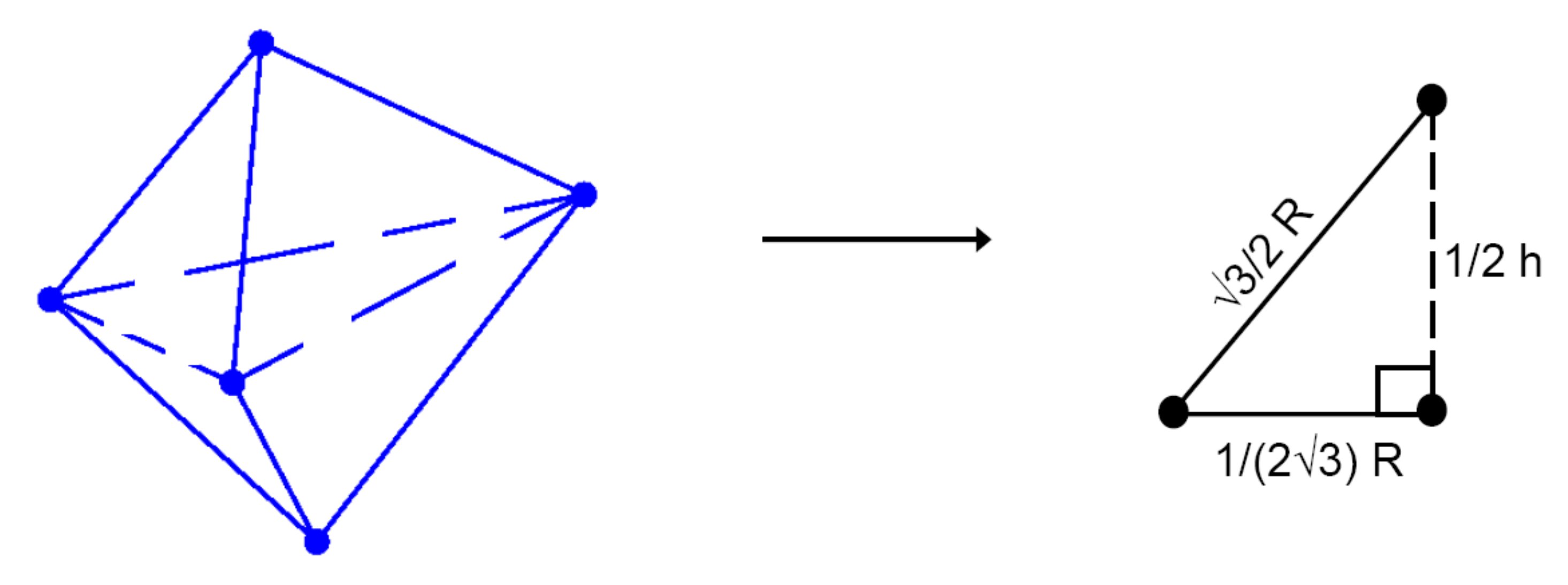}
\caption{Calculating the Distance Between the 2 Points of Intersection of the 3 Intersection Circles of a Trimer (or Triplet).} \label{3 int circs to triangles}
 \end{center}
\end{figure}

\begin{eqnarray}
 \nonumber \tan \frac{\pi}{6} &=& \frac{a}{\frac{1}{2} R}\\
\nonumber \Longrightarrow \frac{1}{\sqrt{3}} &=& \frac{a}{\frac{1}{2} R}\\
\nonumber \Longrightarrow a &=& \frac{1}{2 \sqrt{3}} R
\end{eqnarray}
Because all of the triangles in the pentamer are equivalent, we can use $a$ to construct the triangle in figure \ref{3 int circs to triangles}b, where the $(\sqrt{3}/2) R$ edge corresponds to the radius of one of the in-plane intersection circles.  Thus, the $(\sqrt{3}/2) R$ edge connects one of the in-plane triangle edges to the point of intersection of the 3 in-plane intersection circles.  The $1/2h$ edge corresponds to the line segment connecting the point of intersection of the 3 circles of intersection to the in-plane projection of this intersection point; and therefore represents $1/2$ the length between the 2 intersection points (as the in-plane trimer lies at the midpoint of the 2 intersection points).  This forms a right triangle, and once again we can use right triangle math to solve for $h$:
\begin{eqnarray}
 \nonumber \left( \frac{1}{2 \sqrt{3}} R\right)^2 + \left(\frac{1}{2} h\right)^2 = \left( \frac{\sqrt{3}}{2} R\right)^2\\
\Longrightarrow \label{appendix 2sqrt(2/3) dist rule} \boxed{h = 2 \sqrt{\frac{2}{3}} R}
\end{eqnarray}
\par \indent Thus, the solution to an adjacency matrix corresponding to a 5 particle packing, written in terms of a distance matrix, is
\begin{eqnarray}
\nonumber \left( \begin{matrix} 0	&1	&1	&1	&1\\
1	&0	&1	&1	&1	\\
1	&1	&0	&1	&1	\\
1	&1	&1	&0	&0	\\
1	&1	&1	&0	&0	\end{matrix} \right)
\longrightarrow \left( \begin{matrix} 0	&1	&1	&1	&1\\
1	&0	&1	&1	&1	\\
1	&1	&0	&1	&1	\\
1	&1	&1	&0	&2 \sqrt{\frac{2}{3}}	\\
1	&1	&1	&2 \sqrt{\frac{2}{3}}	&0	\end{matrix} \right)
\end{eqnarray}
\\
\underline{This rule can be identified by the following adjacency matrix pattern:}\\
$\A_{ij} = \A_{ik} = \A_{kj} = 1$, and $\exists$ 2 $p$ for which $\A_{pi} = \A_{pj} = \A_{pk} = 1$.  This rule solves for the distance between those 2 $p$, for \begin{frenchspacing}ex.\end{frenchspacing} if $p = l, m$, then $\D_{lm} = 2\sqrt{\frac{2}{3}}R$.
\\
\\
\textbf{\underline{Rule 3:}}

This follows from rule 2, namely as we have calculated that the distance between the 2 intersection points of 3 mutually intersecting intersection circles is $2\sqrt{2/3}R$, then that distance can not $=R$.  Any adjacency matrix implying that distance $=R$ is unphysical.\\
\\
\underline{This rule can be identified by the following adjacency matrix pattern:}\\
$\A_{ij} = \A_{ik} = \A_{kj} = 1$, $\exists$ 2 $p$ for which $\A_{pi} = \A_{pj} = \A_{pk} = 1$ (\begin{frenchspacing}i.e.\end{frenchspacing} the same pattern as in rule 2), \textit{and} the $\A$ element associated with those 2 $p$'s is 1 (\begin{frenchspacing}i.e.\end{frenchspacing} if $p = q,z$ then $\A_{qz} = 1$).
\\
\\
\textbf{\underline{Rule 4:}}

\begin{figure}
 \begin{center}
 \includegraphics[width = 6in]{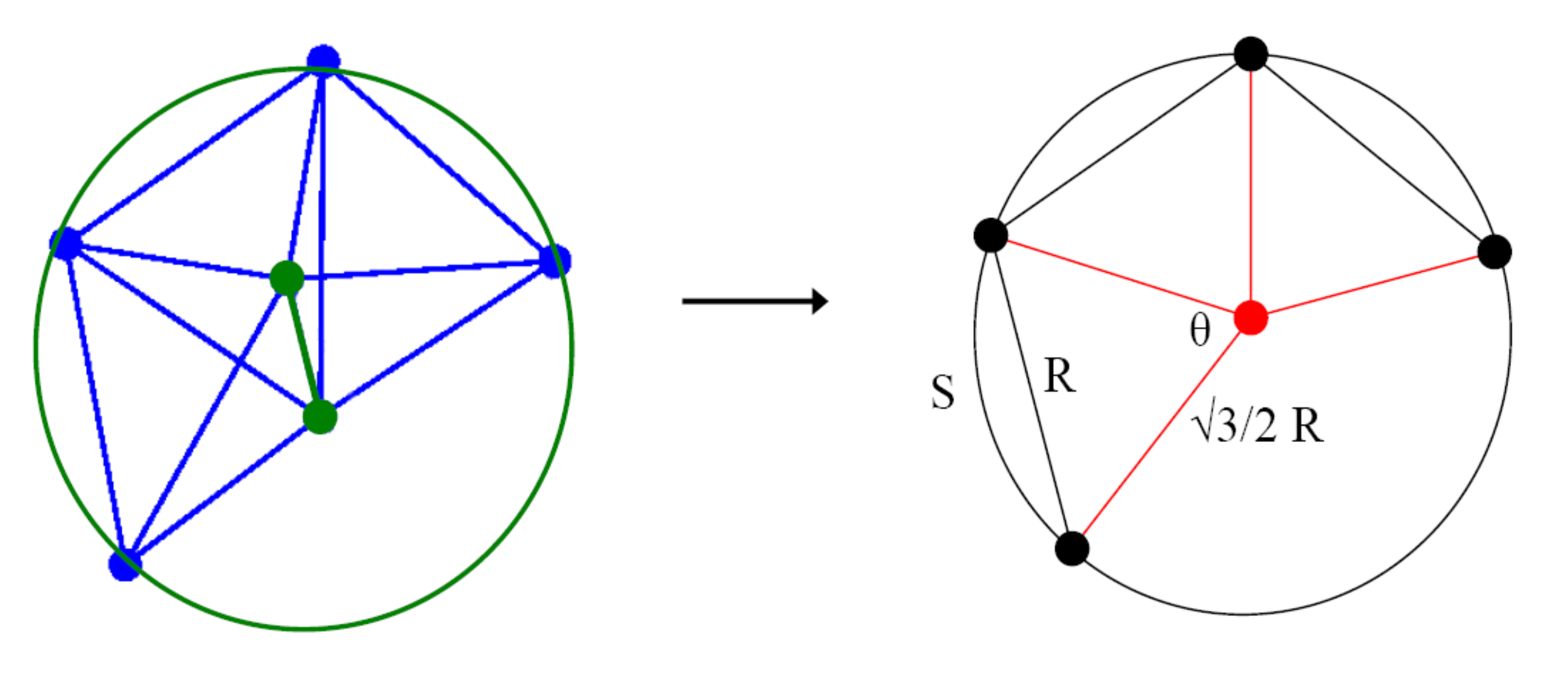}
 \caption{4 Touching Points on an Intersection Circle}\label{4 parts on int circs 1}
\end{center}
\end{figure}

\begin{figure}
 \begin{center}
 \includegraphics[width = 5in]{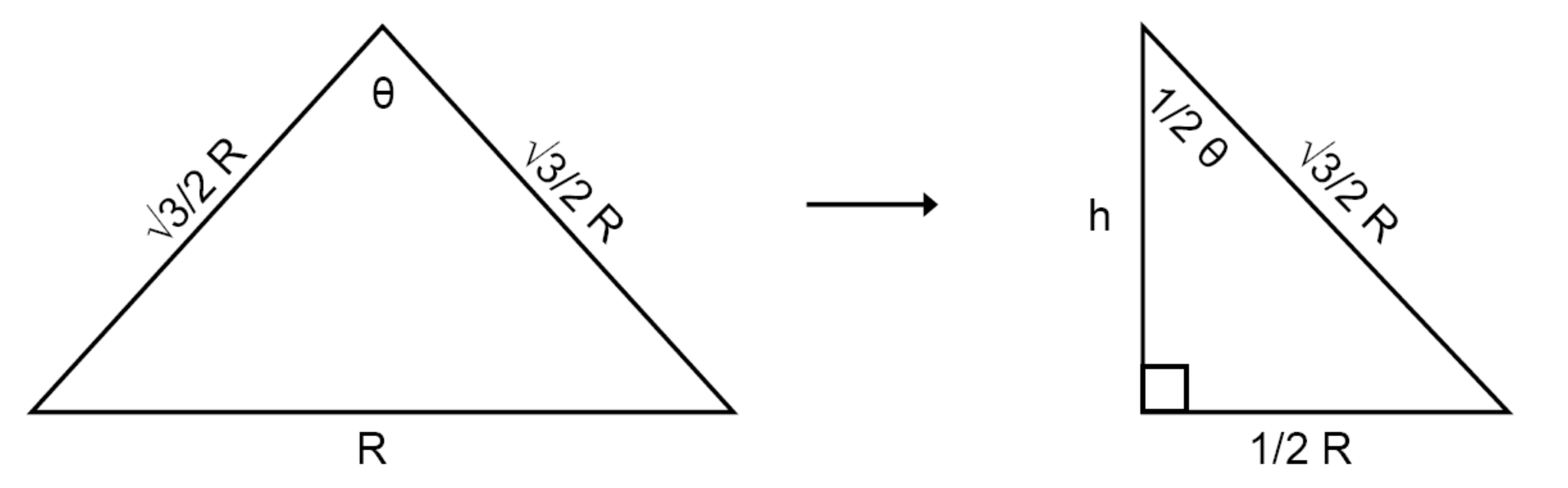}
  \includegraphics[width = 5.9in]{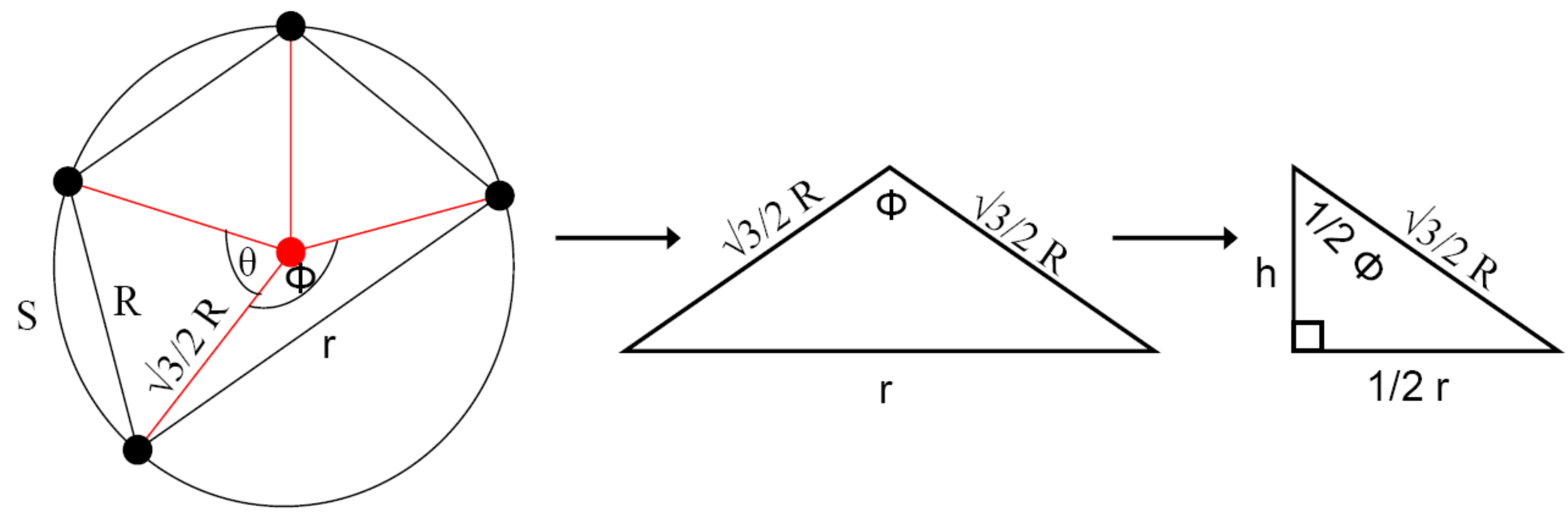}
 \caption{Solving the Long Distance Between 4 Touching Points on an Intersection Circle}\label{4 parts on int circs 3}
\end{center}
\end{figure}

This rule calculates the distance between the two end particles of a total of 4 particles that lie on a circle of intersection, where particle $i+1$ touches particle $i$ ($1 \geq i \leq 4$).  (See figure \ref{4 parts on int circs 1}).
From the figure, we can see that this scenario forms 3 equivalent isoscoles triangles, which we will be able to use to calculate the distance, $r$, between the 2 end particles (see figure \ref{4 parts on int circs 3}).  We can calculate $\theta$ from this triangle, either via the law of cosines
\begin{eqnarray}
 \nonumber \theta &=& \cos^{-1}\left( \frac{\left(\frac{\sqrt{3}}{2} R\right)^2 + \left(\frac{\sqrt{3}}{2} R\right)^2 - R^2}{2 \frac{\sqrt{3}}{2} R \frac{\sqrt{3}}{2} R} \right)\\
\nonumber \Longrightarrow \theta &=& \cos^{-1}\left( \frac{1}{3} \right)
\end{eqnarray}
or via right triangle math:
\begin{eqnarray}
 \nonumber \sin(\frac{1}{2} \theta) &=& \frac{\frac{1}{2} R}{\frac{\sqrt{3}}{2} R}\\
\nonumber \Longrightarrow \theta &=& 2 \sin^{-1}\left( \frac{1}{\sqrt{3}} \right)
\end{eqnarray}
and both of these expressions for $\theta$ are equivalent (both $\approx 1.23096$).

Using $\theta$, we can calculate the desired distance, $r$.  The new angle, $\phi$, depicted in figure \ref{4 parts on int circs 3} is given by $2\pi - 3\theta$:
\begin{eqnarray}
 \nonumber \phi = 2\pi - 3\cos^{-1}\left(\frac{1}{3} \right) = 2\pi - 6\sin^{-1}\left( \frac{1}{\sqrt{3}} \right)
\end{eqnarray}
and thus $r$ is given by
\begin{eqnarray}
 \nonumber \sin\left(\frac{1}{2} \phi\right) &=& \frac{\frac{1}{2} r}{\frac{\sqrt{3}}{2} R}\\
\nonumber \Longrightarrow r &=& 2 \frac{\sqrt{3}}{2} R \sin\left( \pi - 2 \sin^{-1} \left( \frac{1}{\sqrt{3}} \right)\right)\\
\nonumber &=& \sqrt{3}R\sin\left(3 \sin^{-1}\frac{1}{\sqrt{3}}\right)
\end{eqnarray}
and as
\begin{eqnarray}
 \nonumber \sin\left( 3 \sin^{-1}\left( \frac{1}{\sqrt{3}}\right) \right) = \frac{5}{3 \sqrt{3}}
\end{eqnarray}
$r$ is given by
\begin{eqnarray}
\label{appendix 5/3 dist rule} \boxed{r = \frac{5}{3} R}
\end{eqnarray}
Therefore, the distance between particle 1 and 4, where particle $i+1$ touches particle $i$ for $i=1$ to $i=4$, and particles 1 to 4 all touch the same 2 particles is always $5/3 R$.
%\begin{figure}
% \begin{center}
% \includegraphics[width = 7in]{4pointsOnIntCirc_Triangles_Part2}
%\end{center}
%\label{4 parts on int circs 3}
%\end{figure}
\\
\\
\underline{This rule can be identified by the following adjacency matrix pattern:}\\
$\A_{ij} = 1$, and $\exists$ 4 $k$ for which $\A_{ik} = \A_{jk} = 1$, and 3 $\A_{pq} = 1$ amongst the 4 $k$ (\begin{frenchspacing}i.e.\end{frenchspacing} if $k = p, q, l, m$, then $\A_{pq} = \A_{ql} = \A_{lm} = 1$), and the distance $\D_{pm} = \frac{5}{3}R$).
\\
\\
\textbf{\underline{Rule 5:}}

Following from rule 4, we know that any adjacency matrix implying that the above distance $= R$ is unphysical.
\\
\\
\underline{This rule can be identified by the following adjacency matrix pattern:}\\
The same pattern as in rule 4, where $\A_{pm} = 1$ also.
\\
\\
\textbf{\underline{Rule 6:}}

\begin{figure}
 \begin{center}
 \includegraphics[width = 5.9in]{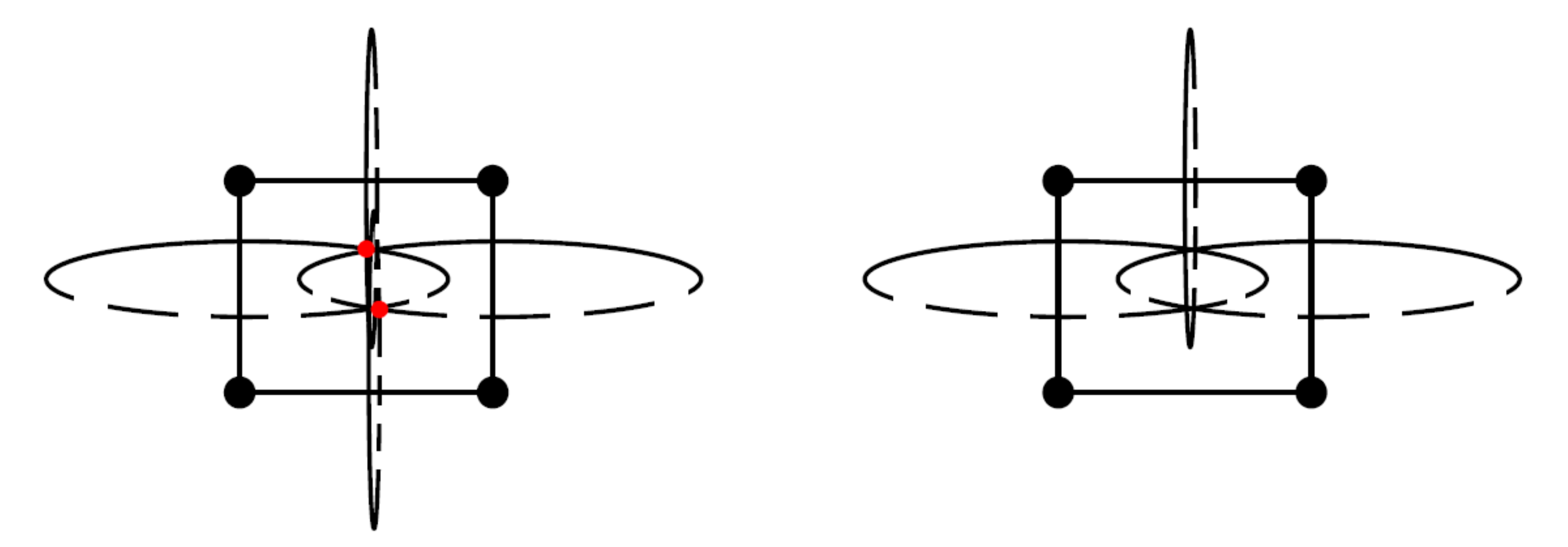}
\caption[c]{4 Intersection Circles Intersecting at 2 Points\\ \footnotesize{The red points in the first figure are the points of intersection of the 4 intersecting circles.  The dashed parts of the circle indicate that part of the circle goes into the page, while the solid lines come out of the page.  The second figure, drawn with one intersection circle not shown, is added for clarity, so that the intersection points of the intersection circles can become more apparent.}}
\label{4 int circs inting at 2 points}
\end{center}
\end{figure}

This rule derives the distance between the 2 intersection points of the mutual intersection of 4 intersection circles associated with 4 points (fig \ref{4 int circs inting at 2 points}).  As mentioned before, $m$ intersection circles can intersect at 1, 2, or 0 points.  As we can see from figure \ref{4 int circs inting at 2 points}, when 4 intersection circles intersect at 2 points, they must be associated with a 2-dimensional square (in fact, we will see later that this is generally true - if it is possible for $m$ intersection circles to intersect at 2 points, they will only intersect at 2 points when associated with a regular $m$-gon lying in a 2 dimensional plane.  And $2 \leq m \leq 5$, if $m \geq 6$ the $m$ associated intersection circles can intersect in at most 1 point (see rule 13).  If 4 intersection circles are not associated with a square, then they can only intersect at 1 or at 0 points.

\begin{figure}
 \begin{center}
 \includegraphics[width = 5.9in]{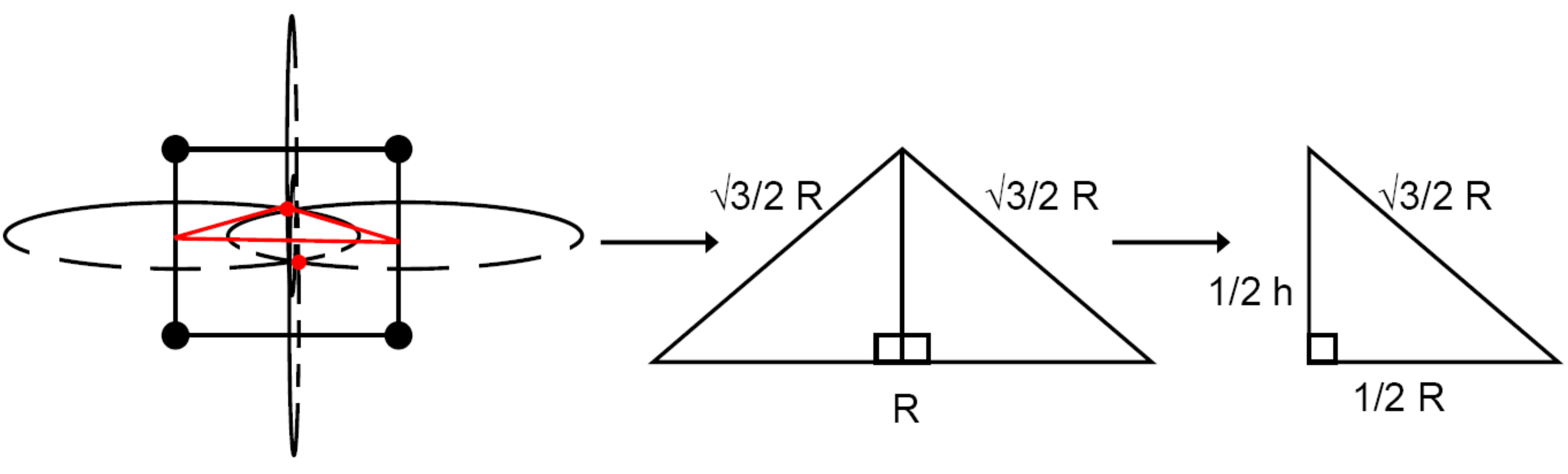}
 \caption{Solving for the Distance Between 2 Intersection Points of 4 Mutually Intersecting Intersection Circles Associated with 4 Points.}\label{4 int circs inting at 2 points triangles}
\end{center}
\end{figure}

We can calculate the distance between the 2 points of intersection of 4 mutually intersecting circles by constructing an isosceles triangle that connects one of the intersection points to the center of 2 of the intersection circles - such a triangle is shown in figure \ref{4 int circs inting at 2 points triangles}.  We know the lengths of all of the sides of this triangle, as the 2 equivalent lengths are the radius of the intersection circle, $(\sqrt{3}/2) R$, and the base length is the distance between 2 touching particles (or equivalently the radius of the neighbor sphere), $R$. 
The height of this triangle corresponds to half the distance between the 2 points of intersection, thus we can use right triangle math to calculate the desired distance:
\begin{eqnarray}\nonumber \left( \frac{1}{2} R\right)^2 + \left( \frac{1}{2} h\right)^2 &=& \left( \frac{\sqrt{3}}{2} R\right)^2\\
\nonumber \frac{1}{4}R^2 + \frac{1}{4}h^2 &=& \frac{3}{4}R^2\\
\nonumber \Longrightarrow h &=& \sqrt{4 \left( \frac{3}{4} - \frac{1}{4}\right)} R\\
\nonumber \\
\label{appendix sqrt(2) dist rule} \Longrightarrow &&\boxed{h = \sqrt{2} R}
\end{eqnarray}
Thus, the distance between 2 points of intersection of 4 intersection circles must be $\sqrt{2} R$, if any adjacency matrix violates this property, by setting this distance $=R$, then it is physically impossible.
\\
\\
\\
\underline{This rule can be identified by the following adjacency matrix pattern:}\\
$\exists$ a closed 4 ring, $\A_{ij} = \A_{jk} = \A_{kp} =\A_{pi} = 1$, where the cross particles don't touch, $\A_{ik} = \A_{jp} = 0$, and 2 particles touch all points in the 4 ring, $\A_{mj} = \A_{mk} = \A_{mp} = \A_{mi} = 1$ for two $m$'s.  This rule then solves for the distance between those 2 $m$'s, \begin{frenchspacing}i.e\end{frenchspacing} if $m = z,s$ then $\D_{zs} = \sqrt{2}R$.\\
\\
The following adjacency matrix contains this pattern:
\begin{eqnarray}
\nonumber \left( \begin{matrix}
0&0&1&1&1&1\\
0&0&1&1&1&1\\
1&1&0&0&1&1\\
1&1&0&0&1&1\\
1&1&1&1&0&0\\
1&1&1&1&0&0
\end{matrix} \right)
\end{eqnarray}
\\
\\
\textbf{\underline{Rule 7:}}

As just mentioned, following from rule 6, any adjacency matrix that implies the distance between the 2 intersection points of 4 mutually intersecting intersection circles $= R$ is unphysical.
\\
\\
\underline{This rule can be identified by the following adjacency matrix pattern:}\\
The same pattern as in rule 6, \textit{and} the adjacency matrix element between those 2 $m$'s is 1 (\begin{frenchspacing}i.e.\end{frenchspacing} if $m = q,l$, then $\A_{ql} = 1$ also).

\section{Rules Sufficient for 7 Particle Packings}

The following geometrical rules, in addition to the rules used to solve the 6 particle packings, are sufficient to solve for all 7 particle packings.\\
\\
\textbf{\underline{Rule 8:}}

\begin{figure}
 \begin{center}
 \includegraphics[width = 6in]{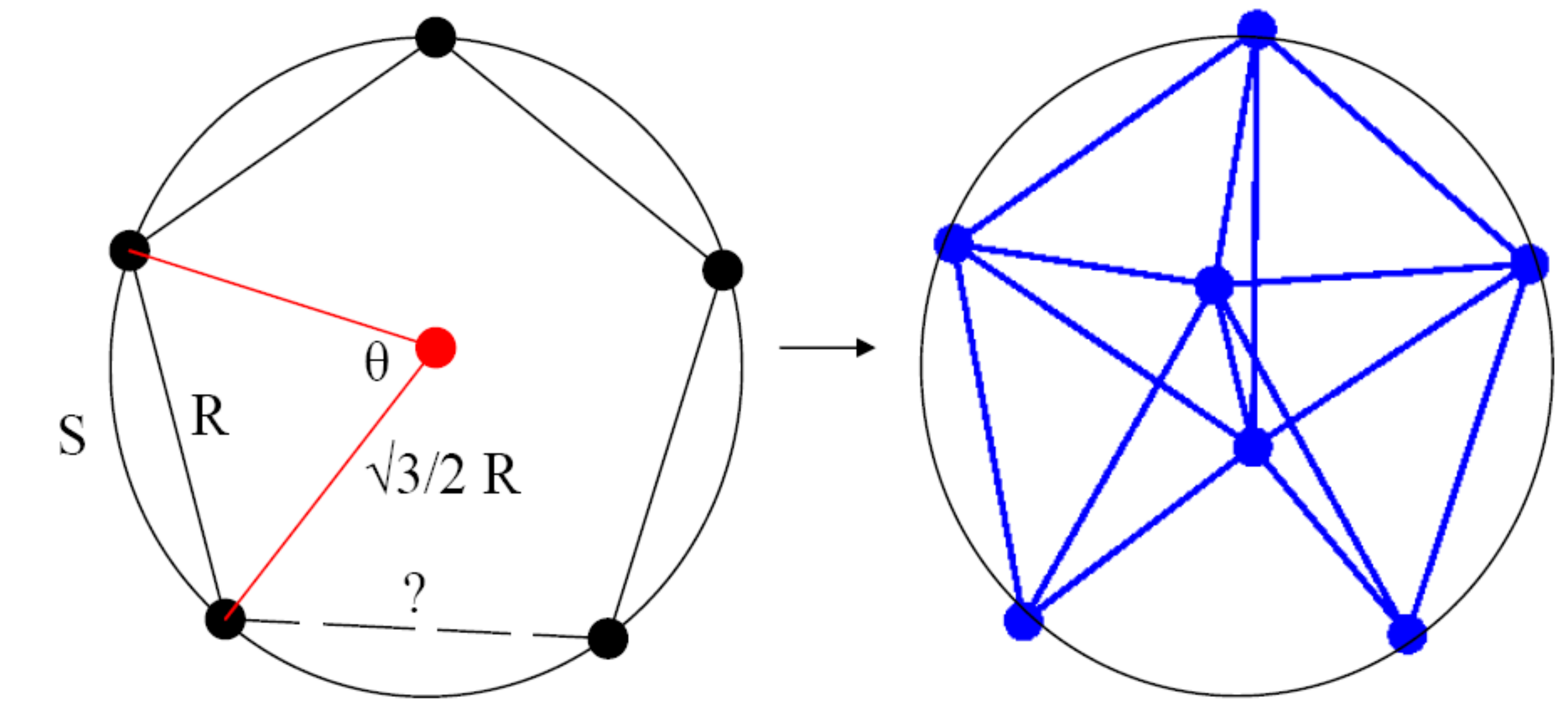}
 \caption{5 Points on an Intersection Circle. (The `?' distance = the unknown distance $r$.)}\label{5PointsOnIntCirc}
\end{center}
\end{figure}

We have calculated the long distance between 4 and between 3 touching points on an intersection circle (rules 2 and 4); for 7 particle packings, we must also know the long distance between 5 touching points on an intersection circle (i.e. the distance between points 1 and 5, where point $i+1$ touches point $i$ for $1 \leq i \leq 5$) - see figure \ref{5PointsOnIntCirc}.

We make this calculation as we did for the one with 4 touching points on an intersection circle, but here the relevant angle of the triangle that will allow us to calculate the distance $r$ is
\begin{eqnarray}
 \nonumber \theta = 2\pi - 4 \cos^{-1}\left( \frac{1}{3} \right) = 2\pi - 8 \sin^{-1}\left( \frac{1}{\sqrt{3}} \right)
\end{eqnarray}
and thus $r$ is given by
\begin{eqnarray}
 \nonumber \sin \left(\frac{1}{2}\theta \right) &=& \frac{\frac{1}{2} r}{\frac{\sqrt{3}}{2} R}\\
\nonumber \sin \left(\pi - 4 \sin^{-1} \left( \frac{1}{\sqrt{3}} \right) \right) &=& \frac{\frac{1}{2} r}{\frac{\sqrt{3}}{2} R}\\
\nonumber \frac{4 \sqrt{2}}{9} &=& \frac{\frac{1}{2} r}{\frac{\sqrt{3}}{2} R}\\
\label{appendix 4sqrt(6)/9 dist rule} \Longrightarrow&&  \boxed{r = \frac{4 \sqrt{6}}{9} R}
\end{eqnarray}\\
\\
\\
\underline{This rule can be identified by the following adjacency matrix pattern:}\\
$\A_{ij} = 1$, and $\exists$ 5 $k$ for which $\A_{ik} = \A_{jk} = 1$, and 4 $\A_{pq} = 1$ amongst the 5 $k$ (\begin{frenchspacing}i.e.\end{frenchspacing} if $k = p, q, l, m, z$, then $\A_{pq} = \A_{ql} = \A_{lm} = \A_{mz} = 1$).  This rule solves $\D_{pz} = \frac{4\sqrt{6}}{9}R$.
\\
\\
\textbf{\underline{Rule 9:}}

Following from rule 8, we can see that 5 particles can not touch the same 2 particles and all touch each other, as the distance between the 4th and 5th points lying on an intersection circle is $= \frac{4\sqrt{6}}{9}R \approx 1.08866 R$, and for 5 particles to touch the same 2 particles, this distance would have to equal exactly $R$.  Thus any $\A$ implying this distance $=R$ is unphysical.\\
\\
\\
\underline{This rule can be identified by the following adjacency matrix pattern:}\\
The same pattern as in rule 8, but 5 $A_{pq} = 1$ amongst the 5 $k$ -- \begin{frenchspacing}i.e.\end{frenchspacing} $\A_{zp} = 1$ also.
\\
\\
\textbf{\underline{Rule 10:}}

Also following from rule 8, we can see that no more than 5 points can fit on an intersection circle (\begin{frenchspacing}i.e.\end{frenchspacing} no more than 5 points can mutually touch a dimer), as only slightly more than 5 points can fit on an intersection circle, and 6 points can not fit.
\par \indent We can calculate exactly how many points a distance $R$ apart can fit on an intersection circle by dividing the circumference of the entire intersection circle by the arc length swept out by one of the isosceles triangles created by 2 particles lying on the intersection circle. (See figure \ref{5PointsOnIntCirc}). The formula for the arc length is
\begin{eqnarray}
\nonumber S = r\theta
\end{eqnarray}
where $r$ is the radius of the circle, $\theta$ the angle between 2 radial line segments, and $S$ is the arc length.  Thus the number of points a distance $R$ apart that can fit on an intersection circle is
\begin{eqnarray}
 \nonumber \frac{2\pi \frac{\sqrt{3}}{2} R}{\frac{\sqrt{3}}{2} R \cos^{-1}\left( \frac{1}{3} \right)} &=& \frac{\pi}{\cos^{-1}\left( \frac{1}{3} \right)}\\
&\approx& 5.1043
\end{eqnarray}
(As an aside, $\cos^{-1}(1/3) = 2 \sin^{-1}(1/\sqrt{3}) = 2 \tan^{-1}(\sqrt{2}/2)$, and we could have expressed the angle by any one of these).

Thus any $\A$ implying that $> 5$ points lie on an intersection circle is unphysical.\\
\\
\\
\noindent \underline{This rule can be identified by the following adjacency matrix pattern:}\\
$\A_{ij} = 1$, and $\exists >$ 5 $k$ for which $\A_{ik} = \A_{jk} = 1$.
\\
\\
\textbf{\underline{Rule 11:}}

\begin{figure}
 \begin{center}
 \includegraphics[width = 5.9in]{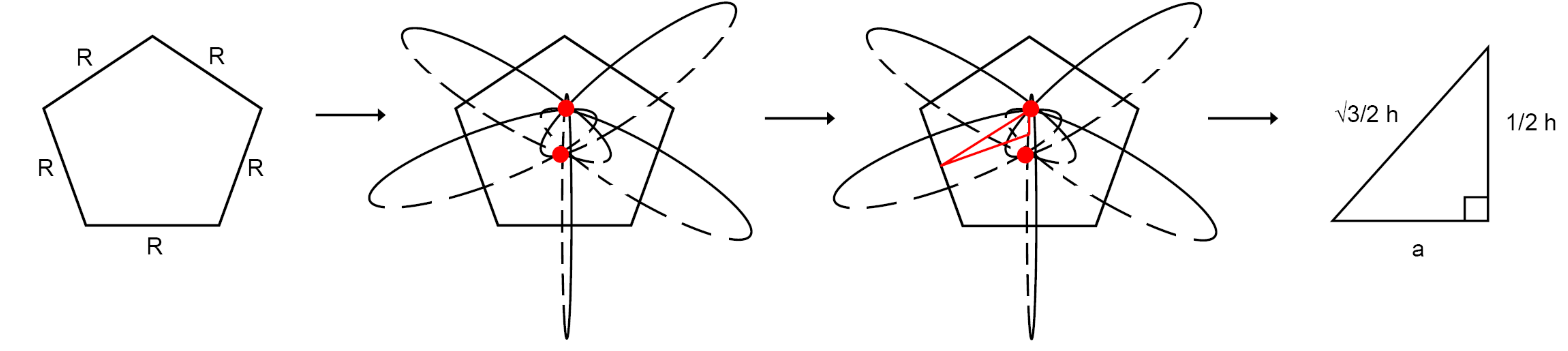}
 \caption{Calculating the Distance Between 2 Intersection Points of 5 Mutually Intersecting Intersection Circles Associated with 5 Points.}\label{5 intersecting int circs}
\end{center}
\end{figure}

\begin{figure}
 \begin{center}
 \includegraphics[width = 2in]{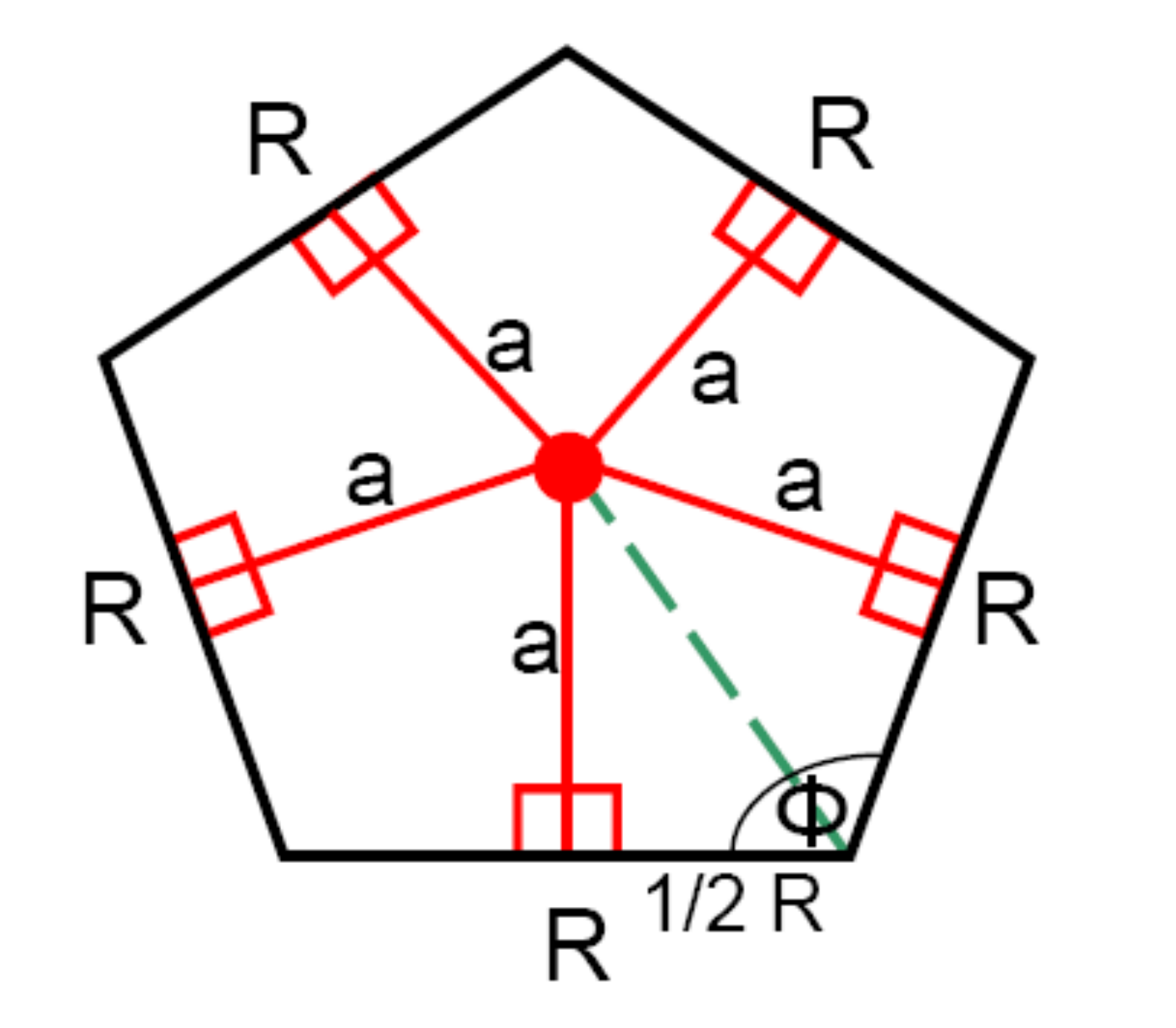}
 \caption{The Apothem and Interior Angle of the Pentagon.}\label{pentagon for 5 intersecting int circs}
\end{center}
\end{figure}

As we did previously with the mutual intersection of 4 intersection circles (rule 6), we will now calculate the distance between the 2 intersection points of 5 mutually intersecting intersection circles.  5 circles of intersection can only intersect at 2 points if they are associated with a 2-dimensional pentagon (see figure \ref{5 intersecting int circs}).  To calculate the distance between these 2 points of intersection, we can construct the triangle that connects one of the points of intersection to the centers of one of the intersection circles (shown in the figure).  The base of this constructed triangle, $a$, is the apothem of a regular pentagon.
We can calculate the apothem of the regular pentagon via the right triangle depicted in figure \ref{pentagon for 5 intersecting int circs} (the formula for the apothem is also well known, so one could alternatively simply look it up).  The interior angle of the pentagon, $\phi$ (fig \ref{pentagon for 5 intersecting int circs}), is given by
\begin{eqnarray}
 \nonumber \phi = \frac{(n-2) \pi}{n} = \frac{3}{5} \pi
\end{eqnarray}
and the angle associated with the right triangle used to calculate the apothem is $\phi/2$.  Therefore, the apothem, calculated via right triangle math, is given by
\begin{eqnarray}
 \nonumber \tan\left( \frac{3}{10} \pi \right) &=& \frac{a}{\frac{1}{2} R}\\
\nonumber \Longrightarrow a &=& \frac{1 + \sqrt{5}}{2 \sqrt{10 - 2 \sqrt{5}}} R
\end{eqnarray}

Now that we know $a$, we can calculate the distance between the 2 points of intersection, $h$, via right triangle math:
\begin{eqnarray}
 \nonumber a^2 + \left( \frac{1}{2} h\right)^2 &=& \left( \frac{\sqrt{3}}{2} R \right)^2\\
\nonumber \left( \frac{1 + \sqrt{5}}{2 \sqrt{10 - 2 \sqrt{5}}} R \right)^2 + \frac{1}{4} h^2 &=& \frac{3}{4} R^2\\
\nonumber \frac{1}{4}\frac{1 + 2 \sqrt{5} + 5}{10 - 2 \sqrt{5}} R^2 + \frac{1}{4}h^2 &=& \frac{3}{4} R^2\\
\nonumber \frac{1}{4}\left(1 + \frac{2}{\sqrt{5}} \right) R^2 + \frac{1}{4}h^2 &=& \frac{3}{4} R^2\\
\nonumber \Longrightarrow h &=& \sqrt{4 \left( \frac{3}{4} - \frac{1}{4}\left(1 + \frac{2}{\sqrt{5}} \right) \right)} R\\
\label{7 part new seed rule bipyramid dist} &\Longrightarrow& \boxed{h = \sqrt{2 - \frac{2}{\sqrt{5}}} R}
\end{eqnarray}
\\
\\
\noindent \underline{This rule can be identified by the following adjacency matrix pattern:}\\
$\exists$ a closed 5 ring, $\A_{ij} = \A_{jk} = \A_{kp} =\A_{pz} = \A_{zi} = 1$, where the cross particles don't touch, $\A_{ik} = \A_{jp} = \A_{kz} = \A_{pi} = \A_{jz} = 0$, and 2 particles touch all points in the 5 ring, $\A_{mj} = \A_{mk} = \A_{mp} = \A_{mz} = \A_{mi} = 1$ for two $m$'s.  This rule identifies the distance between the 2 $m$'s (\begin{frenchspacing}i.e.\end{frenchspacing} if $m = l,y$ this rule gives $\D_{ly} = \sqrt{2 - \frac{2}{\sqrt{5}}} R$).
\\
\\
\textbf{\underline{Rule 12:}}

\begin{figure}
 \begin{center}
 \includegraphics[width = 2in]{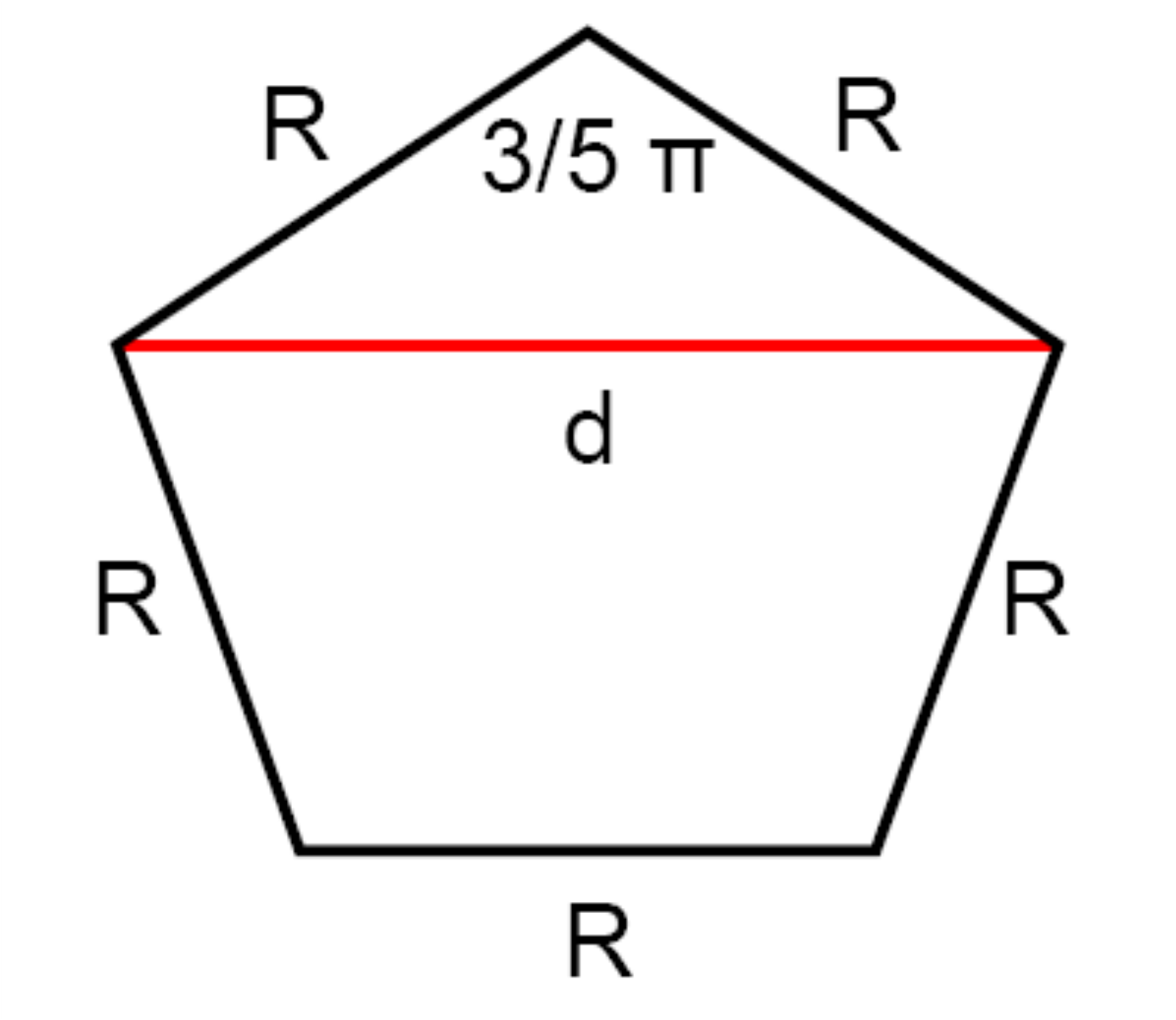}
 \caption{A Triangle Within the Pentagon}\label{pentagon triangle}
\end{center}
\end{figure}

Related to rule 11, the distance between the vertices of the pentagon will be the distance between the non-touching pairs amongst the 5 particles that have intersection circles mutually intersecting at 2 points ($\D_{ik}, \D_{jp}, \D_{kz}, \D_{pi}, \D_{jz}$ from rule 11).  This distance can be calculated from the triangle in figure \ref{pentagon triangle}.  We can solve for the distance, $d$, by using right triangle math or by using the law of cosines, here we'll use the law of cosines:
\begin{eqnarray}
 \nonumber d &=& \sqrt{R^2 + R^2 - 2R^2\cos\left( \frac{3}{5} \pi \right)}\\
\nonumber &=& \sqrt{R^2 + R^2 - 2R^2\left(-\frac{1}{4}\left( -1 + \sqrt{5} \right) \right)}\\
\label{7 part new seed dist rule pentagon} &\Longrightarrow& \boxed{d = \sqrt{\frac{3 + \sqrt{5}}{2}} R}
\end{eqnarray}

\noindent \underline{This rule can be identified by the following adjacency matrix pattern:}\\
The same pattern as in rule 11, but this rule identifies the distances between the cross particles; $\D_{ik} = \D_{jp} = \D_{kz} = \D_{pi} = \D_{jz} = \sqrt{\frac{3 + \sqrt{5}}{2}} R$
\\
\\
\textbf{\underline{Rule 13:}}

6 or more mutually intersecting intersection circles can only intersect at one point.  A regular $m$-gon is the limiting case where $m$ intersection circles will intersect at 2 points, if they can.  For 3, 4, and 5 particles, this is possible.  For $\geq 6$ particles however, this is not.  The longest distance between the non-touching particles of a regular hexagon, whose sides are length $R$, is $2 R$.  (See figure \ref{hexagon}).  
\begin{figure} \label{hexagon}
 \begin{center}
 \includegraphics[width = 4in]{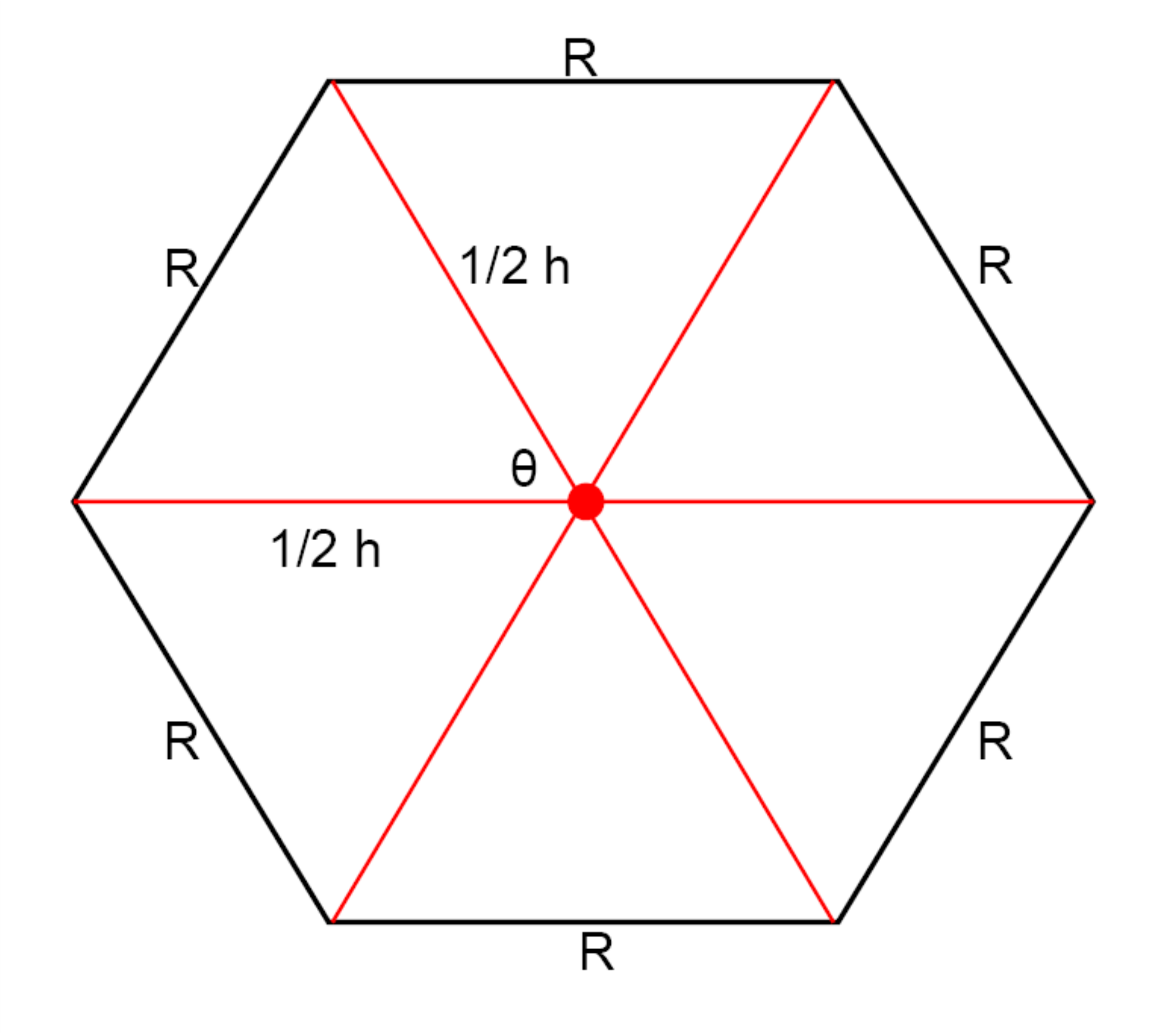}
 \caption{The Hexagon}\label{hexagon}
\end{center}
\end{figure}
The interior angle of a regular $n$-gon is
\begin{eqnarray}
 \nonumber \phi = \frac{(n-2)\pi}{n}
\end{eqnarray}
Thus, $\phi$ here is
\begin{eqnarray}
 \nonumber \phi = \frac{2}{3} \pi
\end{eqnarray}
The angles associated with the equivalent sides of the isosceles triangles formed by the $n$-gon is $((n-2)/2n) \pi$, which implies that $\theta$ is given by
\begin{eqnarray}
 \nonumber \theta &=& \pi - \frac{(n-2)\pi}{n}\\
\nonumber &=& \frac{1}{3}\pi
\end{eqnarray}
(see fig \ref{hexagon}) and we can see that $h$ is indeed equal to $2 R$, from right triangle math (we could also have done this, as usual, from the law of cosines):
\begin{eqnarray}
 \nonumber \sin \left( \frac{1}{2} \theta \right) &=& \frac{\frac{1}{2} R}{\frac{1}{2} h}\\
\nonumber \sin \left( \frac{1}{6} \pi \right) &=& \frac{R}{h}\\
\nonumber \frac{1}{2} &=& \frac{R}{h}\\
\Longrightarrow h &=& 2R
\end{eqnarray}
This distance is equal to twice the distance of 2 touching particles (\begin{frenchspacing}i.e.\end{frenchspacing} to the diameter of the neighbor sphere), and thus we can see that one particle fits right at the center of the regular hexagon, showing us that 6 mutually intersecting intersection circles can only intersect at one point (as only 1 point can touch all 6 points of the hexagon).

Put another way, we can calculate that the height of one of these isoscoles triangles is $= (\sqrt{3}/2) R$, thus implying that the distance from the center of an intersection circle associated with one of the hexagon edges is a distance $\sqrt{3} R$ away from the center of an intersection circle associated with the opposite edge.  Therefore, the distance between the center of the intersection circle and the intersection point of the intersection circles is $(\sqrt{3}/2) R$, and thus the `height between the 2 intersection' points must be 0.  On the other hand, we can see that the distance between the longest non-touching vertices of a regular septagon is $> 2 R$ (not shown here), and thus 7 \textit{in-plane} intersection circles can not all intersect each other.  Of course, once we allow 7 or more intersection circles to arrange themselves in 3 dimensional space, then they can mutually intersect at one point.  

Thus, so far we have established that 6 or more intersection circles can mutually intersect at no more than 1 point.  13 intersection circles can not mutually intersect - this is the \textit{kissing number} limit.  Thus, 2-5 intersection circles can mutually intersect at 2 points, 6-12 intersection circles can mutually intersect at 1 point, and $> 12$ intersection circles can not mutually intersect.

In conclusion, this rule states that any adjacency matrix implying that $\geq 6$ intersection circles mutually intersect at 2 points is unphysical.
\\
\\
\underline{This rule can be identified by the following adjacency matrix pattern:}\\
$\exists$ 2 $m$ for which $\A_{mj} = 1$ for $\geq 6$ $j$ where $\geq 6$ adjacency matrix elements amongst the $j = 1$.  For example, $j = i,l,p,q,z,y$ and $\A_{il} = \A_{lp} = \A_{pq} = \A_{qz} = \A_{zy} = \A_{yi} = 1$ (\begin{frenchspacing}i.e.\end{frenchspacing} there are 6 $j$ and a 6 ring is formed).
\\
\\
\textbf{\underline{Rule 14:}}

\begin{figure}
 \begin{center}
(a)\includegraphics[width = 2in]{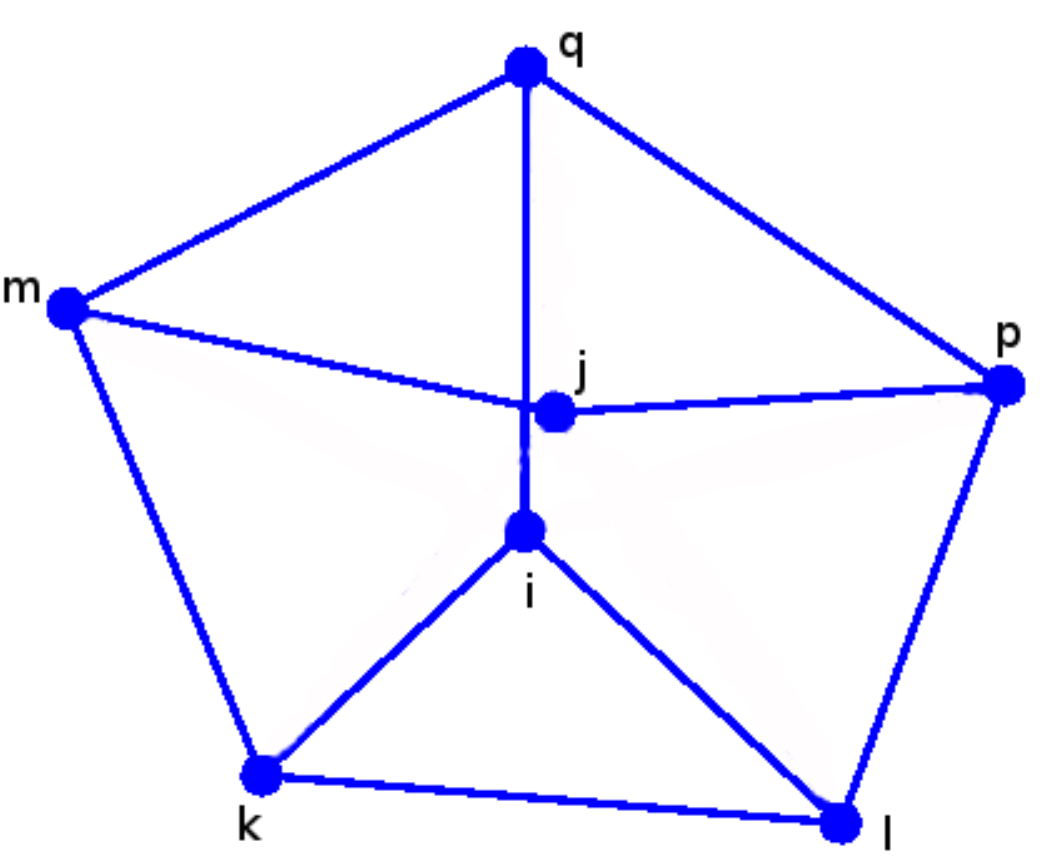}
(b)\includegraphics[width = 2in]{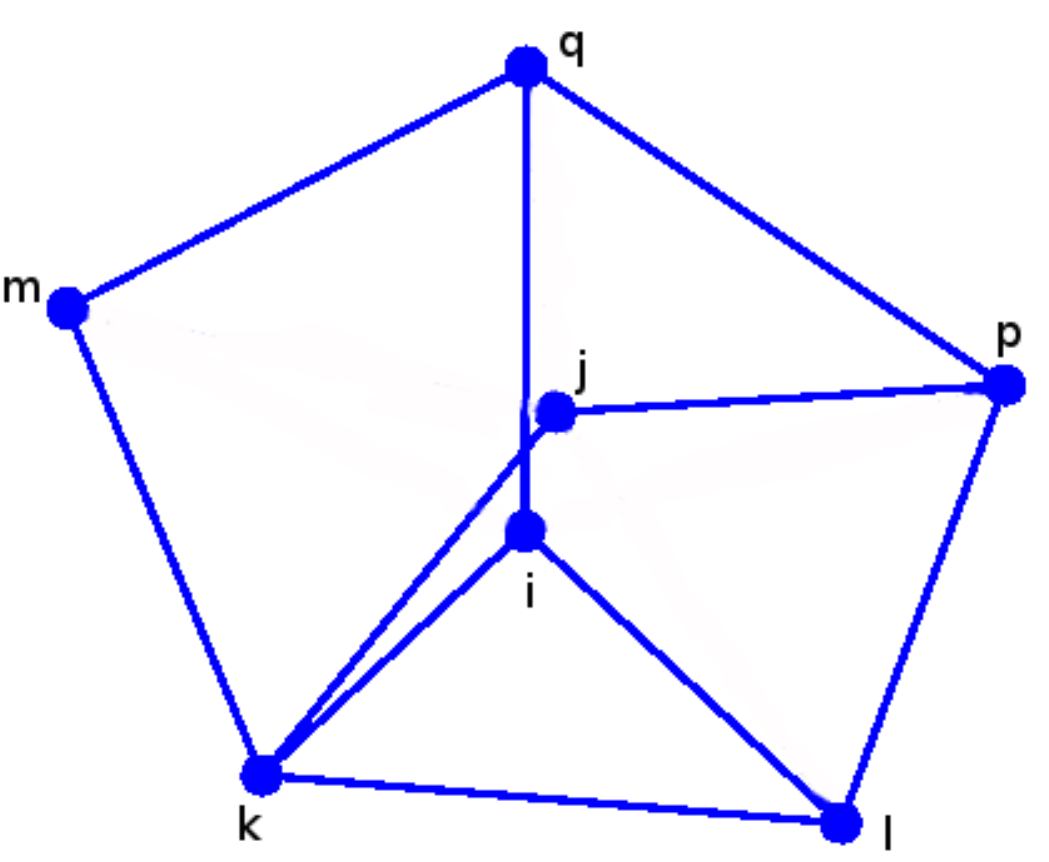}
(c)\includegraphics[width = 2in]{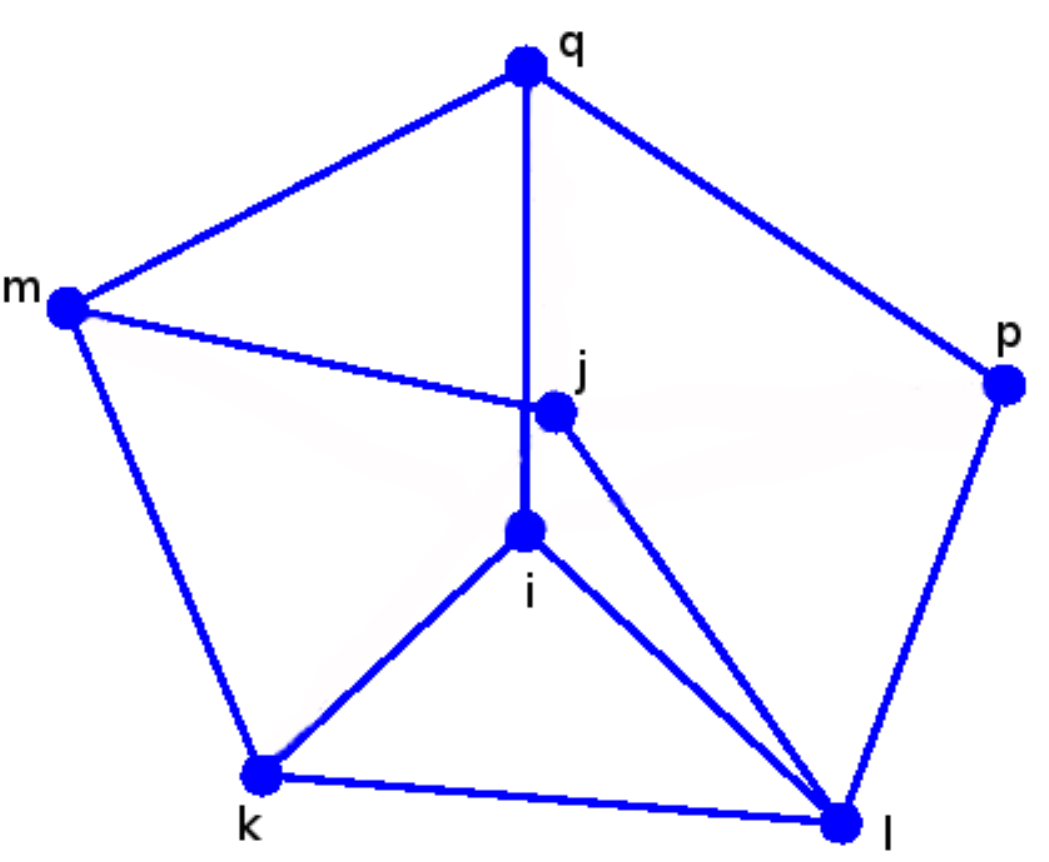}
\caption[c]{A Closed 5 Ring Constrained to Surround a Dimer}\label{5 ring surrounding dimer}
\end{center}
\end{figure}

This rule states that a closed 5 ring can not be confined to surround a dimer.  It is a more general form of the rule stating that the 5 points of a closed 5 ring can not all touch the same dimer (rule 9).  The difference is that this rule does not require that all 5 points be touching the dimer, simply that it make enough contacts with the dimer such that it is confined to surround it; therefore forcing one distance to be $< R$, and thus implying particle overlap (see fig \ref{5 ring surrounding dimer}).\\
\\
\\
\underline{This rule can be identified by the following adjacency matrix pattern:}\\
A violation occurs if
\begin{eqnarray}
 \nonumber \A_{kl} = \A_{lp} = \A_{pq} = \A_{qm} = \A_{mk} &=& 1 \; \; \; \text{(closed 5 ring)}\\
\nonumber \text{\textit{and}}\\
\nonumber \A_{ij} &=& 1 \; \; \; \text{(the dimer)}
\end{eqnarray}
\textit{and} one point in the dimer forms 3 contacts with the closed 5 ring, where 2 of the contacts are adjacent: $\A_{iq} = \A_{ik} = \A_{il} = 1$.  \textit{And} the other point in the dimer (here $j$) touches the 2 points of the 5 ring not touched by the other point (here $i$), $\A_{jm} = \A_{jp} = 1$; or touches one point not touched by $i$, and touches one of the adjacent contacts $i$ makes, $\A_{jm} = \A_{jl} = 1$ or $\A_{jk} = \A_{jp} = 1$.  (See figure \ref{5 ring surrounding dimer}).  If these conditions are satisfied, then the 5 ring is constrained to surround the dimer, and the conformation is unphysical as it implies a distance $< R$.
\\
\\
\textbf{\underline{Rule 15:}}

\begin{figure}
 \begin{center}
 \includegraphics[width = 3in]{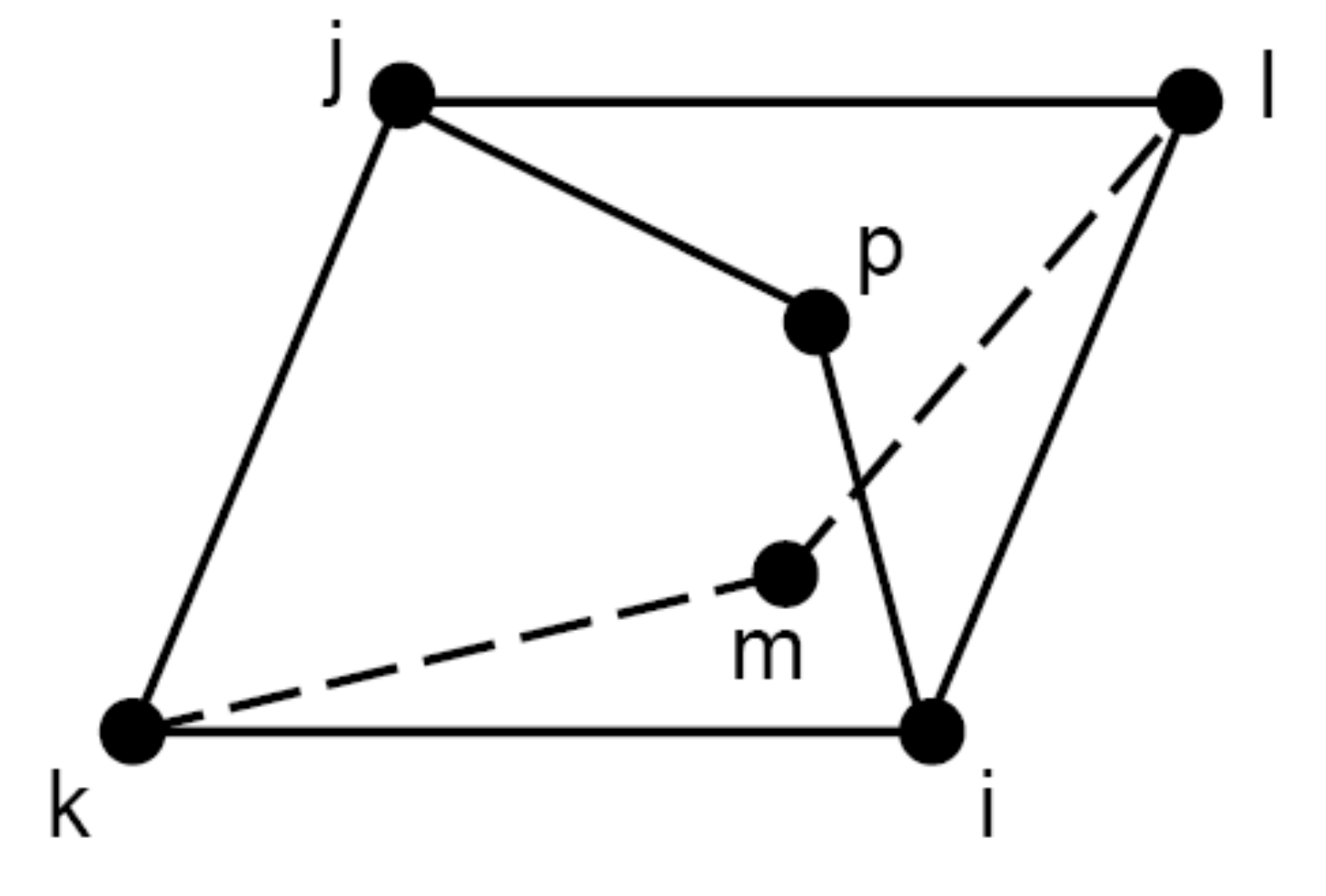}
\caption[c]{2 Points Constrained to Opposite Sides of a Closed 4 Ring}\label{2 points on opp sides of 4 ring}
\end{center}
\end{figure}

2 points confined to opposite sides of a closed 4 ring can not touch (see figure 
\ref{2 points on opp sides of 4 ring}).\\
\\
\\
\underline{This rule can be identified by the following adjacency matrix pattern:}
\begin{small}
\begin{eqnarray}
 \nonumber \text{If} \; \; \A_{ki} = \A_{il} = \A_{lj} = \A_{jk} &=& 1 \; \; \; \text{(a closed 4 ring)}\\
\nonumber \text{and} \; \; \A_{km} = \A_{ml} = \A_{jp} = \A_{pi} &=& 1 \; \; \; \text{(2 points, $m$ and $p$, are constrained to opposite sides of the 4 ring)}\\
\nonumber \Longrightarrow \D_{mp} &>& R \; \; \; \text{(i.e. $\A_{mp} \neq 1$ -- the particles can not touch)}
\end{eqnarray}
\end{small}
If $\A_{mp} = 1$, this adjacency matrix is unphysical.
\\
\\
\textbf{\underline{Rule 16:}}

\begin{figure}
 \begin{center}
 \includegraphics[width = 6in]{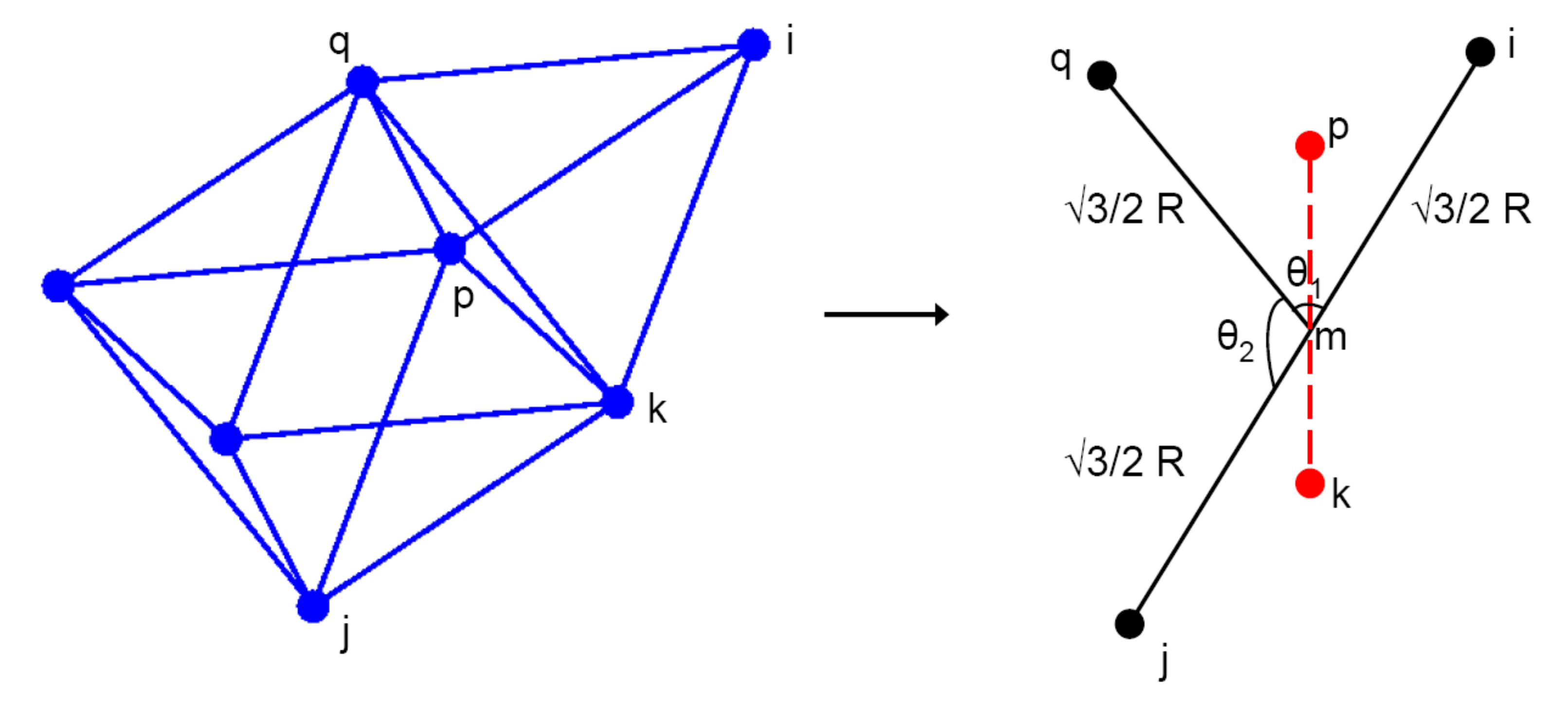}
\caption[c]{A Particle Added to an Octahedron}\label{An Octahedron With an Added Particle}
\end{center}
\end{figure}

This rule calculates the non-touching distances between one point added to an octahedron (See figure \ref{An Octahedron With an Added Particle}).  This is the distance, for example, between points $i$ and $j$, $r_{ij}$.  Particles $i$, $q$, and $j$ all lie on the $p$,$k$ intersection circle, $I_{pk}$.  Thus, we can calculate $r_{ij}$ by determining the angle between $r_{im}$ and $r_{jm}$, where $m$ is the center of $I_{pk}$.  We can determine this angle, $\theta$, simply by summing $\theta_1$ and $\theta_2$.  (We can simply sum the angles here because all of the relevent line segments, $r_{im}$, $r_{qm}$, $r_{jm}$, and $r_{ij}$, necessarily lie in the same plane, as they all lie on the plane defined by the $p,k$ intersection circle).
\par \indent $\theta_1$ is, as before, given by
\begin{eqnarray}
 \nonumber \theta_1 = \cos^{-1}\left( \frac{1}{3} \right)
\end{eqnarray}
(Note that this is the dihedral angle of a tetrahedron).  $\theta_2$ is the dihedral angle of an octahedron, and is given by
\begin{eqnarray}
 \nonumber \theta_2 = \cos^{-1}\left( -\frac{1}{3} \right)
\end{eqnarray}
Thus, $\theta$ is given by
\begin{eqnarray}
 \nonumber \theta &=& \theta_1 + \theta_2 = \cos^{-1}\left( \frac{1}{3} \right) + \cos^{-1}\left( -\frac{1}{3} \right)\\
\nonumber &=& \pi
\end{eqnarray}
Thus, $r_{im}$ and $r_{mj}$ form a straight line, and $r_{ij}$ is simply the sum of these 2 line segments:
\begin{eqnarray} \label{sqrt(3) dist rule}
r_{ij} = r_{im} + r_{mj} = \frac{\sqrt{3}}{2} R + \frac{\sqrt{3}}{2} R = \sqrt{3} R
\end{eqnarray}
Note that, due to the symmetry of this structure, the distance between $i$ and any point that $i$ does not touch is $= \sqrt{3}$ (i.e. $r_{ij} =$ the distance between $i$ and both of the unlabeled points in the octahedron also).\\
\\
\\
\underline{This rule can be identified by the following adjacency matrix pattern:}\\
The same pattern as in rule 6 (keeping in mind that the dummy variables used there do not correspond to the dummy variables shown in fig \ref{An Octahedron With an Added Particle}), and $\exists$ 1 particle, $i$, that touches any trimer within the octahedron -- for example $\A_{iq} = \A_{ik} = \A_{ip} = 1$ (where $q,p,k$ can be identified as a trimer because $\A_{qp} = \A_{qk} = \A_{pk} = 1$).  In this case $\D_{iz} = \sqrt{3}R$, where $z$ are the 3 points that are not part of the trimer to which $i$ is attached, \begin{frenchspacing}i.e.\end{frenchspacing} $z = j$ and the other 2 unmarked points in figure \ref{An Octahedron With an Added Particle}.
\\
\\
\textbf{\underline{Rule 17:}}

\begin{figure}
 \begin{center}
 \includegraphics[width = 4in]{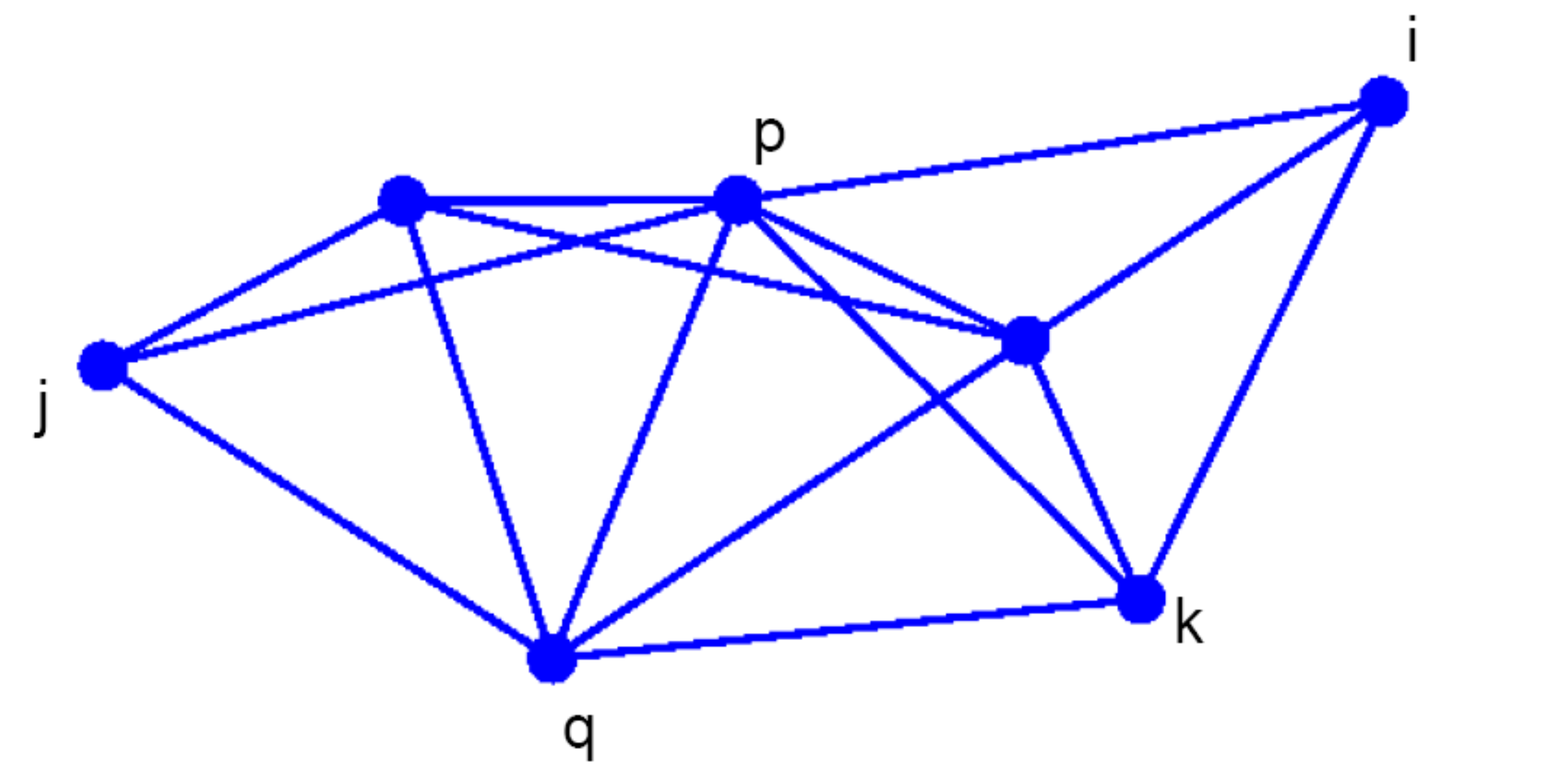}
\caption[c]{A Particle Added to the Side of 3 Connected Tetrahedron}\label{A Polytetrahedron With an Added Particle to the Side}
\end{center}
\end{figure}

The long distance, $r_{ij}$, between a particle added to the side of 3 connected tetrahedron (see figure \ref{A Polytetrahedron With an Added Particle to the Side}) can be derived via spherical trigonometry.
\begin{figure}
 \begin{center}
 \includegraphics[width = 3.2in]{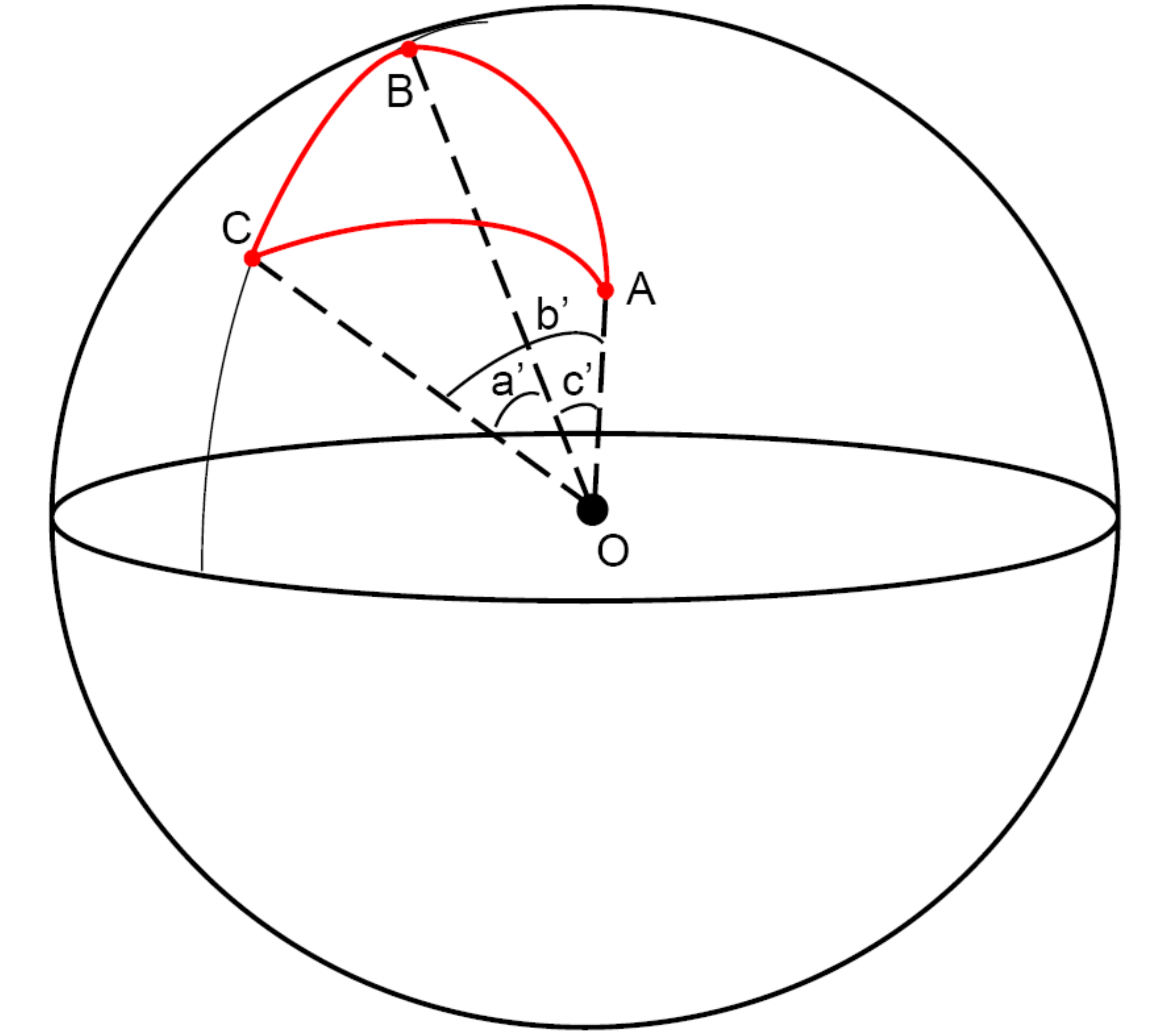}
\includegraphics[width = 2.5in]{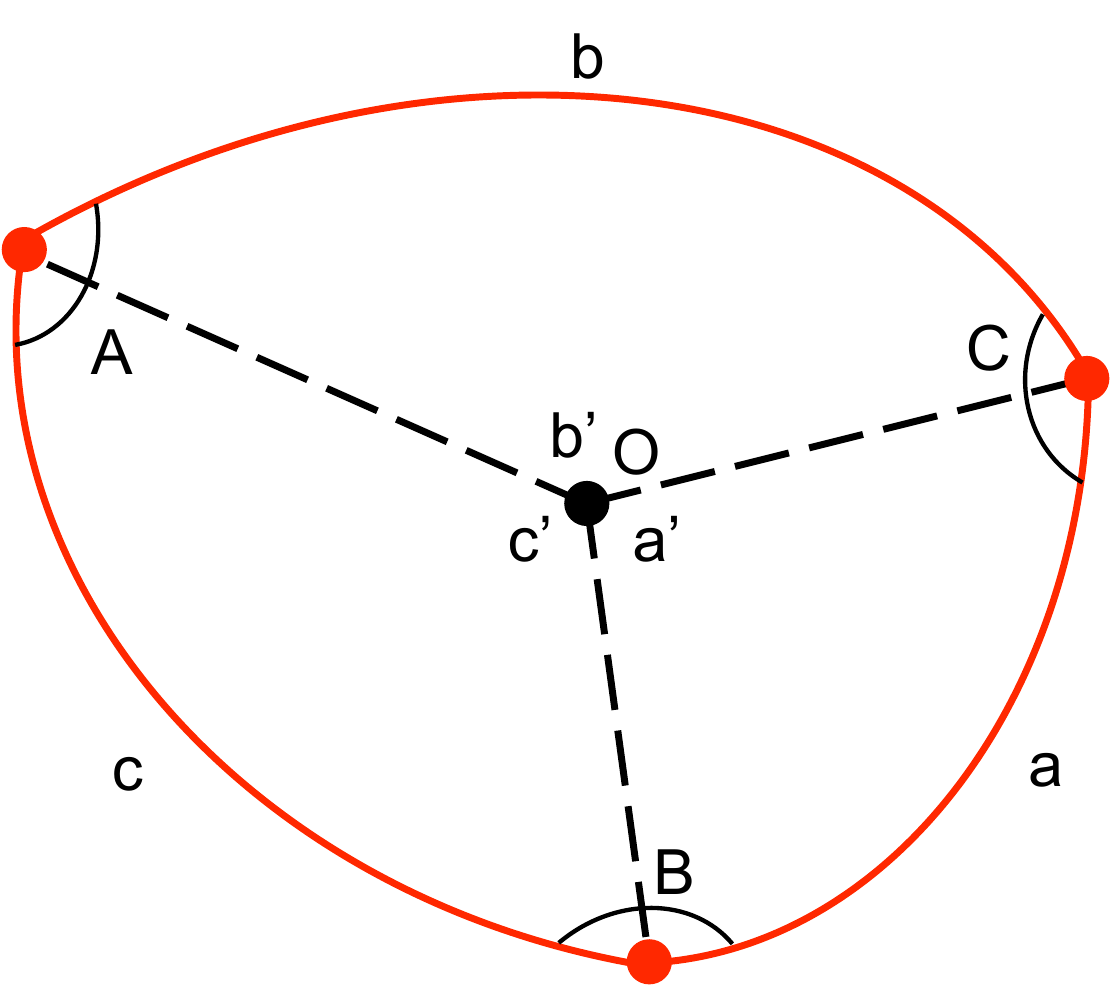}
\caption[c]{A Spherical Triangle}\label{spherical triangle}
\end{center}
\end{figure}
A spherical triangle is shown in figure \ref{spherical triangle}.  Just as in standard trigonometry, there are spherical analogues to the law of cosines and the law of sines.  In the figure, we can see that triangles $OAC$, $OAB$, and $OBC$ form the spherical triangle $ABC$.  $a$ is the arc length opposite the angle $a'$ (\begin{frenchspacing}i.e.\end{frenchspacing} the arc length associated with the side $BC$), $b$ is the arc opposite the angle $b'$ (\begin{frenchspacing}i.e.\end{frenchspacing} associated with the side $AC$, and $c$ is the arc length opposite $c'$ (associated with the side $AB$).  $a'$ is the angle $BOC$ (opposite vertice $A$), $b'$ is the angle $AOC$ (opposite vertice $B$), and $c'$ is the angle $BOA$ (opposite vertice $C$).  The arc length, $S$, is by definition given by
\begin{eqnarray}
 \nonumber S = r\theta
\end{eqnarray}
where $r$ is radius of the sphere and $\theta$ is the angle swept out by the arc length.  Therefore
\begin{eqnarray}
 \nonumber a &=& Ra'\\
\nonumber b &=& Rb'\\
\nonumber c &=& Rc'
\end{eqnarray}
where $R$ is the radius of the sphere, and $O$ is the center of the sphere.  Thus, $R$ is equal to the line segments $OA$, $OB$, and $OC$.  (Note that, by convention, $R$ is often taken to $=1$ without loss of generality, which sets the arc length equal to the angle -- so that $a = a'$ and so on...).  $A$, $B$, and $C$ are taken both to be the vertices of the spherical triangle, as well as the dihedral angle between the relevant triangles, so that
\begin{eqnarray}
\nonumber && A \text{ is the dihedral angle between triangles } OAC \text{ \& } OAB \; (\triangle OAC \text{ and } \triangle OAB)\\
\nonumber && B \text{ is the dihedral angle between } \triangle OAB \text{ \& } \triangle OBC\\
\nonumber && C \text{ is the dihedral angle between } \triangle OBC \text{ \& } \triangle OAC
\end{eqnarray}
\begin{figure}
 \begin{center}
 \includegraphics[width = 4in]{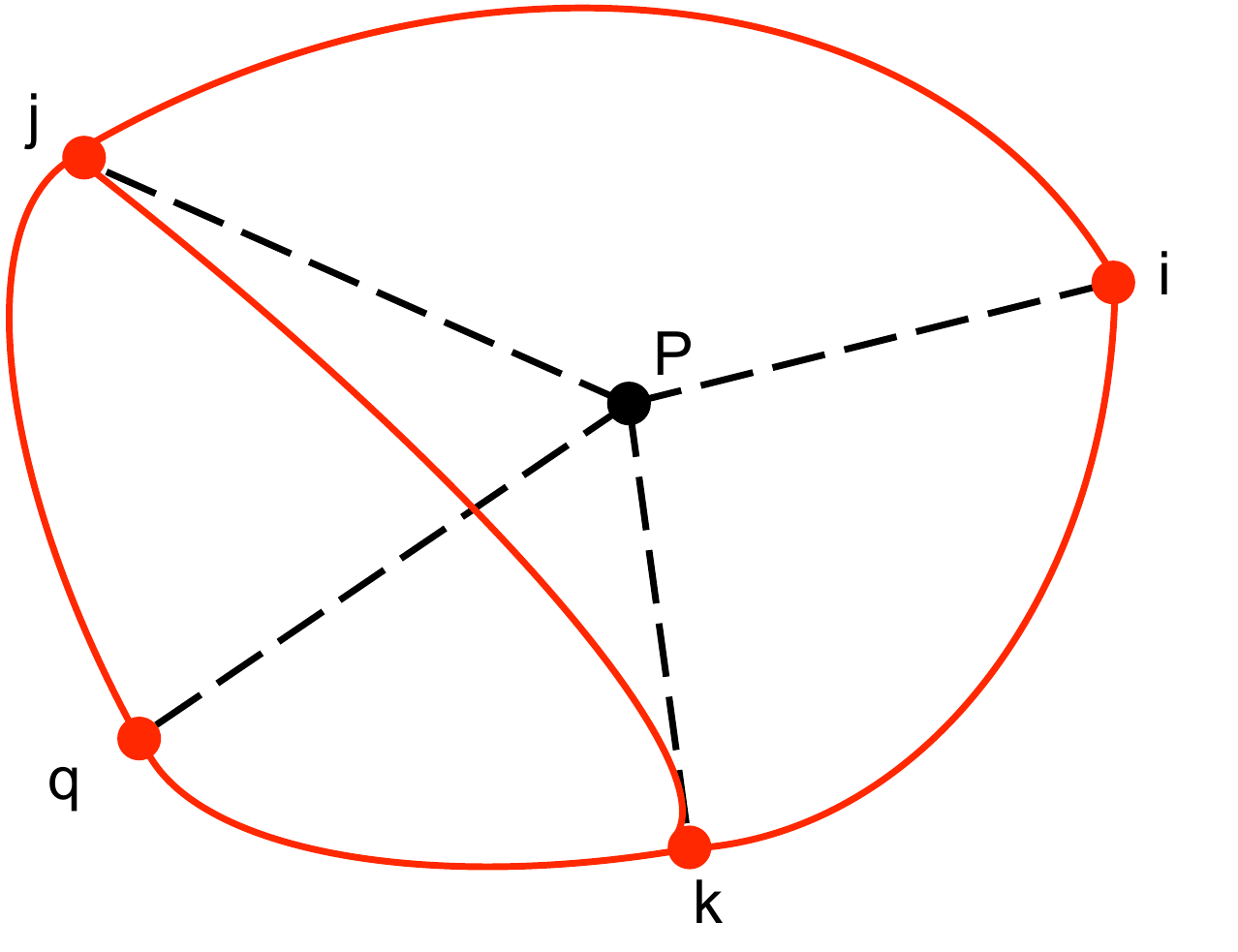}
\caption[c]{The Spherical Triangles Associated with the Conformation in Figure \ref{A Polytetrahedron With an Added Particle to the Side}}\label{Spherical Triangles For Polytetrahedron With an Added Particle to the Side}
\end{center}
\end{figure}
\par \indent The conformation we are considering here forms 2 adjacent spherical triangles (see figure \ref{Spherical Triangles For Polytetrahedron With an Added Particle to the Side}).  We know the following arc lengths of the spherical triangles and their associated line segments: $iq$, $qk$, $ik$, $jk$.  The only line segment we don't know is $ij$, for which we aim to solve.  We also know the following angles ($\angle$): $\angle ipq$, $\angle qpk$, $\angle kpj$.  The only angle we do not know is $\angle ipj$, which we will need in order to solve for $r_{ij}$.
%In addition, we know the dihedral angles between the following triangles: $\triangle ipq$ \& $\triangle qpk$, $\triangle qpk$ \& $\triangle kpj$, $\triangle ipk$ \& $\triangle pkq$.  We do not know the dihedral angles between $\triangle ipk$ and any other triangle, or between $\triangle kpj$ \& $\triangle ipj$.
\par \indent We can use the spherical law of cosines to calculate the associated dihedral angles.  We will use the following form of the spherical law of cosines
\begin{eqnarray}
\label{spher rul cos} \sin b' \sin c' \cos A = \cos a' - \cos b' \cos c'
\end{eqnarray}
Ultimately, we need to calculate the dihedral angle between $\triangle ipk$ and $\triangle jpk$, as this will allow us to back out what $\angle ipj$ is, which we can then use to calculate the unknown distance $r_{ij}$.  In spherical $\triangle ikj$, the dihedral angle between between $\triangle ipk$ and $\triangle jpk$, $\angle ipj$, and $r_{ij}$ are unknown.  This corresponds to both $A$ and $a'$ being unknown in the spherical rule of cosines (eqn \ref{spher rul cos}).  Thus we can not immediately apply this equation to solve for $r_{ij}$.  We can however use spherical $\triangle jqk$ to calculate the dihedral angle between $\triangle ipk$ and $\triangle jpk$, as this dihedral angle will $=$ the dihedral angle between $\triangle ipk$ and $\triangle kpq$ $+$ the dihedral angle between $\triangle kpq$ and $\triangle jpk$, and spherical $\triangle jqk$ will allow us to calculate the dihedral angle between $\triangle kpq$ and $\triangle jpk$.  (We already know the dihedral angle between $\triangle ipk$ and $\triangle kpq$; this is just twice the dihedral angle of a tetrahedron).

Thus, let us consider spherical $\triangle jqk$; here the vertices of the spherical triangle (fig \ref{spherical triangle}) correspond to the following points, $A = k$, $B = j$, $C = q$.  Thus, $A$ is the dihedral angle between $\triangle jpk$ and $\triangle kpq$, $a'$ is $\angle jpq$, $b'$ is $\angle kpq$, and $c'$ is $\angle jpk$.  We know all of these values except for $A$, they are:
\begin{eqnarray}
\nonumber a' &=& \angle jpq = \frac{\pi}{3} \\
\nonumber b' &=& \angle kpq = \frac{\pi}{3} \\
 \nonumber c' &=& \angle jpk = \cos^{-1} \left( \frac{-7}{18} \right)
\end{eqnarray}
the first two angles are $\pi/3$ as they are associated with equilateral triangles, and the last angle is derived from the isosceles $\triangle jpk$, where $r_{jk} = (5/3) R$ has been calculated earlier in rule 4, and $r_{jp} = r_{pk} = R$.  So, from the spherical law of cosines for spherical $\triangle jqk$, we have
\begin{eqnarray}
 \nonumber \sin \left( \frac{\pi}{3} \right) \sin \left(\cos^{-1} \left( \frac{-7}{18} \right)\right) \cos A &=& \cos \left( \frac{\pi}{3} \right) - \cos \left( \frac{\pi}{3} \right) \cos \left( \cos^{-1} \left(\frac{-7}{18} \right)\right)\\
\nonumber \frac{\sqrt{3}}{2} \frac{5 \sqrt{11}}{18} \cos A &=& \frac{1}{2} - \frac{1}{2} \frac{-7}{18}\\
\nonumber \frac{5 \sqrt{33}}{36} \cos A &=& \frac{1}{2} - \frac{-7}{36}\\
\nonumber \Longrightarrow A &=& \cos^{-1} \left( \frac{25 \cdot 36}{36 \cdot 5 \sqrt{33}} \right)\\
\nonumber A &=& \cos^{-1} \left( \frac{5}{\sqrt{33}} \right)\\
\nonumber &\approx& .514806
\end{eqnarray}
This then implies that the dihedral angle between $\triangle ipk$ and $\triangle jpk$, $\phi$, is given by
\begin{eqnarray}
 \nonumber \phi = 2 \cos^{-1}\left( \frac{1}{3} \right) + \cos^{-1}\left( \frac{5}{\sqrt{33}} \right)
\end{eqnarray}
where $2 \cos^{-1}(1/3)$ is the dihedral angle between $\triangle kpq$ and $\triangle ipk$ (which is double the dihedral angle of a tetrahedron).  Now, we can calcualte $\angle ipj$ from the spherical law of cosines for the spherical $\triangle ikj$, where the vertices of the spherical triangle (fig \ref{spherical triangle}) correspond to the following points, $A = k$, $B = j$, $C = i$.  Thus,
\begin{eqnarray}
 \nonumber b' &=& \angle ipk = \frac{\pi}{3}\\
\nonumber c' &=& \angle jpk = \cos^{-1}\left( \frac{-7}{18} \right)\\
\nonumber A &=& \text{ dihedral angle between } \triangle ipk \text{ and } \triangle jpk = \phi = 2 \cos^{-1}\left( \frac{1}{3} \right) + \cos^{-1}\left( \frac{5}{\sqrt{33}} \right)
\end{eqnarray}
and $a' = \angle ipj$ is unknown, and we will solve for it:
\begin{small}
\begin{eqnarray}
 \nonumber \sin \left( \frac{\pi}{3} \right) \sin \left( \cos^{-1}\frac{-7}{18} \right) \cos \left( 2 \cos^{-1} \left( \frac{1}{3} \right) + \cos^{-1} \left( \frac{5}{\sqrt{33}} \right) \right) &=& \cos a' - \cos \left( \frac{\pi}{3} \right) \cos \left( \cos^{-1} \frac{-7}{18} \right)\\
\nonumber \frac{\sqrt{3}}{2} \left( \frac{5 \sqrt{11}}{18} \right) \cos \left( 2 \cos^{-1} \left( \frac{1}{3} \right) + \cos^{-1} \left( \frac{5}{\sqrt{33}} \right) \right) &=& \cos a' + \frac{7}{36}\\
\nonumber \frac{5 \sqrt{33}}{36} \cos \left( 2 \cos^{-1} \left( \frac{1}{3} \right) + \cos^{-1} \left( \frac{5}{\sqrt{33}} \right) \right) &=& \cos a' + \frac{7}{36}\\
\nonumber \Longrightarrow a' &=& \cos^{-1} \left( - \frac{53}{54} \right)\\
\nonumber &\approx& 2.94884
\end{eqnarray}
\end{small}
and so, now that we have $\angle ipj$, we can use the regular law of cosines to calculate $r_{ij}$:
\begin{eqnarray}
 \nonumber \Longrightarrow r_{ij} &=& \sqrt{R^2 + R^2  - 2R^2 \cos \left( \cos^{-1} \left( - \frac{53}{54} \right) \right)}\\
\nonumber &=& \sqrt{2R^2 - 2R^2 \left( - \frac{53}{54} \right)}\\
\nonumber &=& \sqrt{2 + \frac{53}{27}}R\\
\nonumber &=& \sqrt{\frac{107}{27}} R\\
\label{1/3sqrt(107/3) dist rule} \Longrightarrow && \boxed{r_{ij} = \frac{1}{3} \sqrt{\frac{107}{3}} R}
\end{eqnarray}
\\
\underline{This rule can be identified by the following adjacency matrix pattern:}\\
$\A_{ij} = 1$, and $\exists$ 4 points that touch both $i$ and $j$; $\A_{pi} = \A_{pj} = \A_{qi} = \A_{qj} = \A_{li} = \A_{lj} = \A_{mi} = \A_{mj} = 1$.  Where 3 consecutive contacts are made amongst these 4 points; $\A_{pq} = \A_{ql} = \A_{lm} = 1$ (\begin{frenchspacing}i.e.\end{frenchspacing} the same pattern as in rule 4).  \textit{And} $\exists$ another particle that touches the top or bottom of the left or right faces (\begin{frenchspacing}i.e.\end{frenchspacing} that touches either trimer $ilm$, $ipq$, $jlm$, or $jpq$.  Take, for example, the particle $z$, where $\A_{iz} = \A_{lz} = \A_{mz} = 1$, then this rule identifies $\D_{zp} = \frac{1}{3}\sqrt{\frac{107}{3}}R$.\\
\\
Note, this packing is an enantiomer: \begin{frenchspacing}i.e.\end{frenchspacing} there exist left and right handed configurations of this structure, which are different via chiral symmetry, but are not different in terms of their distance matrices.
\\
\\
\textbf{\underline{Rule 18:}}

Following from rule 17, any adjacency matrix that implies $r_{ij} = R$ is unphysical.
\\
\\
\underline{This rule can be identified by the following adjacency matrix pattern:}\\
The same pattern as in rule 17, but where $\A_{zp} = 1$ also.
\\
\\

\section{Some Comments (Non-Uniqueness of Rules)}
\begin{itemize}
\item Rules 1 - 7 are sufficient to solve for packings of $n \leq 6$.  For $n = 7$, we also need rules 8-18.  And in fact, only rules 1, 2, 4, and 6 were actually used to calculate the 6 particle packings (as the application of these rules alone were sufficient to either completely solve or eliminate as unphysical all adjacency matrices).  And for 7 particles, only rules 1, 2, 3, 4, 5, 6, 8, 9, 11, 12, 14, 16, and 17 were used (where some combination of rules 3, 5, and 9 were used -- as the way we've implemented this in the computer code, it lumps these rules together in the output as `matrices were unphysical because they implied that there were $> 2$ points on an intersection circle
that all touched each other').

\item Rule 17 calculates the new distance generated by adding 1 particle to a packing of 6 particles that has already been solved.  From derivations of this nature, it seems that it should be possible (and it would certainly be advantageous) to derive 1 general rule that solves for all unknown distances -- \textit{at least} in the most simple case of adding 1 particle to a packing of $n$ particles that has already been solved.  The \textit{triangular bipyramid rule} derived in the paper is a general rule that can solve all iterative adjacency matrices (where an iterative matrix is one that contains only already encountered sub-matrices, \begin{frenchspacing}i.e.\end{frenchspacing} contains no new structure).  Thus, all iterative rules derived up until this point can be replaced by this rule - this is one example of the non-uniqueness of rules that can be used to derive packings; however, the non-iterative rules (those applied to physical as well as unphysical \textit{new seeds}) must still be applied.

\item  For $n \geq 8$ the triangular bipyramid rule derived in the paper was used to solve for all iterative adjacency matrices and individual geometrical rules were derived for non-iterative adjacency matrices.  The following table shows that iterative matrices account for most of the adjacency matrices, and that the number of non-iterative adjacency matrices does not become `too large' until $n = 10$.\\
\\
\\
\begin{centering}
 \begin{tabular}{|c|c|c|c|}
  \hline
& & &\\
Number of & Number of & Iterative & Non-Iterative\\
Particles & Adjacency Matrices & Adjacency Matrices & Adjacency Matrices\\
\hline
%& & &\\
6 & 4 & 3 & 1\\
\hline
7 & 29 & 26 & 3\\
\hline
8 & 438 & 437 & 1\\
\hline
9 & 13,828 & 13,823 & 5\\
\hline
10 & 750,352 & 750,258 & 94\\
\hline
 \end{tabular}
 \end{centering}
 \\
 \\
 \\
 At $n = 10$, there are 94 non-iterative adjacency matrices.  We have solved for the packings that correspond to these matrices numerically, which is why the new seed list at $n = 10$ is only a preliminary list.  Future work will derive a complete list of $n = 10$ new seeds, as well as go up to higher $n$.  One way of accomplishing this is to apply the triangular bipyramid rule to new seeds.

\item New seed rules for $n = 8$ and 9 can be derived similarly.  Please refer to the computer code for the relevant rules.  The code can be downloaded from\\
\begin{verbatim}
http://people.seas.harvard.edu/~narkus/pubs.html
\end{verbatim}
and is entitled `EvalAdjMats.c.'  In the code, solutions for 8 and 9 particle new seeds were input from Newton iterations to a precision of 16 decimal places.

\end{itemize}

%\noindent \textbf{\underline{Rule 19:}}

%n = 8 seed
%\\
%\\
%\textbf{\underline{Rule 20:}}

%n = 9 seed1
%\\
%\\
%\textbf{\underline{Rule 21:}}

%n =9 seed2
%\\
%\\
%\textbf{\underline{Rule 22:}}

%n =9 seed3
%\\
%\\
%\textbf{\underline{Rule 23:}}

%n =9 seed4
%\\
%\\
%\textbf{\underline{Rule 24:}}

%n =9 seed5
%\\
%\\

\section{Derivation of the Triangular Bipyramid Rule}

We derive this rule as follows:
In 3 dimensions, let the distance, $r_{ij}$, between 2 points, $i$ and $j$, be unknown.  The points $i$ and $j$ share a common 3 particle base, $p,k,q$.  The 5 points $i,j,k,p,q$ together form a (potentially irregular) ditetrahedron or triangular bipyramid (see figure \ref{triang dipyramid}).  This triangular bipyramid can be decomposed into 3 related tetrahedra (see figure \ref{3 spher triangles}).  Let us first derive a formula that relates these dihedral angles ($A_l, B_l, C_l$) to the angles ($a_l, b_l, c_l$); $l = 1,2,3$.

Consider the general (potentially irregular) tetrahedron depicted in figure \ref{GenTetFig}.  The dihedral angle, $A$, between planes $AOB$ and $AOC$ can be calculated using the dot product of the normals to the planes.  The normals to the planes are given by the cross products of the vectors to the vertices.  Thus, this dot product is given by
\begin{eqnarray}
\nonumber (OA \times OB) \cdot (OA \times OC) &=& (|OA||OB|\sin c)(|OA||OC| \sin b)\cos A\\
\label{dot prod eqn 1} &=& |OA|^2|OB||OC|\sin c\sin b\cos A.
\end{eqnarray}

Using a well known vector identity, we also know that
\begin{eqnarray}
\nonumber (OA \times OB) \cdot (OA \times OC) &=& OA \cdot [OB \times (OA \times OC)]\\
\nonumber &=& OA \cdot [OA (OB \cdot OC) - OC (OA \cdot OB)]\\
\nonumber &=& (OB \cdot OC) - (OA \cdot OC) (OA \cdot OB)\\
\nonumber &=& |OB||OC| \cos a - |OA||OC|\cos b |OA||OB|\cos c\\
\label{dot prod eqn 2} &=& |OB||OC|\cos a - |OA|^2|OC||OB|\cos b \cos c
\end{eqnarray}
Setting eqn \ref{dot prod eqn 1} = eqn \ref{dot prod eqn 2}, we thus have

\begin{eqnarray}
\nonumber |OA|^2|OB||OC|\sin c\sin b\cos A = |OB||OC|\cos a - |OA|^2|OC||OB|\cos b \ cos c
\end{eqnarray}
We can divide by $|OB|$ and $|OC|$, leaving us with
\begin{eqnarray}
\nonumber |OA|^2 \sin c \sin b \cos A = \cos a - |OA|^2 \cos b \cos c
\end{eqnarray}
and without loss of generality, we can always rescale things so that $|OA| = 1$, leaving us with
\begin{eqnarray}
\label{spher cos A} \sin c \sin b \cos A = \cos a - \cos b \cos c
\end{eqnarray}
Note that this formula is simply the \textit{spherical rule of cosines}; however, in this derivation we do not assume that all radial distances are equal, as is customary in the derivation of the spherical rule of cosines.
This can analogously be derived for dihedral angles $B$ and $C$, in which case we obtain
\begin{eqnarray}
\sin c \sin a \cos B &=& \cos b - \cos a \cos c\\
\sin a \sin b \cos C &=& \cos c - \cos a \cos b
\end{eqnarray}

\begin{figure}
 \begin{center}
  \includegraphics[width = 6.5in]{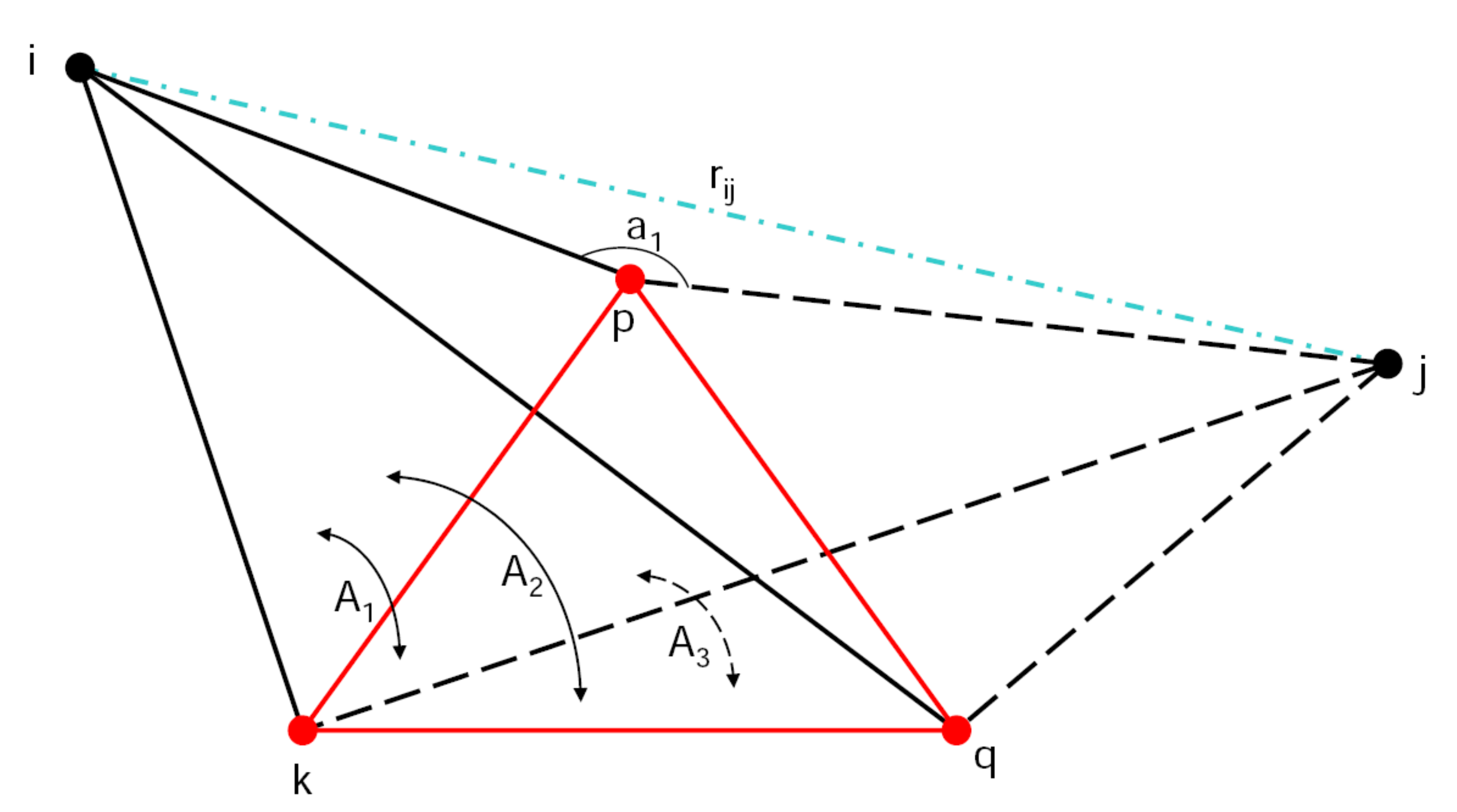}
  \caption[c]{\textbf{The Triangular Bipyramid.}\newline The triangular bipyramid (or ditetrahedron) constructed in the triangular bipyramid rule.  The center triangle ($kpq$), shown in red, corresponds to the common 3 particle base.  Particles $i$ and $j$ are related to one another through the common base.  The distance between $i$ and $j$, $r_{ij}$, shown as the dash-dot blue line, is unknown.  $a_1$ corresponds to $\angle jpi$.  $A_1$ is the dihedral angle between $\triangle ipk$ and $\triangle jpk$, $A_2$ is the dihedral angle between $\triangle ipk$ and $\triangle kpq$, and $A_3$ is the dihedral angle between $\triangle kpq$ and $\triangle jpk$.  Points $i$ and $j$ can either both lie on the same side of the base $kpq$ or each lie on opposite sides of the base (indicated by the dashed lines, that can either go into or come out of the plane).  If $i,j$ lie on the same side, then $A_1$ is equal to the difference of $A_2$ and $A_3$, and if $i,j$ lie on opposite sides of the base then $A_1$ is equal to the sum of $A_2$ and $A_3$.  When all distances other than $r_{ij}$ are known, then an analytical formula can be derived to solve $r_{ij}$.}\label{triang dipyramid}
  \end{center}
 \end{figure}
 
 %\begin{figure}
% \begin{center}
%  %\includegraphics[width = 1.9in]{GeneralSpherTriangle1.eps}
%%\includegraphics[width = 1.9in]{GeneralSpherTriangle2.eps}
%%\includegraphics[width = 1.9in]{GeneralSpherTriangle3.eps}
%\caption[c]{The 3 Spherical Triangles Constructed for the `General Rule' Used to Solve Iterative Packings}\label{3 spher triangles}
%\end{center}
%\end{figure}
 
 \begin{figure}
 \begin{center}
  \includegraphics[width = 3.8in]{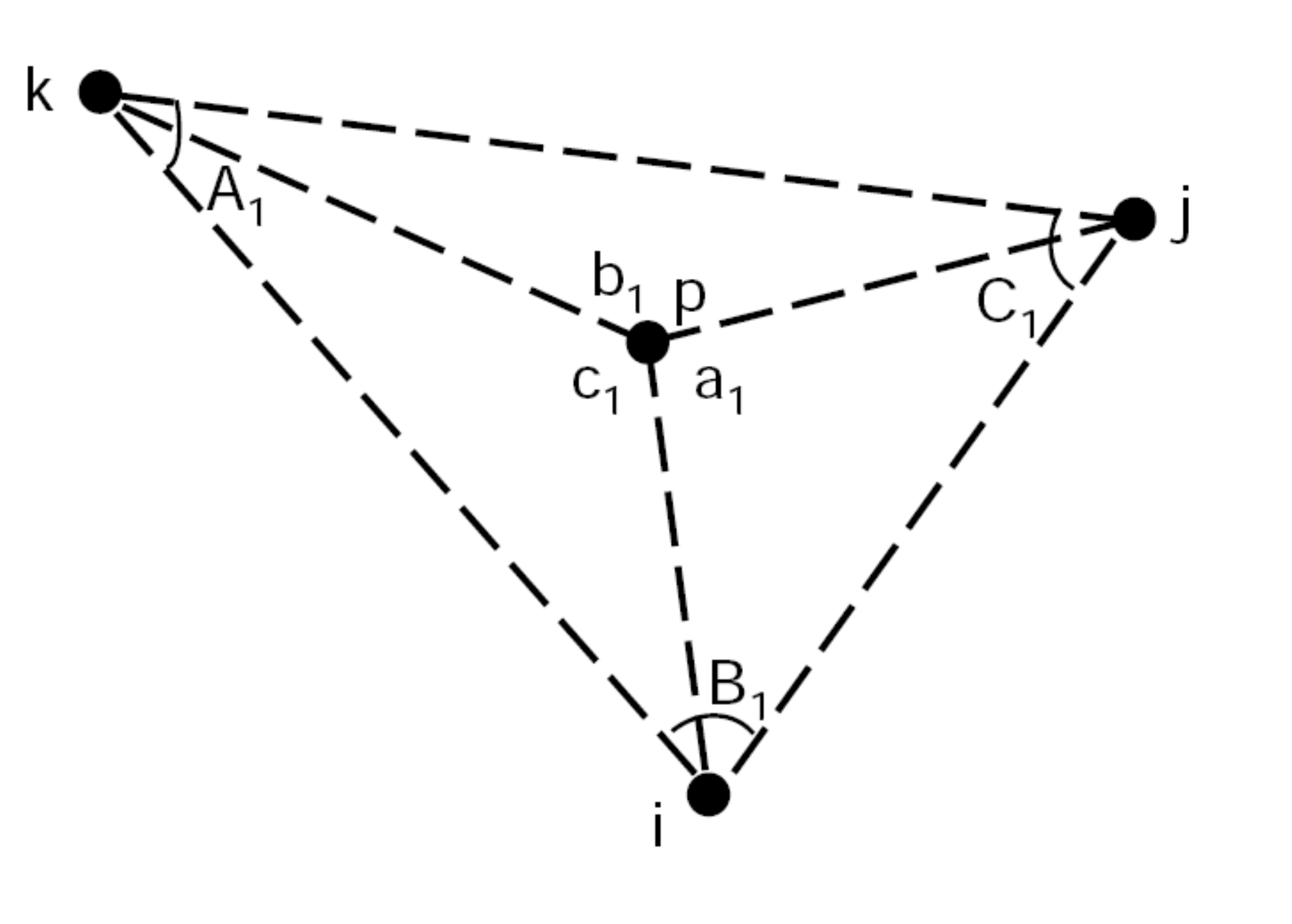}
\includegraphics[width = 3.8in]{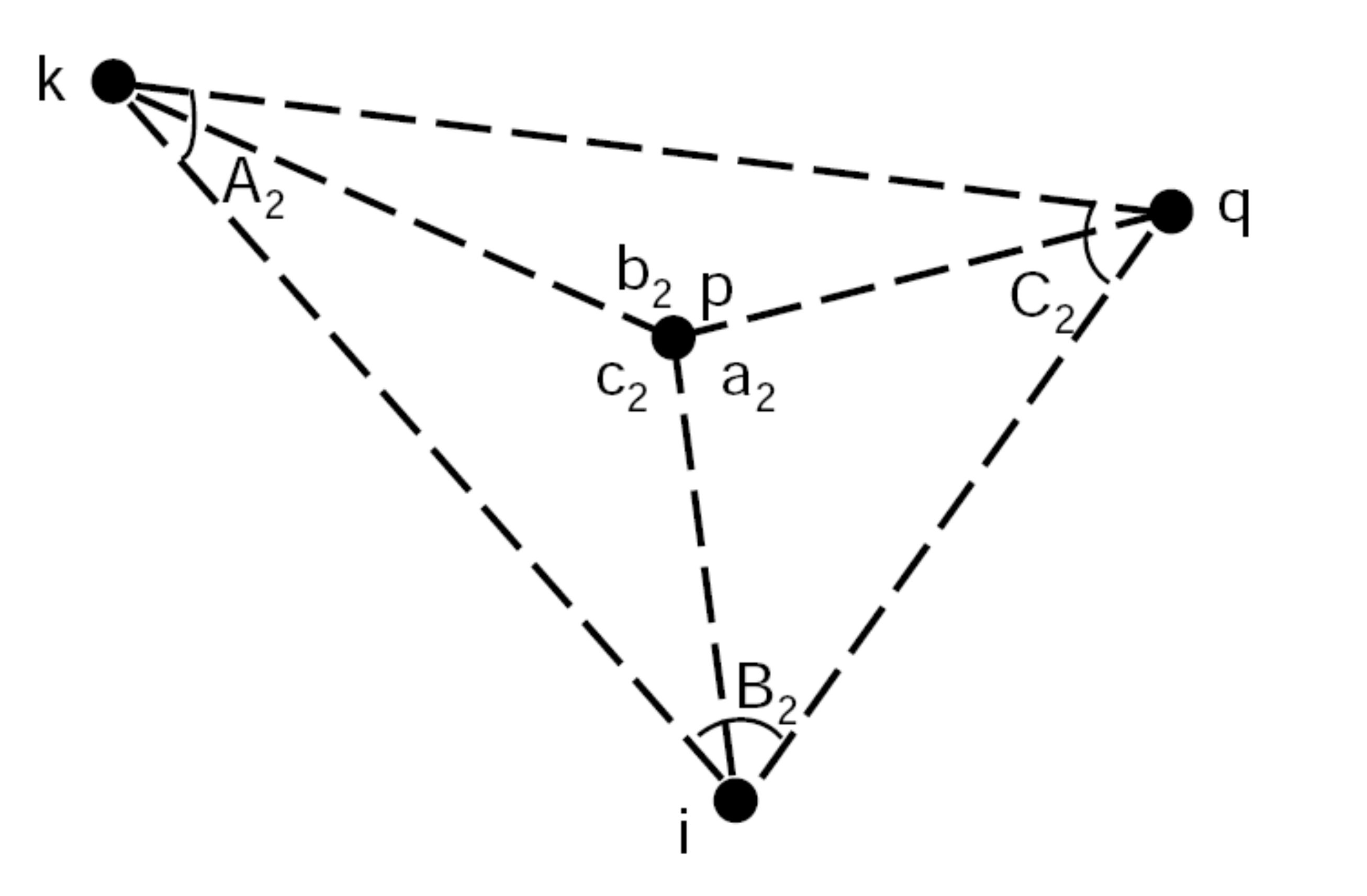}
\includegraphics[width = 3.8in]{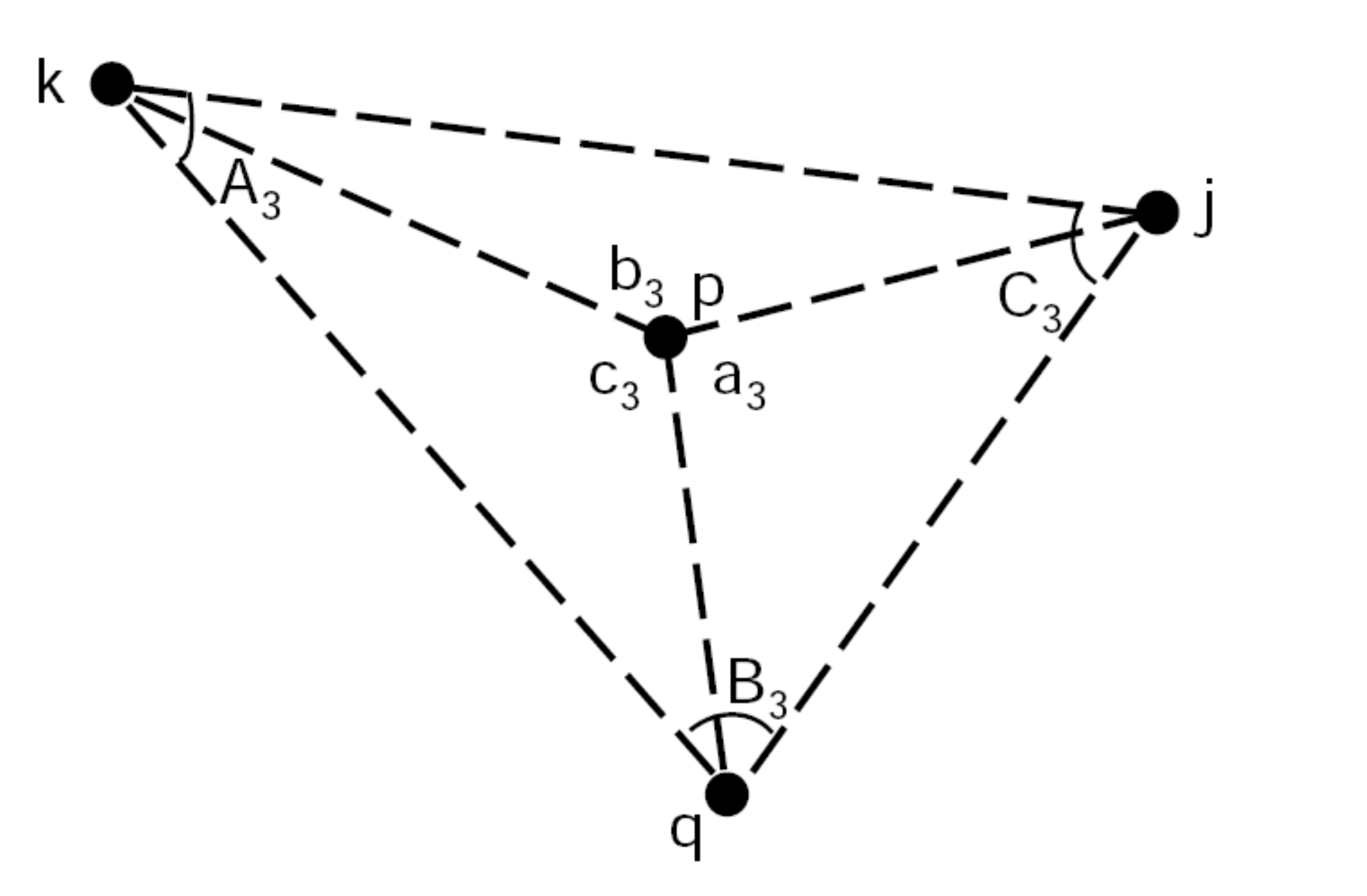}
\caption[c]{\textbf{The Tetrahedra for the Triangular Bipyramid Rule}.  The 3 tetrahedra constructed for the triangular bipyramid rule.  When applied to iterative packings, all of the sides of the tetrahedra are known except for side $ij$ (corresponding to the distance $r_{ij}$).  The known distances can be used to calculate $A_2$ and $A_3$, which in turn allows us to calculate $A_1$, and thus to solve for $r_{ij}$.}\label{3 spher triangles}
\end{center}
\end{figure}

\begin{figure}[htbp]
\begin{center}
\includegraphics[width = 0.5\textwidth]{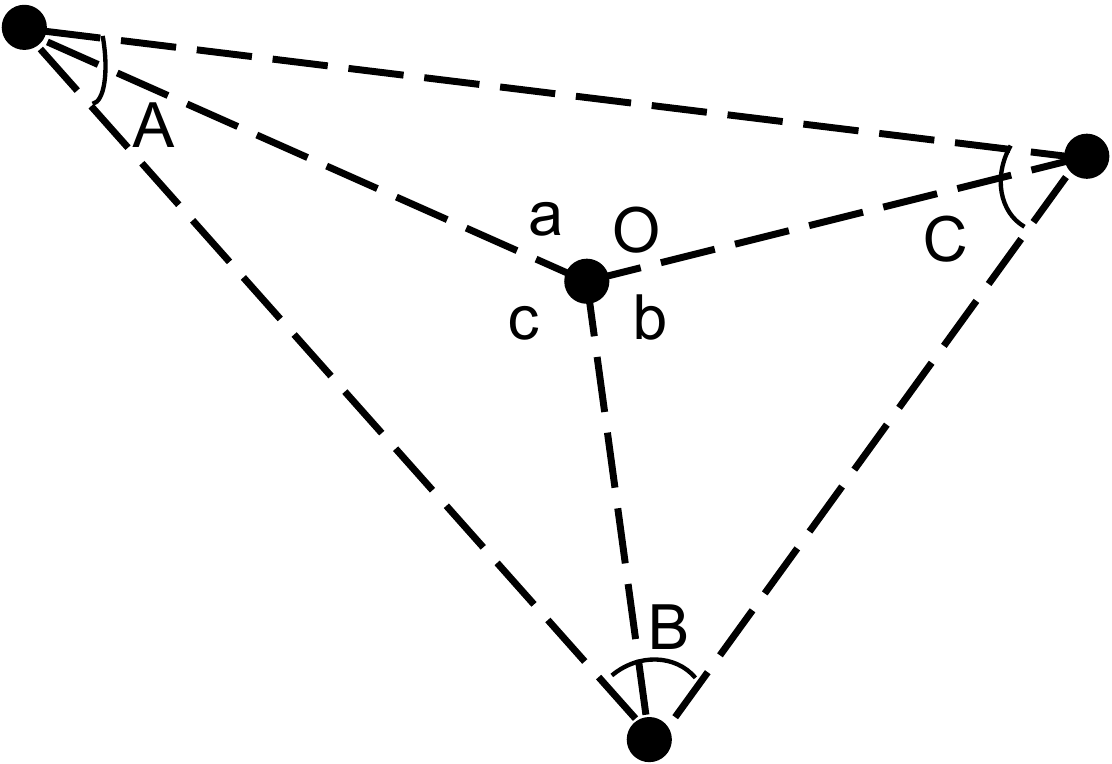}
\caption{\textbf{A General Tetrahedron.}\newline $A,B,C$ and $O$ label the points of the tetrahedron.  $A$ is also used to correspond to the dihedral angle between $\triangle AOB$ and $\triangle AOC$, $B$ corresponds to the dihedral angle between $\triangle AOB$ and $\triangle BOC$, and $C$ corresponds to the dihedral angle between $\triangle BOC$ and $\triangle AOC$.  $a$ corresponds to $\angle COA$, $b$ to $\angle COB$, and $c$ to $\angle AOB$.}
\label{GenTetFig}
\end{center}
\end{figure}

To derive the general formula for $r_{ij}$, we begin with the 3rd tetrahedron of Fig.~\ref{3 spher triangles} (that with angles $A_3, B_3, a_3, b_3,$ \begin{frenchspacing}etc.)\end{frenchspacing} and write down the expression for $A_3$ using eqn \ref{spher cos A}:
\begin{eqnarray}
\nonumber A_3 = \cos^{-1}\left( \frac{\cos a_3 - \cos b_3 \cos c_3}{\sin c_3 \sin b_3} \right)
\end{eqnarray}
Analogously, for the 2nd tetrahedron, we have
\begin{eqnarray}
\nonumber A_2 = \cos^{-1}\left( \frac{\cos a_2 - \cos b_2 \cos c_2}{\sin c_2 \sin b_2} \right)
\end{eqnarray}

$A_1$ will be either the sum or the difference of the dihedral angles $A_2$ and $A_3$ (see fig \ref{triang dipyramid}), depending on whether the points $i,j$ lie on the same or on opposite sides of the base $p,k,q$ (on the same side, $A_1$ corresponds to the difference, and on opposite sides to the sum).

$a_1$ is then given by
\begin{eqnarray}\label{a1 eqn}
a_1 = \cos^{-1}(\sin c_1 \sin b_1 \cos A_1 + \cos b_1 \cos c_1)
\end{eqnarray}
and finally, from the law of cosines, we can calculate $r_{ij}$:
\begin{eqnarray}\label{gen rule eqn rij}
r_{ij} = \sqrt{r_{ip}^2 + r_{pj}^2 - 2r_{ip}r_{pj}\cos a_1}
\end{eqnarray}

Associated with each $r_{ij}$ we have 2 possible $A_1$, and thus 2 possible solutions (similar, in principle, to one having 2 possible solutions to a quadratic equation). 

\subsection{Derivation of the Formula To be Used for New Seeds}

From setting $A_1$ either equal to the sum or the difference of $A_2$ and $A_3$, we have
\begin{eqnarray}
\nonumber \cos^{-1}\left( \frac{\cos a_1 - \cos b_1 \cos c_1}{\sin c_2 \sin b_2} \right) = \cos^{-1}\left( \frac{\cos a_2 - \cos b_2 \cos c_2}{\sin c_2 \sin b_2} \right) \pm \cos^{-1}\left( \frac{\cos a_3 - \cos b_3 \cos c_3}{\sin c_3 \sin b_3} \right)
\end{eqnarray}
writing the angles $a_i,b_i,c_i$ in terms of their relevant distances, using the law of cosines, we then have
\begin{eqnarray}
\nonumber \cos^{-1}\left(\frac{\frac{d_6^2 + d_4^2 - d_2^2}{2d_6d_4} - \frac{d_1^2+d_6^2-d_7^2}{2d_1d_6} \frac{d_1^2+d_4^2-d_5^2}{2d_1d_4}}{\sqrt{1 - \frac{(d_1^2+d_6^2-d_7^2)^2}{4d_1^2d_6^2}} \sqrt{1 - \frac{(d_1^2+d_4^2-d_5^2)^2}{4d_1^2d_4^2}}}\right) &=& \cos^{-1}\left(\frac{\frac{d_6^2 + d_8^2 - d_{10}^2}{2d_6d_8} - \frac{d_1^2+d_6^2-d_7^2}{2d_1d_6} \frac{d_1^2+d_8^2-d_9^2}{2d_1d_8}}{\sqrt{1 - \frac{(d_1^2+d_6^2-d_7^2)^2}{4d_1^2d_6^2}} \sqrt{1 - \frac{(d_1^2+d_8^2-d_9^2)^2}{4d_1^2d_8^2}}}\right) \\
\nonumber && \pm \cos^{-1}\left(\frac{\frac{d_4^2 + d_8^2 - d_3^2}{2d_4d_8} - \frac{d_1^2+d_4^2-d_5^2}{2d_1d_4} \frac{d_1^2+d_8^2-d_9^2}{2d_1d_8}}{\sqrt{1 - \frac{(d_1^2+d_4^2-d_5^2)^2}{4d_1^2d_4^2}} \sqrt{1 - \frac{(d_1^2+d_8^2-d_9^2)^2}{4d_1^2d_8^2}}}\right)
\end{eqnarray}
and solving for $d_2$ then yields
%NOW NOT INCLUDING IN MAIN TEXT - PUT THIS IN SUPP INFO
\begin{eqnarray}\label{GenDistEqnNewSeeds}
\nonumber d_2 = \sqrt{\begin{matrix}-2d_6d_4\sqrt{1 - \frac{(d_1^2 + d_6^2 - d_7^2)^2}{4d_1^2d_6^2}}\sqrt{1 - \frac{(d_1^2 + d_4^2 - d_5^2)^2}{4d_1^2d_4^2}} \cos \left[  \cos^{-1}\left( \frac{\frac{d_6^2 + d_8^2 - d_{10}^2}{2d_6d_8} - \frac{(d_1^2 + d_6^2 - d_7^2)(d_1^2 + d_8^2 - d_9^2)}{4d_1^2d_6d_8}}{\sqrt{1 - \frac{(d_1^2 + d_6^2 - d_7^2)^2}{4d_1^2d_6^2}}\sqrt{1 - \frac{(d_1^2 + d_8^2 - d_9^2)^2}{4d_1^2d_8^2}}}\right) \right. \\
\\
\pm \left. \cos^{-1}\left( \frac{\frac{d_4^2 + d_8^2 - d_3^2}{2d_4d_8} - \frac{(d_1^2 + d_4^2 - d_5^2)(d_1^2 + d_8^2 - d_9^2)}{4d_1^2d_4d_8}}{\sqrt{1 - \frac{(d_1^2 + d_4^2 - d_5^2)^2}{4d_1^2d_4^2}}\sqrt{1 - \frac{(d_1^2 + d_8^2 - d_9^2)^2}{4d_1^2d_8^2}}}\right) \right] - \frac{(d_1^2 + d_6^2 - d_7^2)(d_1^2 + d_4^2 - d_5^2)}{2d_1^2} + d_6^2 + d_4^2\end{matrix}}
\end{eqnarray}
\textbf{{\large \begin{center}APPENDIX B: PACKINGS OF $8 \leq n \leq 10$ SPHERES\end{center}}}
\vspace{.5in}
\setcounter{section}{0}
\setcounter{figure}{0}

\raggedbottom
Here we include the complete set of packings for $8 \leq n \leq 10$ particles, derived via the method specified in the paper.  (Packings of $n \leq 7$ particles can be found in the paper).  Throughout, $R$ corresponds to twice the radius of the particles, and without loss of generality can be normalized to 1, if desired.  In all instances, a `$^*$' appears in front of the 2nd moment that corresponds to the minimum of the 2nd moment of $n$ particles.  If a packing is denoted as chiral, then it has 2 distinct states.

Note that the code used to generate these results as well as the raw output from the code (in which case the results are output to 16 decimal places instead of the 4 \begin{frenchspacing}d.p.\end{frenchspacing} shown here) can be downloaded from the following website: http://people.seas.harvard.edu/$\sim$narkus/.\\
\\
\noindent \underline{Notation:}\\
$\mathcal{A}$: Adjacency Matrix.\\
$\mathcal{D}$: Distance Matrix.\\
$\mathcal{C}$: Coordinates of a packing.\\
$\phi$: Point group of a packing.\\
$\sigma$: Symmetry number of a packing.

%\section{$n \leq 6$}

%\section{$n=7$}

\section{Packings of $n = 8$ Particles}

Here, we include the 13 packings of $n = 8$ particles.  The `Special Properties' column denotes whether the packing is chiral or corresponds to a new seed.  If the special properties column is blank, then the packing is neither.  1 of these packings (packing 13) corresponds to a new seed, and 3 of the packings are chiral.  The minimum of the 2nd moment corresponds to packing 13, graph 408.

\noindent Packing 1 (Graph 27):
\[
\begin{array}{l}
\begin{tiny}\begin{array}{lllll}
&\mathcal{A}:  \left( \begin{matrix} 0&0&0&0&1&1&1&0\\
0&0&0&0&1&1&0&1\\
0&0&0&0&1&0&1&1\\
0&0&0&0&0&1&1&1\\
1&1&1&0&0&1&1&1\\
1&1&0&1&1&0&1&1\\
1&0&1&1&1&1&0&1\\
0&1&1&1&1&1&1&0 \end{matrix} \right)&&&\\
&&&&\\
&\mathcal{D}:\left( \begin{matrix} 0&1.6667&1.6667&1.6667&1&1&1&1.6330\\
1.6667&0&1.6667&1.6667&1&1&1.6330&1\\
1.6667&1.6667&0&1.6667&1&1.6330&1&1\\
1.6667&1.6667&1.6667&0&1.6330&1&1&1\\
1&1&1&1.6330&0&1&1&1\\
1&1&1.6330&1&1&0&1&1\\
1&1.6330&1&1&1&1&0&1\\
1.6330&1&1&1&1&1&1&0 \end{matrix} \right) R  & &&\end{array}
\begin{array}{l}
\mathcal{C}: \left(\begin{matrix}  0\\0\\0\\0\\-1.6667\\0\\-1.4434\\-0.8333\\0\\-0.4811\\-0.8333\\-1.3608\\-0.4811\\-0.8333\\0.2722\\0\\-0.8333\\-0.5443\\-0.7698\\-0.3333\\-0.5443\\-0.7698\\-1.3333\\-0.5443 \end{matrix}\right)R\end{array}\end{tiny} \\
\\
\begin{array}{l}
 \includegraphics[width = 1in]{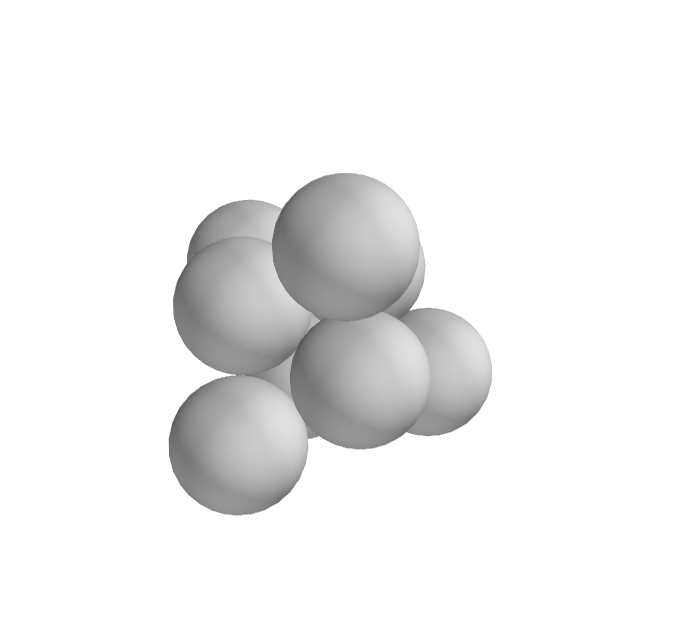} \; \; \; \includegraphics[width = 1in]{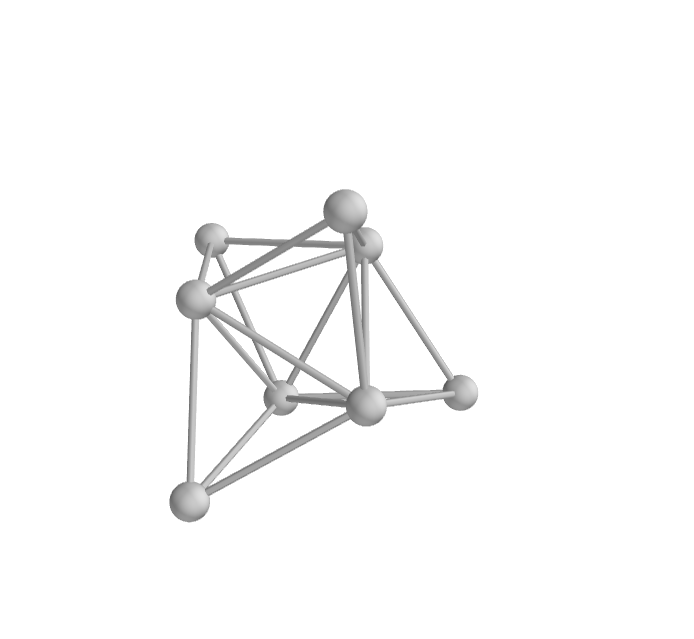}\\
 \begin{tabular}{|c|c|c|c|}
\hline
2nd Moment&$\phi$&$\sigma$&Special Properties\\
\hline
$5.66667 R^2$ & $T_d$ & 12 & \\
\hline
\end{tabular}
\end{array}
\end{array}
\]
\noindent Packing 2 (Graph 77):
\[
\begin{array}{l}
\begin{tiny}\begin{array}{lllll}
&\mathcal{A}:  \left( \begin{matrix} 0&0&0&1&0&1&1&1\\
0&0&0&0&1&1&1&1\\
0&0&0&0&1&0&1&1\\
1&0&0&0&0&1&0&1\\
0&1&1&0&0&0&1&1\\
1&1&0&1&0&0&1&1\\
1&1&1&0&1&1&0&1\\
1&1&1&1&1&1&1&0 \end{matrix} \right)&&&\\
&&&&\\
&\mathcal{D}:\left( \begin{matrix} 0&1.6330&1.0887&1&1.6667&1&1&1\\
1.6330&0&1.6330&1.6667&1&1&1&1\\
1.0887&1.6330&0&1.7249&1&1.6667&1&1\\
1&1.6667&1.7249&0&1.9907&1&1.6330&1\\
1.6667&1&1&1.9907&0&1.6330&1&1\\
1&1&1.6667&1&1.6330&0&1&1\\
1&1&1&1.6330&1&1&0&1\\
1&1&1&1&1&1&1&0 \end{matrix} \right) R  & &&\end{array}
\begin{array}{l}
\mathcal{C}: \left(\begin{matrix}  0\\0\\0\\0\\1.6330\\0\\1.0264\\0.3629\\0\\-0.4811\\0.2722\\0.8333\\0.9623\\1.3608\\-0.0000\\-0.5774\\0.8165\\0\\0.2887\\0.8165\\-0.5000\\0.2887\\0.8165\\0.5000 \end{matrix}\right)R\end{array}\end{tiny} \\
\\
\begin{array}{l}
 \includegraphics[width = 1in]{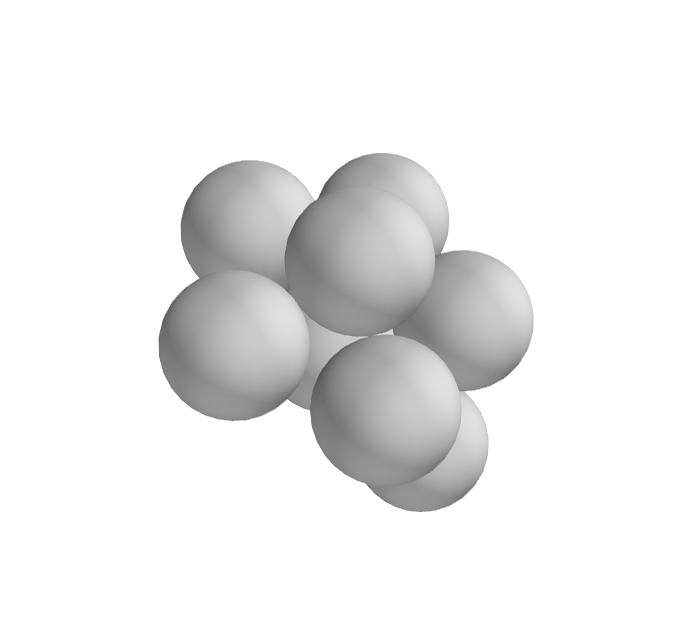} \; \; \; \includegraphics[width = 1in]{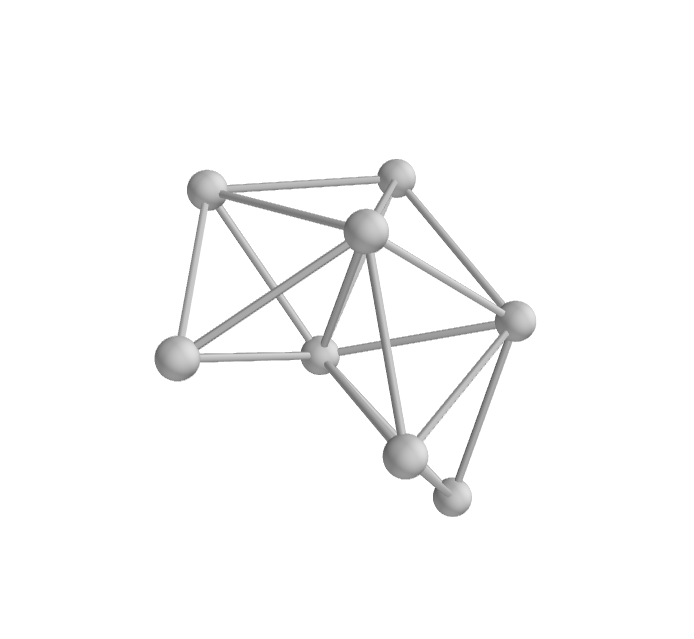}\\
 \begin{tabular}{|c|c|c|c|}
\hline
2nd Moment&$\phi$&$\sigma$&Special Properties\\
\hline
$5.64043 R^2$ & $C_1$ & 1 & chiral \\
\hline
\end{tabular}
\end{array}
\end{array}
\]
\noindent Packing 3 (Graph 82):
\[
\begin{array}{l}
\begin{tiny}\begin{array}{lllll}
&\mathcal{A}:  \left( \begin{matrix} 0&0&0&1&0&1&1&1\\
0&0&0&0&1&1&0&1\\
0&0&0&0&1&0&1&1\\
1&0&0&0&0&0&1&1\\
0&1&1&0&0&1&1&1\\
1&1&0&0&1&0&1&1\\
1&0&1&1&1&1&0&1\\
1&1&1&1&1&1&1&0 \end{matrix} \right)&&&\\
&&&&\\
&\mathcal{D}:\left( \begin{matrix} 0&1.6667&1.6667&1&1.6330&1&1&1\\
1.6667&0&1.6667&1.9907&1&1&1.6330&1\\
1.6667&1.6667&0&1.0887&1&1.6330&1&1\\
1&1.9907&1.0887&0&1.6667&1.6330&1&1\\
1.6330&1&1&1.6667&0&1&1&1\\
1&1&1.6330&1.6330&1&0&1&1\\
1&1.6330&1&1&1&1&0&1\\
1&1&1&1&1&1&1&0 \end{matrix} \right) R  & &&\end{array}
\begin{array}{l}
\mathcal{C}: \left(\begin{matrix}  0\\0\\0\\0\\1.6667\\0\\1.4434\\0.8333\\0\\0.9302\\-0.0556\\-0.3629\\0.7698\\1.3333\\0.5443\\-0.0962\\0.8333\\0.5443\\0.7698\\0.3333\\0.5443\\0.4811\\0.8333\\-0.2722 \end{matrix}\right)R\end{array}\end{tiny} \\
\\
\begin{array}{l}
 \includegraphics[width = 1in]{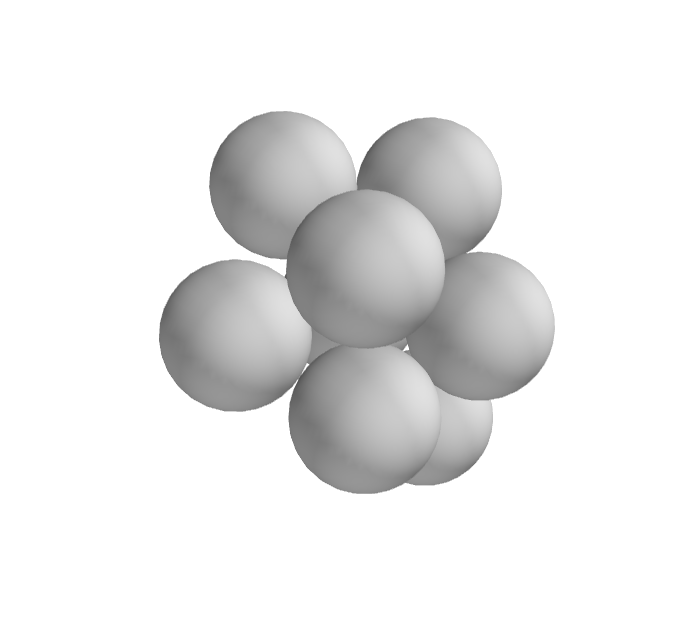} \; \; \; \includegraphics[width = 1in]{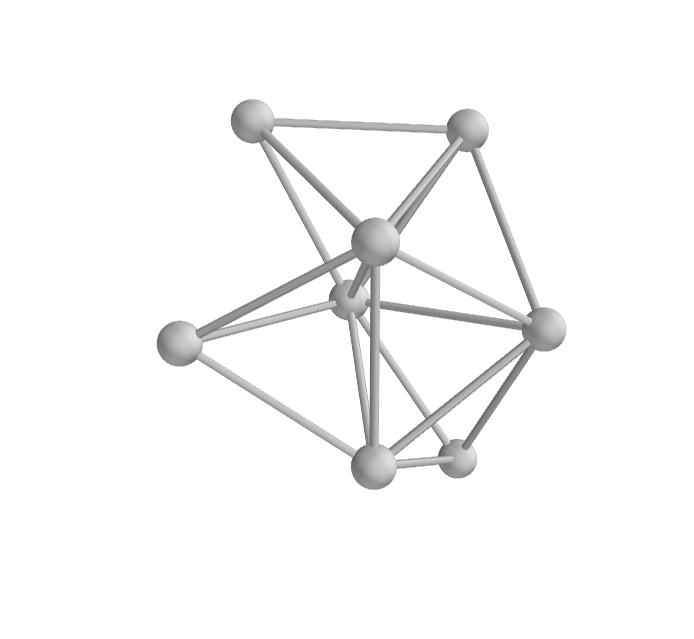}\\
 \begin{tabular}{|c|c|c|c|}
\hline
2nd Moment&$\phi$&$\sigma$&Special Properties\\
\hline
$5.61574 R^2$ & $C_1$ & 1 & chiral \\
\hline
\end{tabular}
\end{array}
\end{array}
\]
\noindent Packing 4 (Graph 107):
\[
\begin{array}{l}
\begin{tiny}\begin{array}{lllll}
&\mathcal{A}:  \left( \begin{matrix} 0&0&0&1&0&1&1&1\\
0&0&0&0&1&1&1&0\\
0&0&0&0&1&1&0&1\\
1&0&0&0&0&0&1&1\\
0&1&1&0&0&1&1&1\\
1&1&1&0&1&0&1&1\\
1&1&0&1&1&1&0&1\\
1&0&1&1&1&1&1&0 \end{matrix} \right)&&&\\
&&&&\\
&\mathcal{D}:\left( \begin{matrix} 0&1.6667&1.6667&1&1.6330&1&1&1\\
1.6667&0&1.6667&1.9907&1&1&1&1.6330\\
1.6667&1.6667&0&1.9907&1&1&1.6330&1\\
1&1.9907&1.9907&0&1.6667&1.6330&1&1\\
1.6330&1&1&1.6667&0&1&1&1\\
1&1&1&1.6330&1&0&1&1\\
1&1&1.6330&1&1&1&0&1\\
1&1.6330&1&1&1&1&1&0 \end{matrix} \right) R  & &&\end{array}
\begin{array}{l}
\mathcal{C}: \left(\begin{matrix}  0\\0\\0\\0\\1.6667\\0\\-1.4434\\0.8333\\0\\0\\-0.0556\\-0.9979\\-0.7698\\1.3333\\-0.5443\\-0.4811\\0.8333\\0.2722\\0\\0.8333\\-0.5443\\-0.7698\\0.3333\\-0.5443 \end{matrix}\right)R\end{array}\end{tiny} \\
\\
\begin{array}{l}
 \includegraphics[width = 1in]{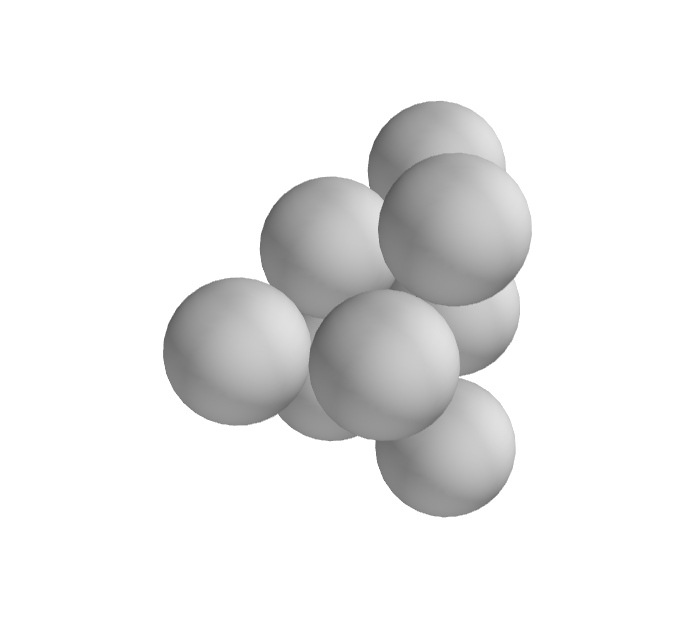} \; \; \; \includegraphics[width = 1in]{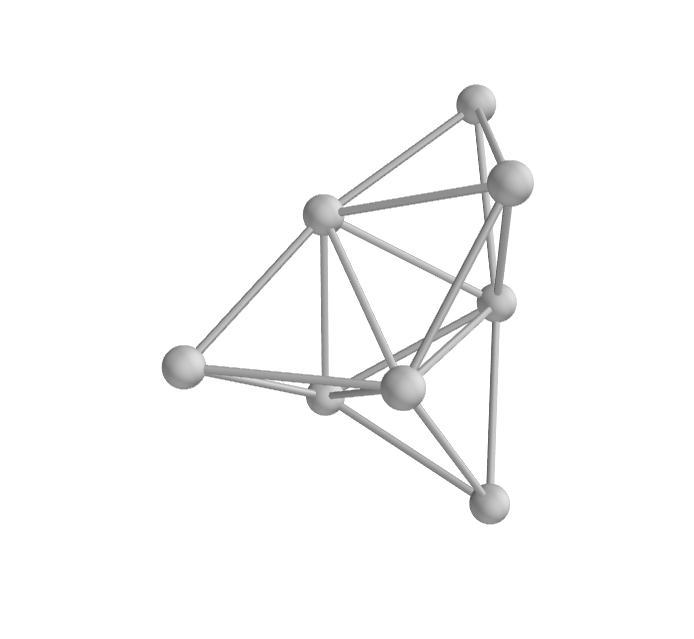}\\
 \begin{tabular}{|c|c|c|c|}
\hline
2nd Moment&$\phi$&$\sigma$&Special Properties\\
\hline
$5.96296 R^2$ & $C_s$ & 1 & \\
\hline
\end{tabular}
\end{array}
\end{array}
\]
\noindent Packing 5 (Graph 204):
\[
\begin{array}{l}
\begin{tiny}\begin{array}{lllll}
&\mathcal{A}:  \left( \begin{matrix} 0&0&0&1&1&1&1&0\\
0&0&0&0&1&0&1&1\\
0&0&0&0&0&1&1&1\\
1&0&0&0&1&1&0&1\\
1&1&0&1&0&0&1&1\\
1&0&1&1&0&0&1&1\\
1&1&1&0&1&1&0&1\\
0&1&1&1&1&1&1&0 \end{matrix} \right)&&&\\
&&&&\\
&\mathcal{D}:\left( \begin{matrix} 0&1.7321&1.7321&1&1&1&1&1.4142\\
1.7321&0&1.4142&1.7321&1&1.7321&1&1\\
1.7321&1.4142&0&1.7321&1.7321&1&1&1\\
1&1.7321&1.7321&0&1&1&1.4142&1\\
1&1&1.7321&1&0&1.4142&1&1\\
1&1.7321&1&1&1.4142&0&1&1\\
1&1&1&1.4142&1&1&0&1\\
1.4142&1&1&1&1&1&1&0 \end{matrix} \right) R  & &&\end{array}
\begin{array}{l}
\mathcal{C}: \left(\begin{matrix}  0\\0\\0\\0\\-1.7321\\0\\1.2910\\-1.1547\\0\\0.1291\\-0.2887\\0.9487\\-0.3873\\-0.8660\\0.3162\\0.9037\\-0.2887\\0.3162\\0.3873\\-0.8660\\-0.3162\\0.5164\\-1.1547\\0.6325 \end{matrix}\right)R\end{array}\end{tiny} \\
\\
\begin{array}{l}
 \includegraphics[width = 1in]{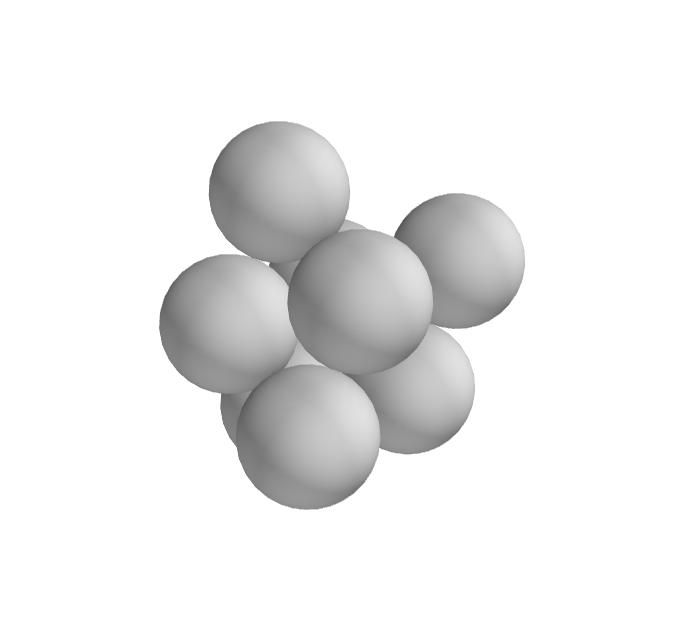} \; \; \; \includegraphics[width = 1in]{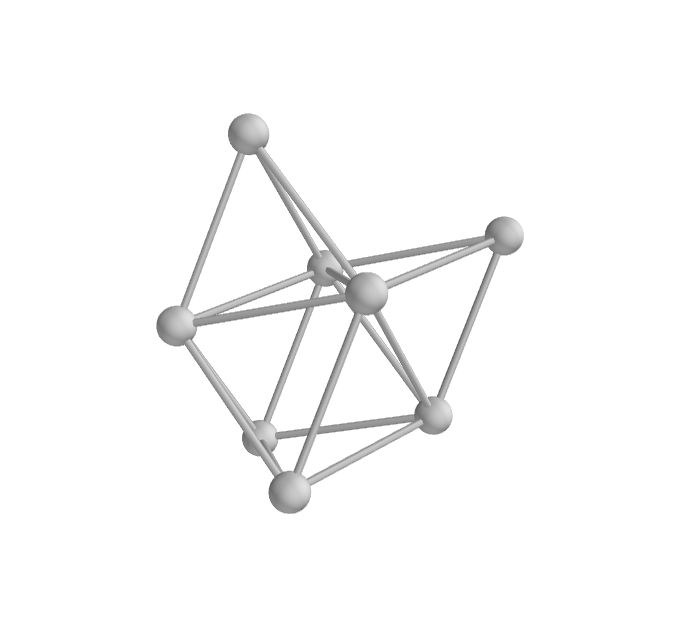}\\
 \begin{tabular}{|c|c|c|c|}
\hline
2nd Moment&$\phi$&$\sigma$&Special Properties\\
\hline
$5.50000 R^2$ & $C_{2 v}$ & 2 & \\
\hline
\end{tabular}
\end{array}
\end{array}
\]
\noindent Packing 6 (Graph 224):
\[
\begin{array}{l}
\begin{tiny}\begin{array}{lllll}
&\mathcal{A}:  \left( \begin{matrix} 0&0&0&1&1&0&1&1\\
0&0&0&1&0&1&1&1\\
0&0&0&0&1&1&1&1\\
1&1&0&0&0&0&1&1\\
1&0&1&0&0&0&1&1\\
0&1&1&0&0&0&0&1\\
1&1&1&1&1&0&0&1\\
1&1&1&1&1&1&1&0 \end{matrix} \right)&&&\\
&&&&\\
&\mathcal{D}:\left( \begin{matrix} 0&1.6330&1.6330&1&1&1.9868&1&1\\
1.6330&0&1.0887&1&1.6667&1&1&1\\
1.6330&1.0887&0&1.6667&1&1&1&1\\
1&1&1.6667&0&1.6330&1.6859&1&1\\
1&1.6667&1&1.6330&0&1.6859&1&1\\
1.9868&1&1&1.6859&1.6859&0&1.6059&1\\
1&1&1&1&1&1.6059&0&1\\
1&1&1&1&1&1&1&0 \end{matrix} \right) R  & &&\end{array}
\begin{array}{l}
\mathcal{C}: \left(\begin{matrix}  0\\0\\0\\0\\1.6330\\0\\-1.0264\\1.2701\\0\\0.5774\\0.8165\\0\\-0.9623\\0.2722\\0\\-0.6077\\1.7189\\-0.7895\\-0.2887\\0.8165\\0.5000\\-0.2887\\0.8165\\-0.5000 \end{matrix}\right)R\end{array}\end{tiny} \\
\\
\begin{array}{l}
 \includegraphics[width = 1in]{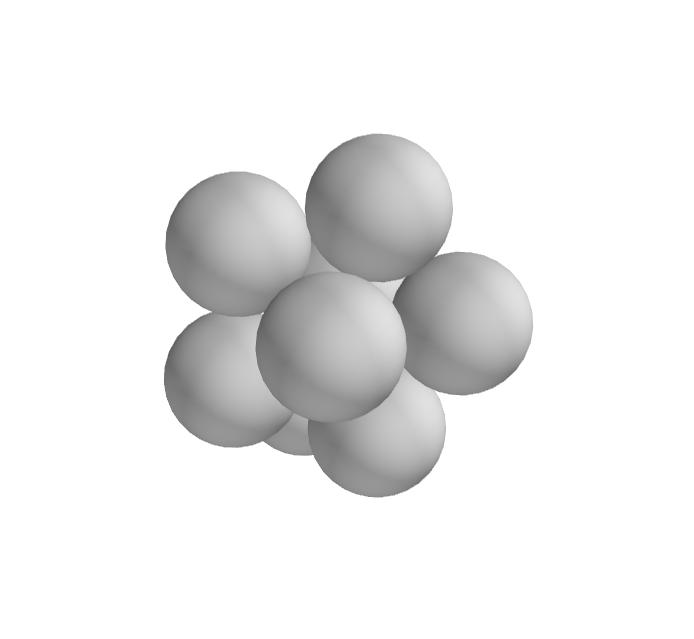} \; \; \; \includegraphics[width = 1in]{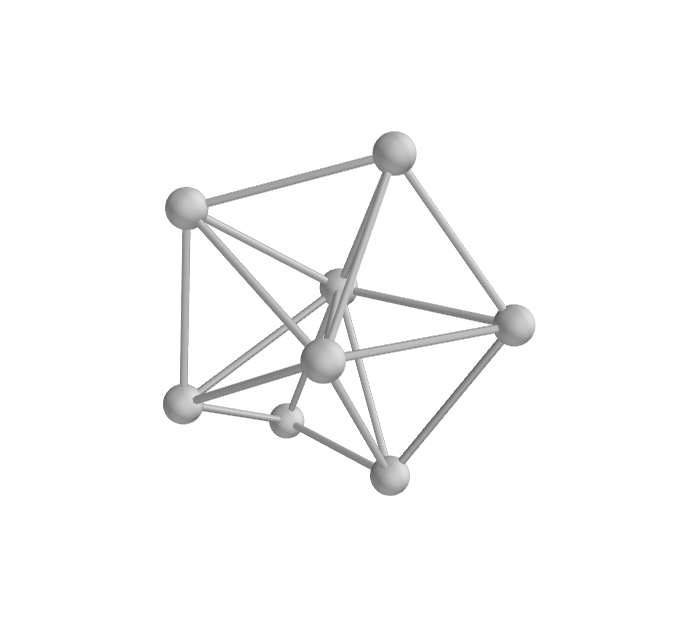}\\
 \begin{tabular}{|c|c|c|c|}
\hline
2nd Moment&$\phi$&$\sigma$&Special Properties\\
\hline
$5.61891 R^2$ & $C_s$ & 1 & \\
\hline
\end{tabular}
\end{array}
\end{array}
\]
\noindent Packing 7 (Graph 271):
\[
\begin{array}{l}
\begin{tiny}\begin{array}{lllll}
&\mathcal{A}:  \left( \begin{matrix} 0&0&0&1&1&0&1&1\\
0&0&0&1&0&1&0&1\\
0&0&0&0&1&1&1&1\\
1&1&0&0&0&1&1&1\\
1&0&1&0&0&0&1&1\\
0&1&1&1&0&0&1&1\\
1&0&1&1&1&1&0&0\\
1&1&1&1&1&1&0&0 \end{matrix} \right)&&&\\
&&&&\\
&\mathcal{D}:\left( \begin{matrix} 0&1.6779&1.6180&1&1&1.6180&1&1\\
1.6779&0&1.6779&1&1.9843&1&1.6532&1\\
1.6180&1.6779&0&1.6180&1&1&1&1\\
1&1&1.6180&0&1.6180&1&1&1\\
1&1.9843&1&1.6180&0&1.6180&1&1\\
1.6180&1&1&1&1.6180&0&1&1\\
1&1.6532&1&1&1&1&0&1.0515\\
1&1&1&1&1&1&1.0515&0 \end{matrix} \right) R  & &&\end{array}
\begin{array}{l}
\mathcal{C}: \left(\begin{matrix}  0\\0\\0\\0\\-1.6779\\0\\-1.4175\\-0.7801\\0\\0.1090\\-0.8390\\-0.5331\\-0.9435\\0\\0.3295\\-0.7671\\-1.3211\\-0.5331\\-0.7459\\-0.3226\\-0.5827\\-0.4617\\-0.8390\\0.2880 \end{matrix}\right)R\end{array}\end{tiny} \\
\\
\begin{array}{l}
 \includegraphics[width = 1in]{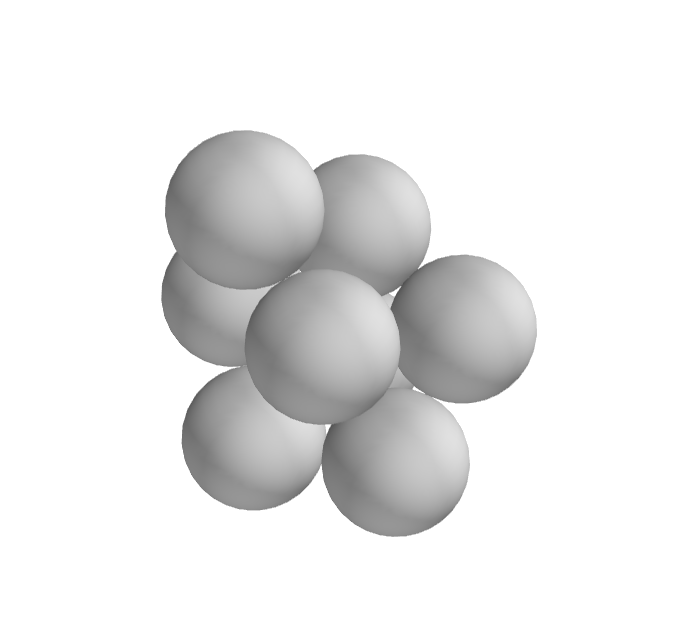} \; \; \; \includegraphics[width = 1in]{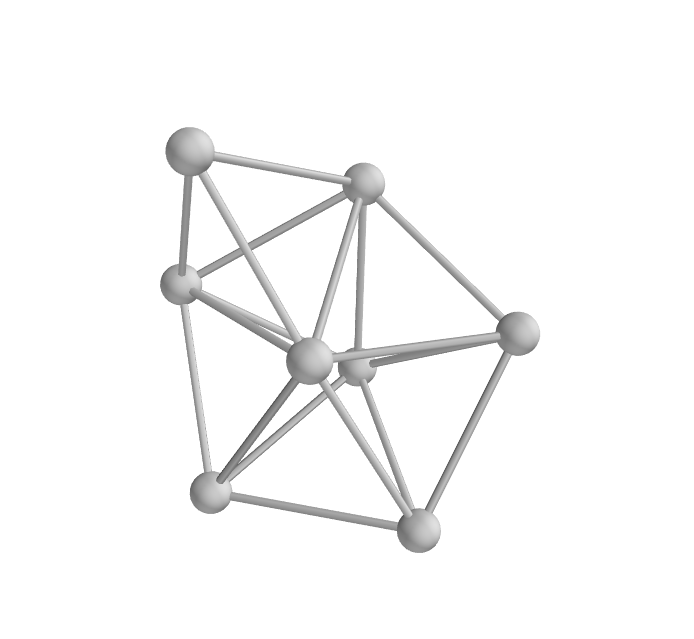}\\
 \begin{tabular}{|c|c|c|c|}
\hline
2nd Moment&$\phi$&$\sigma$&Special Properties\\
\hline
$5.56215 R^2$ & $C_s$ & 1 & \\
\hline
\end{tabular}
\end{array}
\end{array}
\]
\noindent Packing 8 (Graph 294):
\[
\begin{array}{l}
\begin{tiny}\begin{array}{lllll}
&\mathcal{A}:  \left( \begin{matrix} 0&0&0&1&1&1&1&0\\
0&0&0&1&0&1&0&1\\
0&0&0&0&1&0&1&1\\
1&1&0&0&0&1&1&1\\
1&0&1&0&0&1&1&1\\
1&1&0&1&1&0&0&1\\
1&0&1&1&1&0&0&1\\
0&1&1&1&1&1&1&0 \end{matrix} \right)&&&\\
&&&&\\
&\mathcal{D}:\left( \begin{matrix} 0&1.7321&1.7321&1&1&1&1&1.4142\\
1.7321&0&2.0000&1&1.7321&1&1.7321&1\\
1.7321&2.0000&0&1.7321&1&1.7321&1&1\\
1&1&1.7321&0&1.4142&1&1&1\\
1&1.7321&1&1.4142&0&1&1&1\\
1&1&1.7321&1&1&0&1.4142&1\\
1&1.7321&1&1&1&1.4142&0&1\\
1.4142&1&1&1&1&1&1&0 \end{matrix} \right) R  & &&\end{array}
\begin{array}{l}
\mathcal{C}: \left(\begin{matrix}  0\\0\\0\\0\\-1.7321\\0\\-1.6330\\-0.5774\\0\\-0.0000\\-0.8660\\-0.5000\\-0.8165\\-0.2887\\0.5000\\-0.0000\\-0.8660\\0.5000\\-0.8165\\-0.2887\\-0.5000\\-0.8165\\-1.1547\\0 \end{matrix}\right)R\end{array}\end{tiny} \\
\\
\begin{array}{l}
 \includegraphics[width = 1in]{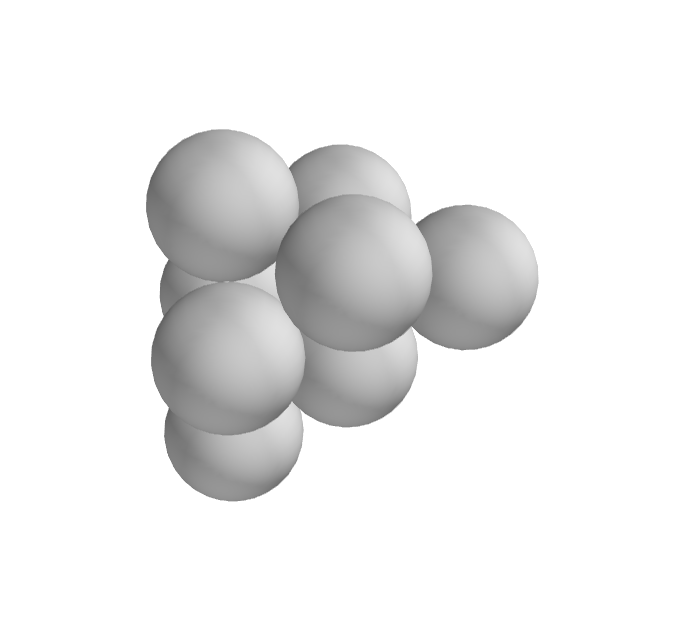} \; \; \; \includegraphics[width = 1in]{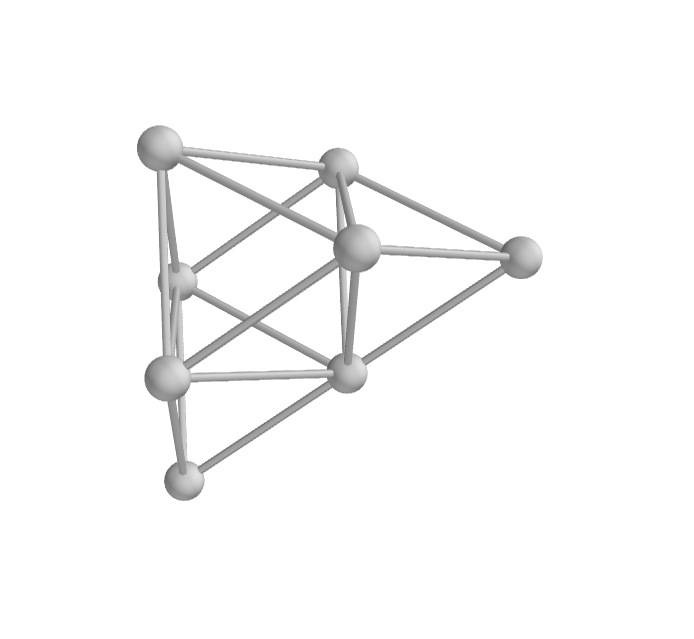}\\
 \begin{tabular}{|c|c|c|c|}
\hline
2nd Moment&$\phi$&$\sigma$&Special Properties\\
\hline
$5.75000 R^2$ & $C_{2 v}$ & 2 & \\
\hline
\end{tabular}
\end{array}
\end{array}
\]
\noindent Packing 9 (Graph 358):
\[
\begin{array}{l}
\begin{tiny}\begin{array}{lllll}
&\mathcal{A}:  \left( \begin{matrix} 0&0&1&0&1&1&1&0\\
0&0&0&1&0&1&1&1\\
1&0&0&0&1&1&0&1\\
0&1&0&0&0&0&1&1\\
1&0&1&0&0&0&1&1\\
1&1&1&0&0&0&1&1\\
1&1&0&1&1&1&0&1\\
0&1&1&1&1&1&1&0 \end{matrix} \right)&&&\\
&&&&\\
&\mathcal{D}:\left( \begin{matrix} 0&1.7321&1&1.9149&1&1&1&1.4142\\
1.7321&0&1.7321&1&1.7321&1&1&1\\
1&1.7321&0&1.9149&1&1&1.4142&1\\
1.9149&1&1.9149&0&1.4142&1.6330&1&1\\
1&1.7321&1&1.4142&0&1.4142&1&1\\
1&1&1&1.6330&1.4142&0&1&1\\
1&1&1.4142&1&1&1&0&1\\
1.4142&1&1&1&1&1&1&0 \end{matrix} \right) R  & &&\end{array}
\begin{array}{l}
\mathcal{C}: \left(\begin{matrix}  0\\0\\0\\0\\1.7321\\0\\-0.9574\\0.2887\\0\\-0.0290\\1.6358\\0.9949\\-0.4352\\0.2887\\0.8528\\-0.2611\\0.8660\\-0.4264\\0.2611\\0.8660\\0.4264\\-0.6963\\1.1547\\0.4264 \end{matrix}\right)R\end{array}\end{tiny} \\
\\
\begin{array}{l}
 \includegraphics[width = 1in]{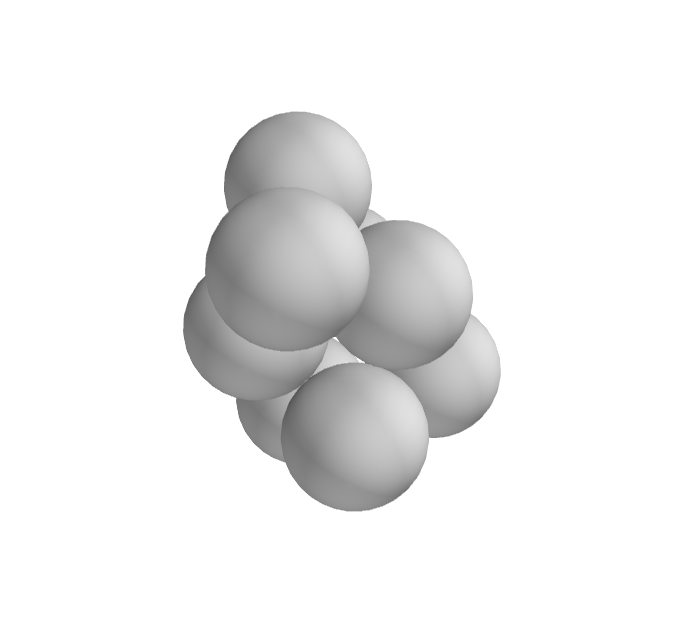} \; \; \; \includegraphics[width = 1in]{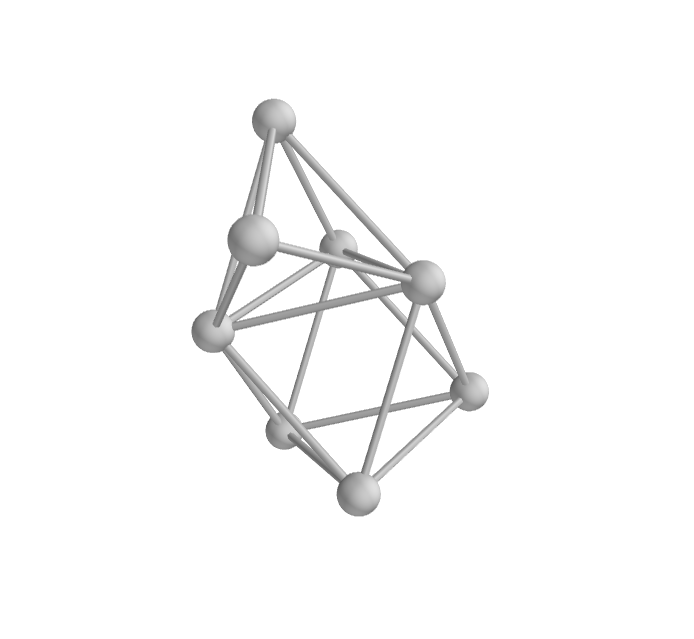}\\
 \begin{tabular}{|c|c|c|c|}
\hline
2nd Moment&$\phi$&$\sigma$&Special Properties\\
\hline
$5.62500 R^2$ & $C_s$ & 1 & \\
\hline
\end{tabular}
\end{array}
\end{array}
\]
\noindent Packing 10 (Graph 366):
\[
\begin{array}{l}
\begin{tiny}\begin{array}{lllll}
&\mathcal{A}:  \left( \begin{matrix} 0&0&1&0&1&1&1&1\\
0&0&0&1&0&1&1&1\\
1&0&0&0&1&0&1&1\\
0&1&0&0&0&1&0&1\\
1&0&1&0&0&0&0&1\\
1&1&0&1&0&0&1&1\\
1&1&1&0&0&1&0&1\\
1&1&1&1&1&1&1&0 \end{matrix} \right)&&&\\
&&&&\\
&\mathcal{D}:\left( \begin{matrix} 0&1.6330&1&1.6667&1&1&1&1\\
1.6330&0&1.6667&1&1.9907&1&1&1\\
1&1.6667&0&1.9907&1&1.6330&1&1\\
1.6667&1&1.9907&0&1.7778&1&1.6330&1\\
1&1.9907&1&1.7778&0&1.6667&1.6330&1\\
1&1&1.6330&1&1.6667&0&1&1\\
1&1&1&1.6330&1.6330&1&0&1\\
1&1&1&1&1&1&1&0 \end{matrix} \right) R  & &&\end{array}
\begin{array}{l}
\mathcal{C}: \left(\begin{matrix}  0\\0\\0\\0\\1.6330\\0\\-0.9623\\0.2722\\0\\0.4811\\1.3608\\0.8333\\-0.5453\\-0.0907\\0.8333\\0.5774\\0.8165\\-0.0000\\-0.2887\\0.8165\\-0.5000\\-0.2887\\0.8165\\0.5000 \end{matrix}\right)R\end{array}\end{tiny} \\
\\
\begin{array}{l}
 \includegraphics[width = 1in]{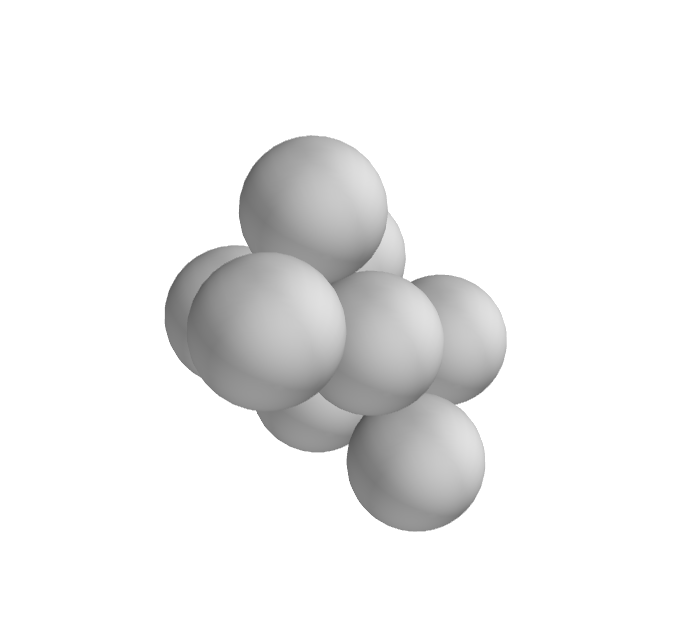} \; \; \; \includegraphics[width = 1in]{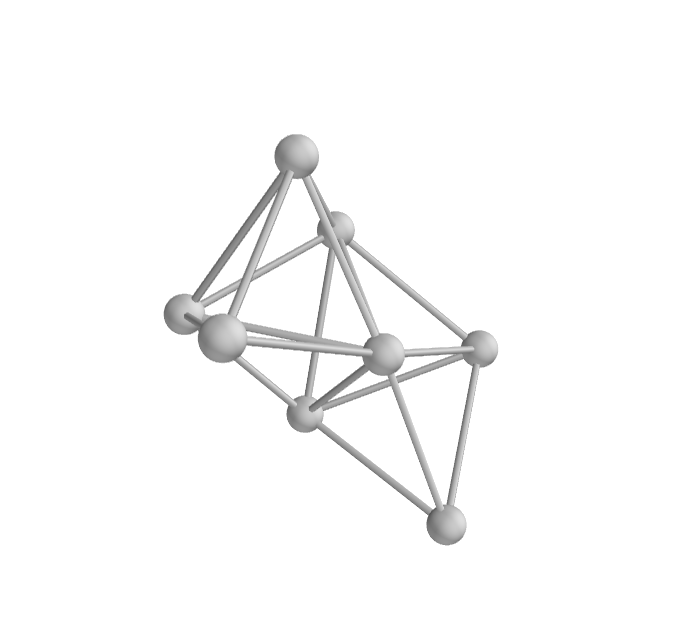}\\
 \begin{tabular}{|c|c|c|c|}
\hline
2nd Moment&$\phi$&$\sigma$&Special Properties\\
\hline
$6.01080 R^2$ & $C_s$ & 1 & \\
\hline
\end{tabular}
\end{array}
\end{array}
\]
\noindent Packing 11 (Graph 374):
\[
\begin{array}{l}
\begin{tiny}\begin{array}{lllll}
&\mathcal{A}:  \left( \begin{matrix} 0&0&1&0&1&1&1&1\\
0&0&0&1&0&1&1&1\\
1&0&0&0&1&0&1&1\\
0&1&0&0&0&1&1&0\\
1&0&1&0&0&0&0&1\\
1&1&0&1&0&0&1&1\\
1&1&1&1&0&1&0&1\\
1&1&1&0&1&1&1&0 \end{matrix} \right)&&&\\
&&&&\\
&\mathcal{D}:\left( \begin{matrix} 0&1.6330&1&1.6667&1&1&1&1\\
1.6330&0&1.6667&1&1.9907&1&1&1\\
1&1.6667&0&1.9907&1&1.6330&1&1\\
1.6667&1&1.9907&0&2.4369&1&1&1.6330\\
1&1.9907&1&2.4369&0&1.6667&1.6330&1\\
1&1&1.6330&1&1.6667&0&1&1\\
1&1&1&1&1.6330&1&0&1\\
1&1&1&1.6330&1&1&1&0 \end{matrix} \right) R  & &&\end{array}
\begin{array}{l}
\mathcal{C}: \left(\begin{matrix}  0\\0\\0\\0\\1.6330\\0\\-0.9623\\0.2722\\0\\0.4811\\1.3608\\-0.8333\\-0.5453\\-0.0907\\0.8333\\0.5774\\0.8165\\0\\-0.2887\\0.8165\\-0.5000\\-0.2887\\0.8165\\0.5000 \end{matrix}\right)R\end{array}\end{tiny} \\
\\
\begin{array}{l}
 \includegraphics[width = 1in]{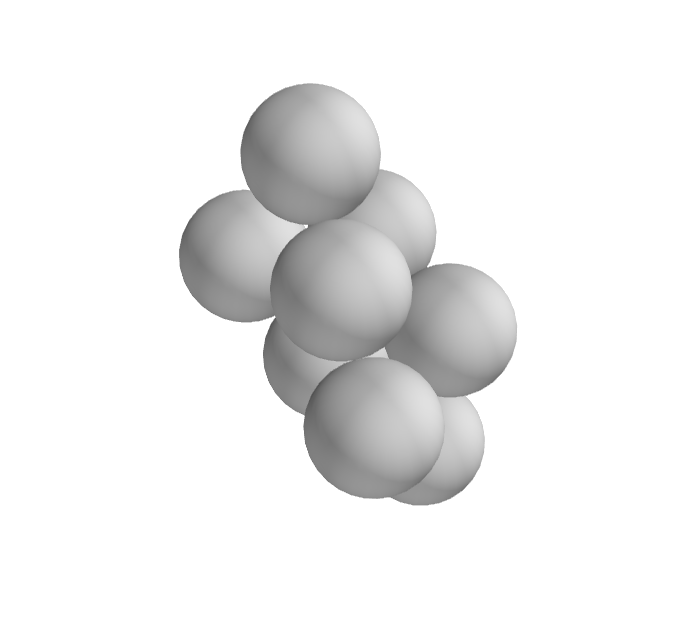} \; \; \; \includegraphics[width = 1in]{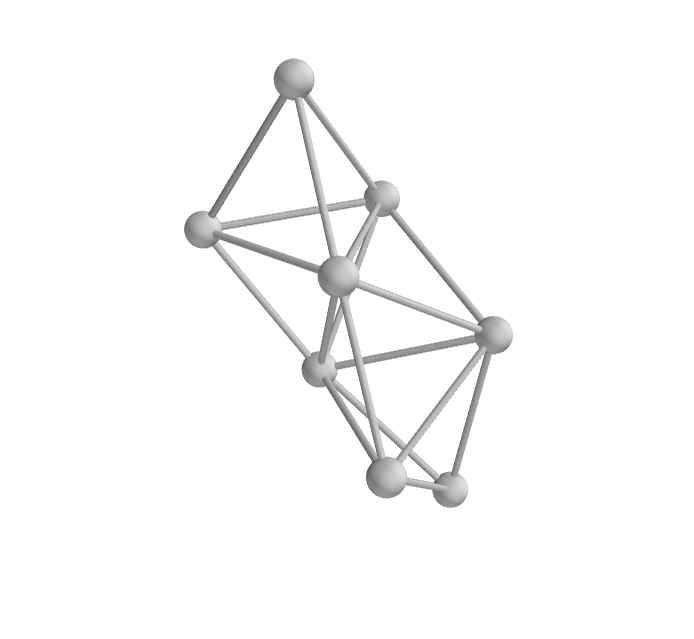}\\
 \begin{tabular}{|c|c|c|c|}
\hline
2nd Moment&$\phi$&$\sigma$&Special Properties\\
\hline
$6.35802 R^2$ & $C_1$ & 1 & chiral \\
\hline
\end{tabular}
\end{array}
\end{array}
\]
\noindent Packing 12 (Graph 397):
\[
\begin{array}{l}
\begin{tiny}\begin{array}{lllll}
&\mathcal{A}:  \left( \begin{matrix} 0&0&1&0&1&1&1&1\\
0&0&0&1&1&1&1&1\\
1&0&0&0&1&0&1&0\\
0&1&0&0&0&1&0&1\\
1&1&1&0&0&0&1&1\\
1&1&0&1&0&0&1&1\\
1&1&1&0&1&1&0&0\\
1&1&0&1&1&1&0&0 \end{matrix} \right)&&&\\
&&&&\\
&\mathcal{D}:\left( \begin{matrix} 0&1.4142&1&1.7321&1&1&1&1\\
1.4142&0&1.7321&1&1&1&1&1\\
1&1.7321&0&2.4495&1&1.7321&1&1.7321\\
1.7321&1&2.4495&0&1.7321&1&1.7321&1\\
1&1&1&1.7321&0&1.4142&1&1\\
1&1&1.7321&1&1.4142&0&1&1\\
1&1&1&1.7321&1&1&0&1.4142\\
1&1&1.7321&1&1&1&1.4142&0 \end{matrix} \right) R  & &&\end{array}
\begin{array}{l}
\mathcal{C}: \left(\begin{matrix}  0\\0\\0\\0\\-1.4142\\0\\1.0000\\0\\0\\-1.0000\\-1.4142\\-0.0000\\0.5000\\-0.7071\\0.5000\\-0.5000\\-0.7071\\-0.5000\\0.5000\\-0.7071\\-0.5000\\-0.5000\\-0.7071\\0.5000 \end{matrix}\right)R\end{array}\end{tiny} \\
\\
\begin{array}{l}
 \includegraphics[width = 1in]{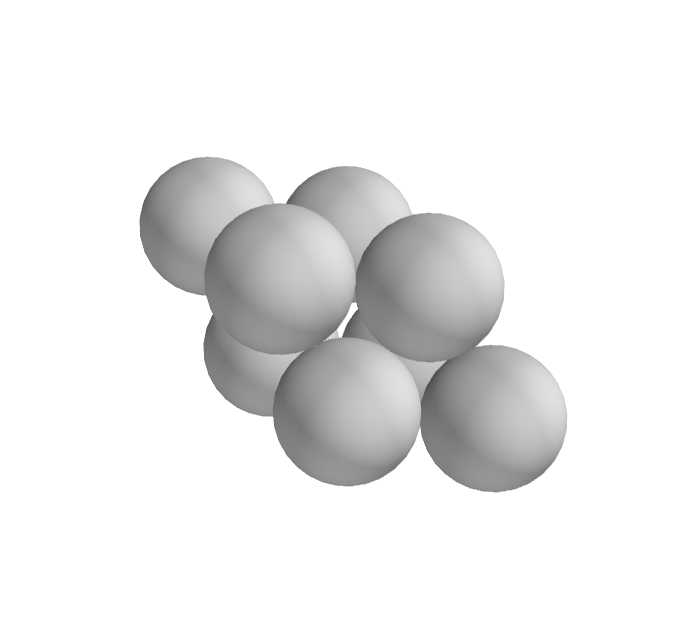} \; \; \; \includegraphics[width = 1in]{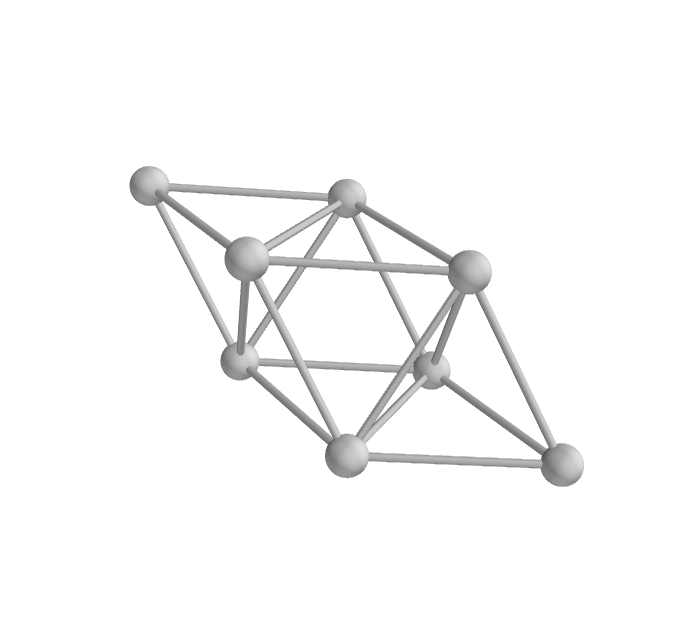}\\
 \begin{tabular}{|c|c|c|c|}
\hline
2nd Moment&$\phi$&$\sigma$&Special Properties\\
\hline
$6.00000 R^2$ & $D_{3 d}$ & 6 & \\
\hline
\end{tabular}
\end{array}
\end{array}
\]
\noindent Packing 13 (Graph 408):
\[
\begin{array}{l}
\begin{tiny}\begin{array}{lllll}
&\mathcal{A}:  \left( \begin{matrix} 0&0&1&0&1&1&1&0\\
0&0&0&1&1&0&1&1\\
1&0&0&0&1&1&0&1\\
0&1&0&0&0&1&1&1\\
1&1&1&0&0&0&1&1\\
1&0&1&1&0&0&1&1\\
1&1&0&1&1&1&0&0\\
0&1&1&1&1&1&0&0 \end{matrix} \right)&&&\\
&&&&\\
&\mathcal{D}:\left( \begin{matrix} 0&1.7199&1&1.7199&1&1&1&1.5130\\
1.7199&0&1.7199&1&1&1.5130&1&1\\
1&1.7199&0&1.7199&1&1&1.5130&1\\
1.7199&1&1.7199&0&1.5130&1&1&1\\
1&1&1&1.5130&0&1.2892&1&1\\
1&1.5130&1&1&1.2892&0&1&1\\
1&1&1.5130&1&1&1&0&1.2892\\
1.5130&1&1&1&1&1&1.2892&0 \end{matrix} \right) R  & &&\end{array}
\begin{array}{l}
\mathcal{C}: \left(\begin{matrix}  0\\0\\0\\0\\1.7199\\0\\0.9568\\0.2907\\0\\0\\1.4292\\-0.9527\\0.2613\\0.8600\\0.4384\\0.3752\\0.4852\\-0.7898\\-0.4124\\0.8600\\-0.3006\\0.8211\\1.2347\\-0.3006 \end{matrix}\right)R\end{array}\end{tiny} \\
\\
\begin{array}{l}
 \includegraphics[width = 1in]{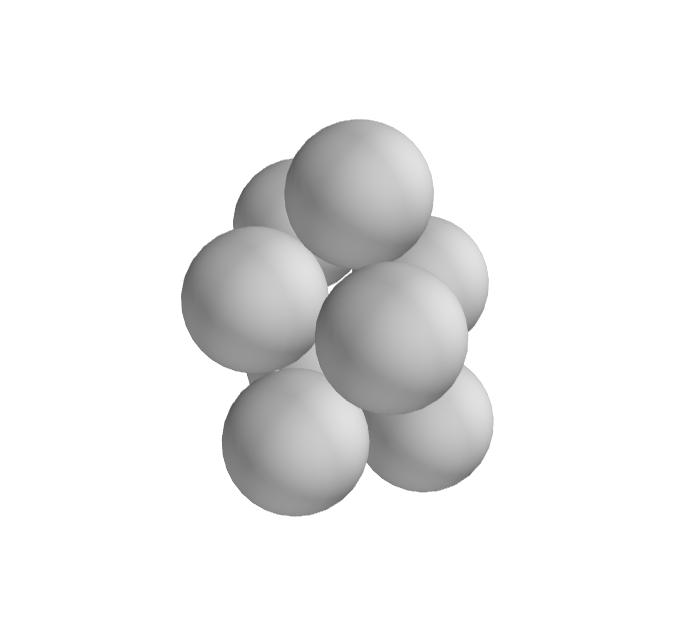} \; \; \; \includegraphics[width = 1in]{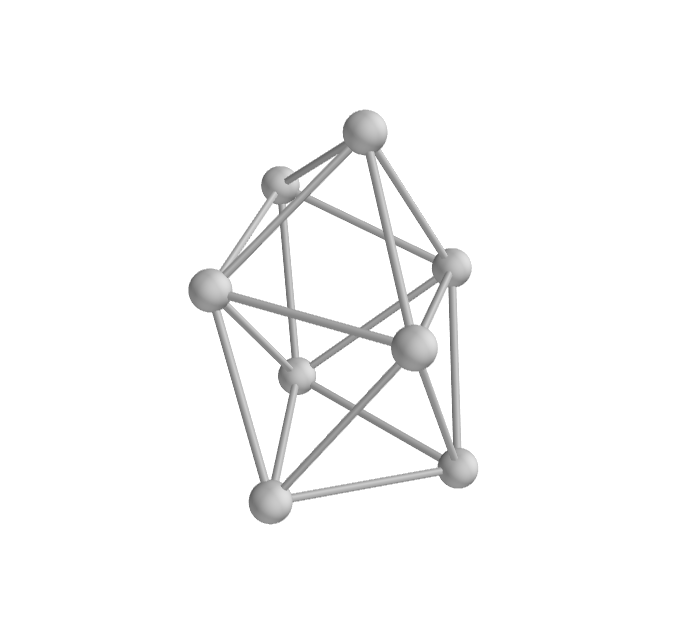}\\
 \begin{tabular}{|c|c|c|c|}
\hline
2nd Moment&$\phi$&$\sigma$&Special Properties\\
\hline
$^*5.28917 R^2$ & $D_{2 d}$ & 4 & new seed \\
\hline
\end{tabular}
\end{array}
\end{array}
\]

\newpage
\section{Packings of $n = 9$ Particles}

Here, we include the 52 packings of $n = 9$ particles.  The `Special Properties' column denotes whether the packing is chiral, non-rigid, or corresponds to a new seed.  If the special properties column is blank, then the packing is none of these.  28 of these packings are chiral, 4 of these packings correspond to new seeds (graphs 8901, 8926, 12811, and 13380), and 1 of them is non-rigid (graph 8901).  The minimum of the 2nd moment corresponds to packing 47, graph 13380.  A star, `*', appears in front of packings that are reported here but were not reported in our last paper (PRL, 103, 118303).\\
\\
\noindent Packing 1 (Graph 406):
\[

\end{array}
\]

\newpage
\section{Packings of $n = 10$ Particles}

Here we include the 262 10 particle packings\footnote{Note that for these 10 particle packings, the packing figures were generated by code that John Lee wrote that interfaces the packing output with the mayavi2 graphing package.  And the point groups and symmetry numbers were also obtained via code that John Lee wrote that interfaces the packing output with the rotational constant calculator website (http://www.colby.edu/chemistry/PChem/scripts/ABC.html) that calculates these numbers.}.  The `special properties' column denotes whether a packing is chiral, corresponds to a new seed, to a non-rigid structure, or to a packing with $> 3n-6$ contacts.  If this column is blank, then that packing has none of these properties.  There are 201 chiral packings, 8 new seeds (graphs 308923, 618718, 622587, 629072, 665570, 679397, 714961, and 749808), 4 non-rigid packings (Graph 665570 corresponds to the packing that is both a new seed and a non-rigid structure, and graphs 286981, 457749, and 593702 (packings 84, 145, and 197) correspond to the 10 particle non-rigid packings that are derived from the 9 particle non-rigid new seed; 9 particle graph 8901)\footnote{Note that while the 3 non-rigid packings based on the 9 particle non-rigid new seed were reported in (PRL, 103, 118303) and the PhD Thesis (http://people.seas.harvard.edu/$\sim$narkus/assets/Thesis.pdf.zip), they were not found by the old code.  We had reported them because we had already solved for their coordinates.  We made a note of this where the packings were listed, and mentioned that we suspected these 3 packings had been missed due to round-off error, which we knew to be an issue.  Thus, while a star does not appear next to these packings because they were previously reported, if one were to compare raw output of the old and new code, these 3 matrices would also lie in the category that the old code misses and the new modified code picks up.}, and 3 packings with $3n-5 = 25$ contacts (packings 259-262).  The minimum of the 2nd moment corresponds to packing 257, graph 749808.

Note that for the 25 contact packings, we have included all graphs that lead to these packings; thus multiple graphs are listed under the same packing header.  Because of this, we have listed all 25 contact packings at the end; these are the only packings that are listed out of chronological graph order.  For these graphs, we have included their associated distance matrix and coordinates, as these can be different up to particle labeling degeneracy (\begin{frenchspacing}i.e.\end{frenchspacing} they can be isomorphic but not identical), but we have only included the table with the 2nd moment, $\phi$, $\sigma$, and the special properties once, as this will always be identical.

A star, `*', appears in front of packings that are reported here but were not reported in our last paper (PRL, 103, 118303).  A double star, `**', appears in front of packings that are reported here but were not reported previously by us or by Hoy and O'Hern (PRL, 105, 068001).  We have also included markings next to the relevant graphs that correspond to 25 contact packings which were not found in these papers; however, note that this did not affect the number of packings reported in either of these papers as multiple graphs lead to the same 25 contact packings, and in both papers all 3 25 contact packings were found.  In front of these graphs, a `*' appears for ones not found in (PRL, 103, 118303), and a `$^+$' appears for ones not found in (PRL, 105, 068001)\footnote{Rob Hoy sent us the raw data he used for the paper (PRL, 105, 068001).  It was comparing this data to our data that allowed us to determine which graphs (ones that uniquely or redundantly correspond to packings) appear here but were not reported in (PRL, 105, 068001).}.

\noindent Packing 1 (Graph 4471):
\[

\]
\noindent The other graphs that lead to the same 25 contact packing (packing 260), with their corresponding distance matrices and coordinates:\\
The table containing the minimum of the 2nd moment, etc., as well as the pictures will  be the same, and thus is not included in the list.\\
\newline
\noindent Graph 458233:
\[
%
\]
\noindent Graph 561797 (note that as a 24 contact $\A$ this is a non-iterative graph, but as a 25 contact packing ($\D$) it is iterative):
\[
%
\]
\noindent Graph 609155 (note that as a 24 contact $\A$ this is a non-iterative graph, but as a 25 contact packing ($\D$) it is iterative):
\[
%
\]
\noindent Graph 609244 (note that as a 24 contact $\A$ this is a non-iterative graph, but as a 25 contact packing ($\D$) it is iterative):
\[
%
\]
\noindent The other graphs that lead to the same 25 contact packing (packing 261), with their corresponding distance matrices and coordinates:\\
The table containing the minimum of the 2nd moment, etc., as well as the pictures will  be the same, and thus is not included in the list.\\
\newline
\noindent Graph 593947:
\[
%
\]
\noindent The other graphs that lead to the same 25 contact packing (packing 262), with their corresponding distance matrices and coordinates:\\
The table containing the minimum of the 2nd moment, etc., as well as the pictures will  be the same, and thus is not included in the list.\\
\newline
\noindent Graph 287036:
\[
%
\]
\noindent $^+$Graph 665605 (note that as a 24 contact $\A$ this is a non-iterative graph, but as a 25 contact packing ($\D$) it is iterative):
\[
%
\end{tiny} 
\end{array}
\]
\textbf{{\large \begin{center}APPENDIX C: PSEUDO CODE\end{center}}}
\vspace{.5in}
\setcounter{section}{0}
\setcounter{figure}{0}

Here we include a list of changes that were made in the code since our last paper (PRL, 103, 118303).  The entire code, both old and new versions of it, can be downloaded from the following website: http://people.seas.harvard.edu/$\sim$narkus/.  We also include the algorithm for the triangular bipyramid rule, as applied to iterative packings, in which case equation 10 of the main text can be applied directly:
%For iterative packings, rule \ref{gen rule eqn rij} can be applied directly.  The algorithm is as follows:
\renewcommand{\labelenumi}{\arabic{enumi}.}

\section{Changes Made to Code}

\begin{enumerate}
\item In rare cases, for some of the triangular bipyramids, there are dihedral angles that are not defined.  This occurs when $>=3$ points lie on a line.  This occurs in certain packings that contain octahedra.  In these cases, the global consistency check that is normally performed and checks for the sum and differences of dihedral angles is not valid.  In the old code, running this check when the dihedral angle was not defined caused some physical packings to be erroneously ruled out as globally inconsistent and thus unphysical.  Now, the global consistency check (in the function `global\_consistency\_check') first checks to see that all dihedral angles are well defined - if they are, the check proceeds as normal, if they're not then a special global consistency check for 3 points lying in a line (as was described in section 4.2 of the main text of the paper) is performed.  Making this correction caused 2 more packings to be realized at $n=9$ and the bulk of the 39 more packings that were found at $n=10$ to be realized.

\item `round\_numi' and `preci,' where `i' is a number, has been entered in various functions as an easier way to control the round-off numbers.  Some of the round-off numbers have been changed so that less error is introduced as a result of these numbers.  This only affects the number of packings found at $n = 10$, and accounts for around 2 more packings being found.

In particular, `prec2' in the function `check\_round' has been changed from -6 to -10.  This means that the precision to which we check that a certain quantity $=0$ is performed to $10^{-10}$ instead of $10^{-6}$.

\item All functions that check for 9 particle physical new seeds now first check to see that the distance between 2 particles, $r_{ij}$, is unknown before writing in the distance corresponding to the appropriate $r_{ij}$ of that seed.  (This prevents any already known distances being overwritten).  In the old code, only 1 of the 9 particle new seed functions performed this check.  For the functions that did not perform this check, occasionally an already known distance was erroneously overwritten and this in some cases caused a physical packing to be erroneously ruled out as unphysical.  Correcting this error also only affects the number of packings found at $n=10$; it also accounts for several more packings being found.
\end{enumerate} 

\section{Algorithm}
\begin{enumerate}
\item Loop through $\A$.  Each $\A_{ij} = 0$ corresponds to an unknown $r_{ij}$.  Keep track of which distances are unknown.

\item For each unknown $r_{ij}$

\begin{enumerate}
\item \label{construct ditet step} Construct each possible ditetrahedron until one with all 9 distances in $\overrightarrow{r}$ is found.

\begin{enumerate}
\item \label{calc rij step} Calculate $r_{ij}$ using the triangular bipyramid rule.

\begin{enumerate}
\item If a solution exists, $r_{ij}$ will fall out immediately.  If $r_{ij} < R$, discard the solution as it is unphysical.

\item If a solution does not exist, this $\A$ is unphysical -- move on to the next $\A$.  A solution will not exist if the triangle inequality is violated -- this will show up as the absolute value of the argument of $\cos^{-1}$ being $> 1$, which is undefined (yielding NaN).
\end{enumerate}

\item Store all physically consistent $r_{ij}$.  (These are locally consistent solutions, global consistency must still be checked).
\end{enumerate}

\item If no ditetrahedron with known $\overrightarrow{r}$ exists, move on to the next $r_{ij}$ and repeat step \ref{construct ditet step}.
\end{enumerate}

\item If any unknown $r_{ij}$ remain, repeat step \ref{construct ditet step}.  (This will be repeated until all $r_{ij}$ are solved).

\item Check for global consistency.

The dihedral angle $A_1$ is given by
\[
A_1 = \left\{
\begin{array}{l}
A_2 + A_3\\
|A_2 - A_3|\\
2\pi - (A_2 + A_3)\\
2\pi - |A_2 - A_3|
\end{array}
\right.
\]
(In calculating $r_{ij}$ in equation 15 of the main text, we need not consider the latter two $A_1$ solutions as $\cos(2\pi - x) = \cos(x)$).

For each potential set of solutions $r_{ij}$, construct all possible ditetrahedra in the structure.  If $\exists$ ditetrahedra where $A_1$ is not satisfied by one of the above formulas, then the solution set is globally inconsistent, and should be discarded. 

\end{enumerate}

For each iterative packing, all $r_{ij}$ can be solved via this algorithm.
\end{document}